\documentstyle[emulateapj,psfig]{article}

\makeatletter

\newenvironment{inlinefigure}{%
\def\@captype{figure}%
\noindent\begin{minipage}{0.999\linewidth}\begin{center}}
{\end{center}\end{minipage}\smallskip}
\makeatother

\lefthead{Cowie \& Barger}

\newcommand{\ha}{H$\alpha$}
\newcommand{\hb}{H$\beta$}

\newcommand{\nii}{[N$\thinspace{\rm II}]$}
\newcommand{\oii}{[O$\thinspace{\rm II}]$}
\newcommand{\oiii}{[O$\thinspace{\rm III}]$}
\newcommand{\sii}{[S$\thinspace{\rm II}]$}

\begin{document}

\slugcomment{Accepted to The Astrophysical Journal}

\title{An Integrated Picture of Star Formation, Metallicity Evolution,
and Galactic Stellar Mass Assembly\altaffilmark{1}}
\author{
L.~L.~Cowie$\!$\altaffilmark{2},
A.~J.~Barger$\!$\altaffilmark{3,4,2}
}

\altaffiltext{1}{Based in part on data obtained at the W.~M.~Keck
Observatory, which is operated as a scientific partnership among
the California Institute of Technology, the University of
California, and NASA and was made possible by the generous financial
support of the W.~M.~Keck Foundation.}
\altaffiltext{2}{Institute for Astronomy, University of Hawaii,
2680 Woodlawn Drive, Honolulu, HI 96822.}
\altaffiltext{3}{Department of Astronomy, University of
Wisconsin-Madison, 475 North Charter Street, Madison, WI 53706.}
\altaffiltext{4}{Department of Physics and Astronomy,
University of Hawaii, 2505 Correa Road, Honolulu, HI 96822.}

\begin{abstract}
We present an integrated study of star formation and
galactic stellar mass assembly from $z=0.05-1.5$ and galactic 
metallicity evolution from $z=0.05-0.9$ using a very large and 
highly spectroscopically complete sample selected by 
rest-frame NIR bolometric flux. 
Our NIR (rest-frame $0.8-2.4~\mu$m) 
sample consists of 2634 galaxies with fluxes in excess of 
$2\times 10^{-15}$~ergs~cm$^{-2}$~s$^{-1}$ in the GOODS-N field. 
It probes to a complete mass limit of $10^{10}$~M$_\odot$ 
for $z=0.05-0.9$ and includes all Milky Way mass 
galaxies for $z=0.05-1.5$. We have spectroscopic 
redshifts and high-quality spectra from $4500-10000$~\AA\ for 
2020 (77\%) of the galaxies. Our 13-band photometric 
redshift estimates show that most of the spectroscopically
unidentified sources in the above redshift ranges are 
early-type galaxies. We assume a Salpeter IMF and fit 
Bruzual \& Charlot (2003) models to the data to compute the 
galactic stellar masses and extinctions. 
We calibrate the star formation diagnostics internally using
our $z=0.05-0.475$ sample.
We then derive the galactic stellar mass assembly and star 
formation histories. We compare our extinction
corrected UV-based star formation rate densities with the
combination of the star formation rate densities that we 
compute from the 24~$\mu$m fluxes and the extinction uncorrected
\oii\ luminosities. We determine the expected 
formed stellar mass density growth rates produced by star formation 
and compare them with the growth rates measured from the 
formed stellar mass functions by mass interval. 
We show that the growth rates match 
if the IMF is slightly increased from the Salpeter IMF at intermediate 
masses ($\sim 10$~M$_\odot$). We investigate the evolution of galaxy 
color, spectral type, and morphology with mass and redshift and the 
evolution of mass with environment. We find that applying extinction
corrections is critical when analyzing the galaxy colors. As an
example, prior to correcting for extinction, nearly all 
of the galaxies in the green valley are 24~$\mu$m sources, but
after correcting for extinction, the bulk of the 24~$\mu$m sources
lie in the blue cloud.
We also compute the metallicities of the sources 
between $z=0.05-0.9$ that have well-detected \hb, \oii~$\lambda3727$, 
and \oiii~$\lambda5007$ emission lines using the R23 diagnostic ratio.
At $z<0.475$ we use the R23, \nii/\oii, and \nii/\ha\ diagnostic ratios.
We find an evolution of the metallicity-mass relation 
corresponding to a decrease of $0.21\pm0.03$~dex between the 
local value and the value at $z=0.77$ in the $10^{10}-10^{11}$~M$_\odot$ 
range. We use the metallicity evolution to estimate the gas mass
of the galaxies, which we compare with the galactic stellar mass 
assembly and star formation histories.
Overall, our measurements are consistent with a galaxy evolution
process dominated by episodic bursts of star formation
and where star formation in the most massive galaxies 
($\gtrsim 10^{11}$~M$_\odot$) ceases at $z<1.5$ because of gas starvation.
\end{abstract}


\section{Introduction}
\label{secintro}

One of the fundamental goals of modern cosmology is
to understand the formation and evolution of the
galaxy population as a whole. We shall refer to
this as the cosmic galaxy formation problem. There has
been spectacular progress in addressing
the cosmic galaxy formation problem over the last
twenty years, beginning with the determination of the
star formation history (e.g., Cowie et al.\ 1995; Lilly et
al.\ 1996; Madau et al.\ 1996; Steidel et al.\ 1999;
Haarsma et al.\ 2000; Barger et al.\ 2000; Le Floc'h et al.\ 2005; 
P{\'e}rez-Gonz{\'a}lez et al.\ 2005;
Hopkins \& Beacom 2006; Wang et al.\ 2006; Reddy et al.\ 2008).
This has been followed more recently by efforts to measure the 
galactic stellar mass assembly history
(e.g., Brinchmann \& Ellis 2000;
Cole et al.\ 2001; Bell et al.\ 2003, 2007;
P{\'e}rez-Gonz{\'a}lez et al.\ 2003, 2008;
Dickinson et al.\ 2003; Rudnick et al.\ 2003, 2006;
Fontana et al.\ 2003, 2004, 2006; Drory et al.\ 2004, 2005;
Bundy et al.\ 2005, 2006; Conselice et al.\ 2005, 2007;
Borch et al.\ 2006; Pannella et al.\ 2006; Elsner et al.\ 2008)
and the evolution of metallicity with galaxy 
mass and redshift (e.g., Kobulnicky et al.\ 2003; Lilly et al.\ 2003;
Kobulnicky \& Kewley 2004; Tremonti et al.\ 2004;
Liang et al.\ 2004; Savaglio et al.\ 2005).
However, ideally what one wants is a 
comprehensive analysis of the history of star formation, the 
growth of galactic stellar mass and metals content, 
and the changes in morphology with redshift, galaxy mass, 
and the environment for a large, mass-selected galaxy
sample that could be compared in detail with local galaxy properties 
and cosmological simulations of galaxy evolution. 
In particular, such an analysis could yield clear explanations
for the migration of star formation to lower mass 
galaxies at later cosmic times and the simultaneous 
quenching of star formation 
in the most massive galaxies (the downsizing of Cowie et al.\ 1996),
as well as for the color bimodality of galaxy populations 
(e.g., Strateva et al.\ 2001; Baldry et al.\ 2004).

Up until now such an analysis has not been possible
since existing data sets are either visually selected, 
have limited color information, and are poorly
suited to a metals analysis because of the spectroscopic
wavelength coverage (e.g., the DEEP2 survey); mass selected
but based on photometric redshifts (e.g., Combo17/GEMS); 
or mass selected and spectroscopically observed but based 
on a relatively small sample (e.g., the Gemini Deep Deep Survey).

In this paper we present, for the first time, an integrated, 
mass-based analysis made possible by the availability of a large,
homogeneous, near-infrared (NIR) selected and spectroscopically 
observed galaxy sample in the Great Observatories Origins Deep 
Survey-North (GOODS-N; Giavalisco et al.\ 2004) field. 
We have obtained extremely deep, wide-field NIR images  
(Keenan et al.\ 2008, in preparation) and highly complete 
spectroscopic identifications of the sources in this field 
(Barger et al.\ 2008, in preparation). We are therefore able 
to use, for the most part, spectroscopic redshifts to make our 
determinations of the galactic stellar mass assembly
and star formation histories, as well as high-quality measurements 
of line fluxes to obtain the metallicity history.

However, we caution that even with such an excellent data set 
there are many complicating factors in relating the star formation 
history to the stellar mass assembly history and the formation of 
metals in galaxies, even at late cosmic times. 
(Here we shall take late cosmic times to be $z<1.5$.)
At the conceptual level, methods of measuring 
star formation rates use diagnostics which are sensitive to the 
high-mass end of the stellar initial mass function (IMF), while 
stellar mass measurements are dominated by lower mass 
stars. Therefore, while the shape of the sub-solar IMF only enters 
as a normalization factor, the shape of the IMF at higher 
masses is critical in relating the star formation rates to the 
stellar masses. Thus, we must be concerned about the uncertainties 
in the IMF shape and the potential variations in the
IMF shape between different types of galaxies. In principle
we could minimize this problem by considering the growth of 
the stellar mass in metals rather than the growth of the total
stellar mass, since the metals are produced by the same 
high-mass stars that are measured by the star formation 
diagnostics (Cowie 1988). However, even this is subject to 
uncertainties in the yields and would require the measurement 
of not only the total stellar mass evolution but also 
the metals evolution in both stars and gas, which would be very 
challenging to do.

Measurements of the star formation rates, stellar masses,
and metals are also complicated by other factors. 
Extinction reradiates
light from the rest-frame UV to the far-infrared (FIR), and we
must determine total star formation rates over a wide range
of galaxies with radically different morphologies and
dust column densities. Conversions even of NIR
light to stellar mass are complicated by ongoing 
active star formation, and there are still major
uncertainties in the stellar modeling of the galaxy populations.
Finally, determinations of the metals throughout the redshift 
range of interest can only be made for the gaseous baryons in 
the star-forming galaxies and depend on the notoriously uncertain 
conversions of the strong oxygen and nitrogen emission lines 
to metallicities.

Cosmic variance is also a significant issue in a field size
as small as the GOODS-N (e.g. Somerville et al. 2004) and
can affect our analysis of the evolution of quantities
such as the galaxy mass density and the universal star
formation rates.

These problems must be borne in mind throughout any work of
the present type, and we attempt at all points to work forward
as self-consistently as possible from the raw information
(NIR luminosities, galaxy line strengths, raw star formation
diagnostics, etc.) to inferences about the evolution of derived
quantities, such as stellar masses, star formation rates,
and metallicities. Also, wherever possible, we have used
multiple independent methods to determine
the sensitivity of the derived quantities to our underlying assumptions.
We attempt to self consistently estimate the effects of cosmic
variance within the data set and also to estimate the effects
which analytic error estimates of the variance could introduce
in our analysis. Finally we compare our results throughout
to other recent work using different, and in some cases
much larger, fields to check for consistency in these portions
of the paper.

The outline of the paper is as follows. In \S\ref{secsample}
and \S\ref{seclum} we describe the basic data and the sample 
selections. In \S\ref{secfit} we fit Bruzual \& Charlot (2003) 
models to the data to determine galactic stellar masses 
and extinctions in the galaxies. In \S\ref{secspectral} we 
measure equivalent widths and line fluxes from the
spectra. In \S\ref{secrelext} we compare measurements of the
continuum and line extinctions. 
In \S\ref{secsfr} we derive self-consistent calibrations
of the various star formation rate diagnostics.
In \S\ref{secha} and \S\ref{seco3} we derive the metallicities
with mass and redshift using various metallicity diagnostics.
In \S\ref{secabs} we consider the galaxies missing from the
metals analysis.
This is a long paper, and some readers may wish to skip much
of the detail and move to the discussion (\S\ref{secdisc})
and summary (\S\ref{seccon}), which we have tried to
make separately readable and which contain the high-level
interpretation of the data, including the 
derivation of the
stellar mass assembly history with redshift, the evolution of the 
mass-metallicity and mass-morphology relations with
redshift and environment, and the use of the metals evolution
to derive an estimate of the baryonic gas mass reservoir
in the galaxies. We find that all of our measurements
provide a broad, self-consistent picture of a galaxy
evolution process dominated by episodic bursts of star formation
and where star formation in the most massive galaxies
is terminated at later cosmic times as a consequence of 
gas starvation.

We adopt the $-1.35$ power-law Salpeter IMF 
(Salpeter 1955) extending from 0.1 to 100~M$_\odot$ 
for ease of comparison with previous results. Most importantly, 
this allows us to compare directly with the local mass function 
computed by Cole et al.\ (2001; hereafter, Cole01) for this IMF. 
The Salpeter IMF only differs significantly from the current 
best IMFs (Kroupa 2001; Chabrier 2003) below 1~M$_\odot$,
and thus these three IMFs differ only in the normalization of the
galactic stellar mass and star formation rate determinations. 
We can convert the total mass formed into stars prior to stellar 
mass loss (which we will refer to as the formed stellar mass to 
distinguish it from the present stellar mass, which is the stellar
mass present at any given time) from the Salpeter 
IMF to the Chabrier IMF by dividing by 1.39 and to the Kroupa IMF 
by dividing by 1.31. The exact conversion when considering the 
present stellar masses rather than the formed stellar masses
depends on the average evolutionary stage of the 
galaxies. However, this dependence is relatively weak, and we may 
approximately convert the present stellar mass from the Salpeter IMF 
to the Chabrier IMF by dividing by 1.70 and to the Kroupa IMF 
by dividing by 1.54. Note that these latter conversion factors 
have been computed for the distribution of ages in our ensemble 
of galaxies. The present stellar mass for the Salpeter IMF is 
roughly 0.74 of the formed stellar mass. 

We assume $\Omega_M=0.3$, $\Omega_\Lambda=0.7$, and
$H_0=70$~km~s$^{-1}$~Mpc$^{-1}$ throughout. 
All magnitudes are given in the AB magnitude system,
where an AB magnitude is defined by
$m_{AB}=-2.5\log f_\nu - 48.60$.
Here $f_\nu$ is the flux of the source in units of
ergs~cm$^{-2}$~s$^{-1}$~Hz$^{-1}$. 
We assume a reference value of the solar
metallicity of $12+\log({\rm O/H})=8.66$ and a conversion to the
mass fraction of metals of $Z=0.0126$ (Asplund et al.\ 2004).
This conversion is weakly dependent on the
assumed chemical composition relative to the oxygen abundance.

%
%
\begin{inlinefigure}
\figurenum{1}
\centerline{\psfig{figure=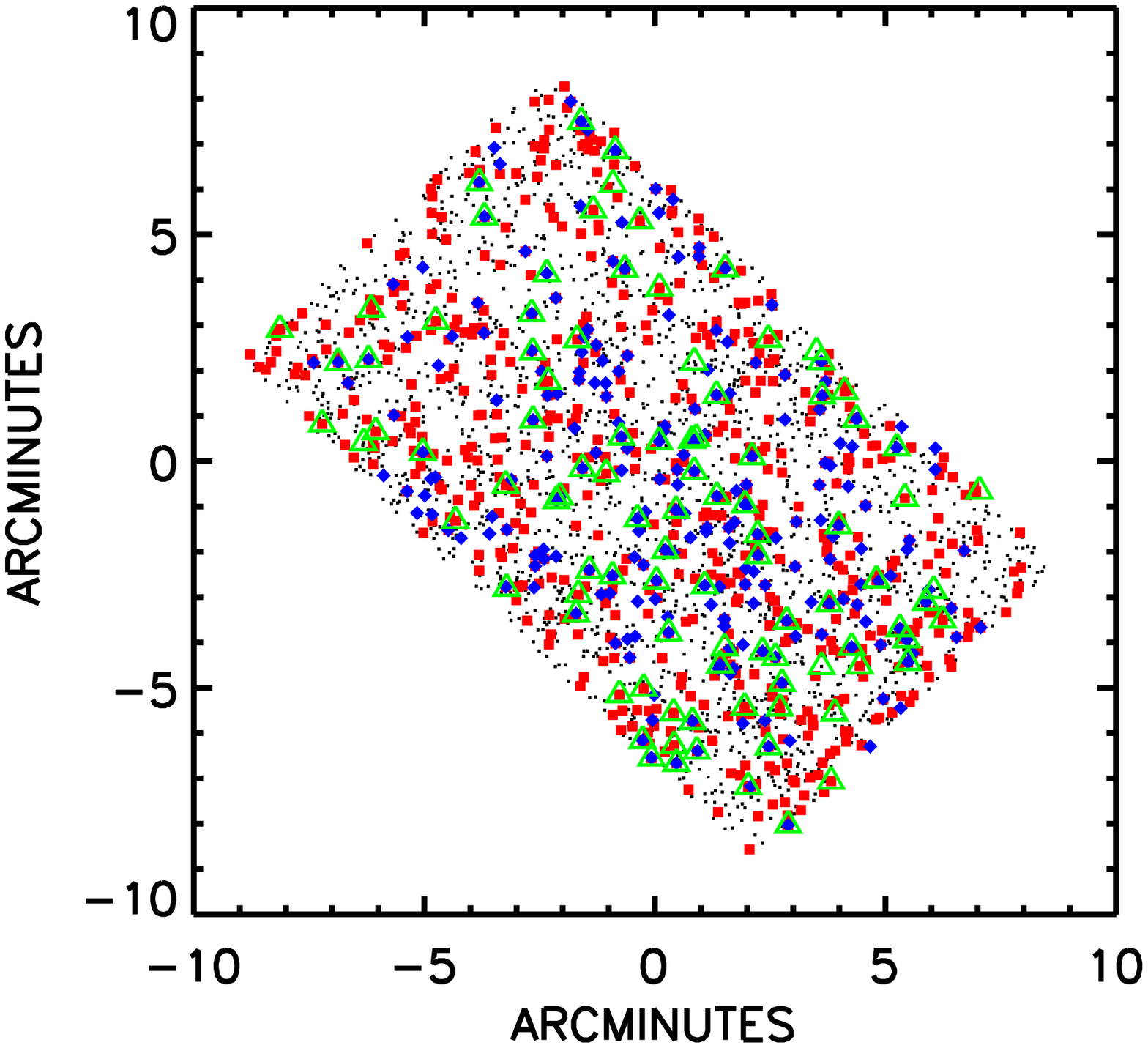,angle=90,width=3.5in}}
\vskip -0.2cm
\figcaption[]{
The observed area in the GOODS-N. The area is centered
on RA(2000) and Dec(2000) coordinates (189.2282, 62.2375)
with corners at (189.5435, 62.2749), (188.9137, 62.2000),
(189.3090, 62.3824), and (189.1482, 62.0909). 
The covered area is 145~arcmin$^2$ ($9\farcm3$ by $15\farcm7$).
The NIR-selected sample is shown with black dots,
the 663 24~$\mu$m detected sources with red squares, 
the 229 X-ray detected sources with blue diamonds, and
the 97 20~cm detected sources with green open triangles.
The concentration of the X-ray sources to the field
center reflects the variation in the sensitivity of
the X-ray image over the field.
\label{acs_nirbol_sample}
}
\end{inlinefigure}

\section{The NIR Bolometric Flux Sample}
\label{secsample}

\subsection{Photometric Selection}
\label{photsample}

The GOODS-N field is one of the most intensively studied 
regions in the sky, and in many bandpasses it has the 
deepest images ever obtained. Thus, it is nearly ideal for 
the present study. In this paper we use photometric 
data taken from existing work. The optical magnitudes are 
from the Subaru 8.2~m SuprimeCam observations of 
Capak et al.\ (2004; $U, B, V, R, I, z'$) 
and from the {\em HST\/} Advanced Camera for Surveys (ACS) 
observations of Giavalisco et al.\ (2004; 
F435W, F606W, F775W, and F850LP).
The NIR magnitudes are from the University of Hawaii
2.2~m ULBCAM ($J, H$) and CFHT WIRCAM and Subaru 8.2~m MOIRCS 
($K_s$) observations of Keenan et al.\ (2008, in preparation). 
To properly match to the optical data, 
the $3.6, 4.5, 5.8$, and $8.0~\mu$m magnitudes were measured 
directly from the IRAC images that Wang et al.\ (2006)
produced. Weighting by exposure time, Wang et al.\ (2006) 
combined the reduced DR1 and DR2 IRAC superdeep
images from the {\em Spitzer\/} Legacy first, interim, and
second data release products (DR1, DR1+, DR2; Dickinson
et al.\ 2008, in preparation).
In all cases, only sources within the well-covered ACS
GOODS-N region were included, as summarized in 
Figure~\ref{acs_nirbol_sample}.
The covered area is 145~arcmin$^2$.

For all of the sources we used corrected aperture magnitudes
to compute the colors and spectral energy distributions (SEDs).
We used $3''$ diameter apertures for the optical and 
NIR data and $6''$ diameter apertures for the MIR data.
We computed the median of the difference  between these magnitudes
and aperture magnitudes computed in  $6''$ diameter apertures
for the the optical and $12''$ diameter apertures for the MIR data
and used this median to correct the smaller aperture magnitude
to an approximate total magnitude.
For the brighter extended sources we also computed
isophotal magnitudes integrated to 0.01\% of
the peak surface brightness and used the difference
between these and the corrected aperture magnitudes
in the $K_s$ band to correct the luminosities and masses.
All of the calibrations are independent, so
it is important to check that we have fully consistent
magnitudes. We will return to this point in \S\ref{secew}.

We computed a rest-frame NIR bolometric flux for all of 
the sources in the region which were significantly detected at any
of the observed wavelength bands by linearly interpolating the observed
mangitudes to form a rest frame SED.
We used spectroscopic redshifts, 
where these were known, or otherwise photometric 
redshifts, which we calculated as in Wang et al.\ (2006) 
using the template method developed by 
P{\'e}rez-Gonz{\'a}lez et al.\ (2005).
We computed the flux over the {\em rest-frame} 
wavelength range 8000~\AA\ to $2.4~\mu$m. We excluded 
from the sample the spectroscopically 
identified stars and all of the sources within $3''$ 
of a brighter object or within $12''$ of the
eleven brightest stars, leaving a final sample of 2634
galaxies with fluxes above 
$2\times10^{-15}$~ergs~cm$^{-2}$~s$^{-1}$.
We take this as our primary NIR sample.
Our exclusion of neighbor sources introduces a 
small, flux-dependent correction to the area, which
we allow for in our determinations of the mass
functions, but the correction never
exceeds 10\%, even at the faintest fluxes.

Our selection by rest frame NIR bolometric flux is compared with the 
more usual selection by observed NIR magnitude ($K_{s, {\rm AB}}$)
in Figure~\ref{nirbol_k} for sources with $z=0.05-1.5$.
The best-fit relation gives
\begin{equation}
K_{s, {\rm AB}} = -11.87 -2.40\log({\rm NIR~ flux}) \,,
\label{eqnnirkrel}
\end{equation}
and the limiting NIR flux corresponds roughly to
$K_{s, {\rm AB}} = 23.4$. This is a shallow sample
compared to the depth of the NIR images. For
the $K_s$ image it corresponds to an $18\sigma$
selection. Thus, there should be no significant 
selection biases.

%
%
\begin{inlinefigure}
\figurenum{2}
\centerline{\psfig{figure=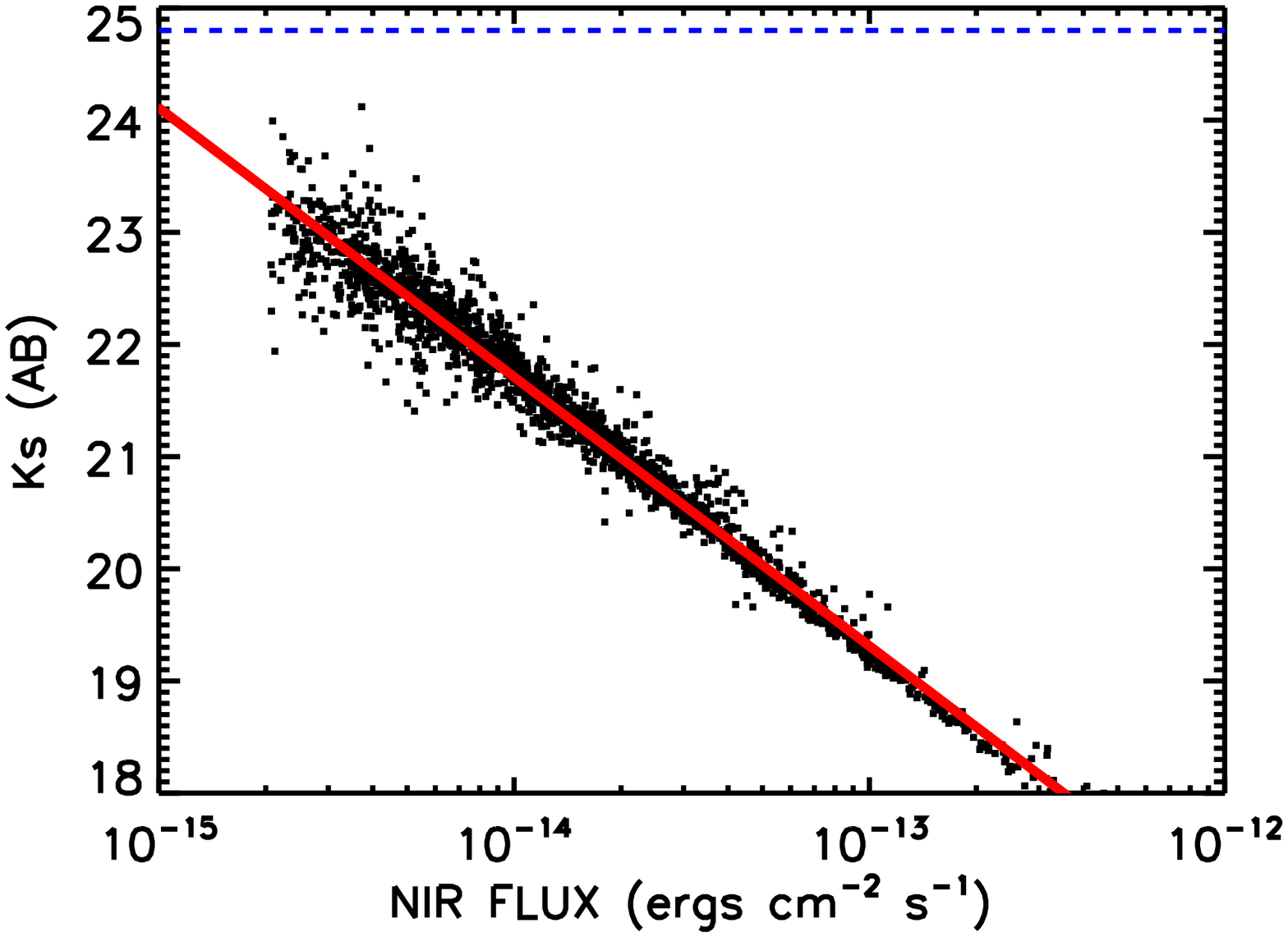,angle=90,width=3.5in}}
\vskip -0.2cm
\figcaption[]{
Comparison of the rest-frame NIR bolometric flux 
with the observed $K_s$ magnitude for sources with
$z=0.05-1.5$. The red diagonal line shows
the least-square polynomial fit of $\log$(NIR flux) to
$K_s$ magnitude. The blue dashed horizontal line shows
the $5\sigma$ flux limit of $24.8$ for the $K_s$ sample.
\label{nirbol_k}
}
\end{inlinefigure}

We identified X-ray counterparts to the NIR sample by 
matching our sample to the sources detected in the 
Alexander et al.\ (2003) catalog of the 2~Ms {\em Chandra\/} 
Deep Field-North (CDF-N) 
using a $2''$ search radius. Near the aim point the 
CDF-N X-ray data reach limiting fluxes of 
$\approx 1.5\times 10^{-17}$ ($0.5-2$~keV) 
and $\approx 1.4\times 10^{-16}$~ergs~cm$^{-2}$~s$^{-1}$ ($2-8$~keV).
We similarly obtained the radio fluxes from the Richards (2000)
1.4~GHz catalog, which reaches a $5\sigma$ limiting flux of 
$40~\mu$Jy and
the $24~\mu$m fluxes are from the DR1+ MIPS $24~\mu$m 
source list and version 0.36 MIPS $24~\mu$m map provided
by the {\em Spitzer\/} Legacy Program. This source catalog
is flux-limited at $80~\mu$Jy and is a subset of a more
extensive catalog (R.~R.~Chary et al.\ 2007, in preparation).

Our NIR-selected sample, together with the X-ray, 20~cm, 
and $24~\mu$m detected sources, are shown in 
Figure~\ref{acs_nirbol_sample}.

\subsection{Spectroscopy}
\label{secspec}

Following the establishment of the Hubble Deep Field-North
(HDF-N) with {\em HST\/}, 
intensive spectroscopic observations of the region
were made by a number of groups, primarily using the
Low-Resolution Imaging Spectrograph (LRIS; Oke et al.\ 1995)
on the Keck~I 10~m telescope (these data are summarized in 
Cohen et al.\ 2000). After the more extended GOODS-N region 
was observed with the ACS camera, a number of groups began
intensive spectroscopic observations with the large-format
Deep Extragalactic Imaging Multi-Object Spectrograph
(DEIMOS; Faber et al.\ 2003) on the Keck~II 10~m telescope.
Wirth et al.\ (2004; Keck Team Redshift Survey or KTRS)
and Cowie et al.\ (2004) presented
large samples of magnitude-selected redshifts,
while Reddy et al.\ (2006) gave a substantial sample of 
$z=2-3$ color-selected redshifts, Chapman et al.\ (2004, 2005)
and Swinbank et al.\ (2004) presented a number of 
radio/submillimeter redshifts, Treu et al. (2005)
measured redshifts for a sample of spheroids,
and Barger et al.\ (2005, 2007)
carried out observations on the X-ray and 
1.4~GHz samples.

We have attempted to make the most complete and homogeneous
spectral database possible by observing all of the missing or 
unidentified galaxies in a variety of flux-limited samples. 
A more extensive description of these samples may be found in 
Barger et al.\ (2007, in preparation). 
In this paper we focus only on the spectroscopic observations of 
our NIR sample. Our observations were made in a number of DEIMOS 
runs between 2004 and 2007. We used the 600 lines per mm grating,
giving a resolution of $3.5$~\AA\ and a wavelength coverage of
$5300$~\AA, which was the configuration used in the KTRS and 
in the Cowie et al.\ (2004)
observations. The spectra were centered at an average wavelength
of $7200$~\AA, though the exact wavelength range for each
spectrum depends on the slit position. Each $\sim 1$~hr exposure
was broken into three subsets, with the objects stepped along
the slit by $1.5''$ in each direction. Unidentified objects
were continuously reobserved, giving maximum exposure times of
up to 7~hrs. The spectra were reduced in the same way as 
previous LRIS spectra (Cowie et al.\ 1996). The dithering 
procedure provides extremely high-precision sky subtraction, 
which is important if we wish to measure accurate equivalent 
widths, as in the present paper. We have only included
spectra in the sample that could be confidently identified 
based on multiple emission and/or absorption lines.

We also reobserved objects where the original spectra 
were of poor quality or where previous redshifts were obtained 
with instruments other than DEIMOS, as well as where
the existing redshift identifications were unconvincing
or where there were conflicting redshifts in the literature
(a small number of sources).
Many of the KTRS spectra have poor sky subtraction. While
these spectra are adequate for redshift identifications, 
they are not suitable for line measurements because of the residual
sky lines. For the fainter objects the absolute sky subtraction 
is often problematic and, in some cases, the spectra even 
have negative continua. Equivalent width measurements made on 
such spectra have very large systematic uncertainties. This
is a substantial problem for previous work 
(e.g., Kobulnicky \& Kewley 2004) that relied on the KTRS 
spectra. We have reobserved most of these sources.

We now have spectroscopic redshift identifications for 2020 
of the 2634 galaxies (77\%) in our NIR sample. 
We show the redshift distribution 
for the spectroscopically identified sample in Figure~\ref{zdist} 
{\em (black histogram)\/}. The photometric redshift analysis of 
the remaining sources {\em (red histogram)\/} implies that 
most of these sources lie outside of our redshift ranges of 
interest ($z=0.05-0.9$ for our metallicity analysis and $z=0.05-1.5$
for our mass assembly and star formation analyses) and hence
that the spectroscopic completeness inside our two redshift 
ranges of interest is extremely high. 

We assigned photometric redshifts to all but 14 of the sources
in our sample. These 14 sources are either extremely faint 
or have peculiar SEDs that could not be adequately fitted by 
the templates. We shall assume that these lie outside of our
redshift ranges of interest. Between $z=0$ and $z=0.9$ there 
are 1260 spectroscopically identified sources, and between
$z=0$ and $z=1.5$ there are 1884 spectroscopically identified 
sources. Photometric redshifts add a further 126 sources 
to $z=0.05-0.9$ and a further 229 sources to $z=0.9-1.5$. 
Based on the SEDs of the sources with only photometric 
redshifts in these ranges, many are red galaxies, 
which are more difficult to identify spectroscopically. 
Of the bluer sources, some may have 
photometric redshifts that have scattered into these redshift 
ranges, even though their true redshifts are higher. 
Of the 126 galaxies with only photometric redshifts in the
redshift range $z=0.05-0.9$, 87 have been spectroscopically 
observed and none of these have strong emission lines. Emission 
lines, if present, would have easily been observed in this 
redshift range.

%
%
\begin{inlinefigure}
\figurenum{3}
\centerline{\psfig{figure=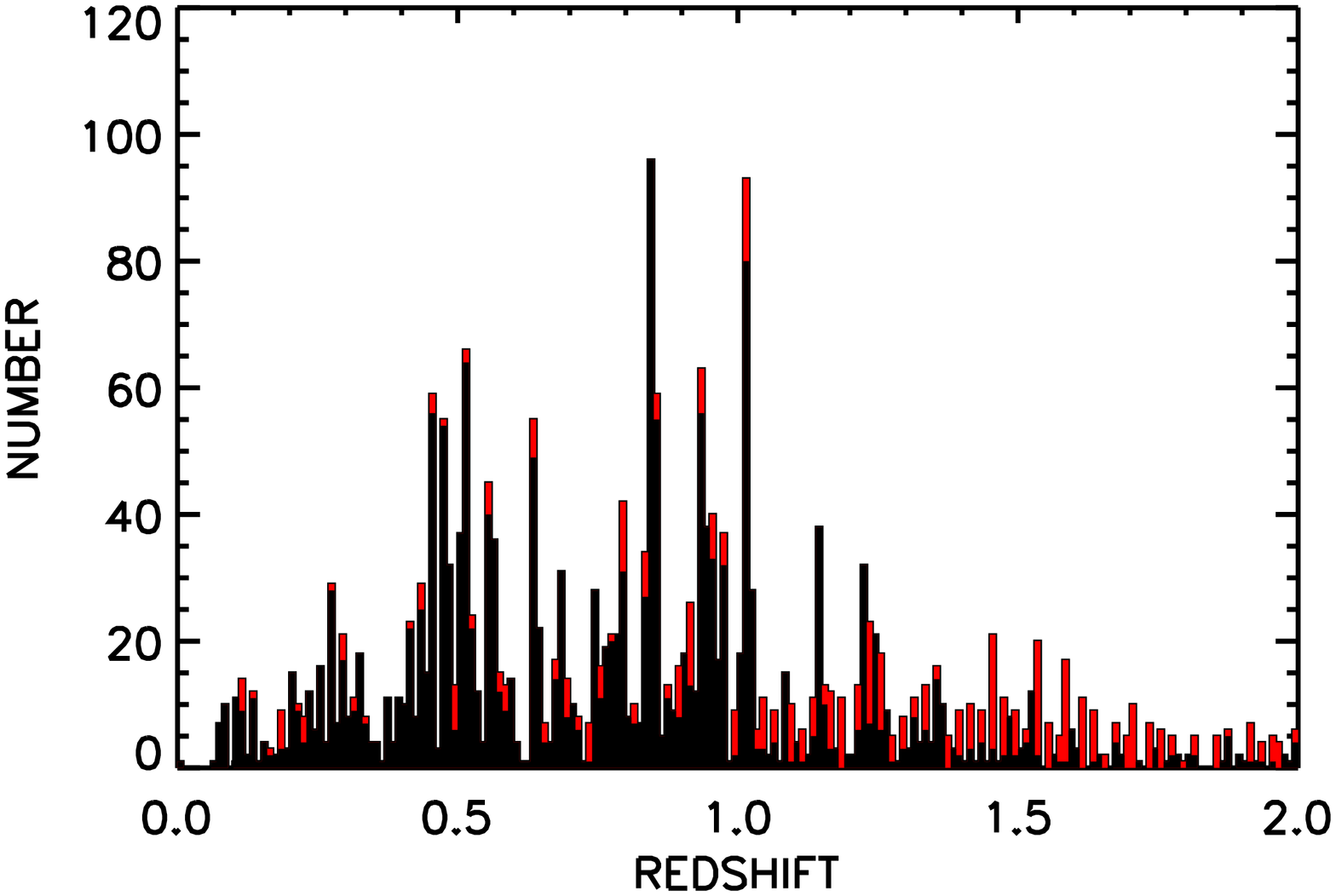,angle=90,width=3.5in}}
\vskip -0.2cm
\figcaption[]{
Redshift distribution of our full NIR sample. 
Black (red) histogram shows spectroscopic 
(photometric) redshifts. The redshift bin size is 0.01. 
\label{zdist}
}
\end{inlinefigure}

We conclude that the spectroscopic sample contains nearly
all of the sources (1260 out of a maximum of 1386, or $>91$\%) 
lying in the redshift range $z=0.05-0.9$ and essentially all 
of the sources with strong emission lines suitable
for measuring emission line metallicities in this redshift
range. Over the $z=0.05-1.5$ interval the spectroscopically 
identified sample contains $>84$\% of the galaxies 
(1884 out of a maximum of 2239).

\subsection{Galaxy Morphologies}
\label{secmorph}

The galaxy morphological types are taken from Bundy et al.\ (2005)
wherever possible. The Bundy et al. catalog is based on
Richard Ellis's visual classification of the sources in the GOODS-N 
according to the following scale:  $-2=$~Star, $-1=$~Compact, 
$0=$~E, $1=$~E/S0, $2=$~S0, $3=$~Sab, $4=$~S, $5=$~Scd, $6=$~Irr,
$7=$~Unclass, $8=$~Merger, and $9=$~Fault.
For galaxies in our sample which were not included in the 
Bundy et al.\ (2005) catalog, we visually classified the sources 
using the {\em HST\/} F850LP images, aiming to reproduce the 
Ellis classifications as closely as possible.

\subsection{Galaxy Environments}
\label{secgalenv}

The local galaxy density can be computed using the 
distance to the $n$th nearest neighbor (Dressler 1980).
In the present work we use the velocity information
only to separate slices; otherwise we use the projected 
distance $d_n$. The surface density is then given by 
$\Sigma=n/(\pi d_n^2)$. An extensive comparison of this 
measure of the density environment with other methods is 
given in Cooper et al.\ (2005), who conclude that the 
projected distance method is generally the most robust for this 
type of work.

Edge effects are important in small field areas, such as the 
GOODS-N region, and can bias the density parameter in low density 
regions where the projected separation extends beyond the edge of 
the field. We correct for this effect by including only that part 
of the area $\pi d_n^2$ which lies within the field 
(e.g., Baldry et al.\ 2006). To further reduce the edge effects, 
we use a low $n$ to minimize the projected distance 
and also exclude regions of the field that, at the redshift
of the galaxy, lie too close to the edge of the field for an 
accurate measurement.

Thus, for each galaxy we computed the projected density based
on the 3rd nearest neighbor having a mass above a uniform
mass cut and lying within 1000~km~s$^{-1}$ of the galaxy.
We exclude galaxies which lie closer than 1~Mpc from the
sample edge. This constraint restricts to galaxies
with $z\gtrsim 0.3$, since all of the lower redshift galaxies 
will lie too close to the edges of the field.

\section{Four Uniform NIR Luminosity Samples}
\label{seclum}

%
%
\begin{inlinefigure}
\figurenum{4}
\centerline{\psfig{figure=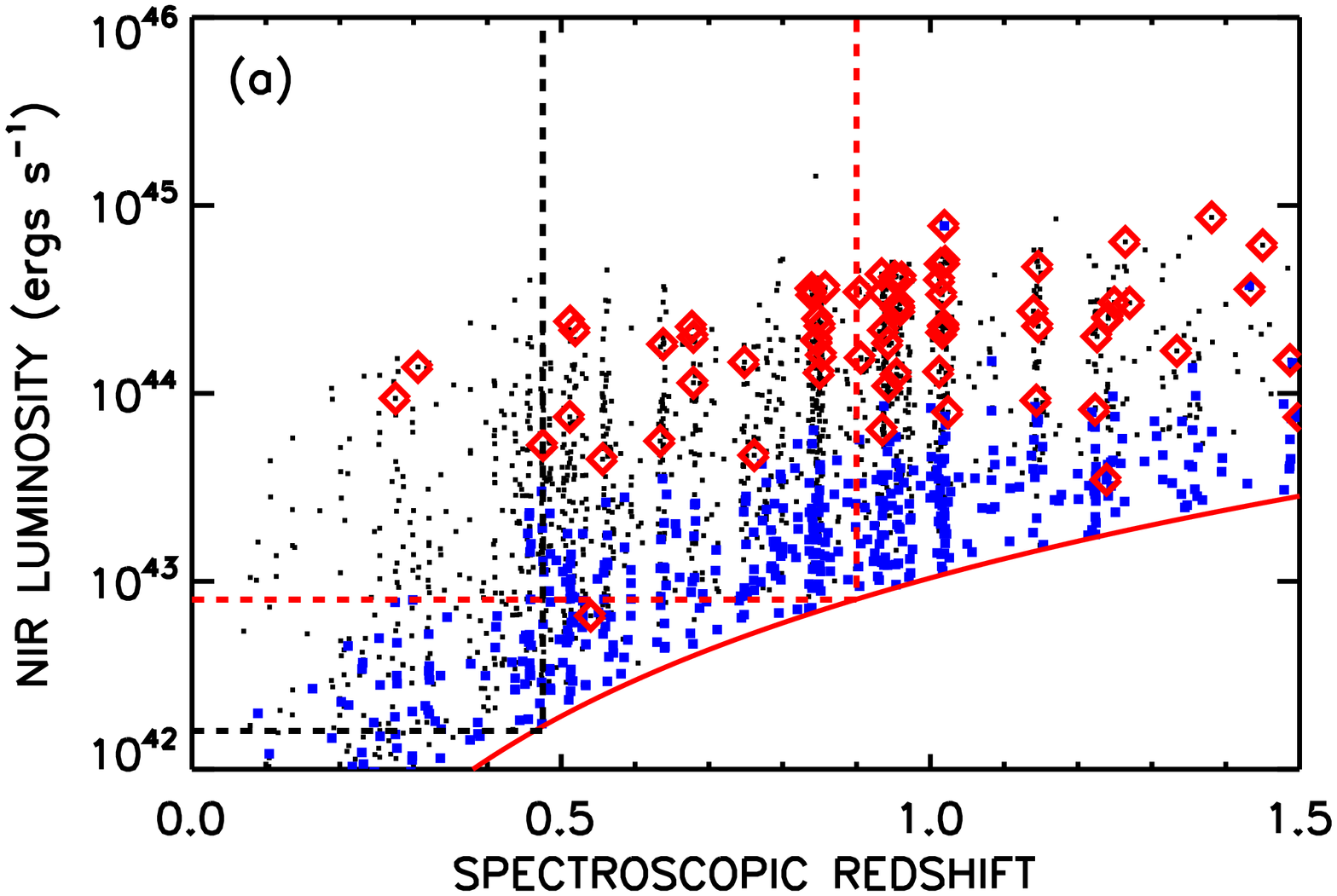,angle=90,width=3.5in}}
\vskip -0.6cm
\centerline{\psfig{figure=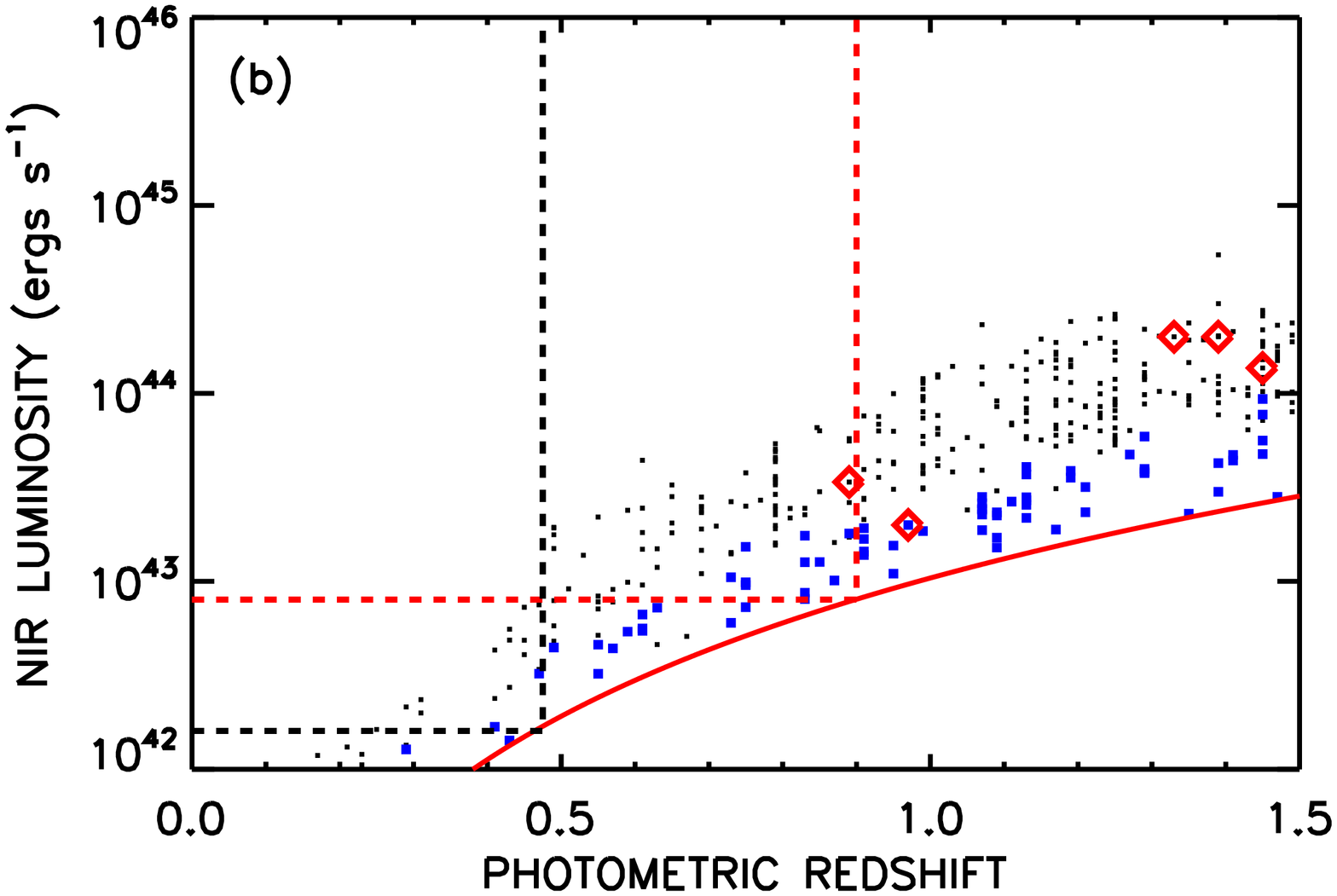,angle=90,width=3.5in}}
\vskip -0.2cm
\figcaption[]{
(a) NIR luminosity vs. spectroscopic redshift for the 
spectroscopically identified sources in the NIR sample, and 
(b) NIR luminosity vs. photometric redshift for the 
spectroscopically unidentified sources in the NIR sample.
The blue solid squares show sources with blue spectra, where
massive stars may still make a substantial contribution to
the NIR luminosity. The red open diamonds show sources 
containing AGNs based on their X-ray luminosities. The red 
solid curve shows the luminosity corresponding to the 
limiting NIR flux of $2\times10^{-15}$~ergs~cm$^{-2}$~s$^{-1}$. 
The red (black) dashed lines mark the region that corresponds 
to the mid-$z$ (low-$z$) uniform NIR luminosity sample, where 
the \oiii~$\lambda5007$ (\ha) line would be in the spectrum. 
\label{nirlum}
}
\end{inlinefigure}

In Figure~\ref{nirlum}a (\ref{nirlum}b) we show NIR 
luminosity versus spectroscopic (photometric) redshift for 
the spectroscopically identified (unidentified) NIR sample. 
We denote sources with blue spectra,
where massive stars may still make a substantial 
contribution to the NIR luminosity, by blue solid squares.
We denote sources that contain AGNs (based on whether 
either their $2-8$~keV or $0.5-2$~keV luminosities 
are $>10^{42}$~ergs~s$^{-1}$) by red open diamonds. 
We show the luminosity corresponding to the limiting NIR 
flux of $2\times10^{-15}$~ergs~cm$^{-2}$~s$^{-1}$
by the solid red curve.

%
%
\begin{inlinefigure}
\figurenum{5}
\centerline{\psfig{figure=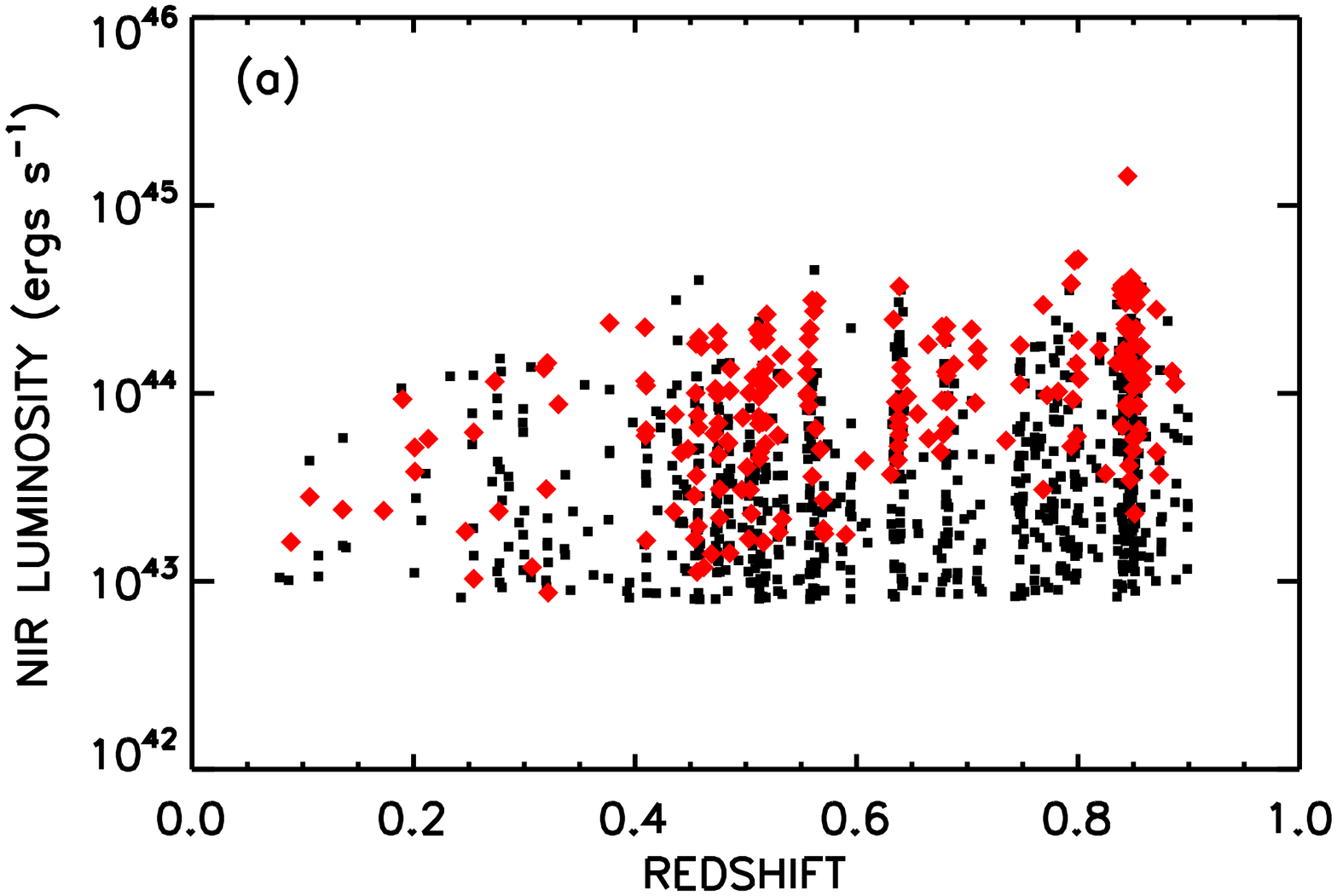,angle=90,width=3.5in}}
\vskip -0.6cm
\centerline{\psfig{figure=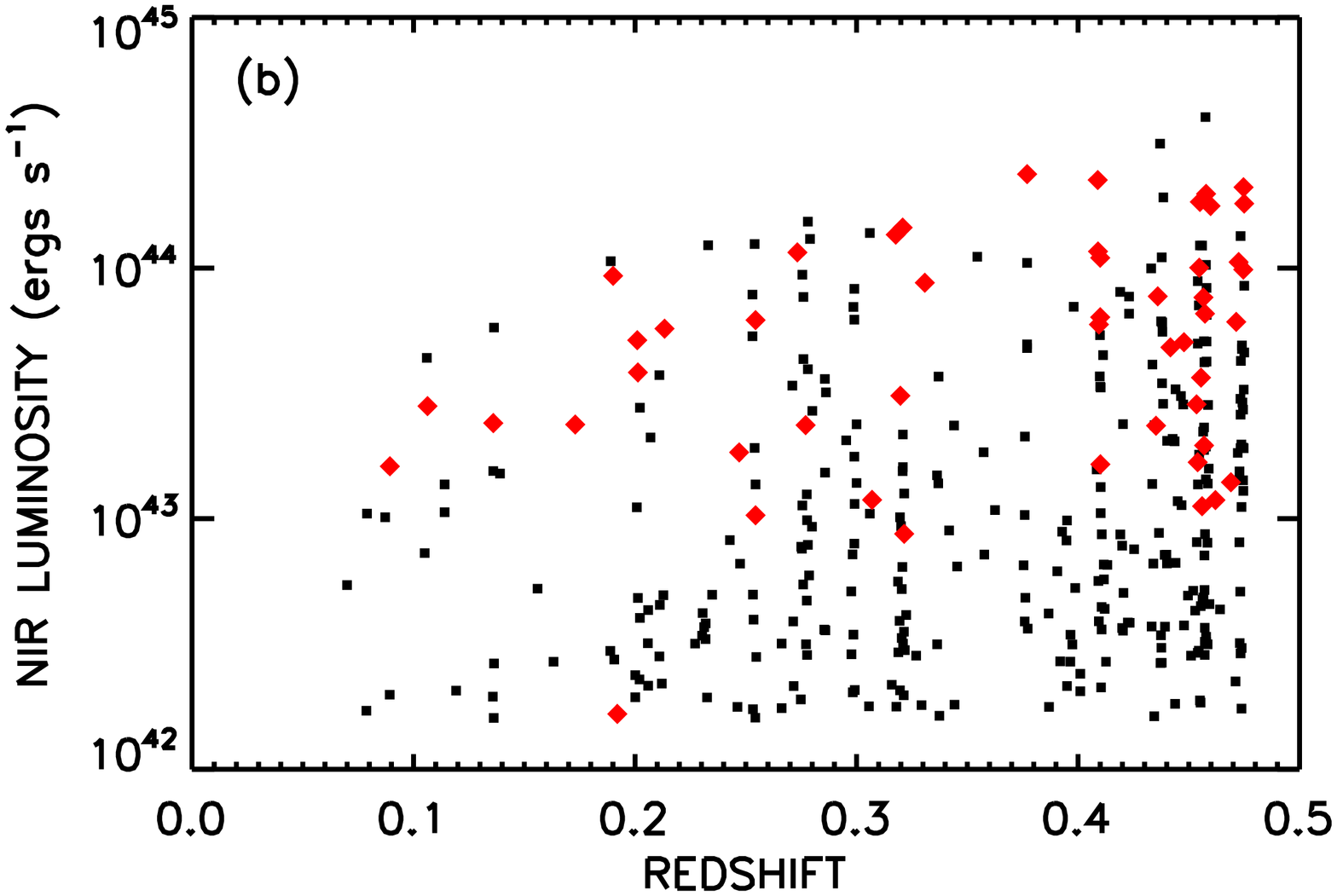,angle=90,width=3.5in}}
\vskip -0.2cm
\figcaption[]{
NIR luminosity vs. redshift for the spectroscopically
identified sources in the (a) mid-$z$ sample and (b) low-$z$ 
sample. The sources with emission (absorption) line redshifts 
are denoted by black squares (red diamonds).
\label{nirlum_sample}
}
\end{inlinefigure}

We construct two uniform NIR luminosity samples for 
our metallicity analysis: a mid-$z$ sample and a 
low-$z$ sample.
Although most of the spectra extend to about $1~\mu$m,
the sensitivity falls rapidly at the reddest wavelengths,
and some spectra are cut off at wavelengths
$<1~\mu$m because of their mask positions. Thus, 
we choose a limiting upper wavelength of 9500~\AA. 
The limiting upper wavelength of 9500~\AA\ corresponds
to $z<0.9$ ($z<0.475$) for the \oiii~$\lambda$5007~\AA\
(\ha) line to be observable if present. In each case this sets a
lower limit on the NIR luminosity, which corresponds to
the NIR flux limit at the maximum redshift. The mid-$z$
sample has $z=0.05-0.9$ and NIR luminosity
$>8\times10^{42}$~ergs~s$^{-1}$,
and the low-$z$ sample has $z=0.05-0.475$ and NIR luminosity
$>1.6\times10^{42}$~ergs~s$^{-1}$. These limits are shown
in Figure~\ref{nirlum} by the red (black) dashed lines
for the mid-$z$ (low-$z$) sample. 
The low-redshift portion of the mid-$z$ sample is contained
as the high-mass subset of the low-$z$ sample.

There are 1009 sources in the mid-$z$ sample, of which 929
(92\%) have spectroscopic redshifts. There are 378 sources
in the low-$z$ sample, of which 354 (96\%) have spectroscopic
redshifts. In Figures~\ref{nirlum_sample}a and 
\ref{nirlum_sample}b we show blow-ups of Figures~\ref{nirlum}a 
and \ref{nirlum}b to more clearly illustrate 
both samples, but here our symbols distinguish 
between absorption line redshifts {\em (red diamonds)\/} 
and emission line redshifts {\em (black squares)\/}. 
Of the 929 (354) spectroscopically identifed redshifts 
in the mid-$z$ (low-$z$) sample, 210 (49) are based on absorption 
line features. The absorbers comprise a much higher fraction of 
the more luminous galaxes. Thus, the smaller
fraction of absorbers in the low-$z$ sample is partly
a consequence of the lower luminosity limit in that sample.

We also construct two higher-redshift uniform NIR luminosity 
samples for studying the evolution of the galaxy masses and
star formation histories. We will 
refer to these as our high-$z$ and highest-$z$ samples.
The high-$z$ sample has $z=0.9-1.2$ and NIR luminosity
$>2\times 10^{43}$~ergs~s$^{-1}$. The highest-$z$ sample has
$z=1.2-1.5$ and NIR luminosity $>3\times 10^{43}$~ergs~s$^{-1}$.

\section{Fitting the Galaxy Spectral Energy Distributions}
\label{secfit}

While NIR luminosities have been the preferred way to estimate
galaxy mass, there is still a wide range in the mass
to NIR luminosity ratios and there are still considerable
uncertainties in the models used to determine 
the masses. The mass ratio for a 
galaxy depends on its star formation history
and on the level of extinction (Brinchmann \& Ellis 2000). 
We may estimate the conversion from NIR luminosity to mass 
by fitting the galaxy SEDs with 
model galaxy types modulated by an assumed extinction law 
(e.g., Brinchmann \& Ellis 2000; Kauffmann et al.\ 2003a; 
Bundy et al.\ 2005). We follow this procedure here to 
estimate the masses of the galaxies and the extinctions. 
We use the Bruzual \& Charlot (2003; hereafter, BC03) models 
for ease of comparison with previous work; however, as we will
discuss, there has been considerable recent debate over
this calibration, which may overestimate galaxy mass.

For every galaxy in each of our four uniform NIR luminosity
samples, we fitted BC03 models 
assuming a Salpeter IMF, a solar metallicity, and a Calzetti 
extinction law (Calzetti et al.\ 2000). 
We included a range of types from single burst models to
exponentially declining models to constant star formation 
models. For each galaxy we varied the age from 
$5\times10^{7}$~yr to the maximum possible age of the 
galaxy at its redshift. In making 
the fits, we calculated the $\chi^2$ values assuming an 
individual error of 0.1~mag in each band together with 
the $1\sigma$ noise in each band, and we fitted the SED 
over the rest-frame wavelength range $0.2-2.4~\mu$m.

Our use of only solar metallicity models does not introduce
significant uncertainties in the inferred masses and extinctions.
As we shall discuss later there is a well-known strong
degeneracy between age and metallicity in the stellar
models. Introducing a range in metallicity in the models
therefore increases the spread in the possible ages while
leaving the other quantities nearly unchanged. 
We have recomputed the results of the present paper using supersolar
and subsolar models and find this does not change any of the
conclusions.

In our subsequent analysis we use the mass ratios
and extinctions corresponding to the minimum $\chi^2$ 
fits. Hereafter, we refer to these best fits as our 
BC03 fits and the 
corresponding masses and extinctions as
our BC03 masses and extinctions. 
However, we note that the mass to NIR luminosity ratio
and extinction probability distributions
(see Kauffmann et al.\ 2003a)
show that there are still substantial uncertainties in 
these quantities (e.g., Papovich et al.\ 2006). 
These uncertainties are the largest for the blue galaxies 
and can range up to 0.3~dex in the mass to NIR luminosity 
ratio. This reflects the ambiguities in the type and extinction
fitting, where models with different star formation histories 
and extinctions can reproduce the same galaxy SED.

In Figure~\ref{bruzual_fit} we show our BC03
fits to two example galaxies. In (a) we show a red galaxy 
at $z=0.850$ which is best fitted by no extinction 
{\em (black squares)\/} and a single burst with an age of 
2.4~Gyr {\em (red SED)\/}. In (b) we show a red galaxy at 
$z=0.433$ which is best fitted with a large extinction 
of $E(B-V)=0.45$ {\em (black squares)\/} and a single burst 
with an age of 0.18~Gyr {\em (red SED)\/}. However, as an example 
of how our distinction between old galaxies and reddened 
younger galaxies relies on the overall shape of the
SED, we also show in (b) a fit with no extinction 
{\em (purple diamonds)\/}
and a 1~Gyr exponential decline with an age of 4.5~Gyr 
{\em (purple SED)\/}. It would not be possible to
differentiate between these two fits with only the
optical data; however, with the NIR data the latter
is a significantly poorer fit. For all the fits shown the 
solid portions of the curves indicate the regions over 
which we made the fits. We did not fit to the rest-frame 
MIR data because of the limitations of the BC03 models at 
the longer wavelengths.

However, there are very serious concerns about
determining the stellar masses from the population
synthesis models and, in particular, from the NIR
fluxes. Maraston (2005) pointed out that an
improved treatment of the thermally-pulsating
asymptotic giant branch (TP-AGB) stars resulted in a substantial 
increase in the NIR light at intermediate ($\sim10^9$~yr) 
ages relative to preceding population synthesis models.
This would reduce the stellar mass estimates 
in the high-redshift galaxies. Bruzual (2007) reports similar
results when an improved treatment of the TP-AGB stars 
is included in a revised version of the Bruzual-Charlot 
code. Kannappan \& Gawiser (2007) have investigated the 
differences in the various models using a local galaxy 
sample and, while not coming to a conclusion about a 
preferred model, they emphasize the uncertainties in the 
mass determination as a function of galaxy type.

Conselice et al.\ (2007; hereafter, Conselice07) 
have used fits to the revised 
Bruzual-Charlot models to argue that the decrease in the 
average masses relative to BC03 is small for galaxies
in the $z=0.4-2$ redshift range when the masses are based 
on rest-frame $0.7-1.5~\mu$m wavelengths. They find an 
average drop in the masses of 0.08~dex relative to BC03 
and a maximum decrease of about 20\%. However, averaging
may obscure the systematic effects of the uncertainties, 
particularly when comparing the higher redshift samples 
with local samples. For example, galaxies with ages of 
$10^{10}$~yr will have BC03 masses that are consistent with 
the Maraston (2005) and revised Bruzual-Charlot codes, while
those with ages of $10^{9}$~yr will have BC03 masses that
are about 25\% too high, and those with ages of 
$10^{8}-10^{9}$~yr will have BC03 masses that are about 40\% 
too high (see Fig.~3 of Bruzual 2007). We will consider the 
possible effects of such systematic uncertainties on the 
stellar mass density growth rates measured from the formed
stellar mass functions and the comparison of those rates
with the expected formed stellar mass density growth rates 
produced by star formation in \S\ref{secsfh}.

%
%
\begin{inlinefigure}
\figurenum{6}
\centerline{\psfig{figure=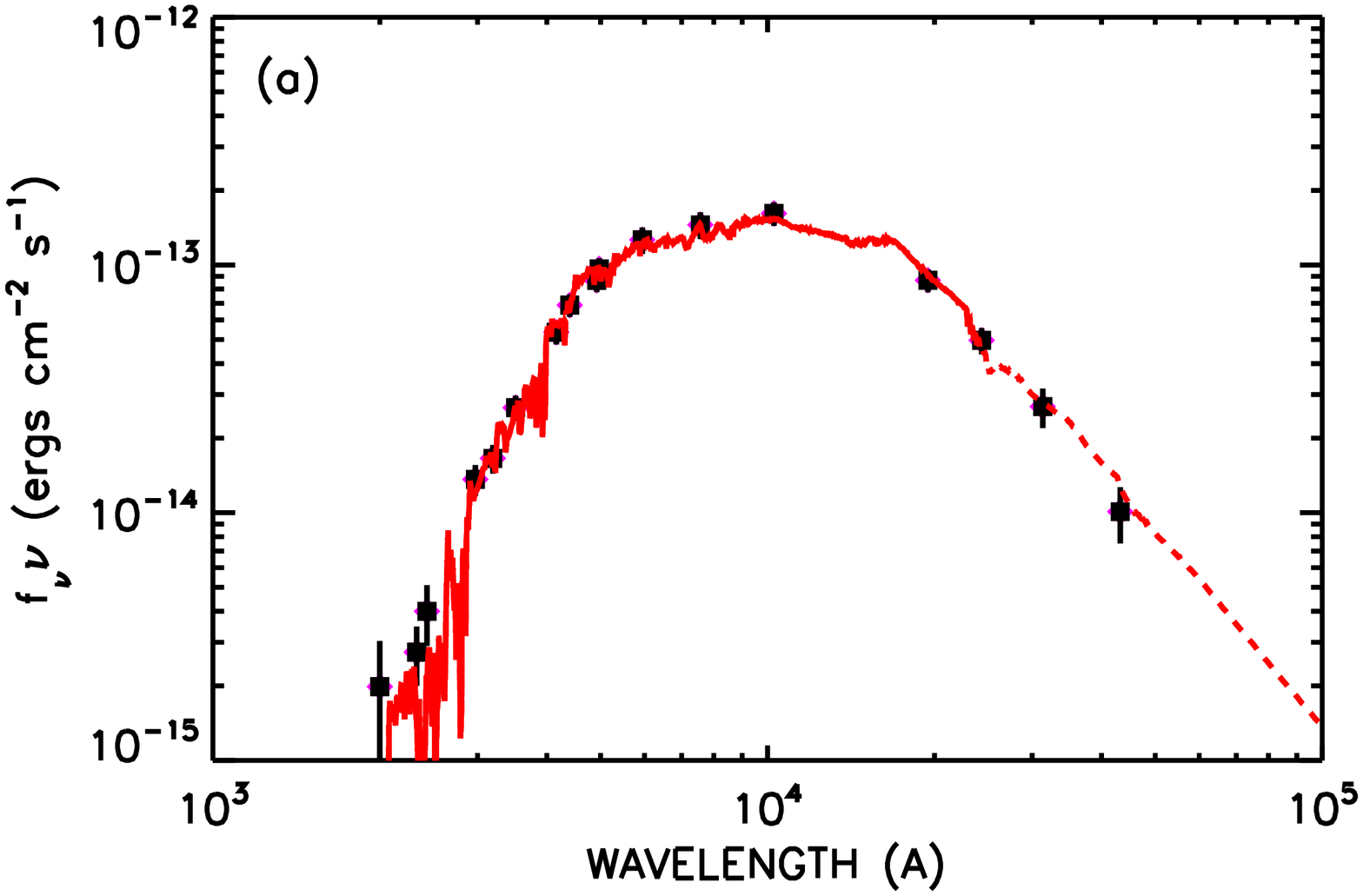,angle=90,width=3.5in}}
\vskip -0.6cm
\centerline{\psfig{figure=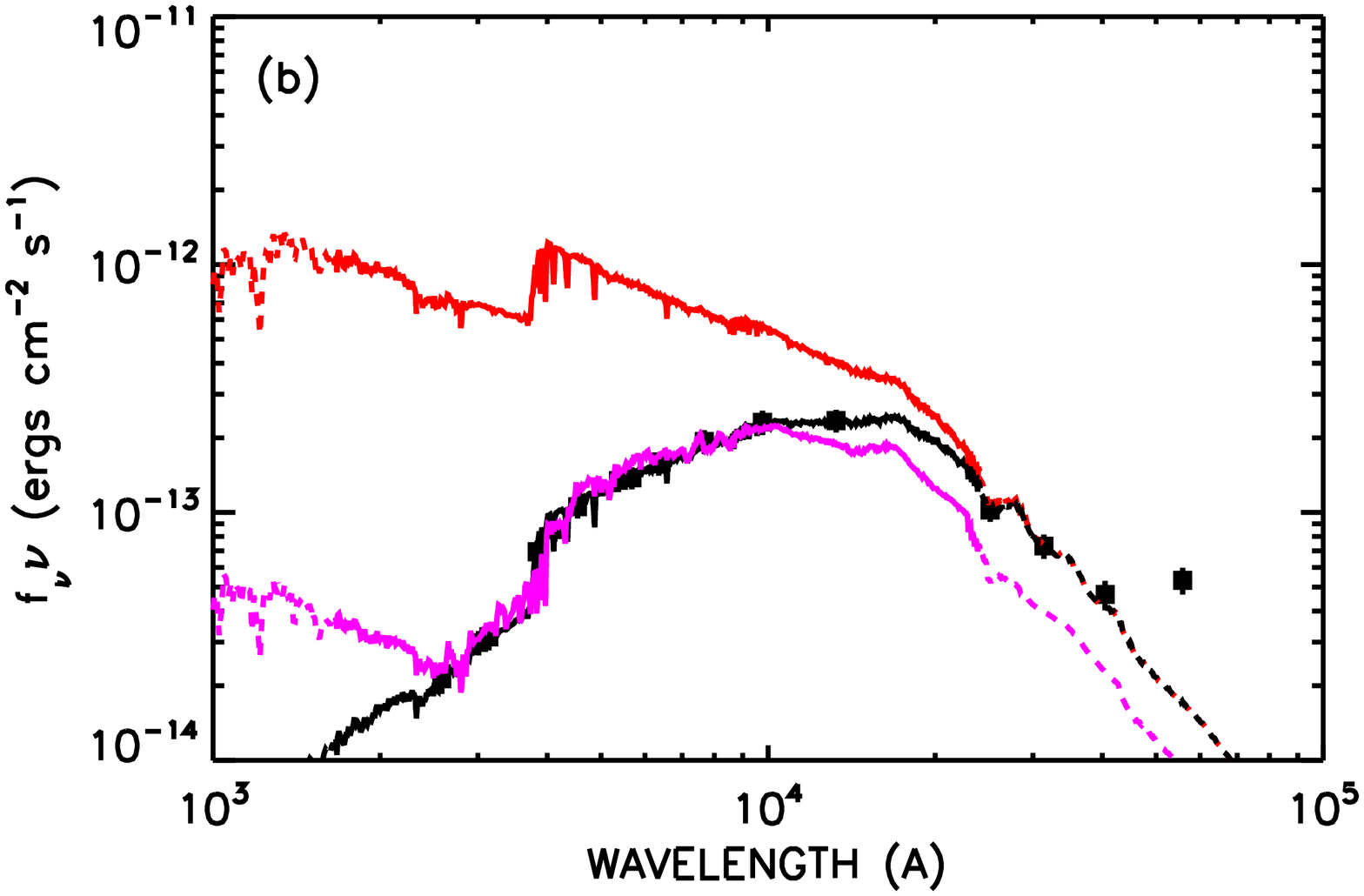,angle=90,width=3.5in}}
\vskip -0.2cm
\figcaption[]{
Sample BC03 fits to two galaxies.
(a) A red galaxy at $z=0.850$ is best
fitted with no extinction {\em (black squares
with the assumed errors from the text)\/} and 
a single burst with an age of 2.4~Gyr {\em (red SED)\/}.
The solid portion of the curve shows the region 
over which we made the fit. 
(b) A red galaxy at $z=0.433$ is best fitted 
with a large extinction of $E(B-V)=0.45$ 
{\em (black squares with the assumed errors from the text
show the observed fluxes)\/} 
and a single burst with an age of 0.18~Gyr 
{\em (the red solid line shows the galaxy SED without
extinction and the black line the SED with the extinction
included)\/}. The NIR data are required to distinguish
between this best-fit extinguished model and
another fit having no extinction {\em (purple 
diamonds with the assumed errors from the text)\/} and a 1~Gyr 
exponential decline with an age of 4.5~Gyr {\em (purple SED)\/}.
The solid portion of each curve shows the region over 
which we made the fit.
Both are good fits to the rest frame optical
and UV but the model with no extinction is a significantly
poorer fit in the near IR. The no-extinction fit would
reduce the mass by a factor of five relative to the
and the inferred star-formation rate by a factor of thirty
relative to the best fit model.
\label{bruzual_fit}
}
\end{inlinefigure}

\subsection{Extinctions}

Given the model uncertainties discussed above, it is critical
to determine how meaningful our BC03 extinctions are.
We may test them in two ways. First we look at how 
they relate to the MIR properties of the galaxies, and then
we compare them with extinctions measured 
from the Balmer lines. We note that the comparison of our BC03
extinctions with the MIR properties is affected by orientation, 
which will add scatter to the comparison. However, for both
of our tests we find reasonable agreement between our BC03 
extinctions and the dust properties measured in other ways.

In Figure~\ref{f24hist} we show the distribution of the 
BC03 extinctions for the mid-$z$ sample {\em (black histogram)\/}.
We find that roughly half of the galaxies in the sample
have weak extinctions of $E(B-V)<0.1$ or no measured extinction, 
while the remainder lie in an extended tail up to our maximum
allowed value of $E(B-V)=0.63$. 
The median extinction of A$_v=0.6$ is twice the local value
given by Kauffmann et al.\ (2003a). This is consistent with 
the high-redshift galaxies having more gas and dust mass. 
We shall return to this point in \S\ref{secgasmass}.

%
%
\begin{inlinefigure}
\figurenum{7}
\centerline{\psfig{figure=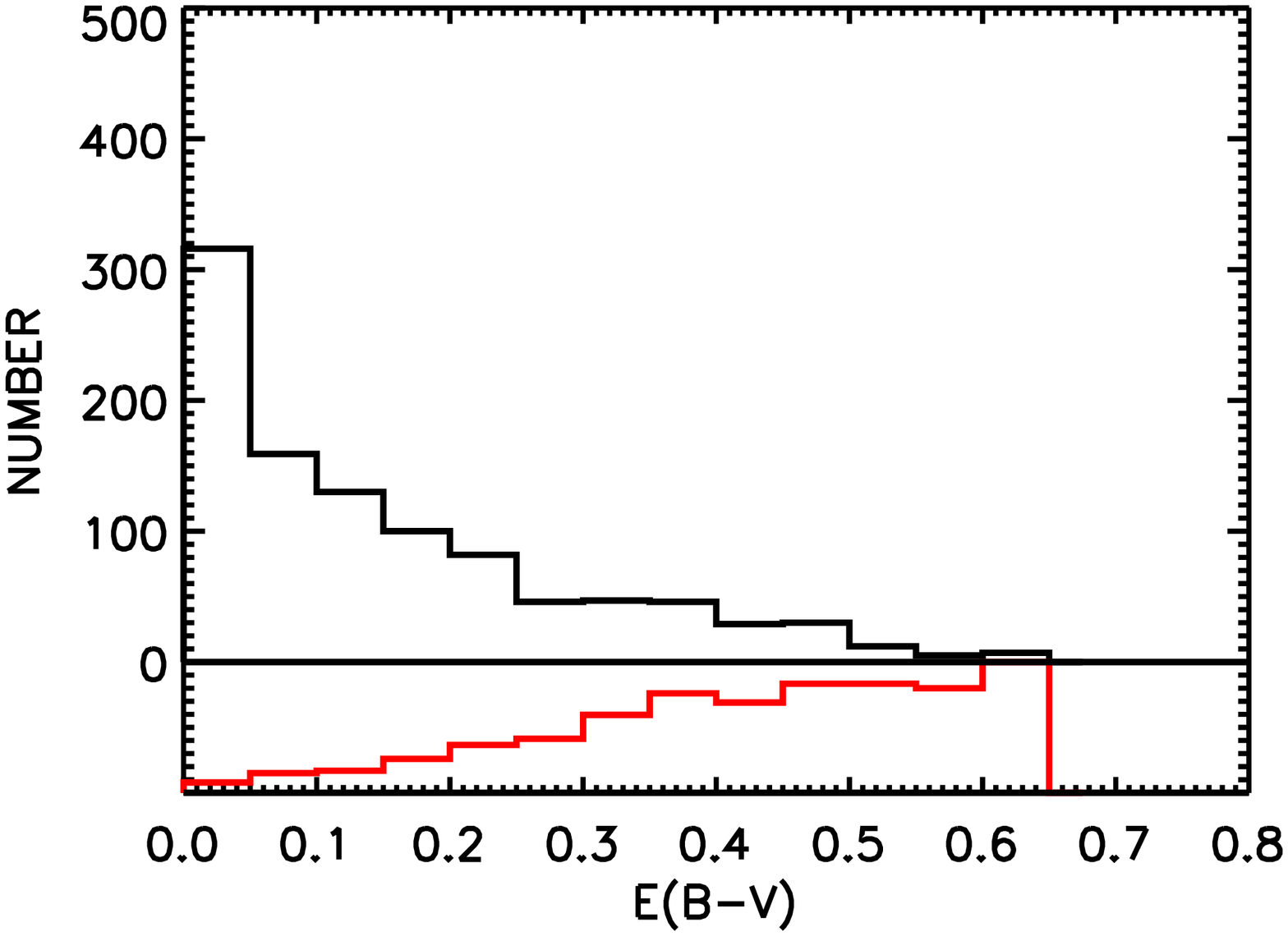,angle=90,width=3.5in}}
\vskip -0.2cm
\figcaption[]{
Distribution of extinctions for the mid-$z$ sample
derived from our BC03 fits {\em (black histogram)\/}.
Roughly half of the galaxies have little or no extinction.
The red histogram shown underneath is the fraction of galaxies
detected at 24~$\mu$m. Only about 11\% of the galaxies with
weak extinctions of $E(B-V)<0.05$ are 24~$\mu$m sources, while
nearly all of the strongly extinguished sources are.
\label{f24hist}
}
\end{inlinefigure}

%
%
\begin{inlinefigure}
\figurenum{8}
\centerline{\psfig{figure=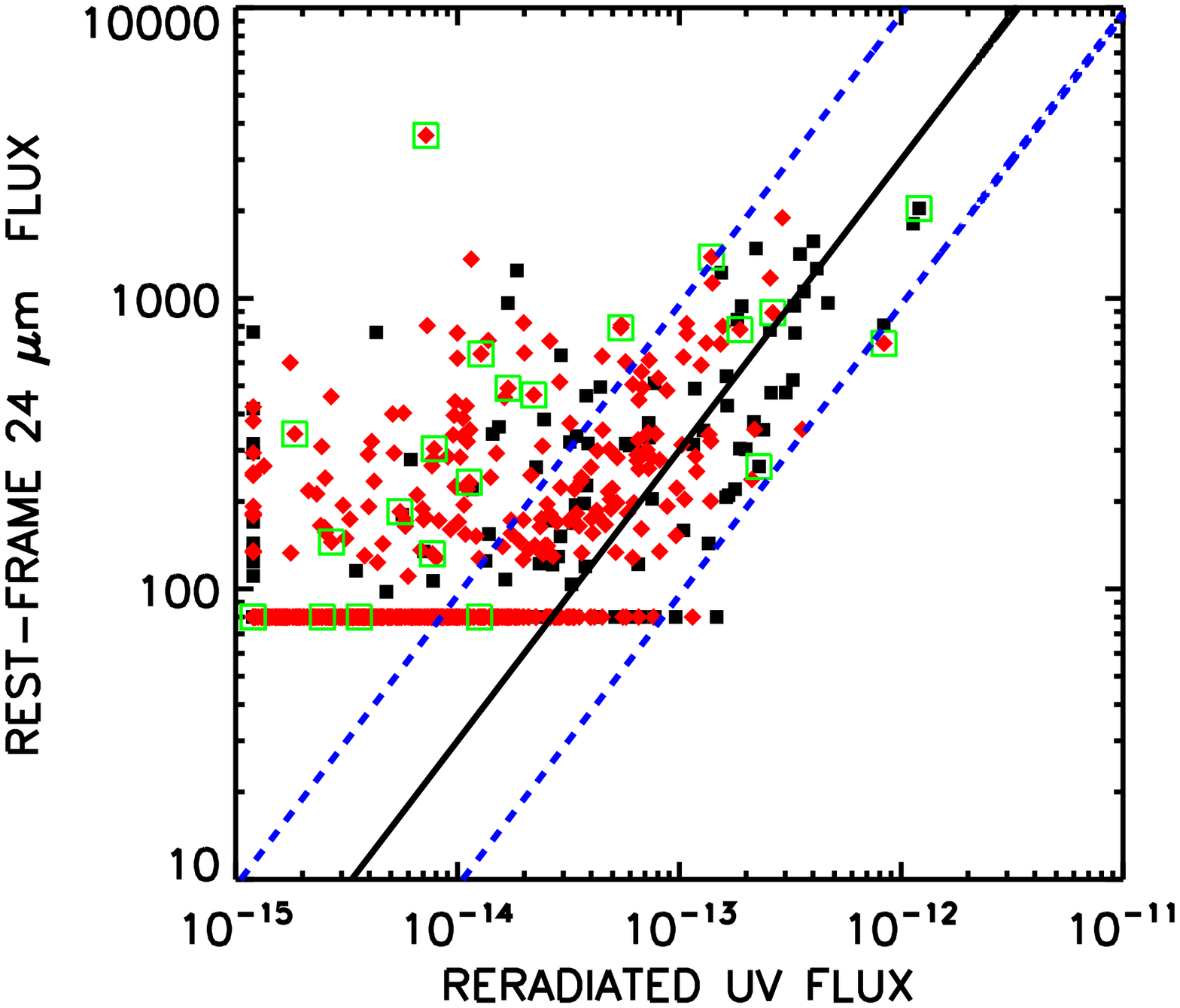,angle=90,width=3.5in}}
\vskip -0.2cm
\figcaption[]{
Comparison of the reradiated light that originated
from rest-frame $2000-3000$~\AA\ (see text for details)
with the rest-frame $24~\mu$m flux for the mid-$z$ sample. 
The $24~\mu$m $K$-correction has been computed assuming 
the M82 SED of Silva et al.\ (1998).
The red diamonds show sources with 
$z=0.475-0.9$, and the black squares show
lower redshift sources. The galaxies without $24~\mu$m
detections are shown at the $80~\mu$Jy limit of the
$24~\mu$m data, and the galaxies with little or no 
reradiated flux are shown at a nominal value of 
$1.2\times10^{-15}$~ergs~cm$^{-2}$~s$^{-1}$. 
The sources containing AGNs based on their X-ray luminosities 
are enclosed in green squares. There is a linear 
correlation {\em (black solid line)\/} between large 
reradiated fluxes and $24~\mu$m detections with a large 
spread ($\pm0.5$~dex, {\em blue dashed lines\/}).
\label{reradiated}
}
\end{inlinefigure}

We may test the extinctions derived from the BC03
fits by comparing the extinguished UV light, which 
is reradiated into the FIR, with other completely independent
measures of the dust reradiated light, such as the $24~\mu$m 
light. This is the longest wavelength light for which 
extremely deep images of the field have been obtained. 
In Figure~\ref{f24hist} we show the
fraction of galaxies detected at 24~$\mu$m {\em (red histogram)\/}.
Interestingly, we see that nearly all of the highly extinguished 
sources are also detected at $24~\mu$m. To further quantify 
this comparison, we computed the reradiated UV light from 
each galaxy in the mid-$z$ sample by subtracting the 
observed SED from the extinction corrected SED.
In Figure~\ref{reradiated} we show the difference
in the rest-frame $2000-3000$~\AA\ wavelength range 
(i.e., the UV flux that got reradiated) versus the 
rest-frame $24~\mu$m flux ($K$-corrected using the M82 SED 
of Silva et al.\ 1998).

The sources with large reradiated fluxes are generally
detected at $24~\mu$m, with a linear relation shown by
the solid black line. There is a large
spread in the relation ($\pm0.5$~dex, shown by the
blue dashed lines), which most likely reflects
the use of only the M82 template to
obtain the $K$-corrections for the $24~\mu$m flux
(see, e.g., Dale et al.\ 2005; Marcillac et al.\ 2006;
Barger et al.\ 2007 for why this is not ideal).
However, there is no substantial
redshift change, with the higher redshift points
{\em (red diamonds)\/} having the same distribution
as the lower redshift points {\em (black squares)\/}.

These results show that the assignment of
substantial extinctions to galaxies by our BC03
fits is confirmed by the MIR measurements. About
60\% of the $24~\mu$m sources in the mid-$z$ sample
to the $80~\mu$Jy flux limit of the $24~\mu$m data
are picked out in this way and lie between the dashed 
lines in Figure~\ref{reradiated}.
However, some of the remaining $24~\mu$m sources in the
mid-$z$ sample have low reradiated UV fluxes.
(A few percent are clearly blended
galaxies, where the $24~\mu$m flux arises from a
different galaxy than the one being fitted in the
UV, but this is a small effect.) Therefore, a critical
question for the present analysis is whether this
implies that we are failing to assign extinctions
to galaxies where there should be extinctions.

We have checked this by inspecting the spectra 
(see \S\ref{secspectral}) of the galaxies in the
mid-$z$ sample with \ha\ in their spectrum which
are detected at $24~\mu$m but for which our BC03 fits 
have assigned a low extinction.
In all cases the $f$(\hb)$/f$(\ha) ratios
are also consistent with little extinction.  
These sources may contain obscured nuclei that
are only seen in the MIR and have little effect
on the measured properties of their host galaxies.
Indeed, the brightest 24~$\mu$m source in the
mid-$z$ sample is not picked out by its reradiated UV 
flux. It is an X-ray source, and it has a Seyfert~2
spectrum, which suggests that it is an obscured AGN.
There is also a higher fraction of X-ray AGNs
among the remaining sources with 24~$\mu$m detections 
but low reradiated UV fluxes.
(In Figure~\ref{reradiated} 
we enclose in green open squares sources containing 
AGNs based on their X-ray luminosities.) However,
many of these sources are not X-ray detected and, if the 
$24~\mu$m light in these sources is produced by AGN 
activity, then the nucleus must be highly obscured 
in the optical and in the X-ray.

\subsection{Masses}
\label{secmass}

In Figure~\ref{mass_ratio} we show the ratio of 
the stellar mass to the observed (i.e., uncorrected
for extinction) NIR luminosity versus the observed
NIR luminosity for our low-$z$ and mid-$z$ samples.
We hereafter refer to this ratio as the mass ratio. 
For both samples the mass ratios range from values of about
$10^{-34}$ (blue galaxies) to $10^{-33}$ (evolved galaxies
with ages comparable to the present age of the universe),
though the more luminous galaxies are primarily at the
high (evolved) end. Multiplying the $10^{-33}$ upper limit 
on the mass ratio {\em (purple line)\/} by the NIR
luminosity limits of the low-$z$ and mid-$z$ samples, 
we find that this upper limit implies
that the low-$z$ and mid-$z$ samples include all galaxies
with masses above $\sim 2\times10^{9}$~M$_\odot$
and $\sim 10^{10}$~M$_\odot$, respectively.
For our high-$z$ and highest-$z$ samples, the corresponding
mass limits are $2\times 10^{10}$~M$_\odot$
and $3\times 10^{10}$~M$_\odot$, respectively.

%
%
\begin{inlinefigure}
\figurenum{9}
\centerline{\psfig{figure=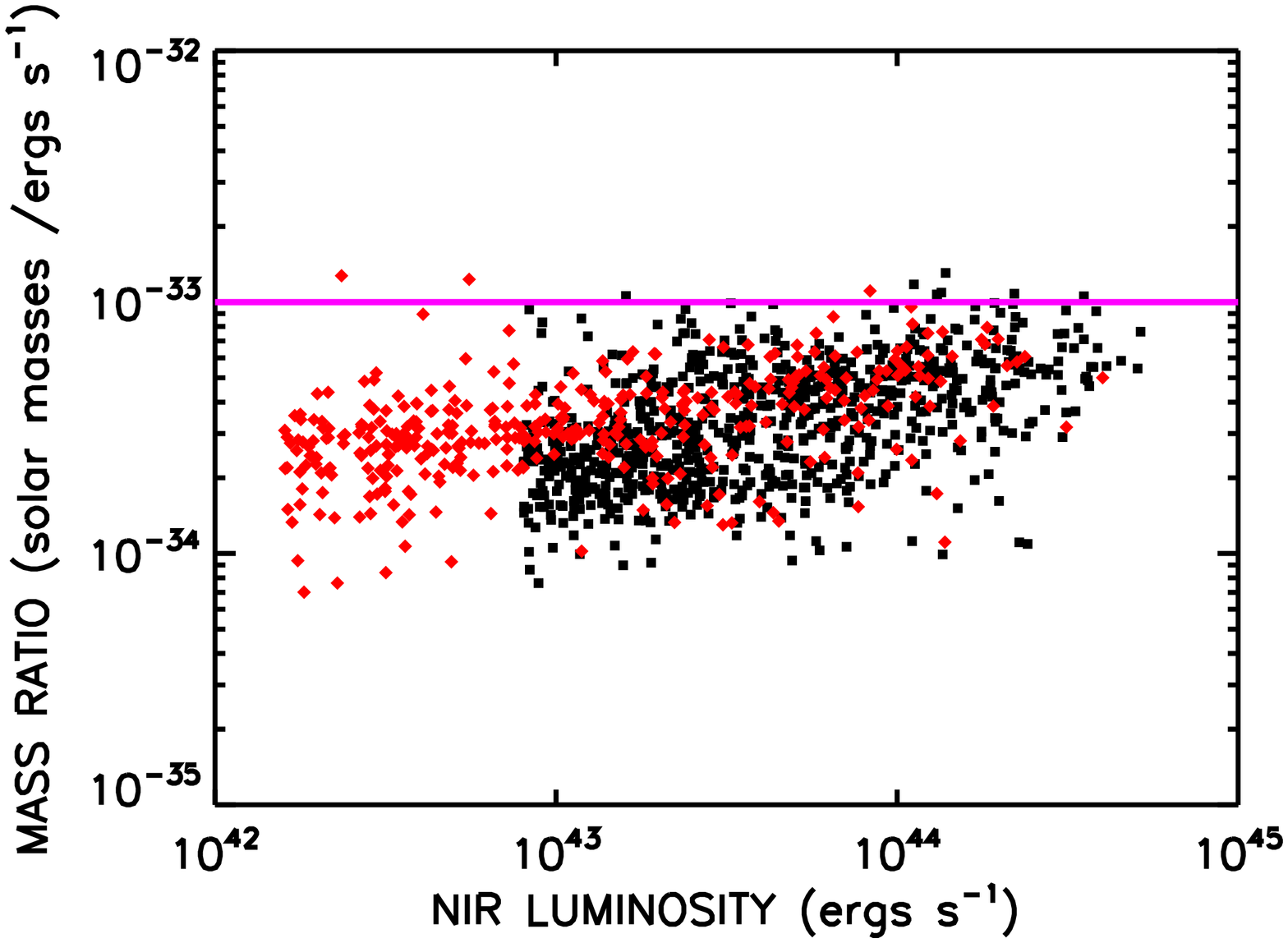,angle=90,width=3.5in}}
\vskip -0.2cm
\figcaption[]{
Ratio of the stellar mass to the observed (i.e., 
uncorrected for extinction) NIR luminosity vs. observed NIR 
luminosity. The mass ratios were computed from our BC03 fits,
which include the effects of extinction. The low-$z$ (mid-$z$) 
sample is shown with the red diamonds (black squares). 
The purple line shows the maximum mass ratio adopted in 
computing the mass limits on the samples.
\label{mass_ratio}
}
\end{inlinefigure}

We also computed the mass to luminosity
ratios ($M/L_{z'}$) for the extinction corrected $z'$-band
luminosity, $L_{z'}$, following the procedures given in 
Kauffmann et al.\ (2004). For $z=0.05-0.9$,
we obtain a median $M/L_{z'}=1.1$ in the $10^{11}-10^{12}$~M$_\odot$
range and a median $M/L_{z'}=0.7$ in the $10^{10}-10^{11}$~M$_\odot$
range. As expected, since the high-redshift galaxies
are younger and have higher star formation rates, these values 
are about a factor of two lower than the local values. After
correcting the Kauffmann et al.\ (2004) local sample's
Kroupa IMF stellar masses to our Salpeter IMF stellar masses, 
their median values are about 2.5 for the high-luminosity galaxies 
and about 1.1 for the lower luminosity galaxies, which we may
roughly compare with our mass-selected values.

In Figure~\ref{mass_byz} we show the stellar masses of 
the galaxies in our NIR sample over the redshift range
$z=0.05-1.5$. The purple solid lines show the mass
limits given above, to which we expect each of the
samples to be complete. (Note that the purple line
for the mid-$z$ sample has been truncated below 
$z=0.475$ where the sample overlaps with the high-mass 
end of the low-$z$ sample.) We mark with red diamonds the 
sources with $2-8$~keV or $0.5-2$~keV luminosities
above $3\times10^{43}$~ergs~s$^{-1}$, where the AGN
luminosity could contaminate the NIR photometry.
However, the number of such sources is too small to
affect the results.
There are only a small number of very high-mass galaxies
in the sample. Between $z=0$ and $z=1.5$ we find only
13 galaxies with masses above $10^{11.5}$~M$_\odot$ 
(shown by the green dashed line in Figure~\ref{mass_byz}).
As can be seen from the figure,
there is a tendency for the more massive galaxies to lie 
in the high-density filaments in the sample
(Cohen et al.\ 1996, 2000). We will return to this point 
in \S\ref{secenv} when we consider the
environmental dependences of the sample.

%
%
\begin{inlinefigure}
\figurenum{10}
\centerline{\psfig{figure=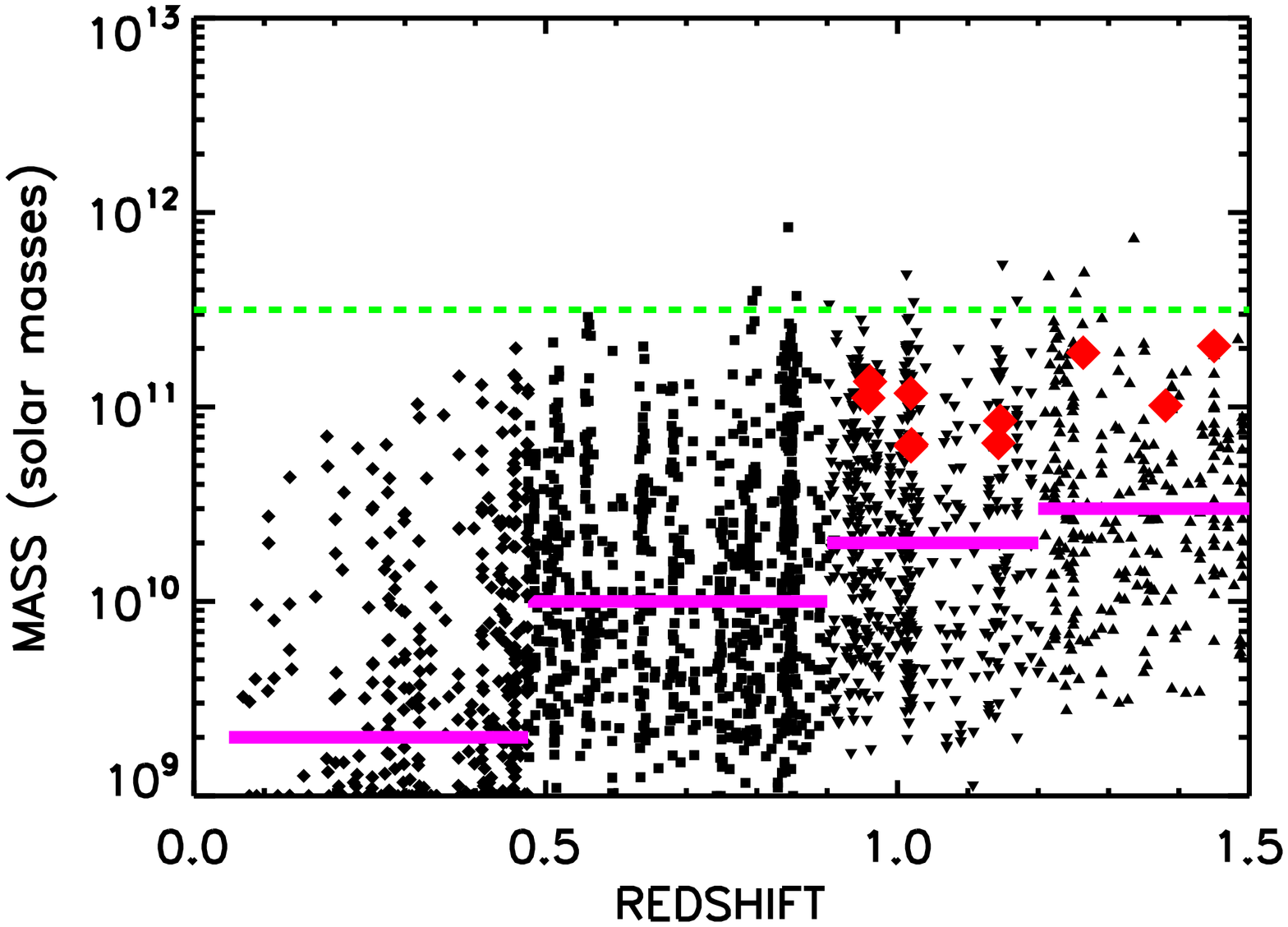,angle=90,width=3.5in}}
\vskip -0.2cm
\figcaption[]{
Stellar mass vs. redshift for
$z=0.05-0.475$ {\em (diamonds)\/},
$z=0.475-0.9$ {\em (squares)\/}, $z=0.9-1.2$
{\em (downward pointing triangles)\/}, and
$z=1.2-1.5$ {\em (upward pointing triangles)\/}.
The purple horizontal lines show the masses above which each
sample (low-$z$, mid-$z$, high-$z$, highest-$z$) is expected to 
be complete. Note that the purple line for the mid-$z$ sample 
has been truncated below $z=0.475$ where that sample overlaps 
with the high-mass end of the low-$z$ sample.
The red diamonds show the small number of X-ray sources with
$2-8$~keV or $0.5-2$~keV luminosities above
$3\times10^{43}$~ergs~s$^{-1}$. The green dashed line
shows a logarithmic mass of 11.5~M$_\odot$, above
which there are relatively few galaxies.
\label{mass_byz}
}
\end{inlinefigure}

\section{Spectral Line Measurements}
\label{secspectral}

\subsection{Equivalent Widths}
\label{secew}

For each spectrum we measured the equivalent widths (EWs) 
of a standard set of lines: \sii~$\lambda\lambda6717, 6731$,
\ha~$\lambda6563$, \nii~$\lambda6584$, \oiii~$\lambda5007$, 
\hb~$\lambda4861$, and \oii~$\lambda3727$. 
For the stronger lines (rest-frame EW$>10$~\AA) 
we used a full Gaussian fit together with a linear fit to the
continuum baseline. For the weaker lines we held the full width
constant using the value measured in the stronger lines, if this
was available, or, if not, then using the nominal width
(i.e., the resolution of the spectrum). For the weaker lines
we also set the central wavelength to the redshifted value. 
We measured the noise as a function of wavelength by 
fitting to random positions in the spectrum and computing 
the dispersion in the results.

In Figure~\ref{ewha}a we show the rest-frame EW(\hb) versus
NIR luminosity for the mid-$z\ (0.05-0.9)$ sample, and in 
Figure~\ref{ewha}b we show the rest-frame EW(\ha) versus 
NIR luminosity for the low-$z\ (0.05-0.475)$ sample.
Both show a strong trend to higher EWs at lower
NIR luminosities (see also Fig.~\ref{ewhb_median}), 
reflecting the higher specific star 
formation rates (the star formation rate per unit galaxy 
stellar mass) in the smaller galaxies. Absorption line 
galaxies {\em (red squares)\/} appear at 
$\gtrsim 10^{43}$~ergs~s$^{-1}$.

Roughly half of the mid-$z$ sample have rest-frame EW(\hb$)>4$~\AA.
At this EW the emission lines may be strong 
enough to make metallicity estimates with line ratios. 
The fraction of galaxies with  a corresponding rest-frame 
EW(\ha$)>12$~\AA\ (assuming the case B Balmer ratio
discussed below) in the low-$z$ sample is closer to 75\%.

%
%
\begin{inlinefigure}
\figurenum{11}
\centerline{\psfig{figure=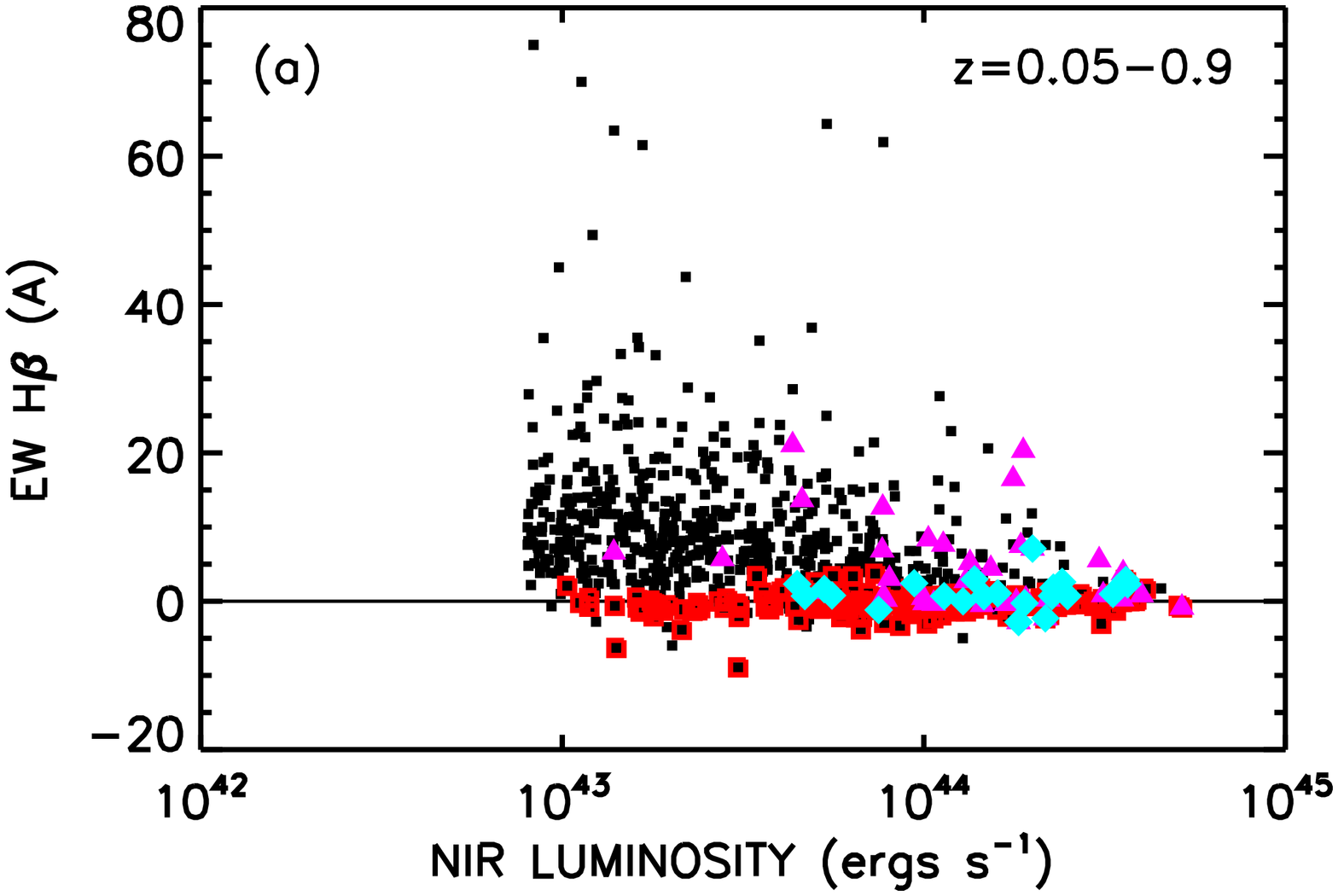,angle=90,width=3.5in}}
\vskip -0.6cm
\centerline{\psfig{figure=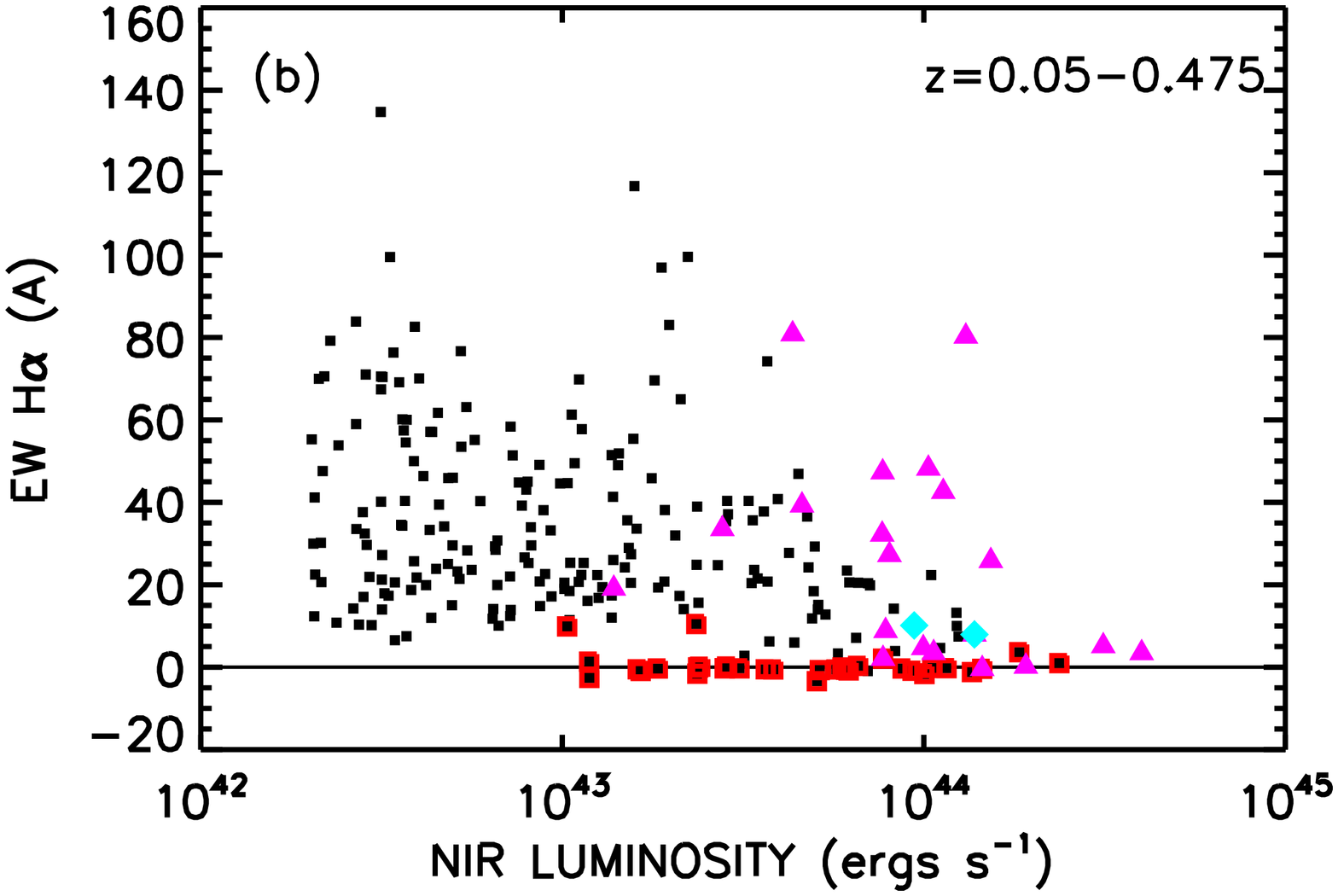,angle=90,width=3.5in}}
\vskip -0.2cm
\figcaption[]{
(a) Rest-frame EW(\hb) vs. NIR luminosity 
for the $z=0.05-0.9$ mid-$z$ sample. (b) Rest-frame EW(\ha) vs. NIR 
luminosity for the $z=0.05-0.475$ low-$z$  sample.
In both panels the cyan diamonds 
show sources containing AGNs based on their X-ray luminosities,
the purple triangles show 20~cm detected sources, 
and the red squares show sources with absorption line redshifts.
\label{ewha}
}
\end{inlinefigure}

%
%
\begin{inlinefigure}
\figurenum{12}
\centerline{\psfig{figure=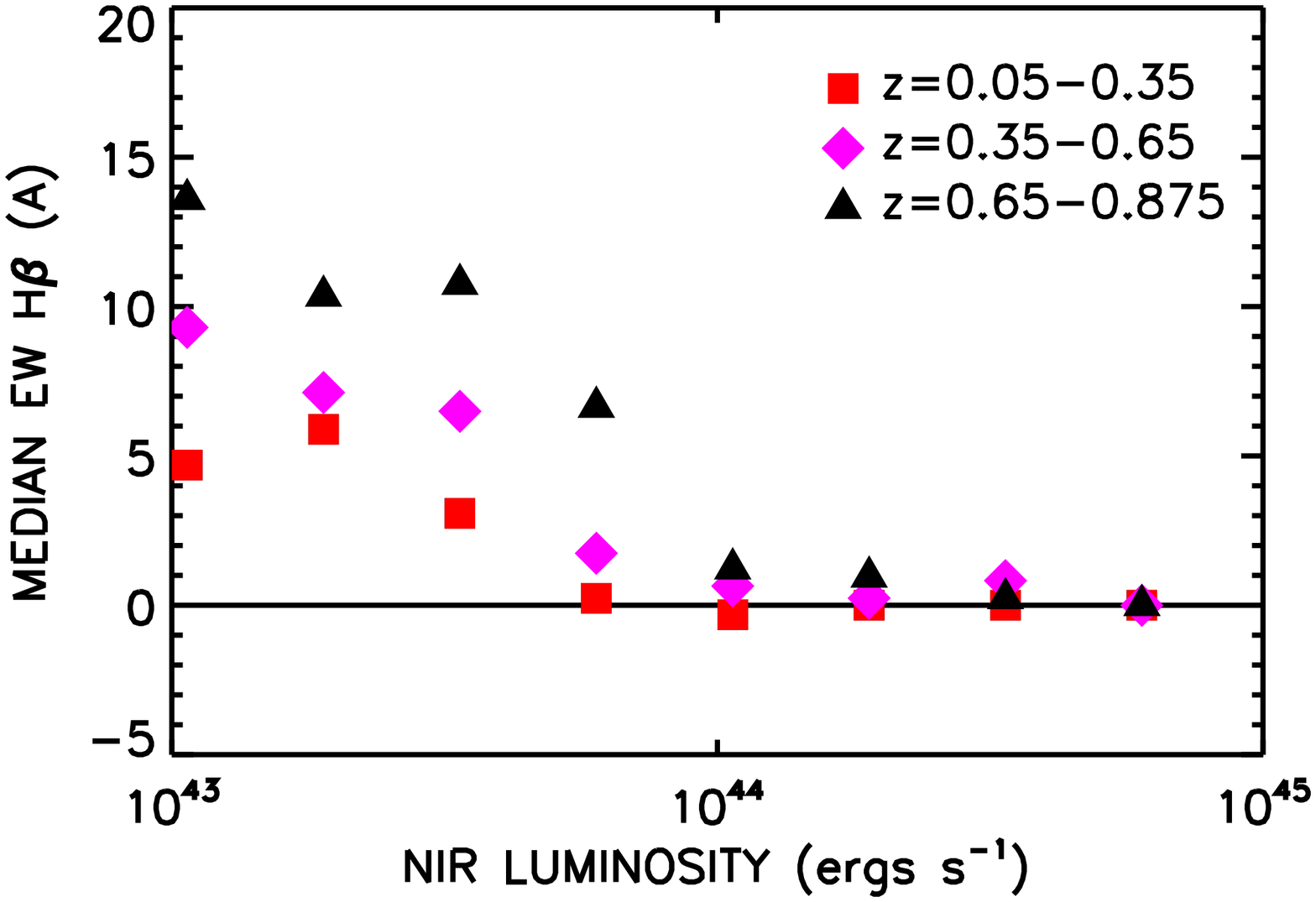,angle=90,width=3.5in}}
\vskip -0.2cm
\figcaption[]{
Median rest-frame EW(\hb) vs.  NIR luminosity for the 
mid-$z$ sample. The symbols correspond to values in the redshift
intervals $z=0.65-0.9$ {\em (black triangles)\/},
$z=0.35-0.65$ {\em (purple diamonds)\/}, and
$z=0.05-0.35$ {\em (red squares)\/}. 
The presence of strong emission lines suitable for 
measuring metallicities is highly dependent on both
NIR luminosity and redshift. 
\label{ewhb_median}
}
\end{inlinefigure}

We also show in Figures~\ref{ewha}a and \ref{ewha}b 
the sources that contain AGNs 
based on their X-ray luminosities {\em (cyan diamonds)\/} 
and the 20~cm detected sources {\em (purple triangles)\/}.
While the 20~cm sources often have quite strong
Balmer emission lines, consistent with them being high-end star 
formers, the AGNs generally have weak Balmer emission lines.
The weakness of the AGN Balmer lines results in most of
them being automatically excluded from our
metallicity analysis, but we also use the X-ray signatures
to identify and exclude any remaining AGNs.

Not only is the presence of strong emission lines a strong 
function of NIR luminosity, but because of 
the decrease in the overall star formation rates with
decreasing redshift, it is also
a strong function of redshift. We show this in
Figure~\ref{ewhb_median}, where we plot the median rest-frame 
EW(\hb) for the mid-$z$ sample versus NIR luminosity for 
three redshift intervals spanning the full redshift range. 
The rapid drop in the EWs with decreasing redshift can be clearly
seen. The change in the emission line strengths with redshift
introduces a strong selection bias when comparing emission
line metallicities in galaxies at different redshifts.
This is an important point, which we will return to in 
\S\ref{secabs}.

The EW(\hb) is reduced by the effects of the underlying 
stellar absorption, and we must correct for this
effect. The simplest procedure is to apply a single offset. 
Kobulnicky \& Phillips (2003) found an offset of $-2$~\AA\
by measuring several Balmer lines in 22 galaxy spectra 
(referred to as the K92+ sample in their paper) and obtaining 
a self-consistent reddening and stellar absorption solution 
for each galaxy. 
However, the correction in our data is smaller than 
this, possibly because of the fitting methods we used. 
In particular, the Gaussian fits to the emission lines are
narrower than the absorption lines, and it is only the
absorption integrated through the Gaussian which we need 
to correct.

We have estimated the correction as follows. First we averaged
the normalized spectra of galaxies in the low-$z$ sample 
grouped by EW(\ha). We show a few of these averaged 
spectra in Figure~\ref{composite} [those with 
EW(\ha$)<5$~\AA, EW(\ha$)=5-15$~\AA, EW(\ha$)=15-30$~\AA, 
and EW(\ha$)=30-60$~\AA], where the bottom spectrum 
[EW(\ha$)<5$~\AA] is the averaged spectrum of the absorption 
line galaxies in the low-$z$ sample (i.e., the absence of \ha\ 
emission guarantees that there is little \hb\ emission). 
We then renormalized the averaged spectrum of the absorption 
line galaxies to match each of the other averaged spectra in 
wavelength regions outside the emission lines. These fits are 
shown in Figure~\ref{composite} in red. The renormalized 
absorption spectrum was then subtracted to form a corrected 
spectrum. Finally, we measured the EW(\hb) in the corrected 
and uncorrected spectra. 
In Figure~\ref{ewdiff} we plot the difference of these two
measurements versus the EW(\ha).
We see no strong dependence of the \hb\ correction on
the galaxy type. Thus, we adopt a fixed offset of 1~\AA\ to
correct for the stellar absorption, regardless of galaxy type.
(Our results are not significantly changed if we use a
2~\AA\ rather than a 1~\AA\ correction.)
In our metals analysis we will restrict to galaxies with
corrected EW(\hb$)>4$~\AA\ to minimize the systematic
uncertainties introduced by this procedure. Hereafter,
we refer to the corrected EW(\hb) as EW(\hb).

%
%
\begin{inlinefigure}
\figurenum{13}
\centerline{\psfig{figure=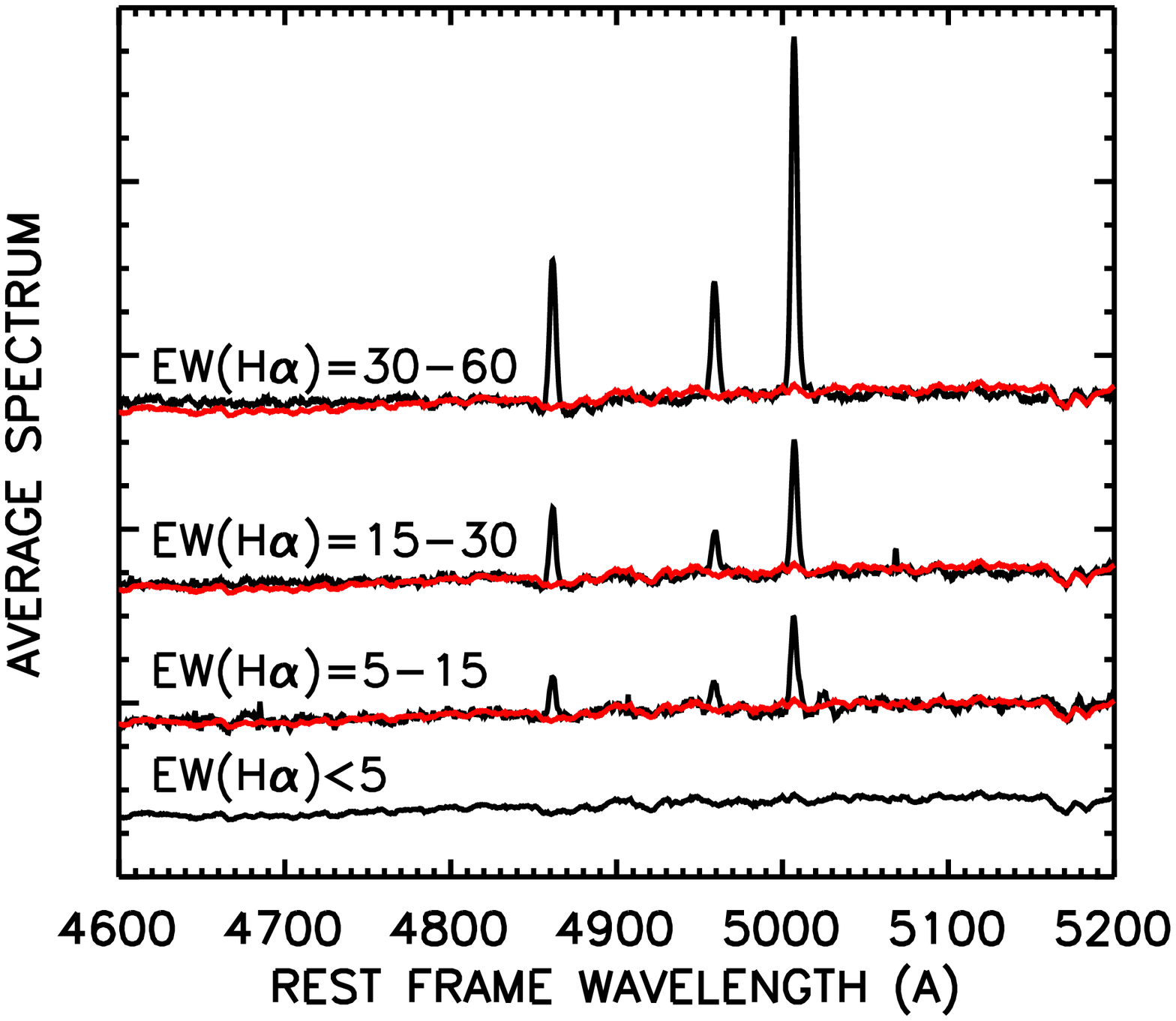,angle=90,width=3.5in}}
\vskip -0.2cm
\figcaption[]{
Averages of the normalized spectra vs. 
rest-frame wavelength for galaxies in the low-$z$ sample 
{\em (black spectra)\/}. The bottom
spectrum shows the averaged spectrum for sources with
rest-frame EW(\ha$)<5$~\AA. This defines an absorption 
line spectrum. The remaining spectra from bottom to top 
correspond to rest-frame EW(\ha$)=5-15$, $15-30$, 
and $30-60$~\AA, respectively. Each spectrum is offset 
in the vertical direction to separate them.  
We also show in red the absorption line spectrum normalized to
each of the other averaged spectra, which we used to remove the 
underlying stellar absorption. 
\label{composite}
}
\end{inlinefigure}

%
%
\begin{inlinefigure}
\figurenum{14}
\centerline{\psfig{figure=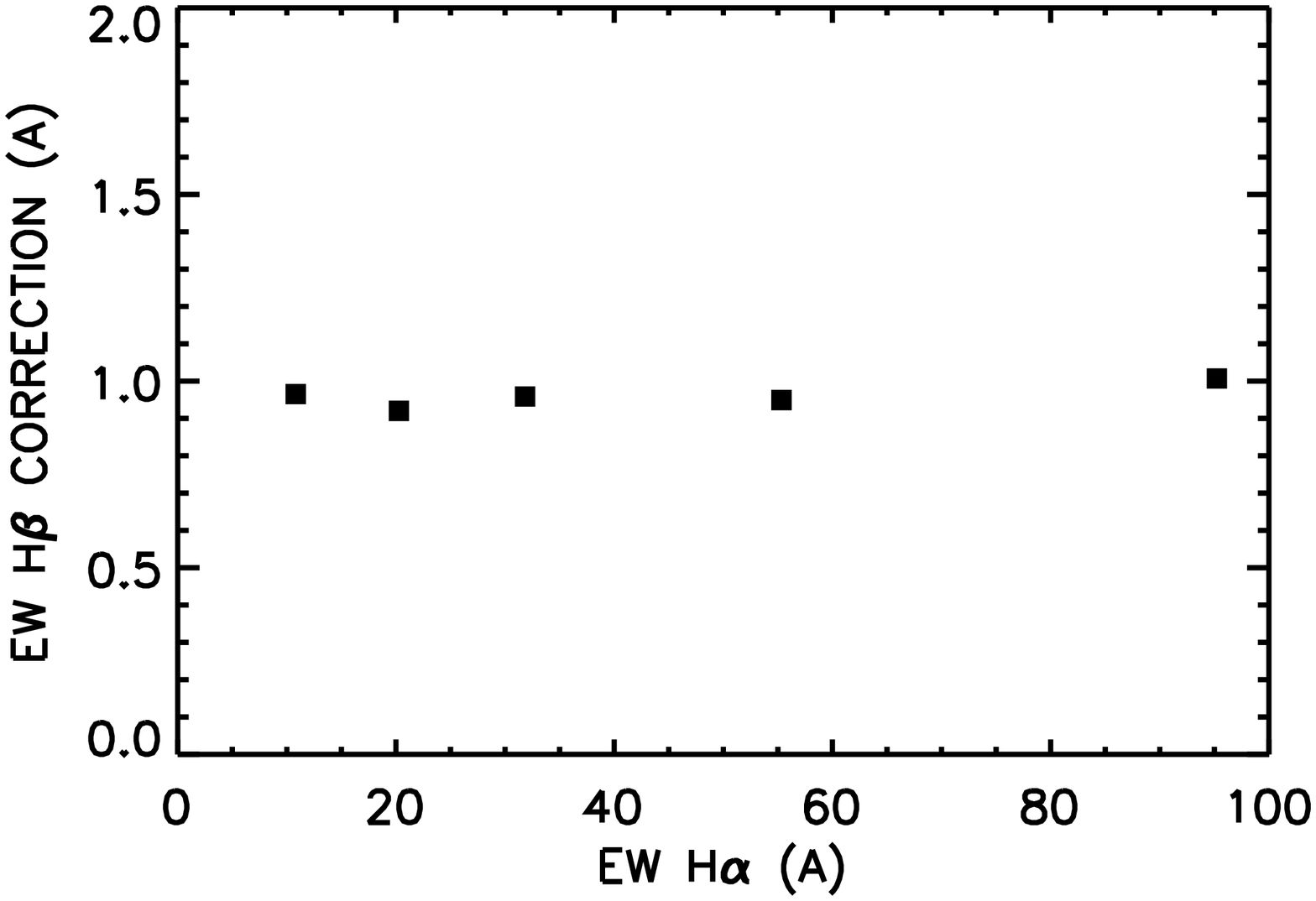,angle=90,width=3.5in}}
\vskip -0.2cm
\figcaption[]{
Measured correction to the rest-frame EW(\hb) vs. the 
rest-frame EW(\ha).
\label{ewdiff}
}
\end{inlinefigure}

\subsection{Line fluxes}
\label{secflux}

Generally the spectra were not obtained at the parallactic 
angle, since this is determined by the DEIMOS mask orientation.
Nor were spectrophotometric standards regularly observed.
Thus, flux calibration is difficult and special care must 
be taken in determining fluxes. Relative line fluxes can be 
measured from the spectra without flux calibration, as long as 
we restrict the line measurements to the short
wavelength range where the DEIMOS response is essentially
constant. For example, one can assume the responses of
neighboring lines (e.g., \oiii$~\lambda$4949 and
\oiii$~\lambda$5007) are the same and then measure
the flux ratio without calibration. However, to measure
quantities such as the R23 metal diagnostic ratio (see \S\ref{secr23}
for definition)
and to estimate the extinction and ionization parameters
in the galaxies, we must calibrate
the line fluxes over much wider wavelength ranges. 
We do this by using the broadband fluxes to calibrate the 
local continuum. We then use the equivalent width to compute
the line flux. This method should work well as long as
the sky subtraction is precise and the spectral continuum
level well determined.

In addition to the line fluxes, we also measured the 
4000~\AA\ break strength in the galaxies. To do this
we measured the ratio of the average flux 
(in units of ergs~cm$^{-2}$~s$^{-1}$~Hz$^{-1}$) at 
$4050-4250$~\AA\ to that at $3750-3950$~\AA\ using
the extinction corrected spectra normalized 
by the instrument throughput, which was measured by 
R.~Schiavon \\(\verb+http://www.ucolick.org/~ripisc/results.html+).
Given the short wavelength span,
this ratio should be fairly insensitive to the flux 
calibration and extinction, but it could be affected 
by the accuracy of the absolute sky subtraction.

We can use the calibrated spectra to search
for any relative offsets in the determination of the
zero points in the imaging data, and inversely 
we can test the spectral shapes by comparing
with the photometric colors from the imaging
data. We carried out these tests by computing
the ratio of the rest-frame $B$ to $U$ bands from
the spectra and comparing this with the photometrically 
determined values for galaxies with masses
$>10^{10}$~M$_\odot$. Over this
wide wavelength range we found that the measured offsets
between the photometric and spectroscopic measurements
showed no dependence on redshift over the $z=0-1.1$ 
range where the values can be measured (see Figure~\ref{b4000_test}).
(Beyond $z=1.1$ the 4500\AA\ band moves above the upper wavelength
limit of the spectra.)
This shows that there are no relative
errors in the photometric calibration of the UV and optical
data. There is a only a small systematic offset throughout
which averages to $-0.05$~mag. 
Translated to the smaller wavelength range
used in the 4000~\AA\ break measurement, we will underestimate
the break strength by a multiplicative factor of 1.02
on average. The measured spread in the offsets translates
to a $1\sigma$ multiplicative error of 1.07 in the individual
4000~\AA\ break strengths.

%
%
\begin{inlinefigure}
\figurenum{15}
\centerline{\psfig{figure=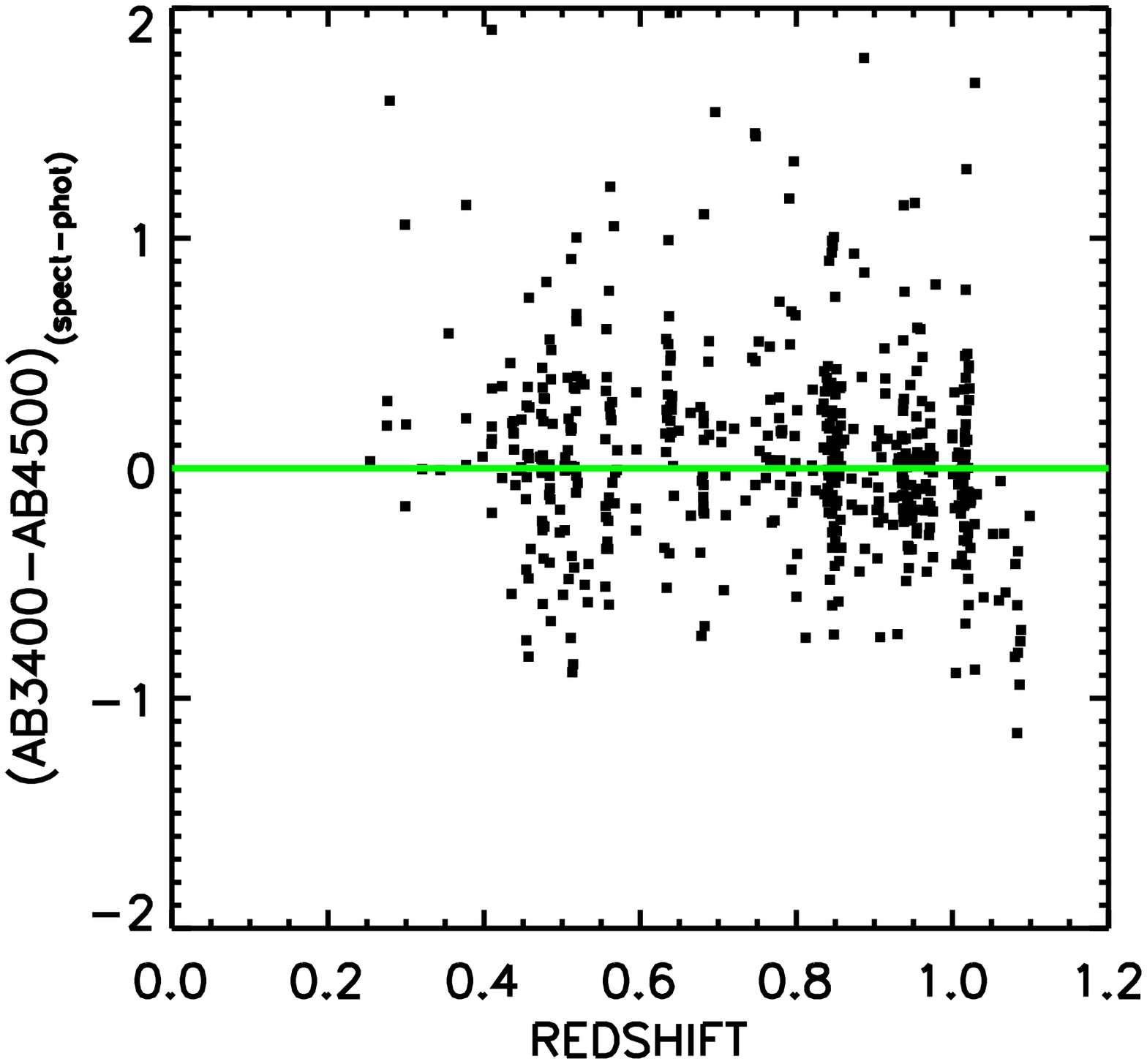,angle=90,width=3.5in}}
\vskip -0.2cm
\figcaption[]{
Magnitude offset between the rest-frame $3400-4500$~\AA\ 
magnitudes measured from the spectra and those measured from the 
photometry vs. redshift for $z=0-1.1$ galaxies with masses 
$>10^{10}~$M$_\odot$ {\em (black squares)\/}.
\label{b4000_test}
}
\end{inlinefigure}

\section{Continuum versus Line Extinction}
\label{secrelext}

Ideally we would measure the extinctions for the line
fluxes from the Balmer ratio $f($\ha$)/f($\hb$)$, since the H~II 
regions producing the emission lines may have different reddening
from the stars providing the continuum. In particular,
we might expect reddening in the H~II region to be systematically
higher than reddening in the continuum, if the H~II regions
lie in regions of higher gas density (Kinney et al.\ 1994).
However, in the present sample we can only measure the Balmer 
ratio extinction in the low-$z$ sample. 
While our analysis of the star formation and stellar mass density
histories can be done without reference to the line fluxes, 
we do require an estimate of the line extinction in computing 
the metallicity diagnostics and in comparing equivalent
widths with galaxy models.

\begin{figure*}
\figurenum{16}
\centerline{\psfig{figure=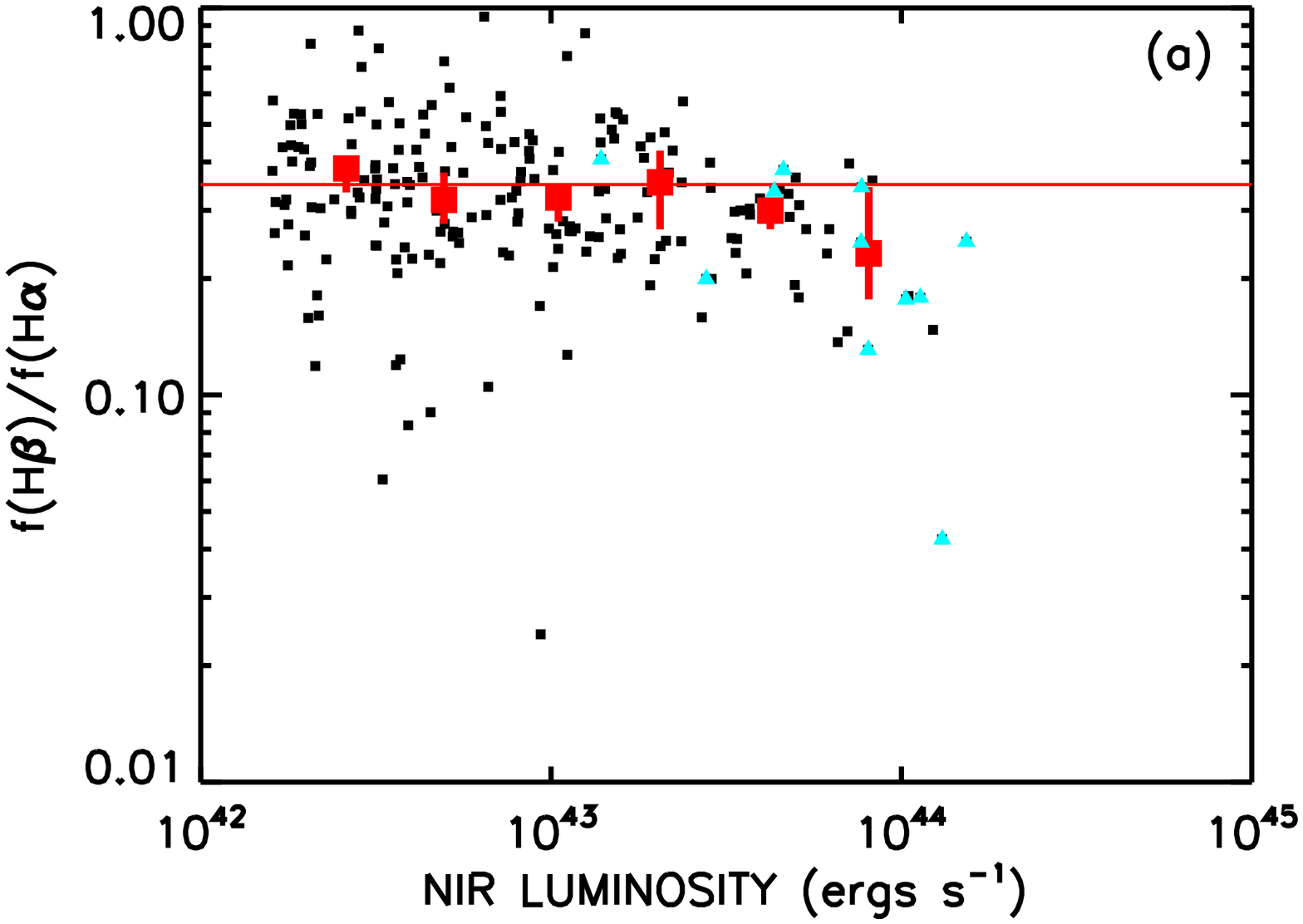,angle=90,width=3.5in}
\psfig{figure=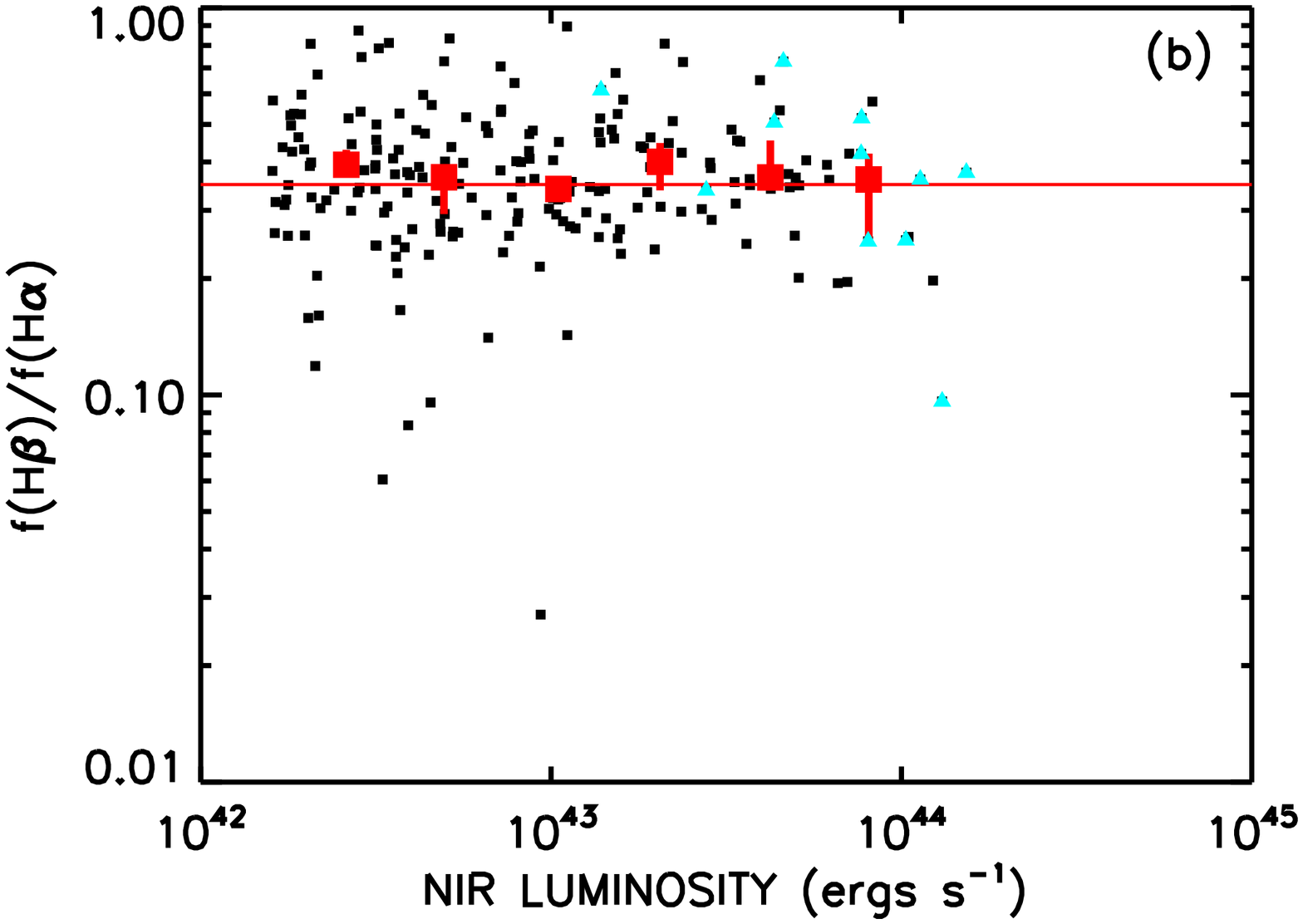,angle=90,width=3.5in}}
\centerline{\psfig{figure=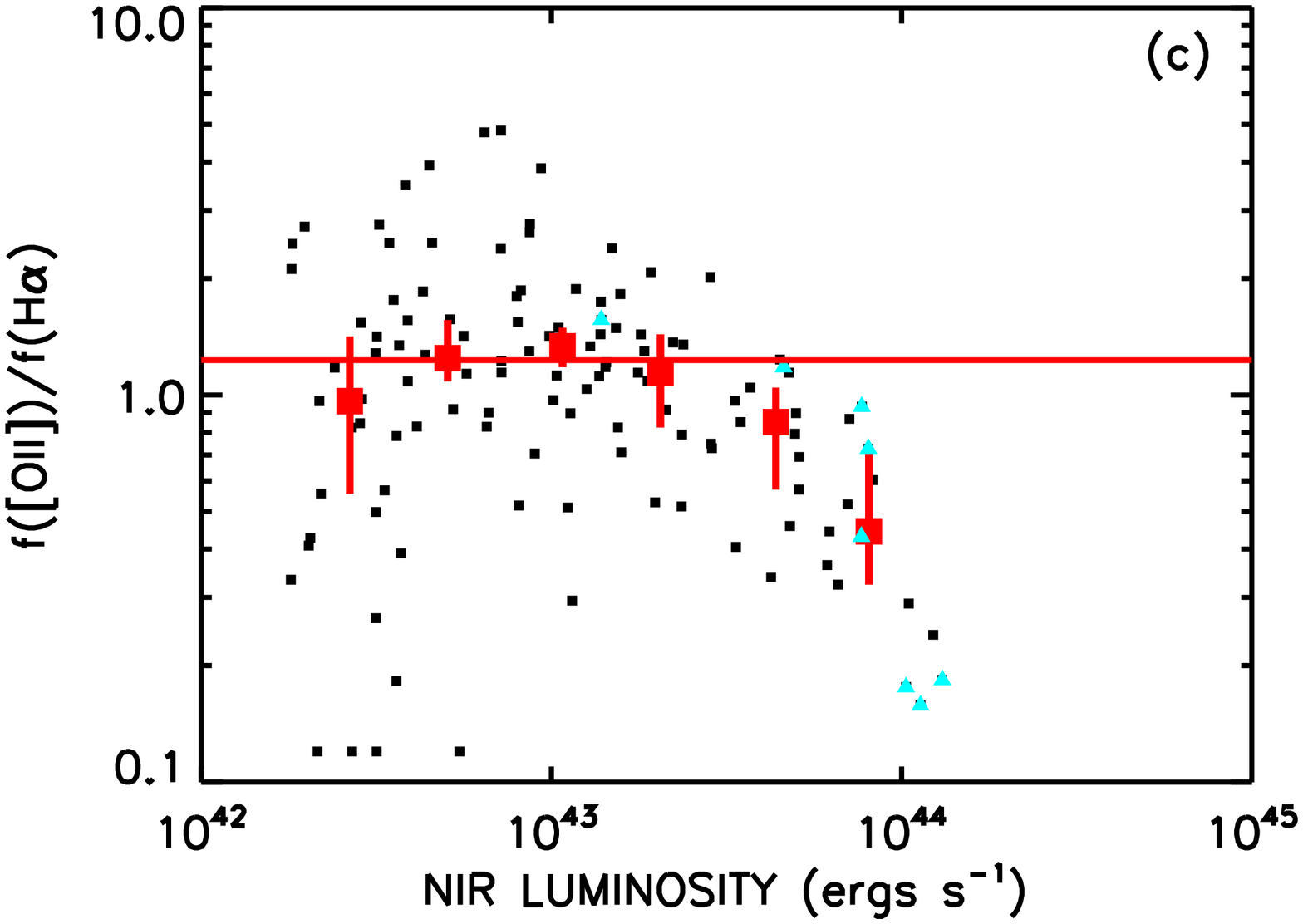,angle=90,width=3.5in}
\psfig{figure=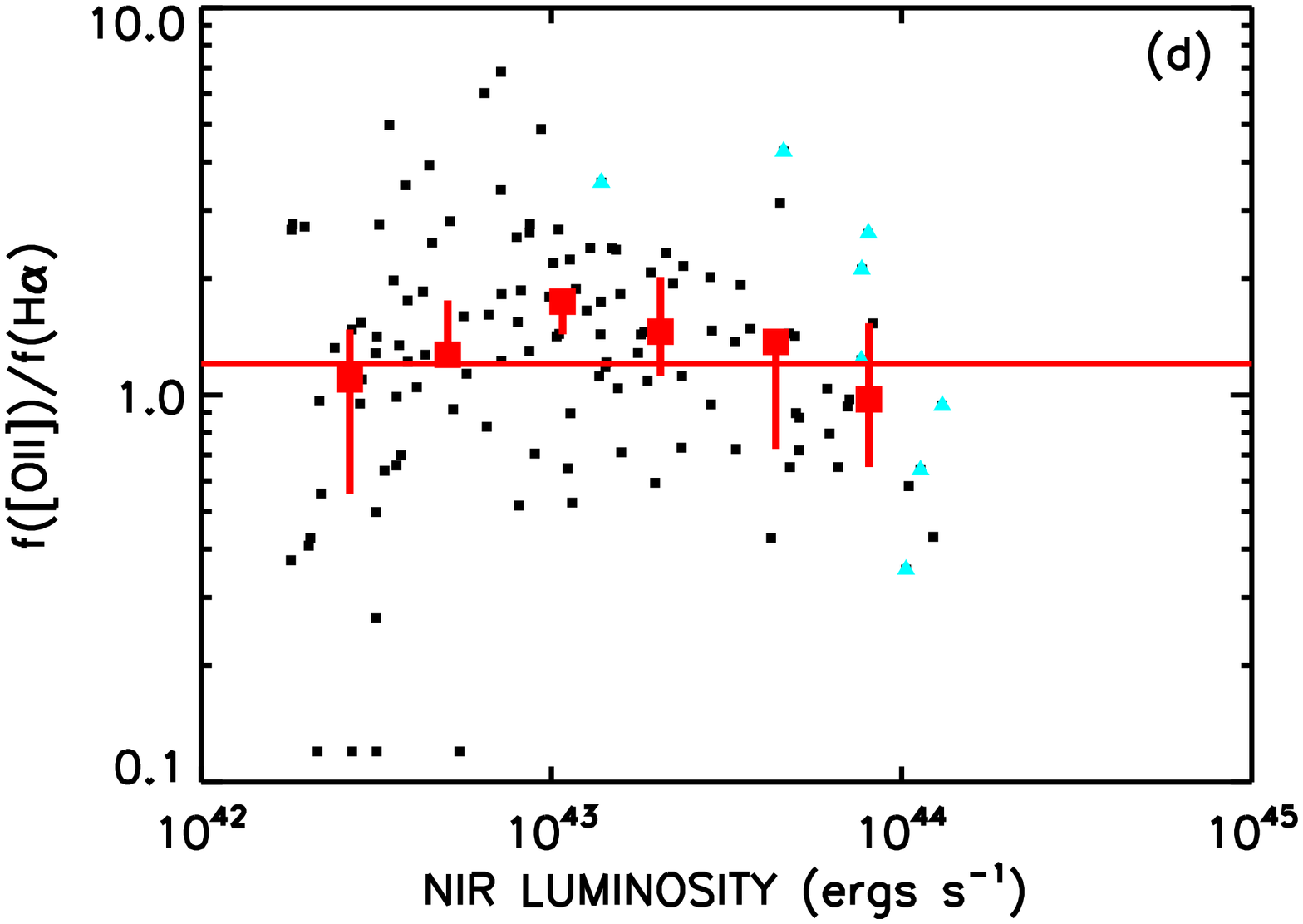,angle=90,width=3.5in}}
\centerline{\psfig{figure=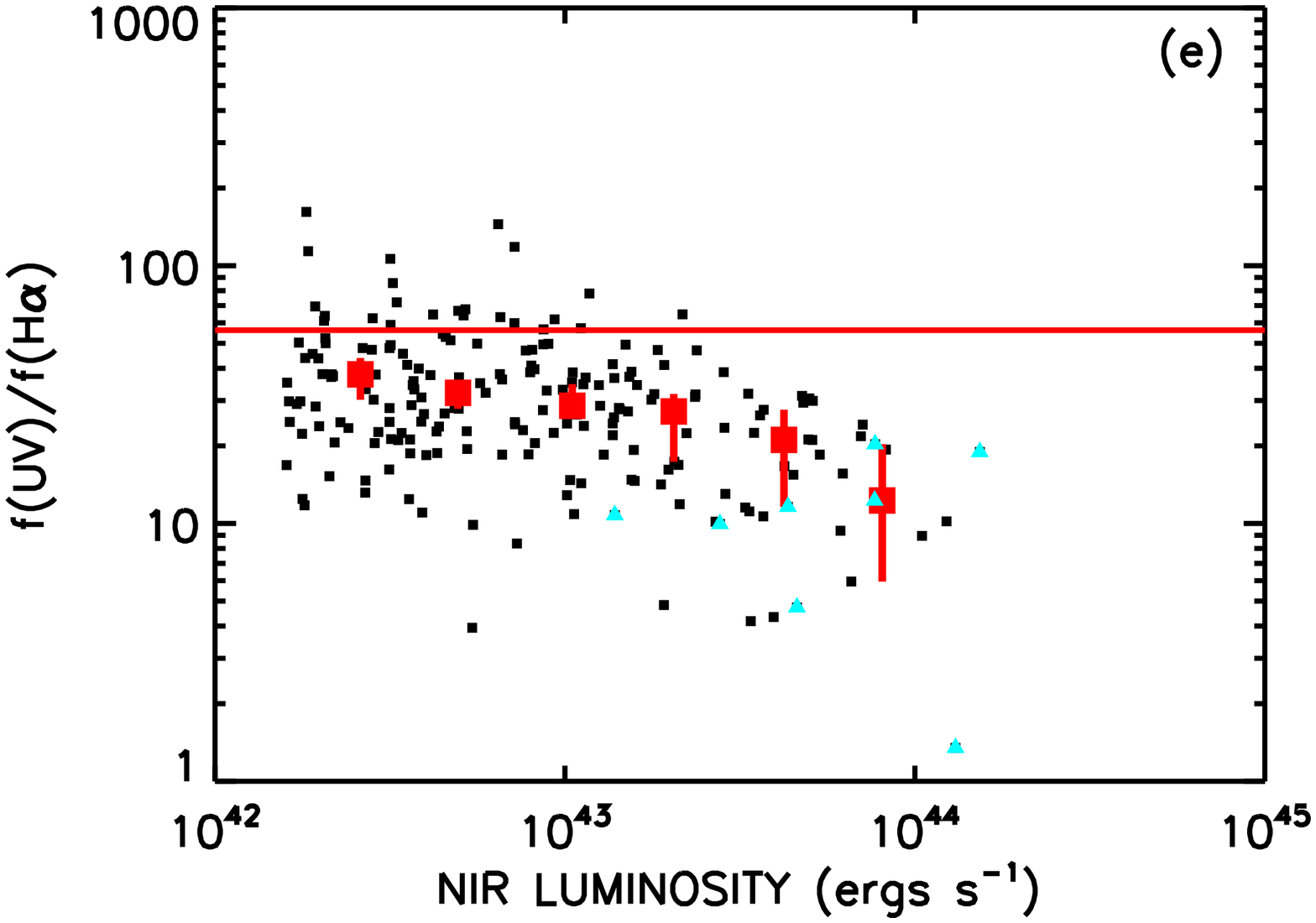,angle=90,width=3.5in}
\psfig{figure=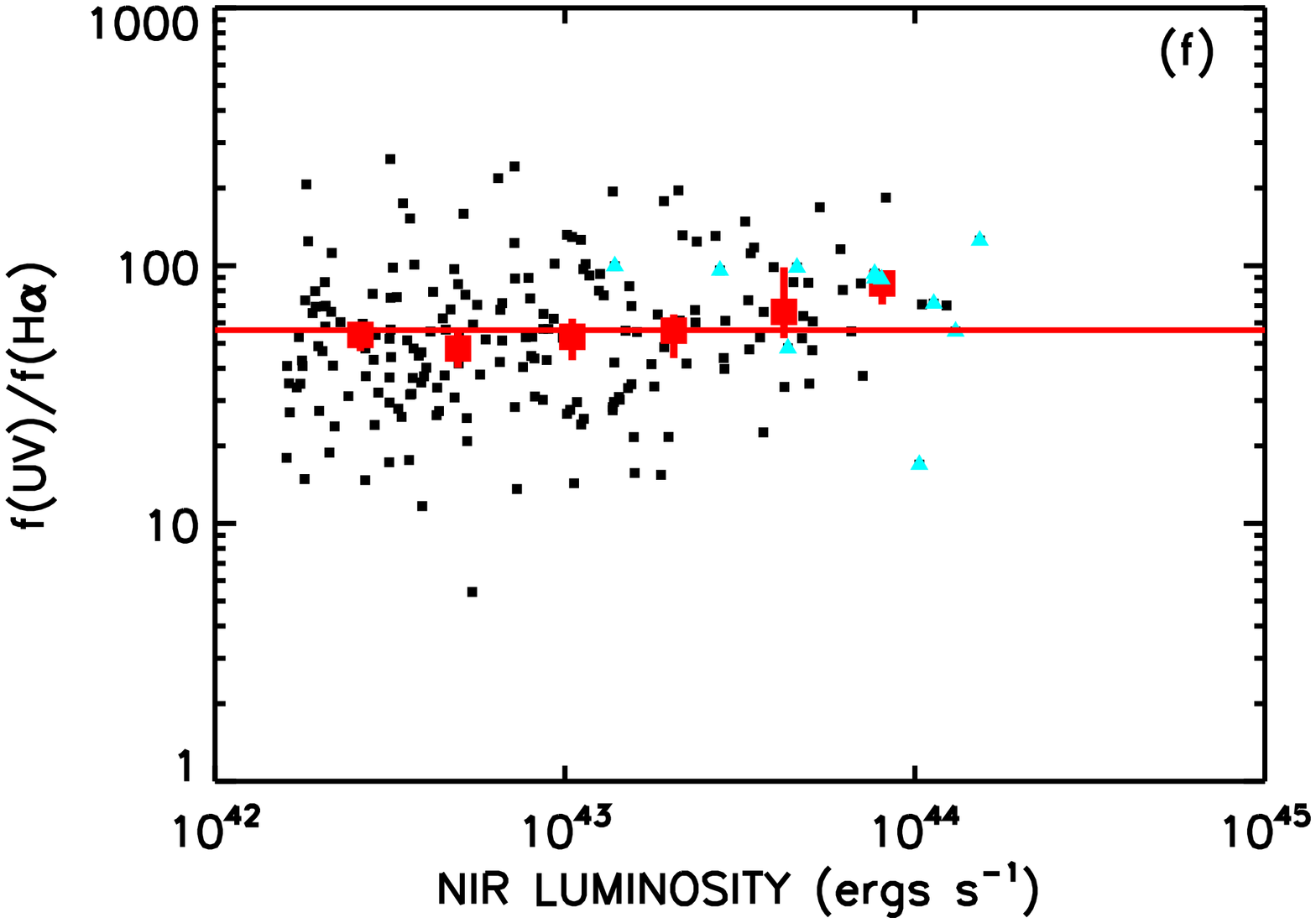,angle=90,width=3.5in}}
\figcaption[]{
In all panels we consider only galaxies in the low-$z$
sample with rest-frame EW(\ha$)>12$~\AA. The large red squares 
denote the median values at each luminosity with $\pm1\sigma$ 
errors computed using the median sign method. 
The 20~cm sources are denoted by cyan triangles.
All ratios are plotted vs. NIR luminosity.
(a) Balmer ratio. (b) Extinction corrected Balmer ratio.
In these two panels the case B ratio in the absence
of extinction is shown by the red horizontal line.
(c) $f($\oii$~\lambda3727)/f($\ha).
(d) Extinction corrected $f($\oii$~\lambda3727)/f($\ha).
In these two panels the $f($\oii$~\lambda3727)/f($\ha) ratio 
used to compute the calibration of the \oii$~\lambda3727$ 
determined SFR is shown by the red horizontal 
line. (e) $f($UV$)/f($\ha) using the rest-frame UV flux
defined in the text.
(f) Extinction corrected $f($UV$)/f($\ha).
In these two panels the $f($UV$)/f($\ha) ratio 
used to compute the calibration of the UV determined SFR
is shown by the red horizontal line.
\label{balmer}
}
\end{figure*}

We therefore checked how well the continuum extinctions derived
from the BC03 fits match to the extinctions derived from
the Balmer ratios in the low-$z$ sample. In
Figure~\ref{balmer}a we show $f$(\hb)$/f$(\ha) versus 
NIR luminosity for the low-$z$ sample 
with rest-frame EW(\ha$)>12$~\AA. 
At low NIR luminosities the median values {\em (red squares)\/}
are very close to an intrinsic $f$(\hb)$/f$(\ha) ratio of
0.35 (the ratio for case B recombination at 
$T=10^4$~K and $n_e\sim 10^2-10^4$~cm$^{-3}$, Osterbrock 1989;
{\em red solid line\/}),
suggesting that there is little extinction. Only in the 
highest luminosity galaxies (NIR luminosities 
$>10^{44}$~ergs~s$^{-1}$) does the ratio fall significantly 
below the case B ratio. The highest NIR luminosity galaxies
pick out the 20~cm detected sources, which are shown with
the cyan triangles. 

In Figure~\ref{balmer}b we show the Balmer ratio after 
applying the extinction corrections derived from our BC03 fits.  
As can be seen from the median values {\em (red squares)\/},
this completely removes the dependence on NIR luminosity.
Thus, the SED derived extinctions can be used to
correct the average line ratios. However, the individual points 
still scatter significantly about the median, implying 
that there are systematic uncertainties in the flux 
determinations.

%
%

We next estimated the extinctions from the Balmer ratios
for the low-$z$ sample with rest-frame EW(\ha$)>12$~\AA\
and made a linear fit against the continuum extinctions.
This gives the relation
\begin{equation}
E(B-V)_{\rm spectral} = 0.0 + (0.87\pm0.18) E(B-V)_{\rm BC03} \,.
\label{eqnebv}
\end{equation}
This suggests that the line extinction is, if anything, smaller 
than the continuum extinction, but that, within the errors, the
two extinction measurements are consistent on average. We
will therefore use the BC03 extinctions to deredden the line
fluxes at higher redshifts. However, we will regularly check
to make sure that this assumption is providing consistent results.

\section{Star Formation Rates}
\label{secsfr}

\subsection{Relative Calibrations}
\label{seccal}

Extensive discussions of the calibrations of the
star formation diagnostics can be found throughout
the recent literature (e.g., Kennicutt 1998;
Rosa-Gonz{\'a}lez et al.\ 2002; Hopkins et al.\ 2003;
Kewley et al.\ 2004; Moustakas et al.\ 2006).
However, for consistency, it is best to internally
calibrate data sets where possible. We can
do this using our low-$z$ sample.

In Figures~\ref{balmer}c and \ref{balmer}e we show, 
respectively, $f$(\oii)$/f$(\ha) and $f$(UV)$/f$(\ha) 
versus NIR luminosity for the low-$z$
sample with rest-frame EW(\ha$)>12$~\AA. 
Here we have defined the rest-frame UV luminosity as the
integral over the frequency range $10^{15}$ to 
$1.5\times10^{15}$~Hz (which corresponds to the wavelength
range 1993 to 2990~\AA) minus 0.03 of the NIR luminosity. 
The latter correction, the value of which was determined
from absorption line galaxies in the sample,
removes the contribution from the older stars in the galaxy.
Since our shortest measured wavelength
is the $U$-band centered at 3700~\AA,
determining the UV luminosity requires an extrapolation 
for the lowest redshift sources. 

Similar to what we saw for the Balmer ratio, there is 
substantial extinction only in the highest NIR luminosity 
galaxies. In Figures~\ref{balmer}d and \ref{balmer}f we 
plot the same ratios shown, respectively, in 
Figures~\ref{balmer}c and \ref{balmer}e, but this time 
after applying our BC03 extinction corrections. Again the 
medians {\em (red squares)\/} are flattened out by the 
extinction corrections.

To obtain the SFR calibrations and to quantify the scatter 
in the flux ratios, in Figure~\ref{calibrate} we show, 
respectively, the distribution of the extinction corrected 
logarithmic (a) Balmer ratios, (b) $f$(\oii)$/f$(\ha) ratios, 
and (c) $f$(UV)$/f$(\ha) ratios {\em (black histograms)\/}, 
together with Gaussian fits to the distributions
{\em (purple dashed curves)\/}. 
The mean value of the extinction corrected Balmer
ratios agrees precisely with the case B ratio
{\em (red line in Fig.~\ref{calibrate}a)\/},
illustrating how well the extinction corrections work.
However, there is a symmetric scatter 
with a $1\sigma$ dispersion of 0.13~dex
in the Balmer ratio about the case B value. We take this
as a rough measure of the systematic and statistical
errors in the extinction corrected flux determinations
at the \hb\ wavelength relative to \ha.

Regarding the $f$(\oii)$/f$(\ha) ratio, it has been shown 
that there is a dependence of this ratio on the oxygen 
abundance (e.g., Kewley et al.\ 2004; Mouhcine et al.\ 2005; 
Moustakas et al.\ 2006). In fact, it can be seen 
from Figures~\ref{calibrate}b and \ref{calibrate}c
that the UV flux has a somewhat 
tighter relation to the \ha\ flux than the \oii\ flux does, 
despite both the extrapolation made to shorter wavelengths 
for the lowest redshift sources to determine the UV flux
and the larger wavelength 
separation of the UV from \ha\ than of \oii\ from \ha\ 
(which means any uncertainties in the extinction corrections 
would have a larger effect). 
In addition, there are no significant outliers in the 
$f$(UV)$/f$(\ha) ratios, unlike the case for the 
$f$(\oii)$/f$(\ha) ratios.

We can now use the flux ratios to determine
the calibrations to star formation rates (SFRs).
We use as our primary calibration the conversion of $L($UV)
to SFR from the BC03 models. 
The calibration for any individual galaxy depends on the 
galaxy star formation history (SFH), but the ensemble average 
is well determined since it is simply the amount of UV light 
produced per unit mass of stars. Since our principal goal 
is to compute the universal SFH, this 
is the appropriate quantity to choose. For our definition 
of the rest-frame UV luminosity given above,
\begin{equation}
\log {\rm SFR} = -42.63 + \log L({\rm UV})  \,.
\end{equation}
Here the SFR is in M$_\odot$ per year and is computed
for the Salpeter IMF used throughout this paper.
$L$(UV) denotes the intrinsic (i.e., corrected for
extinction) rest-frame UV luminosity and is in units of 
ergs~s$^{-1}$. Our UV calibration is slightly
lower than the value of $-42.55$ that would be obtained
from the commonly used relation given by Kennicutt (1998).
The Kennicutt relation is appropriate for an individual 
galaxy undergoing constant star formation, which gives
a lower UV flux per unit SFR.

%
%
\begin{inlinefigure}
\figurenum{17}
\centerline{\psfig{figure=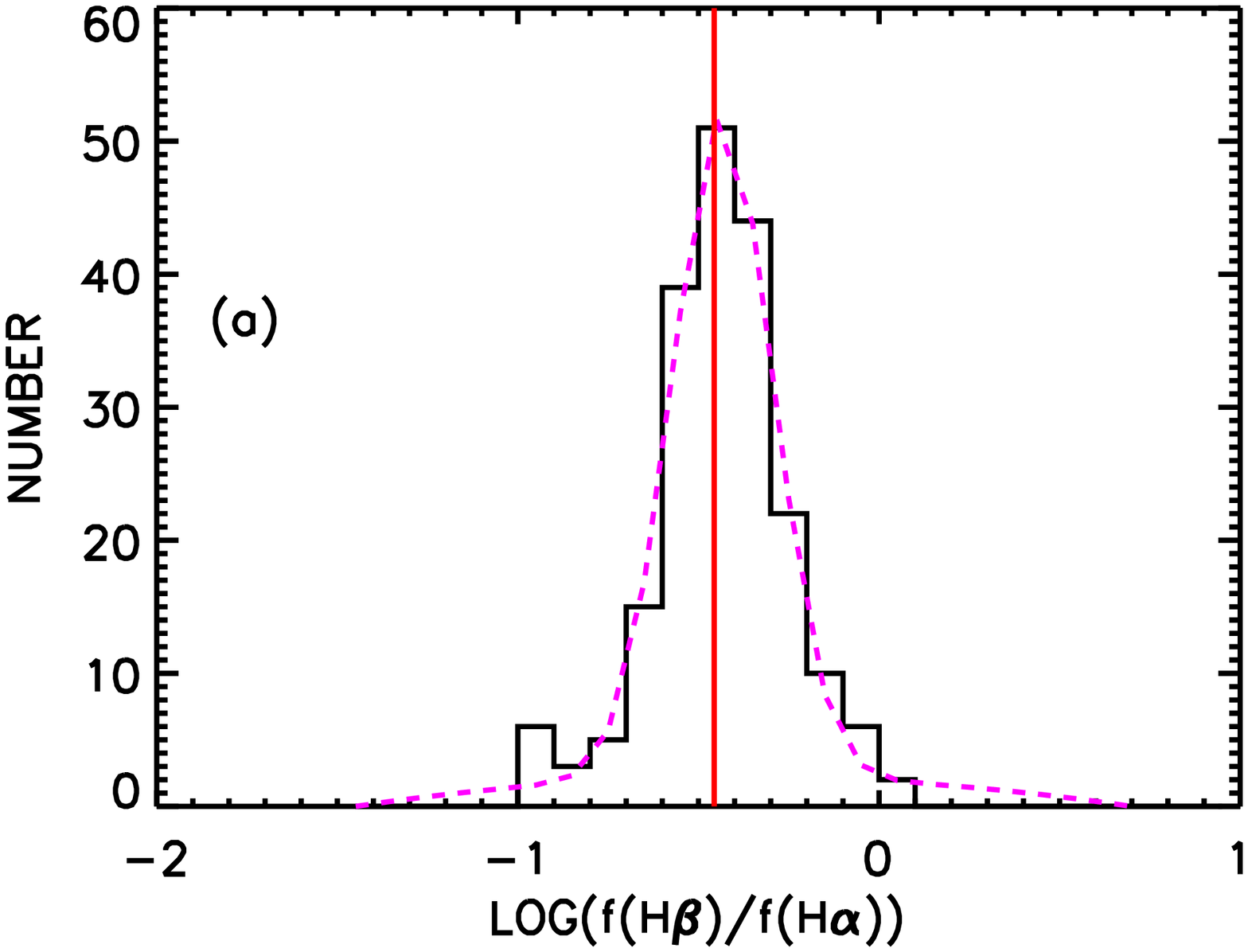,angle=90,width=3.5in}}
\vskip -0.6cm
\centerline{\psfig{figure=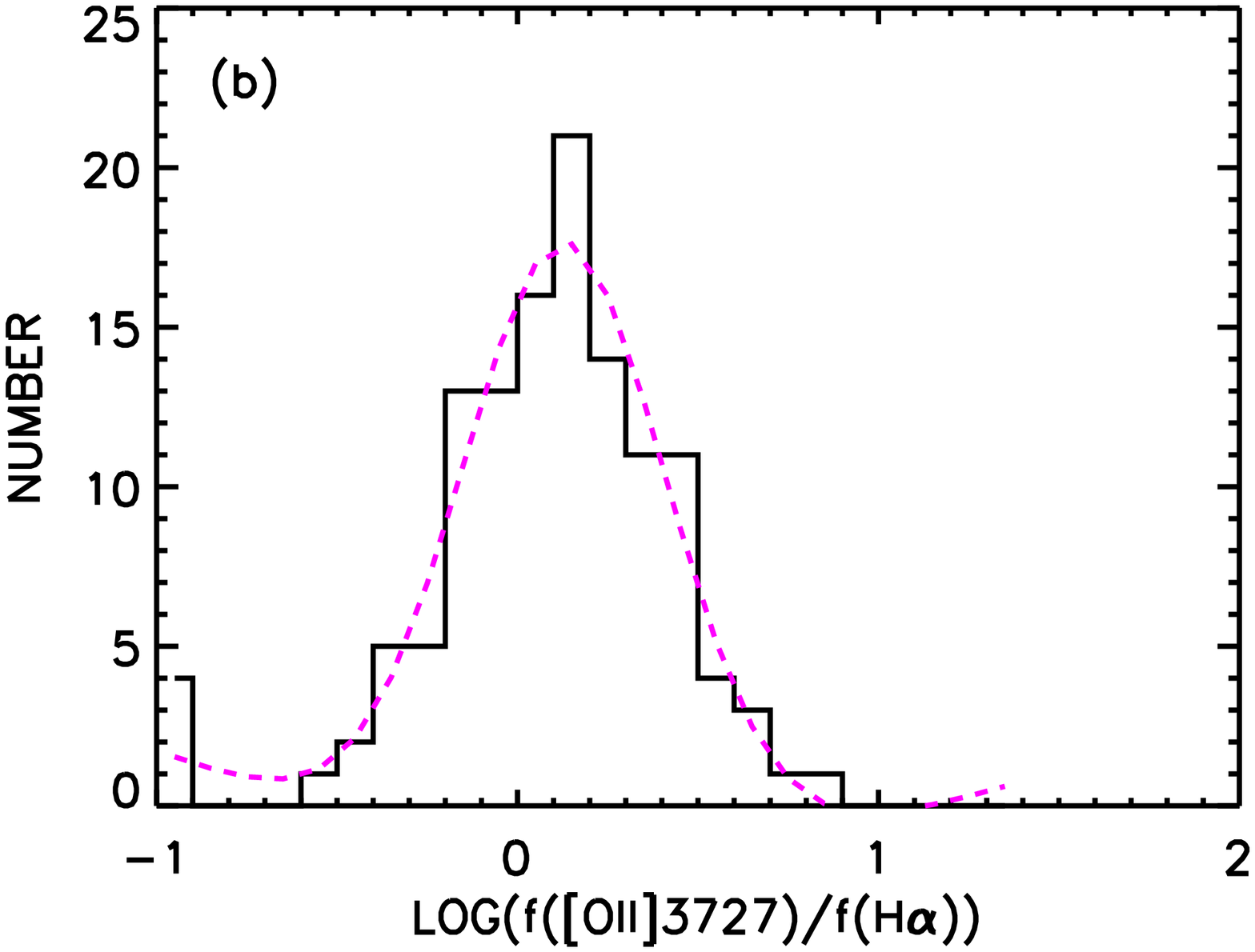,angle=90,width=3.5in}}
\vskip -0.6cm
\centerline{\psfig{figure=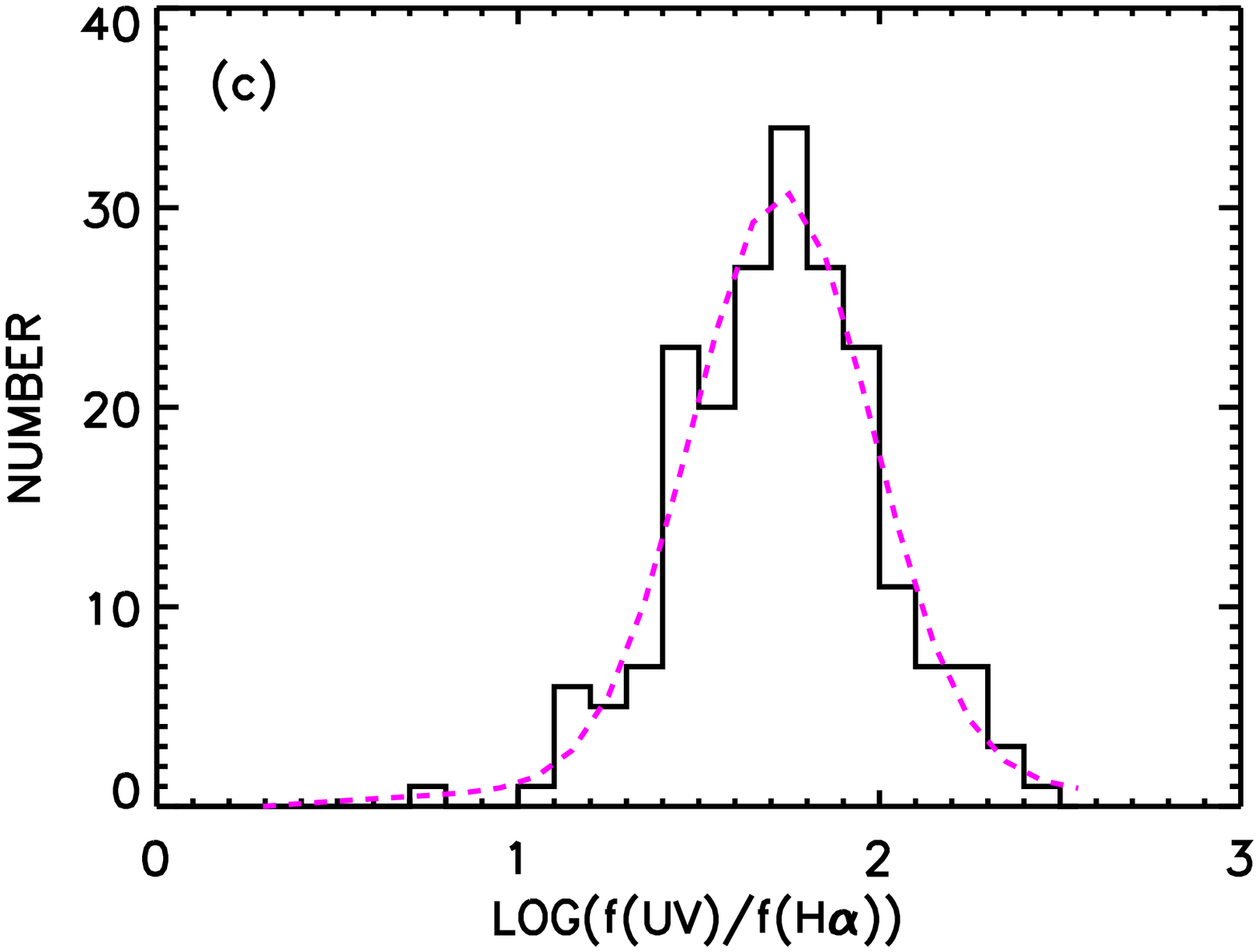,angle=90,width=3.5in}}
\vskip -0.2cm
\figcaption[]{
Distribution of the logarithm of the
extinction corrected ratio of (a) \hb$~\lambda4861$,
(b) \oii$~\lambda3727$, and (c) rest-frame UV flux as defined
in the text to the \ha$~\lambda6563$ flux for the low-$z$ sample
with rest-frame EW(\ha$)>12$~\AA\ {\em (black histograms)\/}. 
In each case the Gaussian fit is shown by the purple dashed curve.
In (a) the red vertical line shows the case B value,
which agrees precisely with the mean measured ratio.
\label{calibrate}
}
\end{inlinefigure}

We next compute the \ha\ calibration relative to
our UV calibration based on the data.  We must
be careful in our object selection here since
there is a potential problem with using objects 
selected to have \ha\ flux. This arises because
the \ha\ is produced by more massive and
younger stars than those responsible for the
UV light. As long as the SFR is smoothly changing
with time, then this is not a large effect. However, 
if the galaxies undergo episodic bursts rather than
a smooth evolution (as we shall argue is the case
in \S\ref{seccolor}), then we will bias this ratio 
since we will not count galaxies which still have
substantial UV light but where \ha\ producing 
stars are no longer present. The different time averaging 
present in the emission line and UV calibrations 
should be borne in mind when considering the SFR 
of individual galaxies, but it should average 
out in the ensemble. 

We can properly calibrate the \ha/UV
ratio by computing the average \ha/UV for a 
UV-selected sample, which will then include the 
all of the  \ha\ emitters as a subsample. Using the 
UV-selected sample we obtain
\begin{equation}
\log {\rm SFR} = -40.90 + \log L({\rm H}\alpha) \,,
\label{eqhaSFR}
\end{equation}
with a $1\sigma$ spread of 0.24~dex. This differs by
0.2 dex from the value of $-41.1$ given in Kennicutt
(1998) and is also higher than the value that would
be derived from the BC03 models. (The H alpha luminosity
for the BC03 models is derived from the ionizing photon
production assuming this is fully absorbed in the galaxy.)
This is not a consequence
of the extinction model, since even using the unreddened
values only reduces our value to $-41.03$. 

This result can be restated as the sample having
too much UV light relative to \ha\ for the adopted IMF.
The \ha\
production is dominated by the most massive stars, and
it is probable that the high UV/\ha\ ratio is a consequence 
of differences in the true high-end IMF relative to that
used in the models. 
As we noted in the introduction all of the currently
used IMFs have similar high mass slopes and so would
not change this result. Rather we need more intermediate
mass stars relative to the very high mass end in the IMF.
Fardal et al.\ (2007) describe this type of IMF as "paunchy"
and as we shall discuss later it can resolve a number of
additional problems in making a consistent interperetation
of the data set.

From the mean value of the logarithmic $f$(\hb)$/f$(\ha)
ratio, which is just the case B value (see Fig.~\ref{calibrate}), 
we find the calibration of Equation~\ref{eqhaSFR} can be 
translated to
\begin{equation}
\log {\rm SFR} = -40.45 + \log L({\rm H}\beta) \,,
\end{equation}
with a $1\sigma$ spread of 0.13~dex. \hb\ is often avoided
as a SFR diagnostic because of concerns about the contamination 
by the underlying stellar absorption. 
However, with such a low $1\sigma$ spread
for our low-$z$ sample, it appears to be quite a good diagnostic. 

From Figure~\ref{calibrate} we saw that the UV flux 
provided a better estimate of the SFR than the \oii\ flux 
did. The average calibration for \oii,
using the mean value of the logarithmic $f$(\oii)$/f$(\ha)
ratio, is
\begin{equation}
\log {\rm SFR} = -41.03 + \log L({\rm [OII]}~\lambda3727) \,,
\end{equation}
with a $1\sigma$ spread of 0.26~dex. This may be compared with
the value of $-40.85$ given in Kennicutt (1998).

%
%
\begin{inlinefigure}
\figurenum{18}
\centerline{\psfig{figure=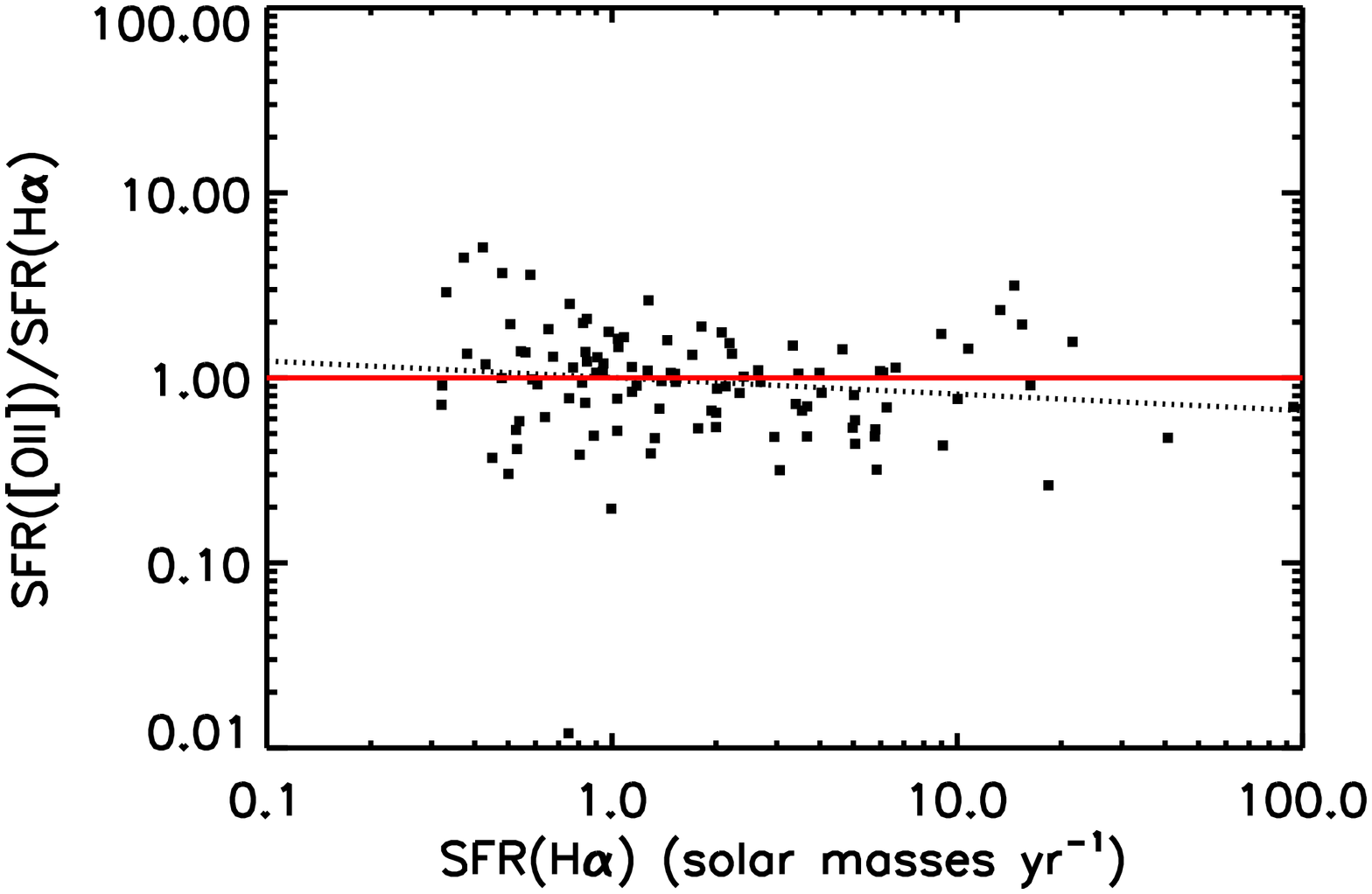,angle=90,width=3.5in}}
\figcaption[]{
Ratio of the \oii\ SFR to the \ha\
SFR versus the \ha\ SFR (all extinction
corrected) for the low-$z$ sample with EW(\ha$)>12$~\AA\ and 
$\log({\rm SFR}({\rm H}\alpha))>-0.5$ {\em (black squares)\/}. 
The black dotted line shows the least-square polynomial fit 
with a gradient of $-0.09$~dex per dex. 
\label{sfrratio_sfr}
}
\end{inlinefigure}

The broader spread in the \oii\ calibration
is reflected in its systematic dependence on
the galaxy properties. In Figure~\ref{sfrratio_sfr}
we show the weak dependence of the ratio of the \oii\ SFR to 
the \ha\ SFR on the \ha\ SFR (all extinction corrected).
There is a $0.09\pm0.05$~dex decline per dex 
increase in the \ha\ SFR, which is slightly larger than  
the locally measured $0.03\pm0.02$~dex decline per dex
increase in the \ha\ SFR measured to somewhat lower \ha\ SFRs
(Kewley et al.\ 2004) but consistent within the errors.
We show in \S\ref{seccompare} that the dependence
of the SFR ratio on the mass of the galaxy is even 
stronger than its dependence on the \ha\ SFR of the galaxy.

Given the larger scatter and systematic dependences of the 
\oii\ calibration, we will adopt the UV calibration as our 
primary calibration.

\subsection{Comparisons Using the Mid-$z$ Sample}
\label{seccompare}

We next tested the relative \oii\ and \hb\ calibrations 
against our primary UV calibration over the wider redshift 
range of the full mid-$z$ sample ($z<0.9$). In Figure~\ref{nuv_o2}a 
we plot the ratio of the \oii\ SFR to the UV SFR (both extinction
corrected) versus galaxy 
mass. We show the lower redshift sources ($z=0.05-0.475$)
with red diamonds and the higher redshift sources
($z=0.475-0.9$) with black squares, and we only show the
results above the mass at which each redshift range is
complete. The ratio changes slowly with both mass
and redshift. The change is a 0.19~dex decline per dex 
increase in mass and a 0.04~dex increase between the low and 
high redshift ranges. This probably primarily reflects 
the higher metallicity in the more massive galaxies and the 
increase of metallicity to lower redshifts, which results
in temperature changes in the H~II regions and
the \oii\ line being stronger relative to the primary \ha\ line.
Thus, if we were to use our \oii\ calibration, we would find 
it difficult to study the evolution of the SFRs as a function 
of redshift, mass, and metallicity.

In contrast, as shown in Figure~\ref{nuv_o2}b (same symbols
as in Fig.~\ref{nuv_o2}a), the ratio of the \hb\ SFR to the 
UV SFR (both extinction corrected) varies more slowly with 
galaxy mass and does not vary significantly with redshift. 
The change with mass is a 0.11~dex decline per dex increase 
in galaxy mass. Thus, \hb\ can also be used over the 
$z=0.05-0.9$ range where it can be measured.

%
%
\begin{inlinefigure}
\figurenum{19}
\centerline{\psfig{figure=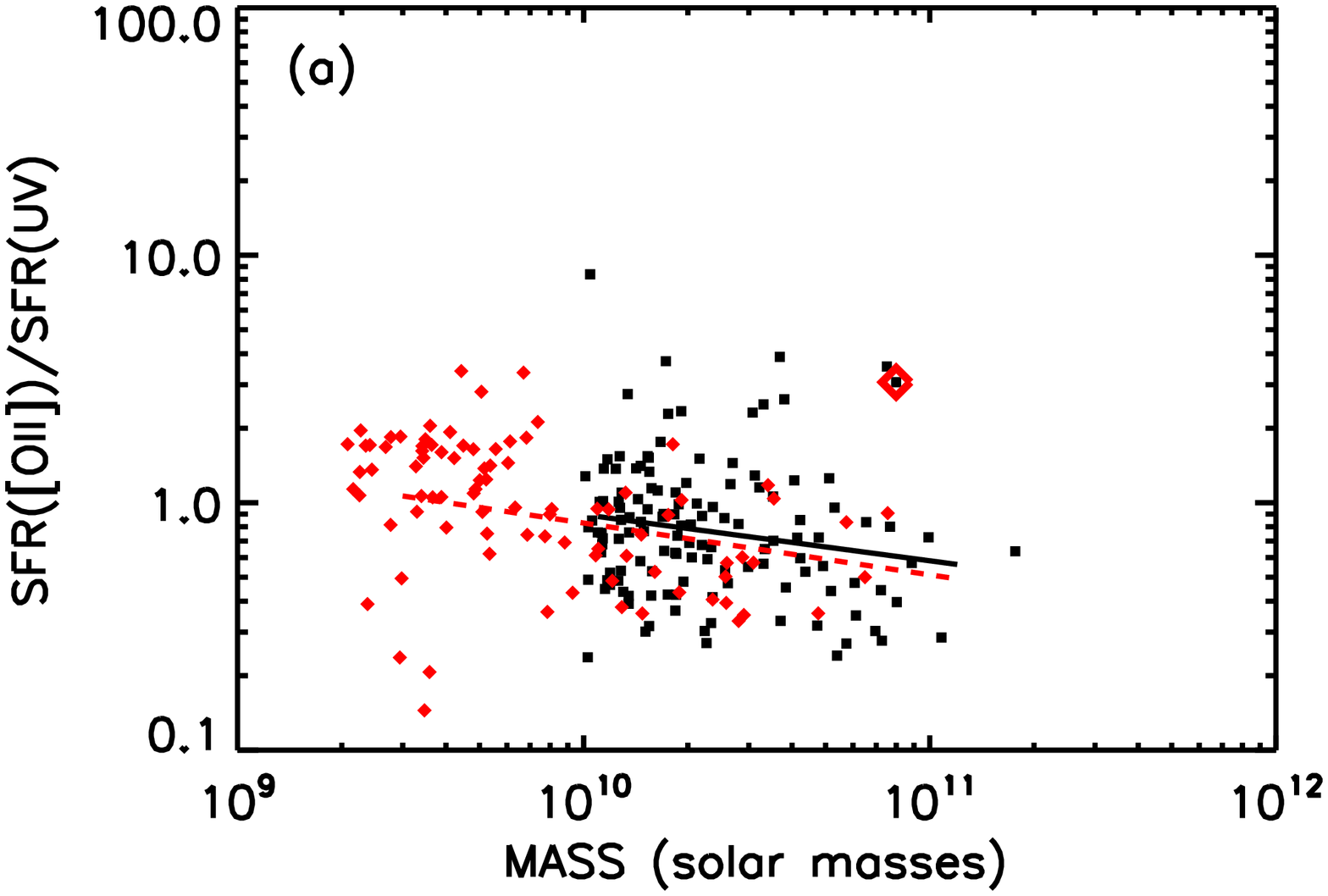,angle=90,width=3.5in}}
\vskip -0.6cm
\centerline{\psfig{figure=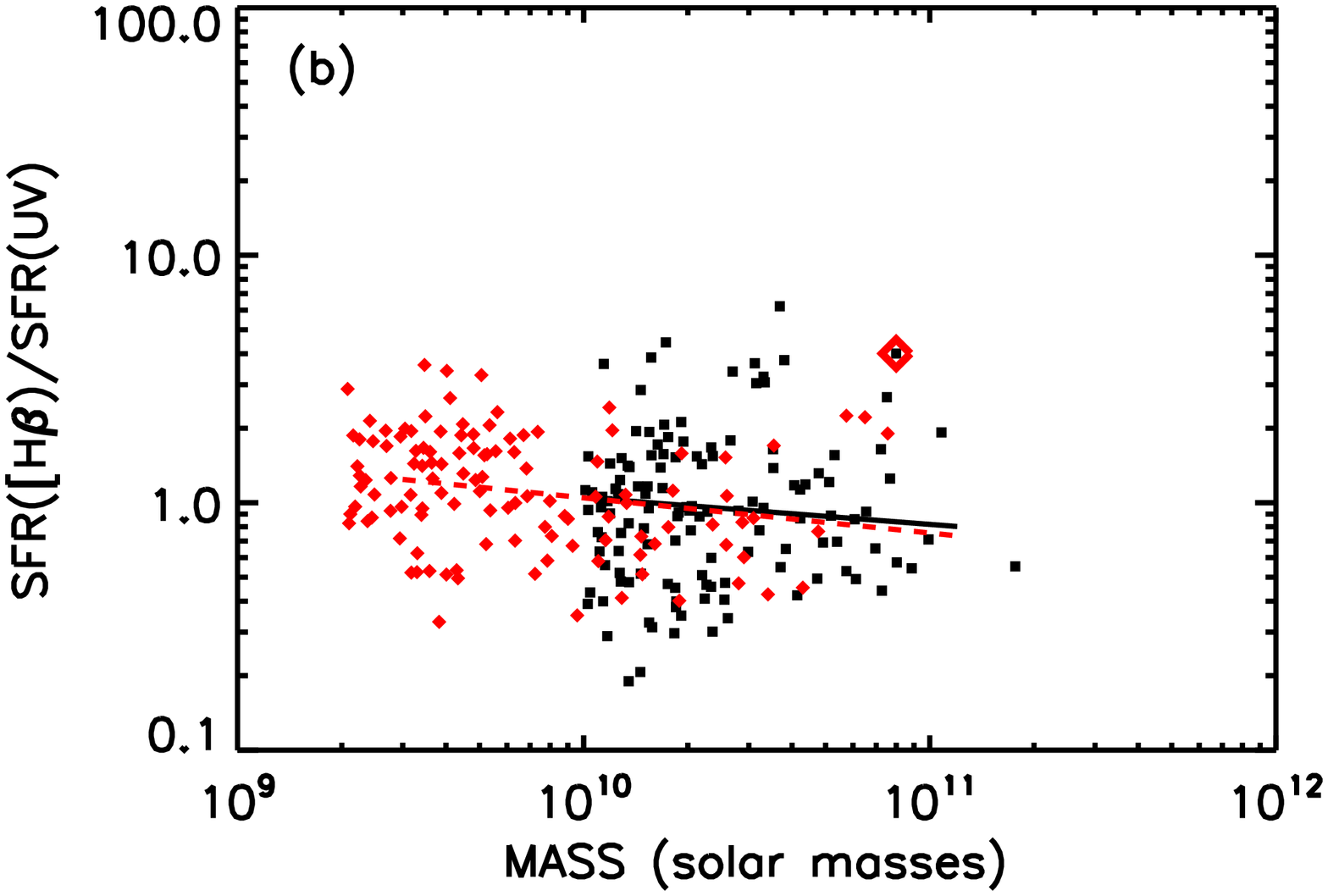,angle=90,width=3.5in}}
\vskip -0.2cm
\figcaption[]{
(a) Ratio of the \oii\ SFR to the UV SFR (both
extinction corrected) vs. galaxy mass.
(b) Ratio of the \hb\ SFR to the UV SFR (both extinction
corrected) vs. galaxy mass. 
Only sources with EW(\hb$)>4$~\AA\ are 
included. The red diamonds (black squares) show the results for 
$z=0.05-0.475$ ($z=0.475-0.9$). 
We show the results only above the mass 
at which each redshift interval is complete. 
The least-square polynomial fits are shown with the dashed red 
(solid black) lines for the $z=0.05-0.475$ ($z=0.475-0.9$) sample.
The one AGN based on its X-ray luminosities is enclosed in the 
large red diamond and is excluded from the fits. 
In (a) there is a significant gradient with mass and a 
small amount of evolution between the two redshift
samples. In (b) there is a smaller change with galaxy mass 
and no significant evolution between the two redshift samples.
\label{nuv_o2}
}
\end{inlinefigure}

\subsection{Comparison with 24~$\mu$m} 
\label{sec24}

Using the observed-frame 24~$\mu$m fluxes of the galaxies is
a relatively poor way to determine the FIR luminosities and
hence SFRs of the galaxies (e.g., Dale et al.\ 2005).
However, because of the ready availability of 24~$\mu$m
data from the {\em Spitzer\/} MIPS observations, this has
become a common way of estimating SFRs, and nearly
all of the papers that we will compare with use some
version of this method to determine their SFRs. We
therefore compare with 24~$\mu$m determined SFRs here.

We compute the $24~\mu$m SFRs following the presciption given 
in Conselice07. We use the Dale \& Helou (2002) 
SEDs to convert the 24~$\mu$m flux to FIR luminosity following 
Figure~7 of Le Floc'h et al.\ (2005). We then use the 
Bell et al.\ (2005) relation between the reradiated SFR and the 
FIR luminosity (converted to the Salpeter IMF of the present paper),
\begin{equation}
\log {\rm SFR} = -43.36 + \log L(FIR) \,,
\end{equation}
to compute the SFRs.

In Figure~\ref{mstar_f24} we compare the SFRs determined from 
the $24~\mu$m fluxes with the reradiated SFRs determined from
the UV luminosities. (Here the reradiated SFR is the
difference between the SFR computed after correcting for
extinction and the SFR computed without an extinction
correction.) In each redshift interval we only show the
sources with sufficiently high SFRs that they will lie
above the 80~$\mu$Jy sensitivity of the $24~\mu$m
observations.

%
%
\begin{inlinefigure}
\figurenum{20}
\centerline{\psfig{figure=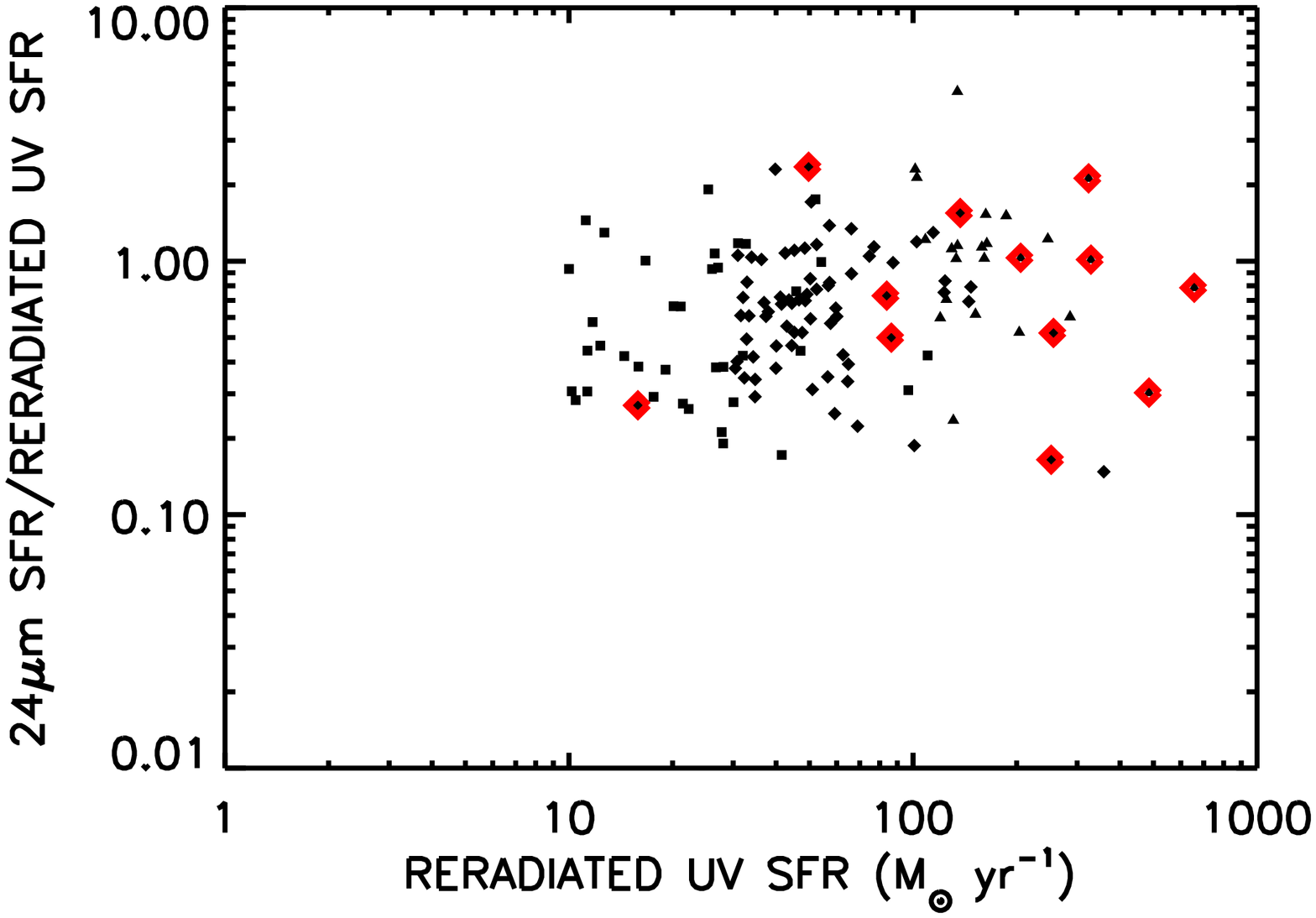,angle=90,width=3.5in}}
\vskip -0.2cm
\figcaption[]{
Ratio of the 24~$\mu$m SFR to the reradiated UV
SFR vs. the reradiated UV SFR. The sources are separated
by redshift interval and are only shown if their SFR is
above the SFR where the galaxies will be detected at the
flux limit for the $24~\mu$m sample. Squares correspond to
$z=0.05-0.475$ and are shown above 10~M$_\odot$~yr$^{-1}$,
diamonds correspond to $z=0.475-0.9$ and are shown above
30~M$_\odot$~yr$^{-1}$, and triangles correspond to $z=0.9-1.5$
and are shown above 100~M$_\odot$~yr$^{-1}$. Sources containing
an AGN based on their X-ray luminosities are enclosed in
larger red diamonds.
\label{mstar_f24}
}
\end{inlinefigure}

While there is a considerable spread in the individual SFR
ratios, there is no clear dependence on redshift or on SFR. 
The normalization difference between the two methods is only
$-0.08$~dex in the ensemble average, which is
well within the uncertainties in the calibrations. The ensemble
distribution is symmetric about the average with a $1\sigma$
multiplicative spread of 0.29~dex. Thus, while there is
a substantial spread in the individual galaxy determinations,
the two methods give good agreement when applied to the galaxy
population as a whole. Excluding AGNs based on their
X-ray properties (i.e., sources enclosed in red diamonds in
Figure~\ref{mstar_f24}) has no effect on the relative calibration.

\section{Metallicities in the Emission Line Galaxies: The
Low-$z$ Sample}
\label{secha}

We may construct a number of emission line diagnostics
for the low-$z$ sample, since many of the spectra
cover all of the emission lines from \oii\ to \sii.
We therefore begin with a comparison of the
metallicity-luminosity and metallicity-mass relations
for these sources before proceeding to the mid-$z$
sample in \S\ref{seco3}. We know that the \sii ratios 
place most of the galaxies in the low-density regime, 
but we do not use this information further.

\subsection{[NII]/[OII] Diagnostic Ratio}
\label{n2o2}

Kewley \& Dopita (2002; hereafter KD02) advocate the use 
of the N2O2 $=f($\nii$)/f($\oii)
diagnostic ratio whenever possible, since this ratio is quite
insensitive to the ionization parameter $q$ and has a
strong dependence on metallicity above $\log$(O/H)$+12\sim8.5$.
However, the downside of this diagnostic ratio is that it compares 
two widely separated lines where the uncertainties in the
flux calibration and extinction are more severe. Throughout this 
section we use the BC03 extinctions of \S\ref{secfit}.

%
%
\begin{inlinefigure}
\figurenum{21}
\centerline{\psfig{figure=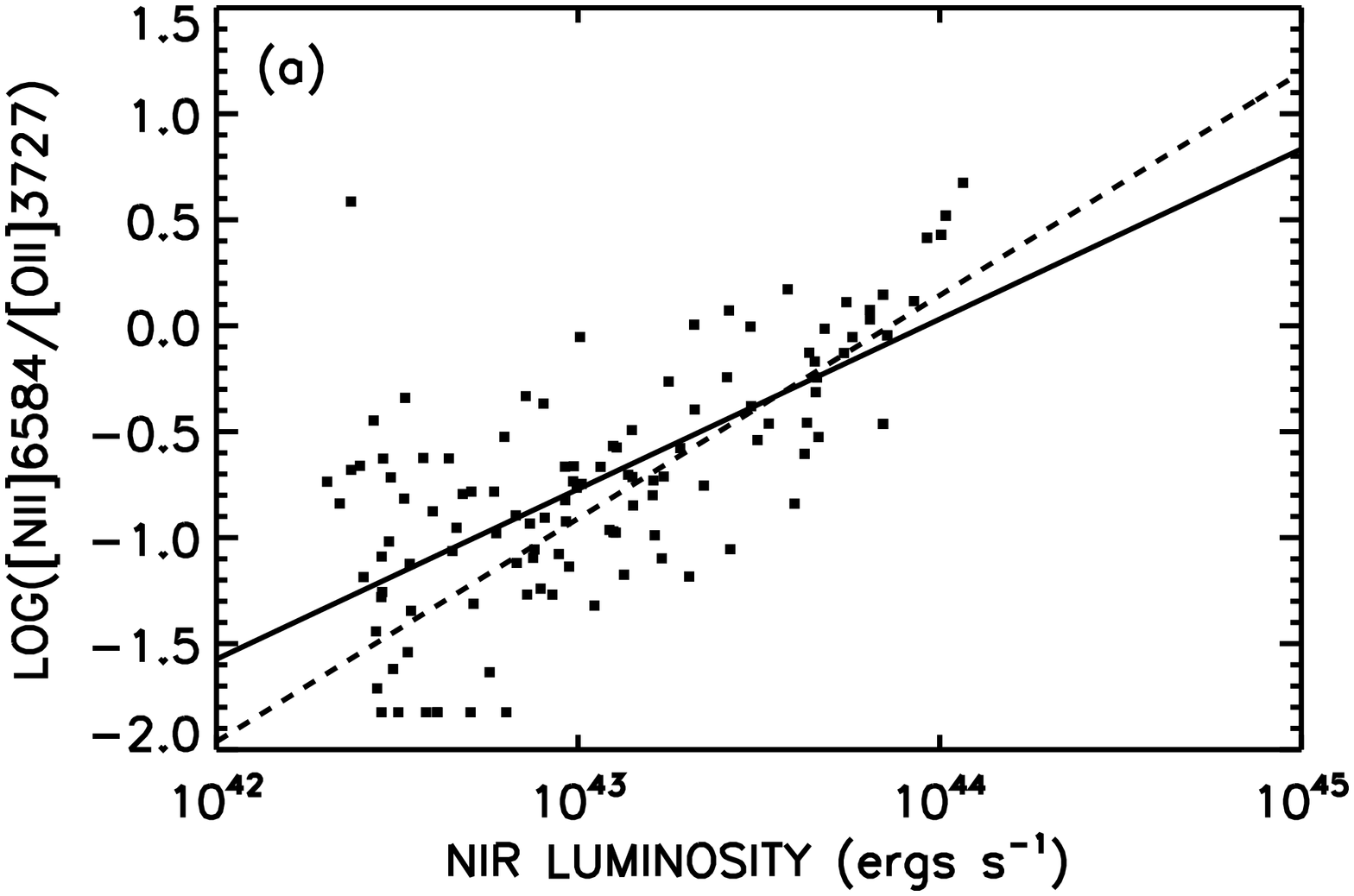,angle=90,width=3.5in}}
\vskip -0.6cm
\centerline{\psfig{figure=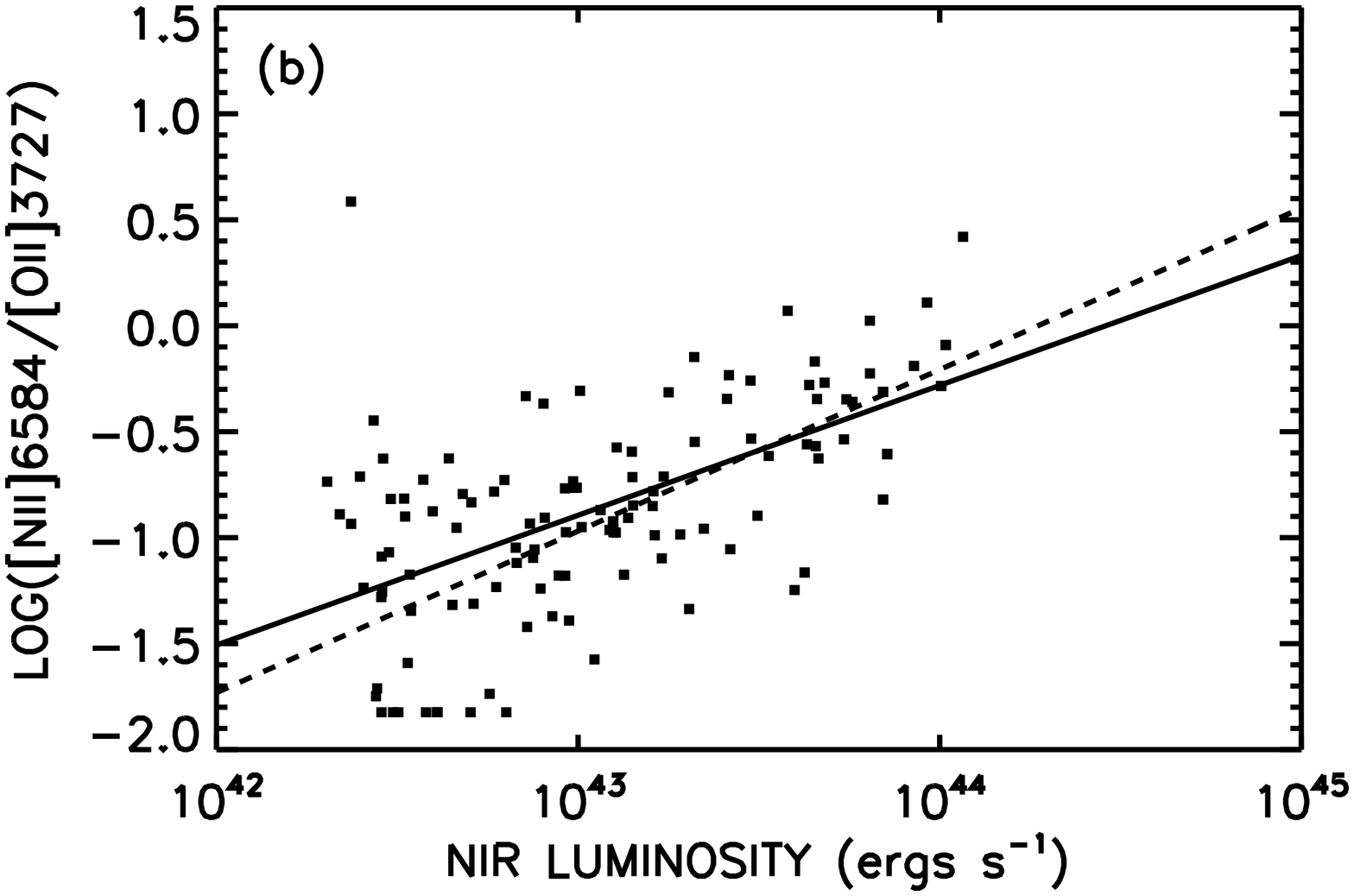,angle=90,width=3.5in}}
\vskip -0.2cm
\figcaption[]{
Logarithmic (a) extinction uncorrected and (b) extinction
corrected N2O2 diagnostic ratio vs. NIR luminosity
for the sources in the low-$z$ sample with EW(\ha$)>12$~\AA\ 
and an \oii\ line that can be measured {\em (black squares)\/}.
The sources with a flux ratio less than 0.015 are 
plotted at that value ($-1.82$ in the logarithm). In each panel
the solid line shows the least-square polynomial fit 
to all the data, and the dashed line shows the least-square 
polynomial fit to only the data with NIR luminosities in the range 
$10^{43}-10^{44}$~ergs~s$^{-1}$.
\label{noplot}
}
\end{inlinefigure}

In Figure~\ref{noplot} we plot (a) extinction uncorrected and 
(b) extinction corrected N2O2 versus NIR luminosity
for the 115 sources in the low-$z$ sample with 
EW(\ha$)>12$~\AA\ and an \oii\ line that can be measured
(it can be negative). These sources cover the redshift 
range $z=0.2-0.475$ and have a median redshift of $z=0.4$. 
We find a strong correlation between N2O2
and NIR luminosity. For each panel 
in Figure~\ref{noplot} we use a solid line to show the 
least-square polynomial fit to all the data. In the faintest 
sources the $f($\nii) and $f($\oii) are weak, which results 
in larger scatter. For example, the extreme up-scattered 
point at low NIR luminosity in Figure~\ref{noplot} is a
source with very weak $f($\nii) and $f($\oii). Thus, for each
panel we also do a fit to the data in the restricted NIR 
luminosity range $10^{43}-10^{44}$~ergs~s$^{-1}$ to minimize 
these effects {\em (dashed line)\/}. The relation for the
extinction corrected data restricted in luminosity is 
\begin{equation}
\log({\rm N2O2}) = -0.94 + (0.63\pm0.08) \log(L_N) \,,
\end{equation}
where $L_N$ is the NIR luminosity in units of 
$10^{43}$~ergs~s$^{-1}$.

%
%
\begin{inlinefigure}
\figurenum{22}
\centerline{\psfig{figure=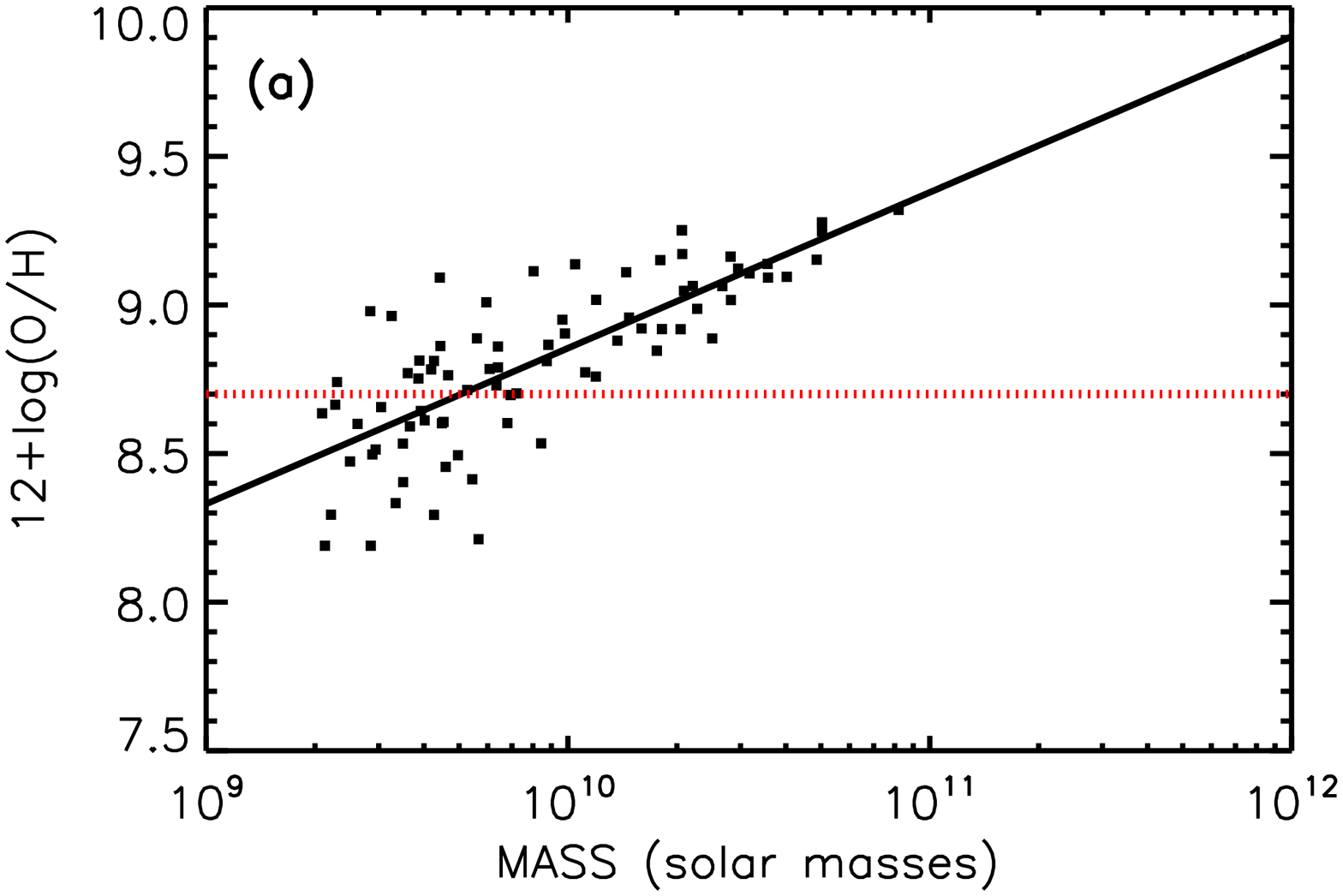,angle=90,width=3.5in}}
\vskip -0.6cm
\centerline{\psfig{figure=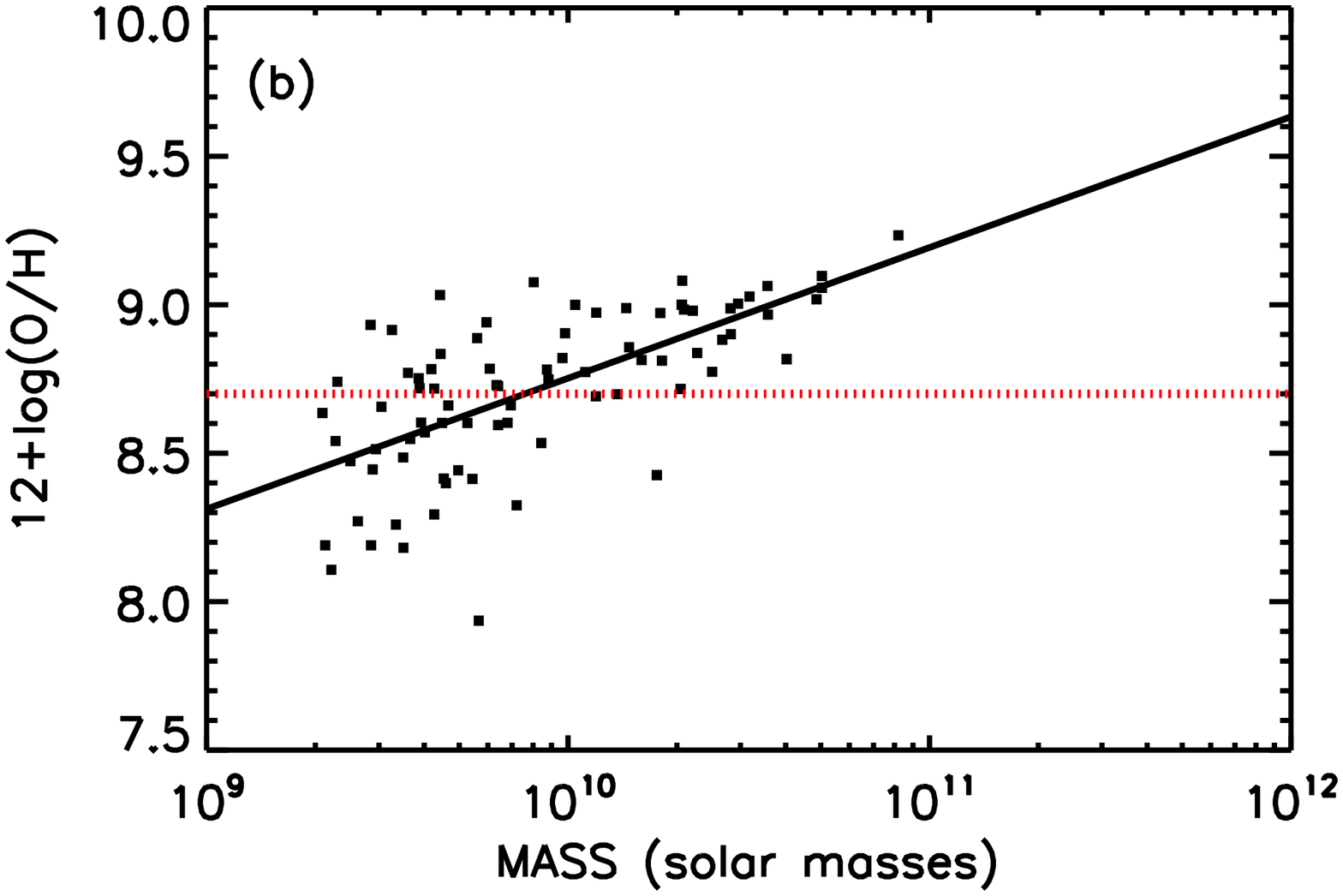,angle=90,width=3.5in}}
\vskip -0.2cm
\figcaption[]{
Metallicity determined from the (a) extinction uncorrected and 
(b) extinction corrected 
N2O2 diagnostic ratio vs. galaxy mass for the sources in the low-$z$ 
sample with EW(\ha$)>12$~\AA\ and an \oii\ line that can be measured 
{\em (black squares)\/}. The conversion is for an ionization parameter 
$q=4\times 10^{7}$~cm~s$^{-1}$, but the conversion is insensitive to 
this choice. In each panel the black solid line shows the 
least-square polynomial fit to all the data, and the red dotted 
ine shows the solar abundance.
\label{n2o2mass}
}
\end{inlinefigure}

We use the KD02 calibration (their Eq.~4 and Table~3) to convert 
our N2O2 values to metallicities, assuming an ionization 
parameter $q=2\times 10^{7}$~cm~s$^{-1}$.
This $q$ value is typical of the ionization parameters in our 
galaxies (see \S\ref{secq}). However, the present conversion 
is extremely insensitive to the choice of $q$. 
In Figure~\ref{n2o2mass} we plot metallicity determined from
(a) extinction uncorrected and (b) extinction corrected 
N2O2 versus galaxy mass.
For each panel we use a solid line to 
show the least-square polynomial fit to all the data. The 
metallicity-mass relation for the extinction corrected case is
\begin{equation}
12+\log({\rm O/H}) = 8.75 + (0.44\pm0.06) \log(M_{10}) \,,
\label{eqnn202rel}
\end{equation}
where $M_{10}$ is the galaxy mass in units of $10^{10}$~M$_\odot$.
The extinction corrections flatten the slope
(the unextinguished slope is $0.52\pm0.06$) since
they reduce the N2O2 values more substantially in the high mass 
(high NIR luminosity) galaxies than they do in the lower
mass galaxies.

\subsection{Ionization Parameter}
\label{secq}

We can now combine the metallicities determined from
extinction corrected N2O2 with extinction
corrected O32 $=1.3\times f$(\oiii)$/f$(\oii)
to determine the ionization parameters $q$. Here we do so 
using the KD02 parameterization of 
the O32 dependence on $q$ and metallicity. We plot 
$q$ versus NIR luminosity in Figure~\ref{no_q_plot}.
The $q$ values lie in a surprisingly small range around
a median value of $2.2\times10^{7}$~cm~s$^{-1}$, with more
than 77\% lying within a multiplicative factor of 2 of
this value. There also appears to be little dependence
of $q$ on NIR luminosity.

%
%
\begin{inlinefigure}
\figurenum{23}
\centerline{\psfig{figure=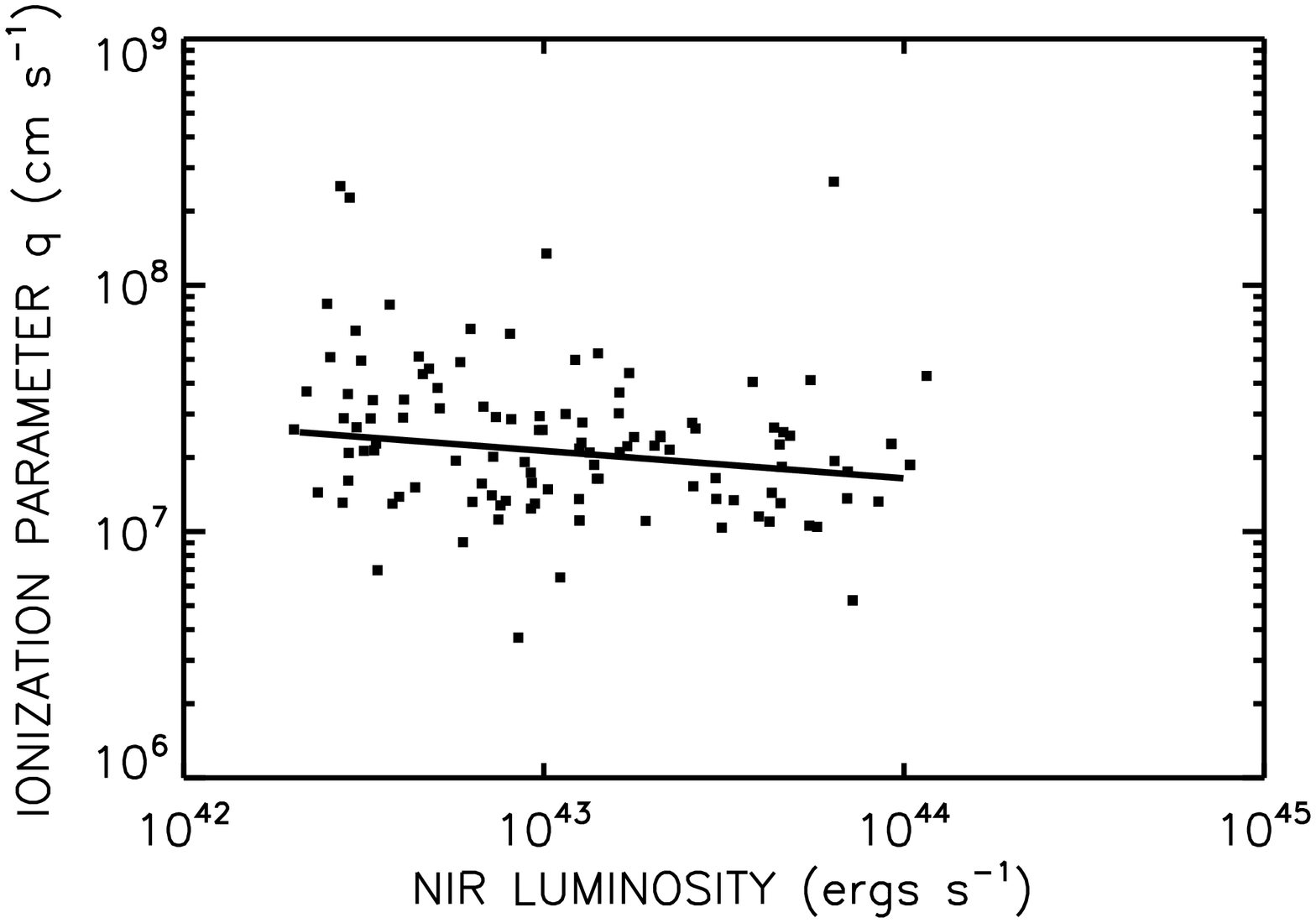,angle=90,width=3.5in}}
\vskip -0.2cm
\figcaption[]{
Ionization parameter obtained by combining the 
metallicity determined from the extinction corrected N2O2 
diagnostic ratio with the extinction corrected O32 diagnostic 
ratio vs. NIR luminosity {\em (black squares)\/}. 
The effects of the extinction correction
are to slightly reduce the average ionization parameter and also
to flatten the dependence on luminosity. Most of the sources 
lie in the $q=1-4\times10^{7}$~cm~s$^{-1}$ range with
an average value of $q=2.2\times10^{7}$~cm~s$^{-1}$. The black
solid line shows the least-square polynomial fit. The ionization 
parameter shows little dependence on NIR luminosity. 
\label{no_q_plot}
}
\end{inlinefigure}

\subsection{[NII]/[H$\alpha$] Diagostic Ratio}
\label{secnh}

The tightly determined $q$ values mean that other 
diagnostic ratios, such as NH = $f$(\nii)$/f$(\ha), which 
normally have too much dependence on ionization parameter 
to be useful, actually work surprisingly well. 
Because \nii\ and \ha\ are extremely close in wavelength,
NH is not dependent on the extinction corrections
nor on the flux calibration methodology and thus provides
a powerful check on the metallicities computed from
N2O2. In Figure~\ref{nhlum} we show 
NH versus NIR luminosity. There is a 
strong correlation, which we fit over the NIR luminosity 
range $10^{43}-10^{44}$~ergs~s$^{-1}$ with the solid line.

We next constructed the metallicity-mass relation from
NH using Eq.~12 of Kobulnicky \& Kewley (2004;
hereafter, KK04) and $q = 2\times10^{7}$~cm~s$^{-1}$.
(We note that the use of the coefficients given in Table~3 of 
KD02 for NH appears to give results that are inconsistent with 
Figure~7 of KD02 and Eq.~12 of KK04.)
The metallicity-mass relation,
\begin{equation}
12+\log({\rm O/H}) = 8.85 + (0.36\pm0.04) \log(M_{10}) \,,
\label{eqnnhorel}
\end{equation}
is shown in Figure~\ref{nhomass} by the black solid line.
Recomputing the data points using higher ($q=4\times 10^7$~cm~s$^{-1}$)
or lower ($q=10^7$~cm~s$^{-1}$) ionization 
parameters and fitting the revised data points
{\em (red dash-dotted lines)\/} does not substantially change 
the slope, but it does significantly change the normalization.

%
%
\begin{inlinefigure}
\figurenum{24}
\centerline{\psfig{figure=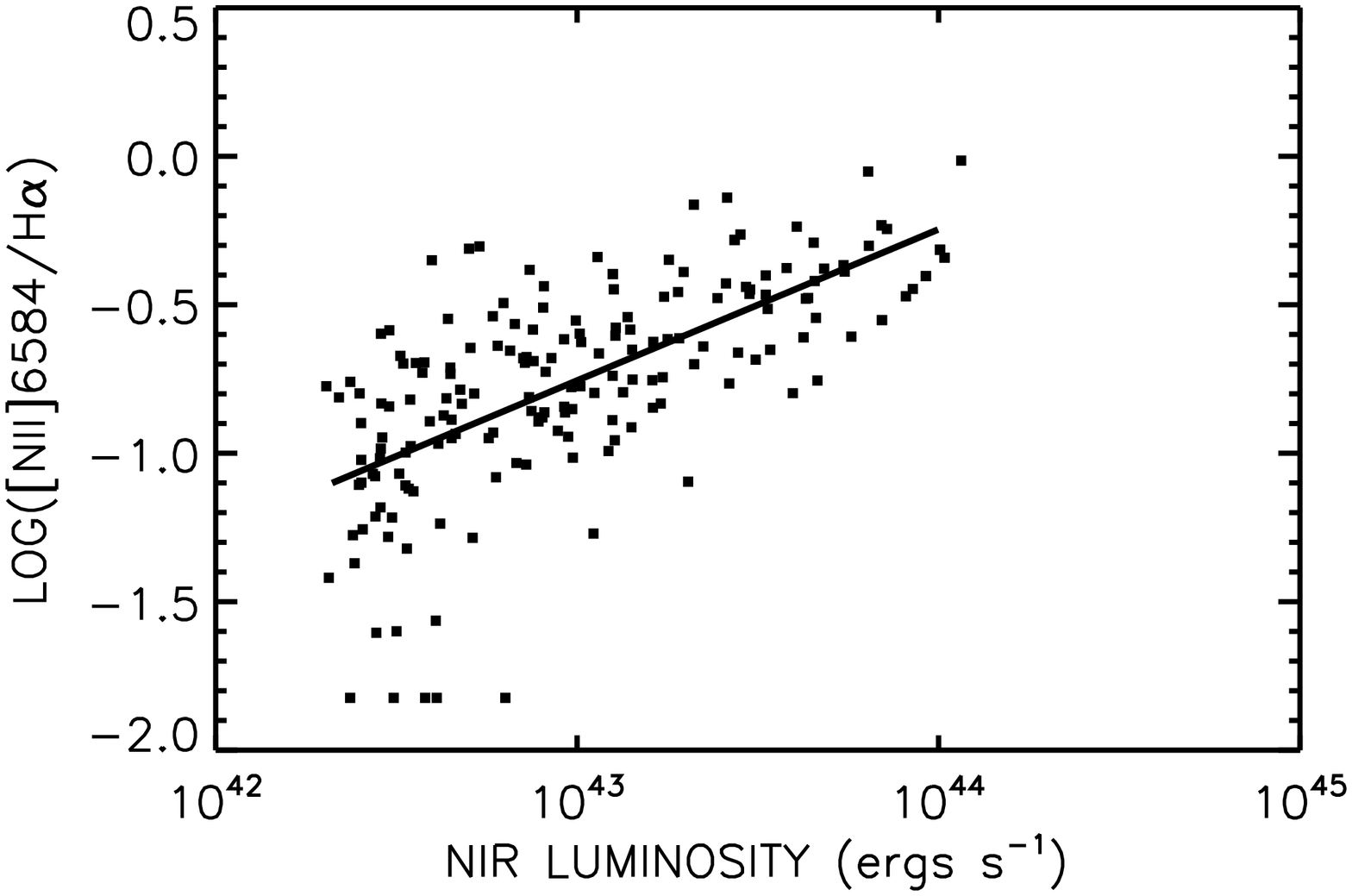,angle=90,width=3.5in}}
\vskip -0.2cm
\figcaption[]{
Logarithmic NH diagnostic ratio vs. NIR luminosity 
for the sources in the low-$z$ sample with EW(\ha$)>12$~\AA\
{\em (black squares)\/}.
The black solid line shows the least-square polynomial
fit to the data. 
\label{nhlum}
}
\end{inlinefigure}

%
%
\begin{inlinefigure}
\figurenum{25}
\centerline{\psfig{figure=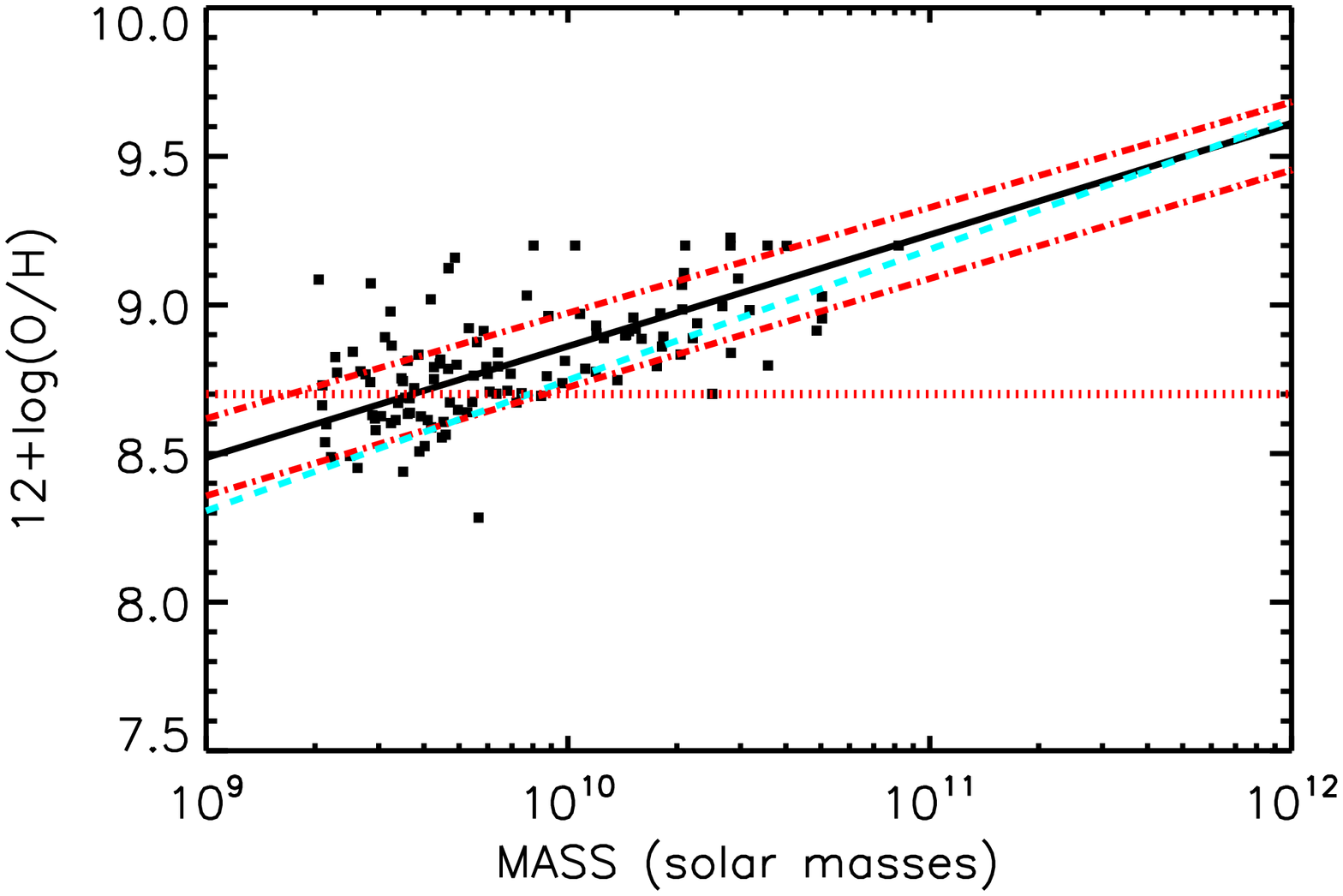,angle=90,width=3.5in}}
\vskip -0.2cm
\figcaption[]{
Metallicity based on the NH 
diagnostic ratio vs. galaxy mass for the sources in the low-$z$ sample 
with EW(\ha$)>12$~\AA\ computed for an 
ionization parameter of $q = 2\times10^{7}$~cm~s$^{-1}$
{\em (black squares)\/}.
The black solid line shows the least-square polynomial
fit to the data. The red dash-dotted lines
show the fits which would be obtained for
$q = 4\times10^{7}$~cm~s$^{-1}$ {\em (upper line)\/}
and $q = 10^{7}$~cm~s$^{-1}$ {\em (lower line)\/}.
The cyan dashed line shows the metallicity-mass relation
derived from the extinction corrected N2O2 diagnostic ratio 
(Eq.~\ref{eqnn202rel}). The red dotted horizontal line shows 
the solar metallicity.
\label{nhomass}
}
\end{inlinefigure}

In the figure we compare the metallicity-mass relation of
Eq.~\ref{eqnnhorel} {\em (black solid line)\/} 
with the metallicity-mass relation of Eq.~\ref{eqnn202rel}
{\em (cyan dashed line)\/} derived from N2O2. 
The agreement is reasonable given the sensitivity
of the NH diagnostic to the ionization parameter. 
The mean metallicity in
the $10^{10}$ to $10^{11}$~M$_\odot$ range is
9.03 from NH and 8.99 from N2O2, which reassuringly shows that 
the metallicity and ionization parameter determinations are 
roughly self-consistent and that our treatment of the extinctions 
is plausible.

\subsection{R23 Diagnostic Ratio}
\label{secr23}

%
%
\begin{inlinefigure}
\figurenum{26}
\centerline{\psfig{figure=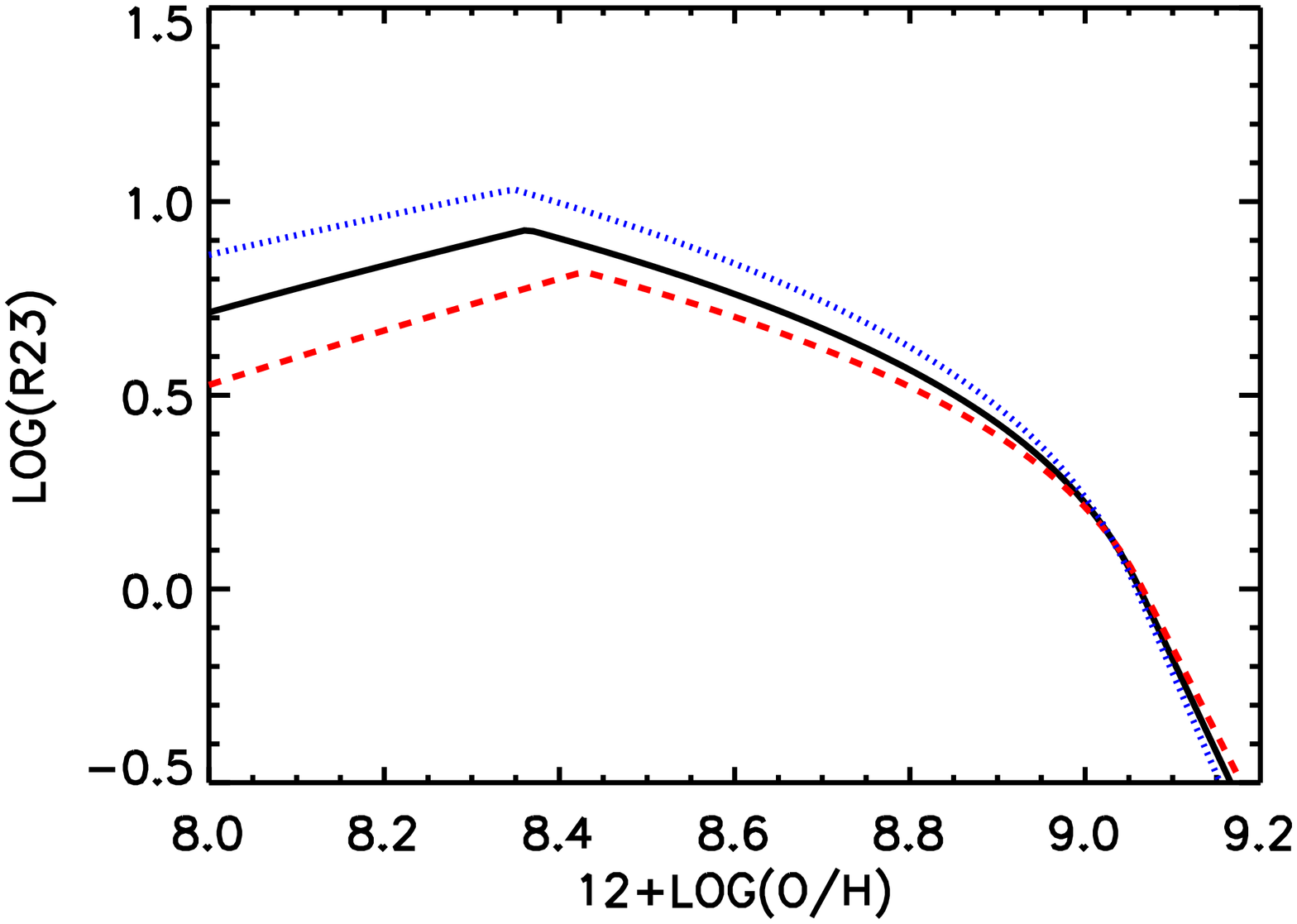,angle=90,width=3.5in}}
\vskip -0.6cm
\centerline{\psfig{figure=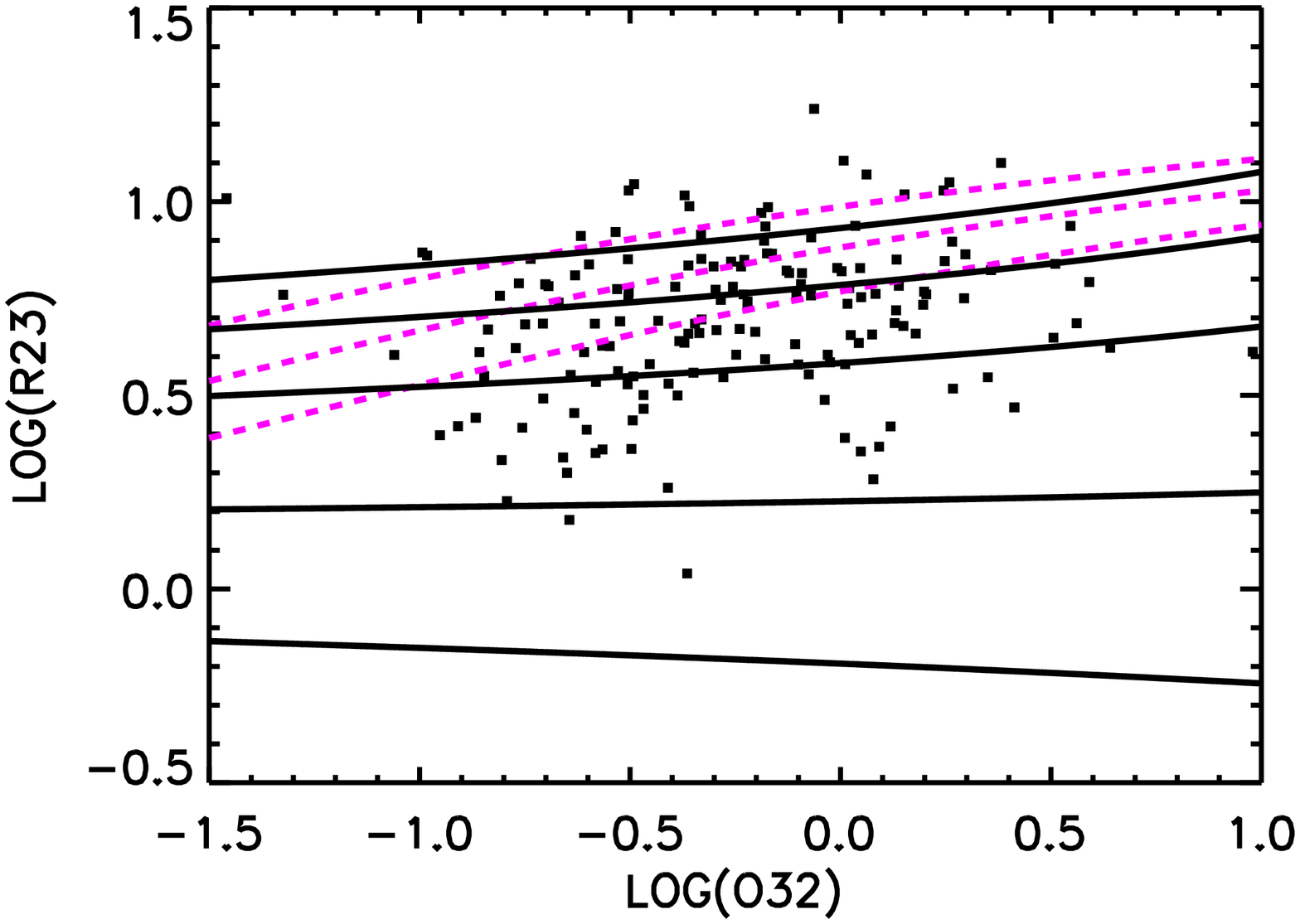,angle=90,width=3.5in}}
\vskip -0.2cm
\figcaption[]{
(a) Logarithmic R23 vs. metallicity for the McGaugh (1991)
models corresponding to $\log({\rm O32})=-1$ {\em (red dashed)\/}, 
$-0.25$ {\em (black solid)\/}, and 0.5 {\em (blue dotted)\/}.
These values cover the range of our data.
(b) Logarithmic extinction corrected R23 vs. logarithmic 
extinction corrected O32 for the sources in the low-$z$ sample with 
EW(\hb$)>4$~\AA\ and detected \oii\ {\em (black squares)\/}.
Tracks give the R23-O32 relations from McGaugh (1991):
purple dashed lines show the relation on the low metallicity 
branch for three values of $12+\log({\rm O/H})$
[8.0 {\em (lowest curve)\/}, 8.2, and 8.4 {\em (highest curve)\/}],
and solid black lines show the relation on the high
metallicity branch for five values of $12+\log({\rm O/H})$
[8.4 {\em (highest curve)\/}, 8.6, 8.8, 9.0, and 9.1 
{\em (lowest curve)\/}]. On the high metallicity branch
there is little dependence of R23 on O32.
\label{r23_o32_ha}
}
\end{inlinefigure}

The R23 diagnostic ratio
[$1.3\times f($\oiii$~\lambda 5007)+f($\oii$~\lambda 3727)]/f($\hb) 
of Pagel et al.\ (1979) is one of the most frequently used 
metallicity diagnostics. It can be measured in 
our data out to $z=0.9$. However, it is unfortunately 
multivalued with both a low metallicity and a high metallicity 
branch. Thus, the same value of R23 can correspond to two 
substantially different metallicities. Moreover, R23
is only weakly dependent on metallicity for 
$12+\log({\rm O/H})\lesssim 8.6$. We can see these traits in 
Figure~\ref{r23_o32_ha}. 
In Figure~\ref{r23_o32_ha}a we plot logarithmic R23 versus 
metallicity for the McGaugh (1991) models corresponding to
$\log({\rm O32})=-1$ {\em (red dashed)\/},
$-0.25$ {\em (black solid)\/}, and 0.5 {\em (blue dotted)\/}.
These cover the range of ionization dependent O32 measurements 
from our data, as can be seen from Figure~\ref{r23_o32_ha}b 
where we plot logarithmic extinction corrected R23 versus 
logarithmic extinction corrected O32 for the 164 sources 
in the low-$z$ sample with EW(\hb$)>4$~\AA\ and detected \oii\
{\em (black squares)\/}.
Here we overlay the models of McGaugh (1991) for the low
metallicity branch {\em (purple dashed)\/} and
the high metallicity branch {\em (black solid)\/} onto the
data.

Unfortunately, much of the data in Figure~\ref{r23_o32_ha}b
lie near the peak of the
R23 ratio [$\log$(R23) in the range 0.5 to 1], where R23 
is a relatively slowly varying function of the metallicity
and where the metallicity could lie on either the upper or lower 
branches (see Fig.~\ref{r23_o32_ha}a). Fortunately, however, 
we can see from Figures~\ref{n2o2mass} and \ref{nhomass} 
(where the metallicities were computed from N2O2 and NH, 
respectively) that the galaxies in the 
low-$z$ sample with masses greater than 
$2\times10^{9}$~M$_\odot$ primarily lie on the high metallicity 
branch of the R23 diagnostic ratio [$12+\log({\rm O/H})>8.4$]
(see also KK04). In \S\ref{seco3} we will assume that the higher redshift
$(z>0.475)$ sources contained in the mid-$z$ sample also predominantly 
lie on the high metallicity branch, though if there is significant
metallicity evolution, this assumption could break down.

With most of the galaxies on the upper branch,
the interpretation is considerably simplified, since 
on this branch there is very little dependence on O32. 
From the McGaugh (1991) 
models in Figure~\ref{r23_o32_ha}, we can see that
at metallicities of $12+\log({\rm O/H})>8.6$, the
R23 parameter is nearly ionization independent and has a 
steep dependence on metallicity. We can therefore
make a simple investigation of the metal evolution by 
analyzing the dependence of R23 on the 
NIR luminosity without considering O32.
In Figure~\ref{r23_nirlum_ha} we show the dependence of 
extinction corrected R23 
on NIR luminosity for the low-$z$ sample with EW(\hb$)>4$~\AA\
and detected \oii. The median R23 is relatively flat at lower
luminosities and then drops at the higher luminosity end.
The least-square polynomial fit is
\begin{equation}
\log({\rm R23}) = 0.61 - (0.08\pm0.04) \log(L_N) \,.
\end{equation}

%
%
\begin{inlinefigure}
\figurenum{27}
\centerline{\psfig{figure=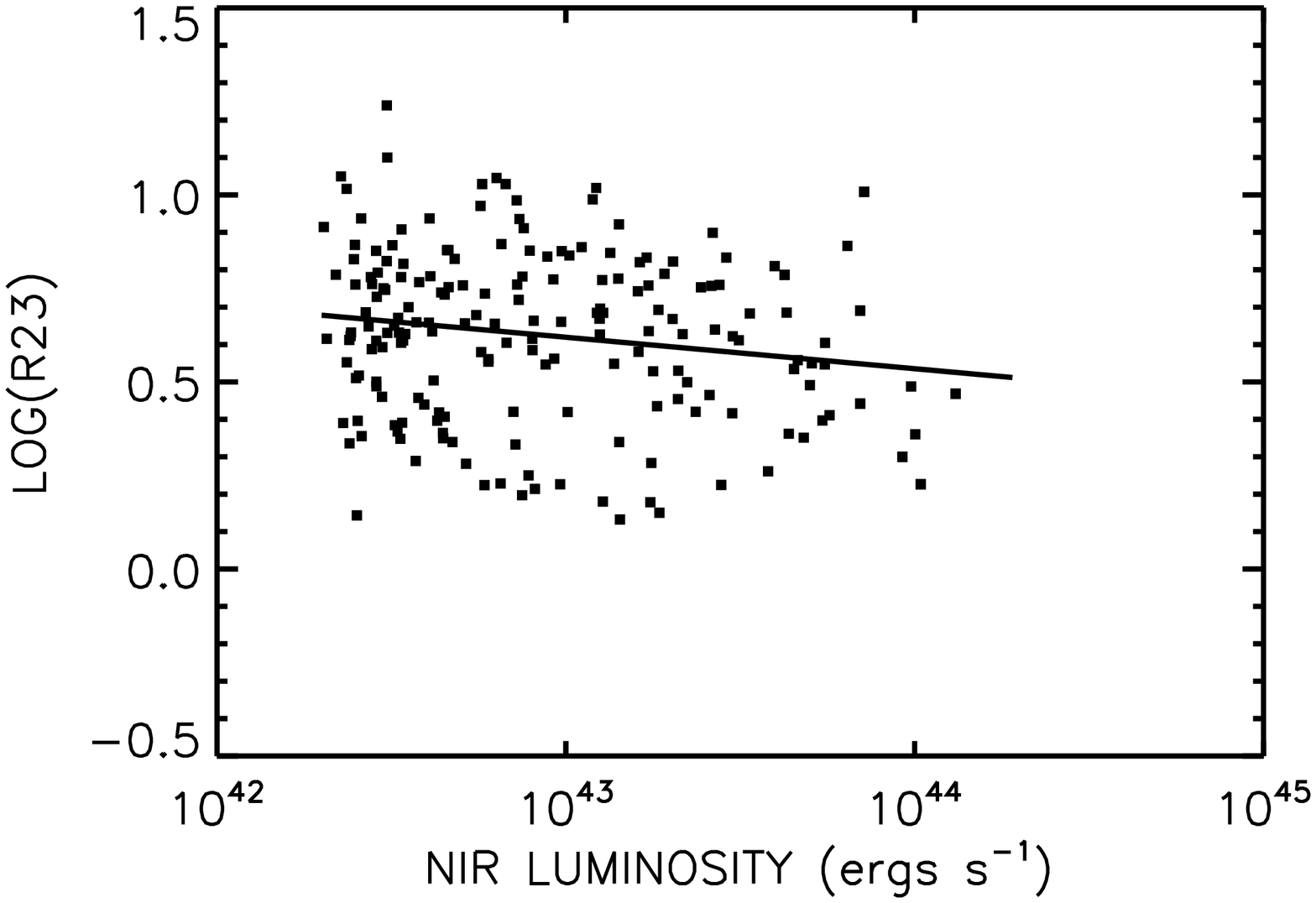,angle=90,width=3.5in}}
\vskip -0.2cm
\figcaption[]{
The extinction corrected R23 diagnostic ratio 
vs. NIR luminosity for the low-$z$ sample with EW(\hb$)>4$~\AA\ 
and detected \oii\ {\em (black squares)\/}.
The black solid line shows the least-square polynomial fit.
\label{r23_nirlum_ha}
}
\end{inlinefigure}

We can now also translate R23 and O32 to metallicity
using the KK04 calibration (their Eq.~18),
which is an average of the McGaugh (1991) 
and the KD02 R23 methods and is only valid for the upper branch. 
The dependence of the metallicity on the mass is then given by
\begin{equation}
12+\log({\rm O/H}) = 8.77 + (0.23\pm0.05) \log(M_{10}) \,,
\label{eqr23rel}
\end{equation}
which we show with a black solid line in Figure~\ref{r23_o_extinct_mass}.
Alternatively, we can translate R23 alone to $12+\log({\rm O/H})$
using Equation~1 of Tremonti et al.\ (2004;
hereafter, Tremonti04), which is an analytical fit to their
R23-metallicity relation and is also only valid for the upper branch.
In this case we find for the dependence of the metallicity on the
mass
\begin{equation}
12+\log({\rm O/H}) = 8.78 + (0.24\pm0.05) \log(M_{10}) \,,
\label{eqr23rel_tr}
\end{equation}
which has a very similar slope to Equation~\ref{eqr23rel} 
and, over the mass range $10^{10}$ to $10^{11}$~M$_\odot$, 
only differs in the intercept by 0.01~dex.

%
%
\begin{inlinefigure}
\figurenum{28}
\centerline{\psfig{figure=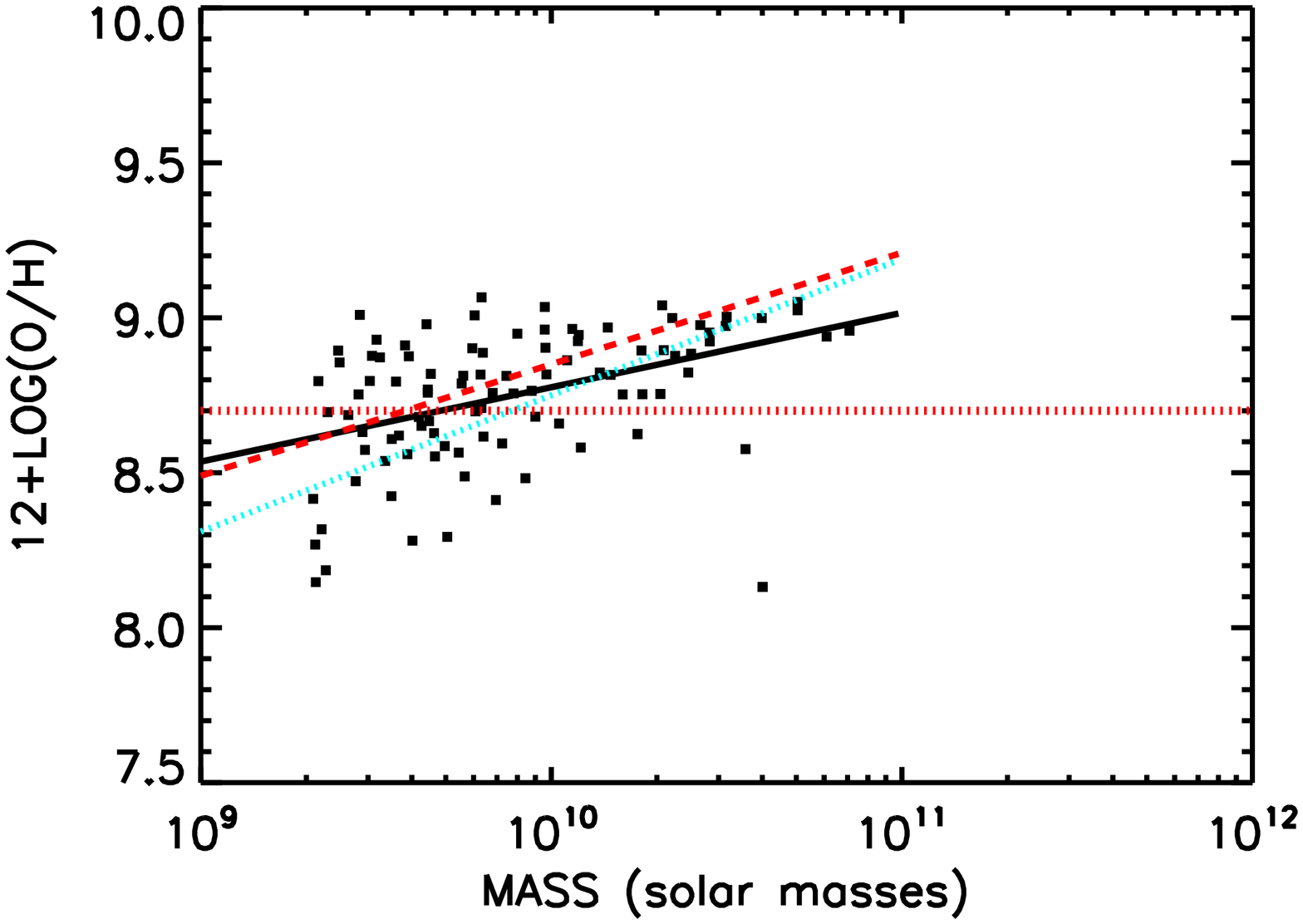,angle=90,width=3.5in}}
\vskip -0.2cm
\figcaption[]{
$12+\log({\rm O/H})$ computed from the extinction 
corrected R23 and O32 diagnostic ratios using the KK04 
calibration vs. galaxy mass for the low-$z$ sample with
EW(\hb$)>4$~\AA\ and detected \oii\ {\em (black squares)\/}.
The black solid line shows the least-square polynomial fit
(Eq.~\ref{eqr23rel}), which is flatter than the fits to the 
metallicities derived from the extinction corrected N2O2
diagnostic ratio (Eq.~\ref{eqnn202rel}; {\em cyan dotted line\/}) 
and the NH diagnostic ratio (Eq.~\ref{eqnnhorel}; 
{\em red dashed line\/}). 
The red dotted line shows the solar abundance.
\label{r23_o_extinct_mass}
}
\end{inlinefigure}

In Figure~\ref{r23_o_extinct_mass} we compare the fit to the
metallicities derived from R23 and O32 using the KK04
calibration {\em (black solid line)\/}
with the fits to the metallicities derived from N202 
{\em (cyan dotted line)\/} and NH {\em (red dashed line)\/}.
While the metallicities are close, using R23 gives a shallower 
slope, as can be seen quantitatively by comparing Equation~\ref{eqr23rel}
with Equations~\ref{eqnn202rel} and \ref{eqnnhorel}. The 
shallower slope is most likely a consequence of
our assumption that the bulk of the sources lie on the
high metallicity branch of R23. This assumption breaks
down at the low mass end, causing some of the low
metallicity branch points to be assigned artificially high 
metallicities.

We also calculated the values of the ionization parameter $q$ 
from R23 and O32 using the new parameterizations of 
KK04 (their Eqs.~13 and 17) for the KD02 models.
(Note that KK04 did
not give a parameterization for $q$ for their averaged KD02
and M91 R23 method, which we used to compute the metallicities 
in Figure~\ref{r23_o_extinct_mass}. However, since $q$ is not 
particularly sensitive to metallicity, the fact that we are 
using the updated KD02 parameterizations here to calculate the 
$q$ values should not be important.)
These $q$ values are very similar to those derived from N2O2. 
The median $q=2.3\times10^{7}$~cm~s$^{-1}$, and 80\% 
of the sample lies between $q=1.1\times10^{7}$~cm~s$^{-1}$
and $q=4.4\times10^{7}$~cm~s$^{-1}$.

\section{Metallicities in the Emission Line Galaxies: 
The [OIII] Sample}
\label{seco3}

At higher redshifts only the R23 diagnostic ratio can be computed
from the optical spectra. As can be seen from 
Figure~\ref{r23_o32_ha} (and as we discussed in \S\ref{secr23}), 
this results in some ambiguities in calculating the metallicities
since R23 is relatively insensitive to
metallicity at sub-solar values, and the translation from R23
to metallicity requires assumptions about which branch 
of the R23 relation the metallicity is on. We therefore first
analyze the raw R23 evolution without making any translation
to metallicity.

In Figure~\ref{r23_bylum} we show the dependence of R23 on
redshift for the mid-$z$ sample with EW(\hb$)>4$~\AA\ and 
detected \oii. We have divided the sample into 
three NIR luminosity intervals:
(a) $(5-20)\times 10^{43}$~ergs~s$^{-1}$;
(b) $(2-5)\times 10^{43}$~ergs~s$^{-1}$;
and (c) $(0.8-2)\times 10^{43}$~ergs~s$^{-1}$.
The purple solid line shows the least-square polynomial
fit to the data in (c). This same fit is also shown as a 
purple line in (a) and (b) for comparison with the fit to the 
data in each of those intervals {\em (black solid line)\/}.
In all three luminosity intervals we see a similar evolution
with redshift, with R23 increasing by about $0.2-0.3$~dex 
between $z=0.3$ and $z=0.9$. The parallel evolution of
the three luminosity intervals indicates that the shape of
the metallicity-mass relation does not evolve rapidly with
redshift over this redshift range. This result is not changed
if we exclude the strong redshift filament seen at $z=0.85$.
Since R23 decreases with increasing metallicity on the high 
metallicity branch (see Fig.~\ref{r23_o32_ha}a), 
we also see from Figure~\ref{r23_bylum} that high 
luminosity galaxies have higher metallicities than lower 
luminosity galaxies at all redshifts. Moreover, in a given
luminosity interval, the metallicity is increasing with decreasing
redshift. The dependence of the metallicity on redshift is
sufficiently strong that high luminosity galaxies at high
redshifts have comparable metallicities to low luminosity
galaxies at lower redshifts.

We may now compute the metallicity evolution with redshift
and luminosity more directly at the expense of both assuming
that the metallicity is on the upper branch of the R23
relation and adopting a specific calibration to metallicity.
In Figure~\ref{metal}
we show $12+\log({\rm O/H})$ computed from R23 and O32 in the
same manner as in \S\ref{secr23} using the KK04 calibration.
The points are separated into the redshift intervals $z=0.05-0.475$ 
{\em (red diamonds)\/} and $z=0.475-0.9$ {\em (black squares)\/} 
with median redshifts of 0.44 and 0.75. Even over this relatively 
narrow redshift range the average metallicity has dropped by about 
0.13~dex. Both distributions show an upper envelope of 
$12+\log({\rm O/H})\sim 9.1$. The spread to lower metallicities
at the higher redshifts should not be taken seriously as it is 
a consequence of applying the high metallicity branch equation 
to galaxies that are on the low metallicity branch.
We find the least-square polynomial fit to the metallicities, 
NIR luminosities, and $(1+z)$ values for all the galaxies together 
to be
\begin{equation}
12+\log({\rm O/H}) = 8.95 +0.18 \log(L_N) - 1.52 \log(1+z) \,,
\label{eqcombinedfit}
\end{equation}
which we show in Figure~\ref{metal} calculated at the median redshifts
of $0.44$ {\em (red solid)\/} and $0.75$ {\em (black solid)\/}, 
respectively. We also show the least-square polynomial fits to the
metallicities and NIR luminosities for $z=0.05-0.475$ 
{\em (red dashed)\/} and $z=0.475-0.9$ {\em (black dashed)\/}.

%
%
\begin{inlinefigure}
\figurenum{29}
\centerline{\psfig{figure=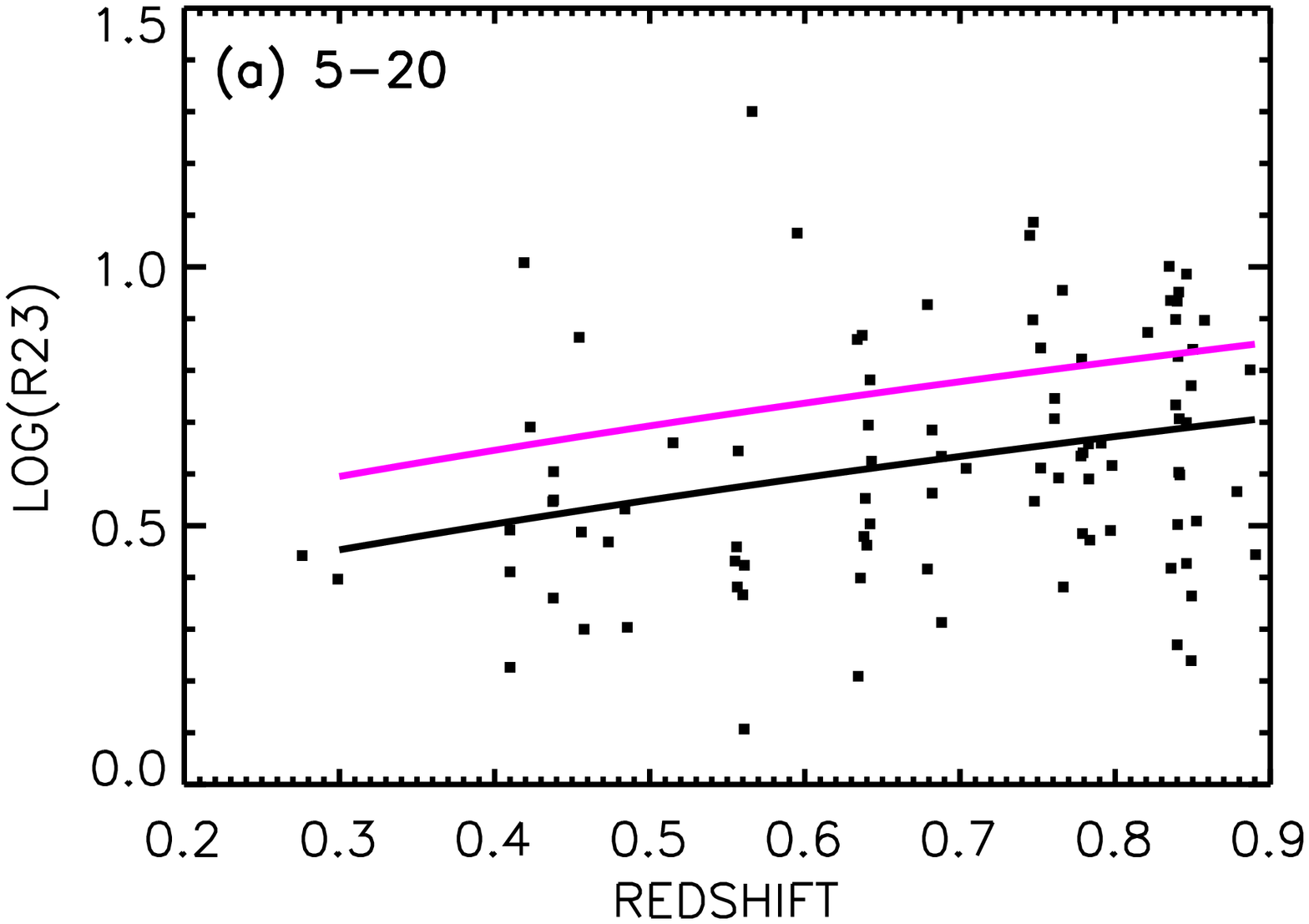,angle=90,width=3.5in}}
\vskip -0.6cm
\centerline{\psfig{figure=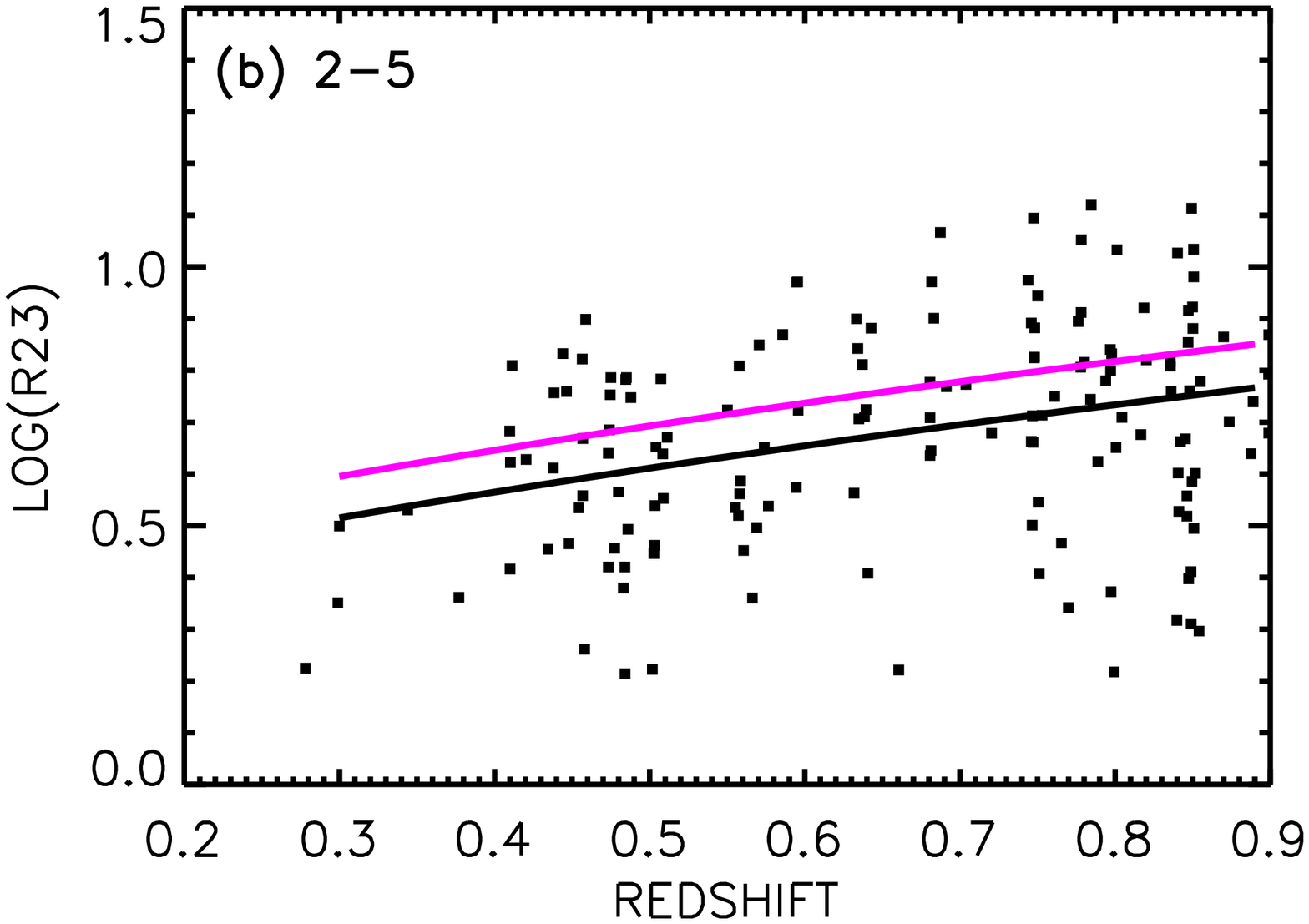,angle=90,width=3.5in}}
\vskip -0.6cm
\centerline{\psfig{figure=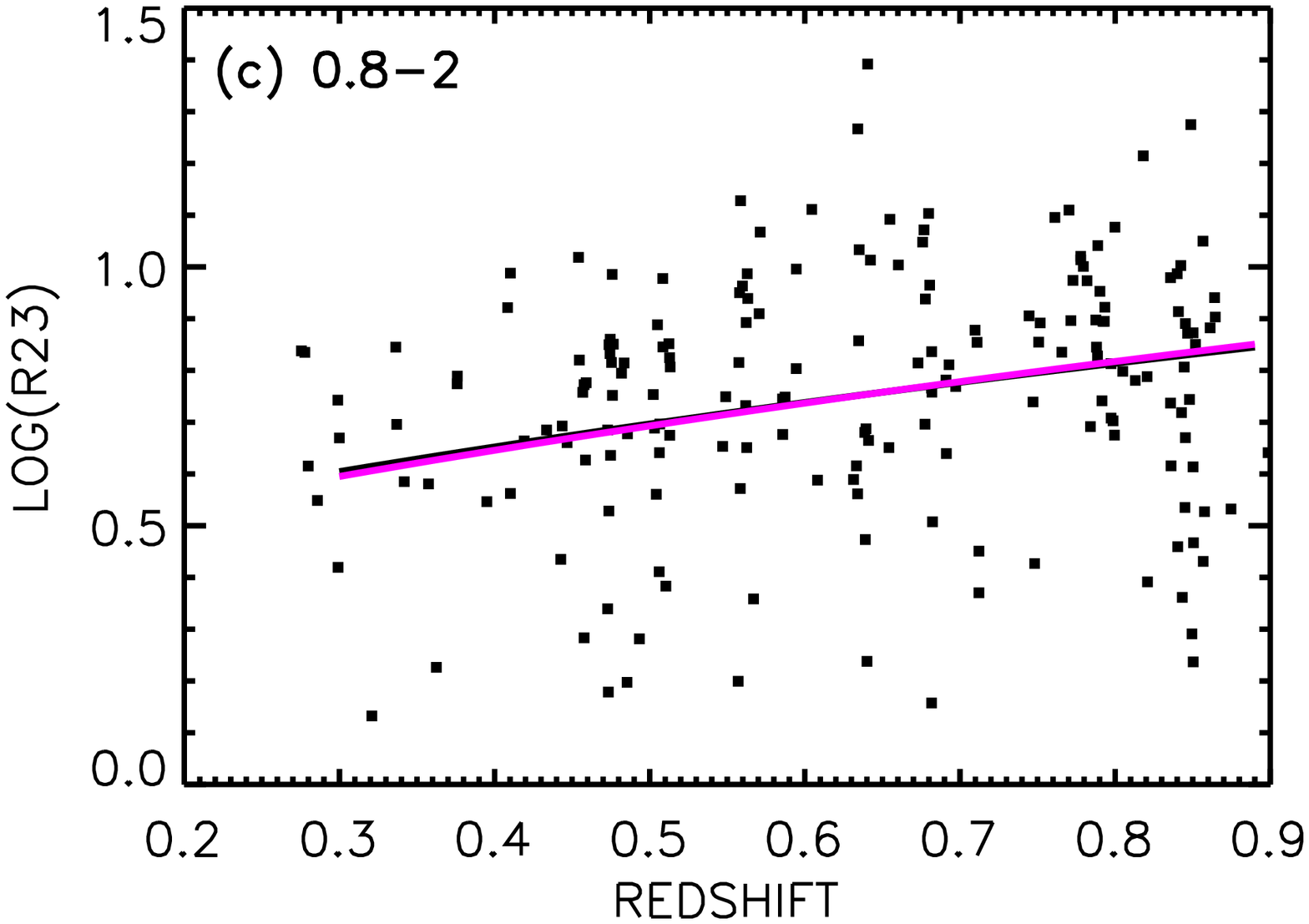,angle=90,width=3.5in}}
\vskip -0.2cm
\figcaption[]{
Logarithmic extinction corrected R23 vs. redshift for
the mid-$z$ sample with EW(\hb$)>4$~\AA\ and detected \oii\ 
for three NIR luminosity ranges: 
(a) $(5-20)\times10^{43}$~ergs~s$^{-1}$,
(b) $(2-5)\times10^{43}$~ergs~s$^{-1}$, 
and (c) $(0.8-2)\times10^{43}$~ergs~s$^{-1}$.
Each panel is marked with the luminosity range 
in units of $10^{43}$~ergs~s$^{-1}$.
In (a) and (b) the black solid line shows the least-square 
polynomial fit to $\log($R23) vs. $\log(1+z)$. In (a), (b), 
and (c) the purple solid line shows the least-square polynomial 
fit to the data in (c).
\label{r23_bylum}
}
\end{inlinefigure}

Using the KK04 calibration, we find the least-square polynomial 
fit to the metallicities and masses for $z=0.475-0.9$ to be
\begin{equation}
12+\log({\rm O/H}) = 8.72 + (0.13\pm0.06) \log M_{10} \,.
\label{eqr23hizrel}
\end{equation}
Using the Tremonti04 calibration, we find
\begin{equation}
12+\log({\rm O/H}) = 8.70 + (0.17\pm0.05) \log M_{10} \,.
\label{eqr23hizrel_tr}
\end{equation}
These two fits only differ by an average of 
0.02~dex over the $10^{10}-10^{11}$~M$_\odot$ range. Moreover,
within the wide errors, the slopes of both fits are 
consistent with the slopes of the fits given in 
Equations~\ref{eqr23rel} and \ref{eqr23rel_tr}, 
which were made to the low-$z$ sample alone over the wider mass 
range using the KK04 and Tremonti04 calibrations, respectively. 
To show the similarity of the slopes for the two redshift intervals, 
in Figure~\ref{hiz_r23_o_extinct_mass} we plot
Equations~\ref{eqr23rel} {\em (red line)\/} and \ref{eqr23hizrel}
{\em (black line)\/} on top of the $z=0.05-0.475$ 
{\em (red diamonds)\/} and $z=0.475-0.9$ {\em (black squares)\/} 
data restricted to masses above $10^{10}$~M$_\odot$, where the 
mid-$z$ sample is complete.

%
%
\begin{inlinefigure}
\figurenum{30}
\centerline{\psfig{figure=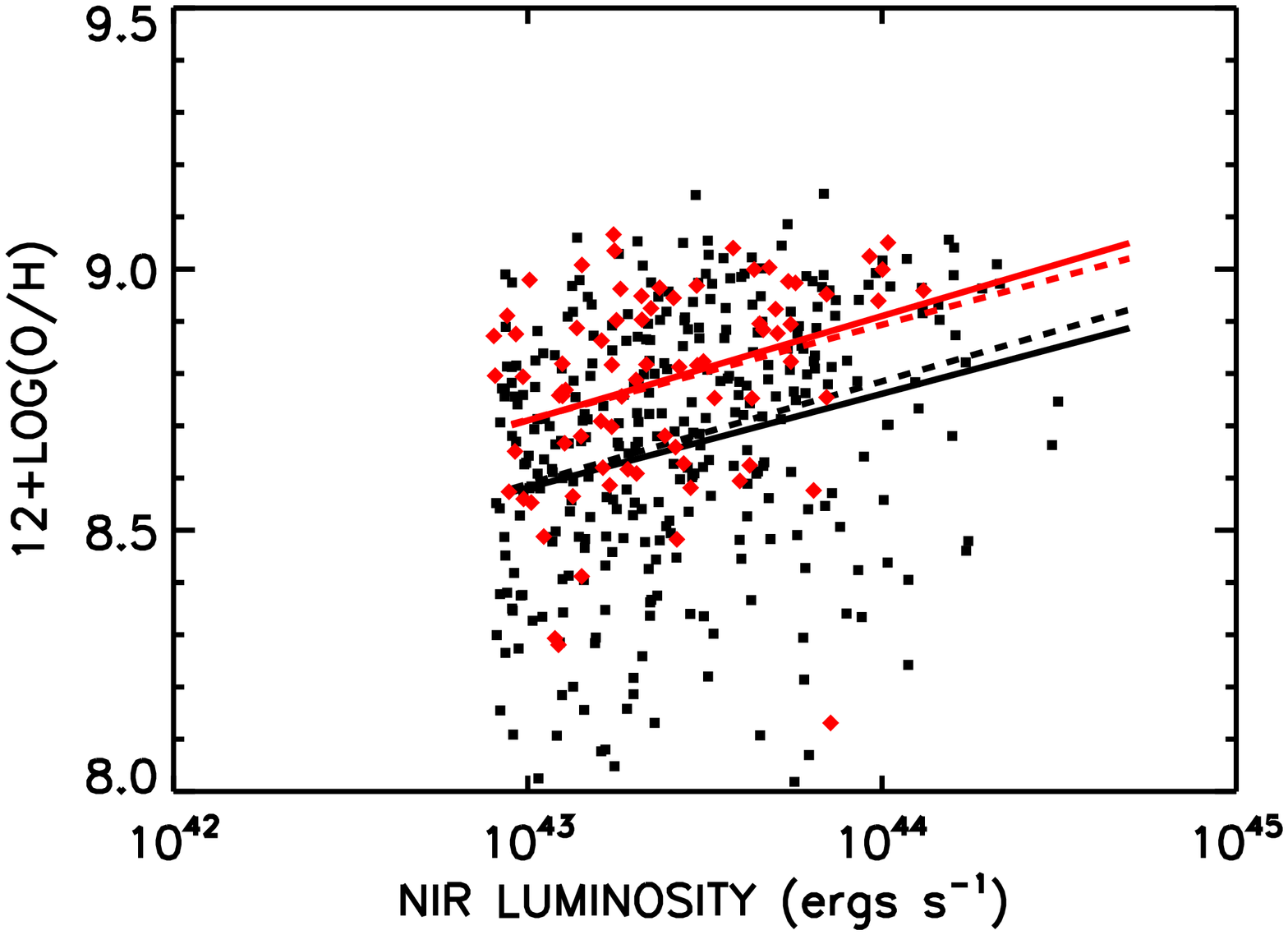,angle=90,width=3.5in}}
\vskip -0.2cm
\figcaption[]{
Metallicity vs. NIR luminosity for the mid-$z$ sample
computed from the extinction corrected R23 and O32 diagnostic ratios
using the KK04 calibration. The red diamonds show the galaxies with 
$z=0.05-0.475$. The black squares show the galaxies with $z=0.475-0.9$. 
Both have an upper limit of about 9.1 for $12+\log({\rm O/H})$. However, 
the mean metallicity has dropped by about 0.13~dex.
The solid lines show the least-square polynomial fit to the
metallicities, NIR luminosities, and $(1+z)$ values for all the
galaxies together (Eq.~\ref{eqcombinedfit}) computed at the 
median redshifts {\em (red for 0.44; black for 0.75)\/}. The dashed 
lines show the least-square polynomial fits to the metallicities and 
NIR luminosities for $z=0.05-0.475$ {\em (red)\/}
and $z=0.495-0.9$ {\em (black)\/}.
\label{metal}
}
\end{inlinefigure}

%
%
\begin{inlinefigure}
\figurenum{31}
\centerline{\psfig{figure=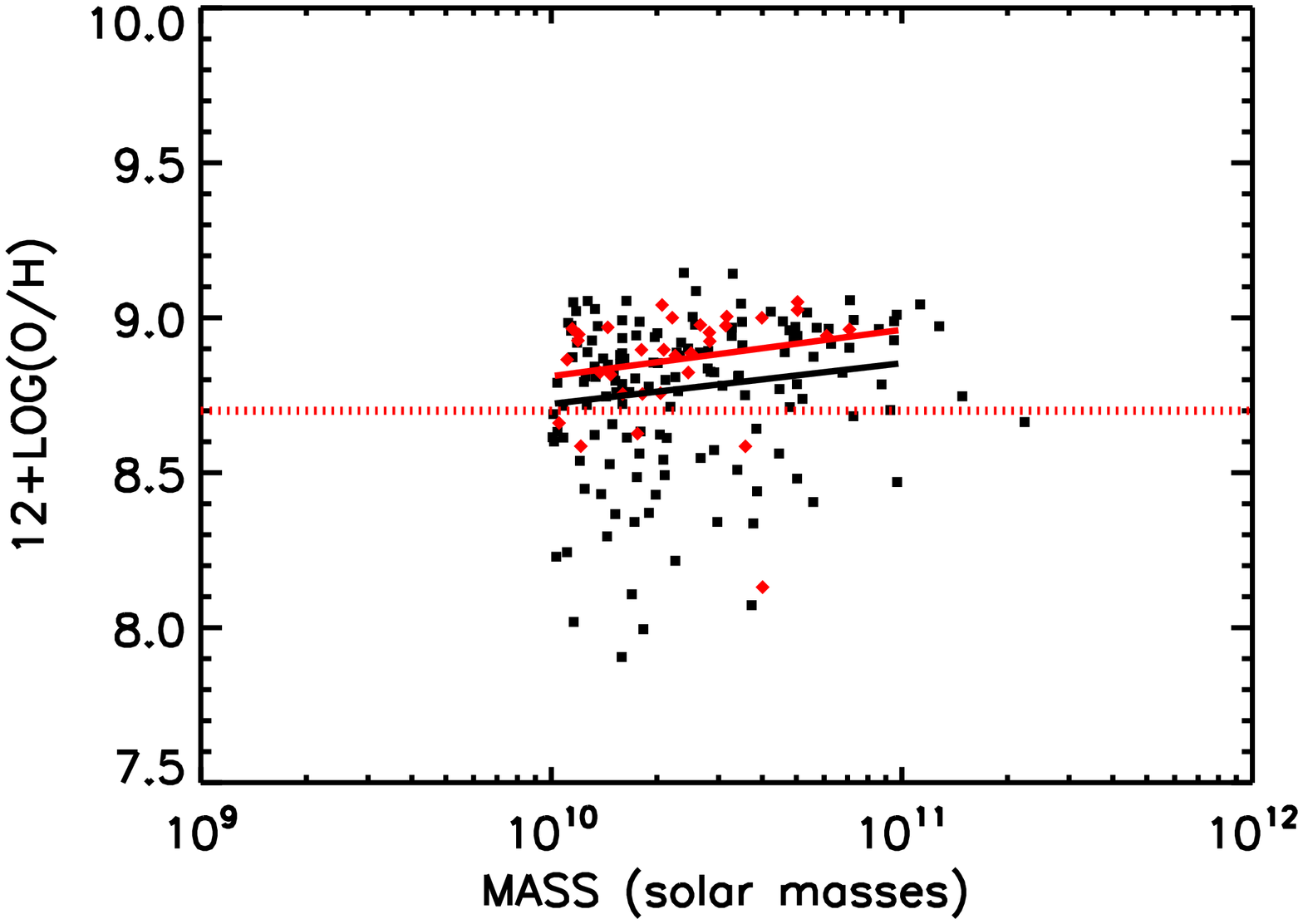,angle=90,width=3.5in}}
\vskip -0.2cm
\figcaption[]{
Metallicity vs. mass for the mid-$z$ sample computed 
from the extinction corrected R23 and O32 diagnostic ratios using
the KK04 calibration. The red diamonds show the galaxies with 
$z=0.05-0.475$.
The black squares show the galaxies with $z=0.475-0.9$. 
The solid lines show the least-square polynomial fits to
each redshift interval from Eq.~\ref{eqr23rel} {\em (red line)\/}
and Eq.~\ref{eqr23hizrel} {\em (black line)\/}.
\label{hiz_r23_o_extinct_mass}
}
\end{inlinefigure}

We can more clearly compare the metallicities
in the low-$z$ and mid-$z$ samples in histogram form.
In Figure~\ref{r23_o_hist} we show the distribution of metallicities 
computed using the KK04 calibration for the galaxies in the 
$10^{10}-10^{11}$~M$_\odot$ range. The $z=0.475-0.9$ ($z=0.05-0.475$)
interval is shown in black (red). The median metallicity in 
the mid-$z$ (low-$z$) interval is $8.79\pm0.03$ ($8.92\pm0.04$), 
where the errors give the 68\% confidence range calculated
using the median sign method. The increase of 0.13~dex in metallicity
from median redshift $0.75$ to median redshift $0.44$ is 
consistent with that found from the fits to the larger luminosity 
selected sample in Equation~\ref{eqcombinedfit}.

%
%
\begin{inlinefigure}
\figurenum{32}
\vskip -0.2cm
\centerline{\psfig{figure=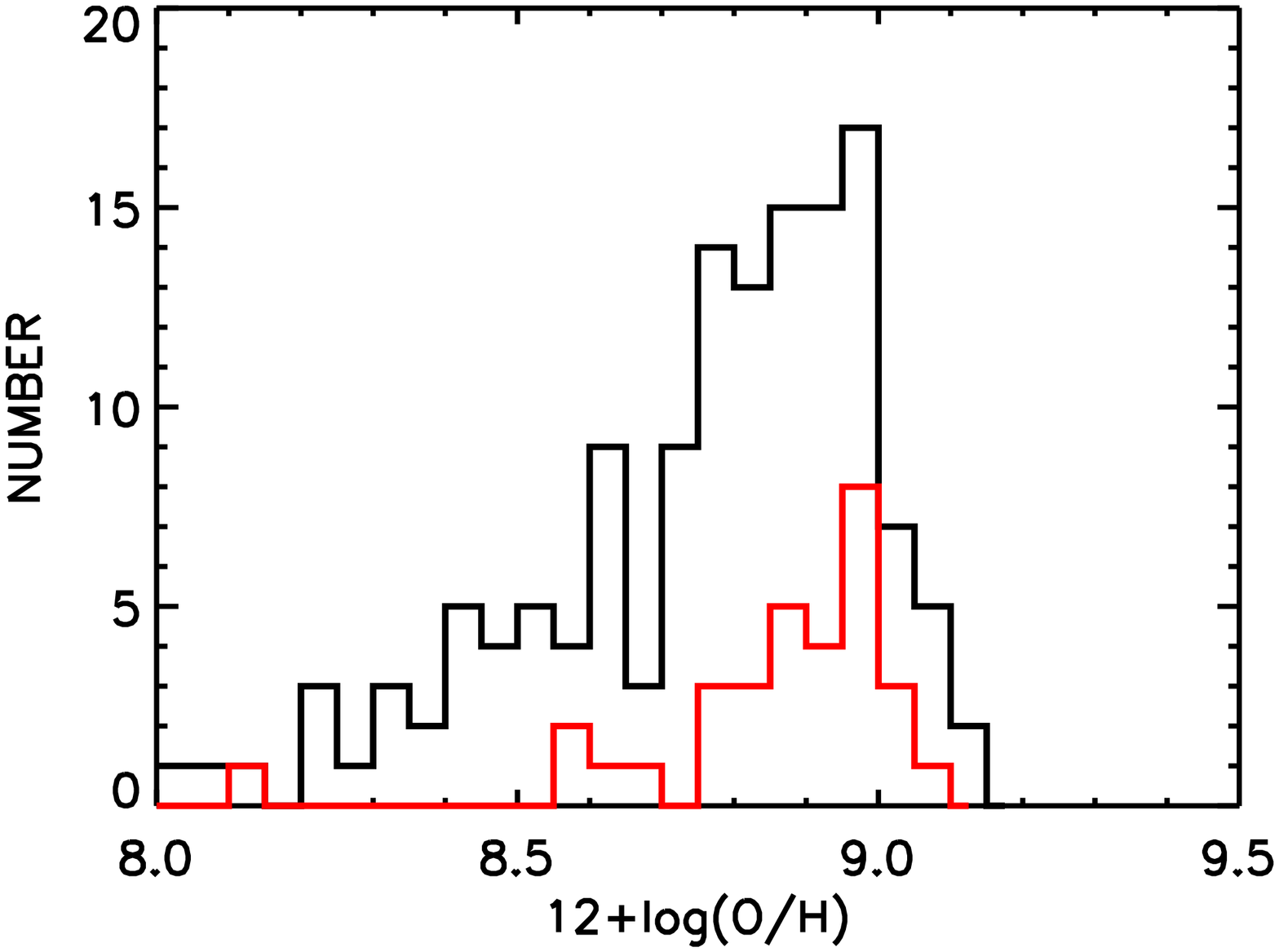,angle=90,width=3.5in}}
\figcaption[]{
Distribution of metallicities computed from
the extinction corrected R23 and O32 diagnostic ratios
using the KK04 calibration for the galaxies in the 
mid-$z$ sample with masses $10^{10}-10^{11}$~M$_\odot$.
The black (red) histogram shows the distribution
of galaxies between $z=0.475-0.9$ ($z=0.05-0.475$).
\label{r23_o_hist}
}
\end{inlinefigure}

\section{Missing Galaxies in the Metals Analysis}
\label{secabs}

The metals analysis of \S\ref{secha} and \S\ref{seco3}
only includes galaxies with strong emission lines, and this 
selection is a strong function of galaxy mass. In 
Figure~\ref{emission_fraction} we show the fraction of 
galaxies with strong enough emission lines to have been 
included in that analysis versus logarithmic mass for 
$z=0.05-0.475$ {\em (red diamonds)\/} and $z=0.475-0.9$ 
{\em (black squares)\/}. The fraction
decreases smoothly with increasing logarithmic mass, 
falling to zero above $\sim10^{11}~$M$_\odot$, depending
on the redshift interval. 

We also show the relative contributions to the mass density 
versus logarithmic mass from all the galaxies in the redshift 
intervals $z=0.05-0.475$ {\em (red histogram)\/} and 
$z=0.475-0.9$ {\em (black histogram)\/}. Combining this 
distribution with the fraction of galaxies in the strong emission
line sample, we find that only about 20\%$-30$\% of the 
total galaxy mass density is included in the strong emission line 
metallicity analysis.  
It is therefore important to understand 
the evolution of the remaining galaxies if we are to understand 
fully the metallicity-mass evolution.

We can divide the galaxies excluded from the strong emission 
line metallicity analysis into two categories: apparently 
passive galaxies with no clear signs of ongoing
star formation, and weakly active galaxies where
the source has EW(\hb$)<4$~\AA\ and either EW(\oii$)>5$~\AA\
or a $24~\mu$m detection. In Figure~\ref{emission_fraction}
we show the fraction of non-strong emission
line galaxies in the redshift intervals $z=0.05-0.475$ {\em (red)\/}
and $z=0.475-0.9$ {\em (black)\/} that are weakly active
{\em (solid curves)\/}. This fraction is quite
substantial. In fact, there are very few completely
passive galaxies below $\sim3\times10^{10}$~M$_\odot$ 
over the entire $z=0.05-0.9$ redshift range. However, 
$\sim40$\% of the total galaxy mass density
is contained in the passive population, and the majority 
of the massive galaxies ($>10^{11}$~M$_\odot$) fall into 
this category.

%
%
\begin{inlinefigure}
\figurenum{33}
\vskip -0.2cm
\centerline{\psfig{figure=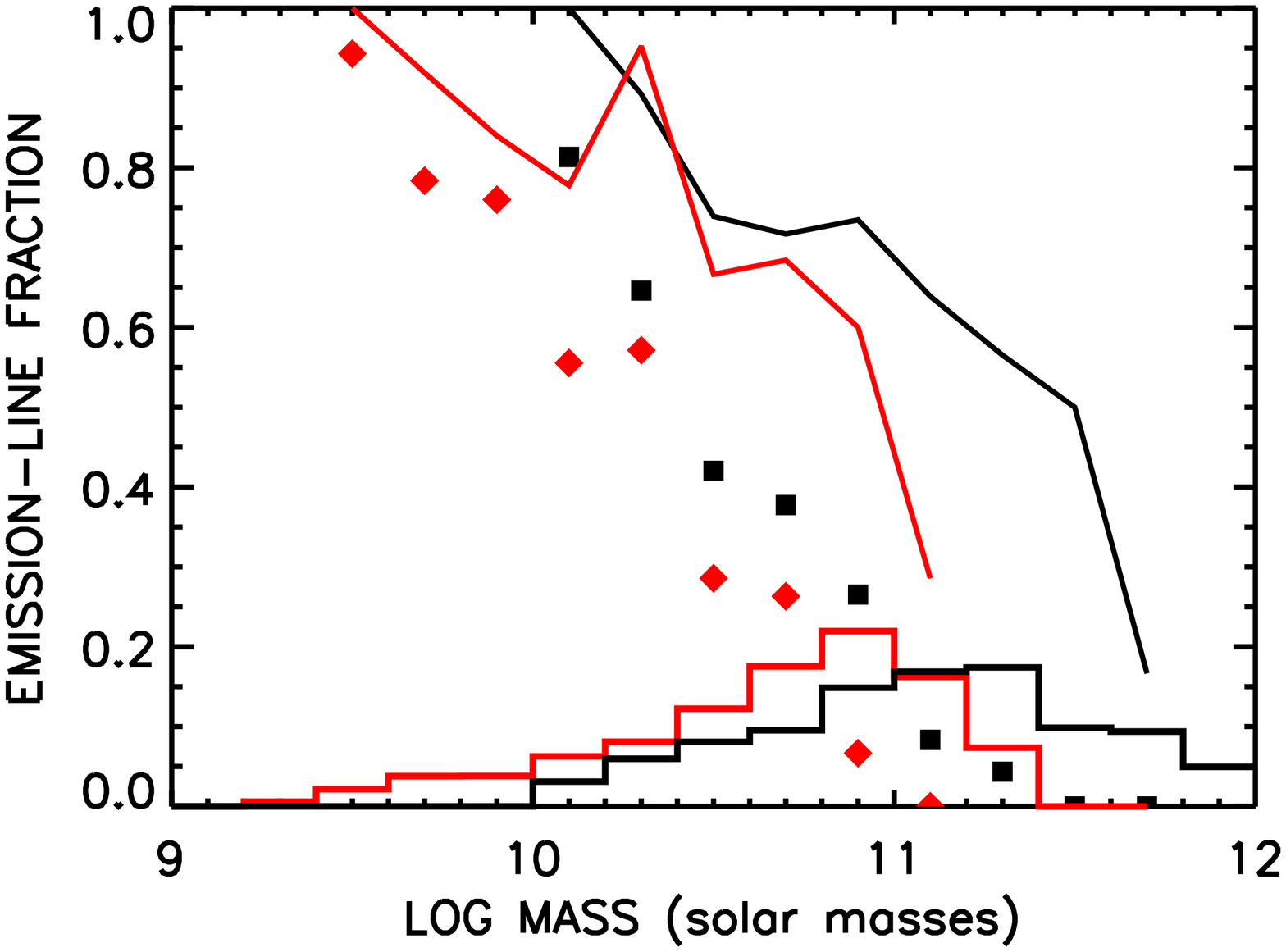,angle=90,width=3.5in}}
\figcaption[]{
Fraction of galaxies in the redshift intervals
$z=0.05-0.475$ {\em (red diamonds)\/} and $z=0.475-0.9$ 
{\em (black squares)\/} that are included in the strong emission
line metallicity analysis of \S\ref{secha} and \S\ref{seco3}
vs. galaxy mass. The red (black) histogram shows the fraction of 
the mass density in each bin that comes from the galaxies in the 
redshift interval $z=0.05-0.475$ 
($z=0.475-0.9$). The solid curves show the fraction of 
non-strong emission line galaxies 
in the redshift intervals $z=0.05-0.475$ {\em (red)\/} 
and $z=0.475-0.9$ {\em (black)\/} that are weakly active.
In all cases we only show the fractions where there are 
more than 4 objects in the bin.
\label{emission_fraction}
}
\end{inlinefigure}

\subsection{Passive Galaxies}

We formed stacked spectra of the passive galaxies in
each of the two redshift intervals by normalizing the
flux calibrated spectra in the $3500-5000$~\AA\ range and 
then averaging the spectra. Regions around the atmospheric
bands and the chip boundaries in the DEIMOS instrument
were blanked out in the individual spectra. 

In Figure~\ref{stack_passive}a we show the stacked spectrum 
{\em (black)\/} for the redshift interval $z=0.475-0.9$.
We fitted the spectrum by assuming a range of metallicities
and then found the best-fit BC03 model for each metallicity. 
Prior to measuring the $\chi^2$ we removed slight shape differences 
by fitting a second-order polynomial to the ratio of the stacked 
spectrum to the model spectrum. Unfortunately, even in the 
stacked spectrum there is almost no distinction between fits 
with very different metallicities because of the metallicity-age 
degeneracy. In the case of the $z=0.475-0.9$ stacked spectrum,
the best-fit solar metallicity model is a single burst 
with an age of 1.9~Gyr {\em (upper red spectrum)\/}. It closely 
matches the stacked spectrum. However, in Figure~\ref{stack_passive}b 
we show the best-fit $2.5\times$ solar metallicity model,
which is a single burst with an age of 1.0~Gyr {\em (upper red spectrum)\/}.
This fit is indistinguishable from the solar metallicity fit. 
In each panel we also show the residual {\em (lower red spectrum)\/}
between the best-fit model and the stacked spectrum. 
There is very weak \oii\ emission visible in the stacked spectrum 
but no \hb\ or \oiii, so there is no possibility of using the 
gaseous emission to estimate the metallicity.

%
%
\begin{inlinefigure}
\figurenum{34}
\centerline{\psfig{figure=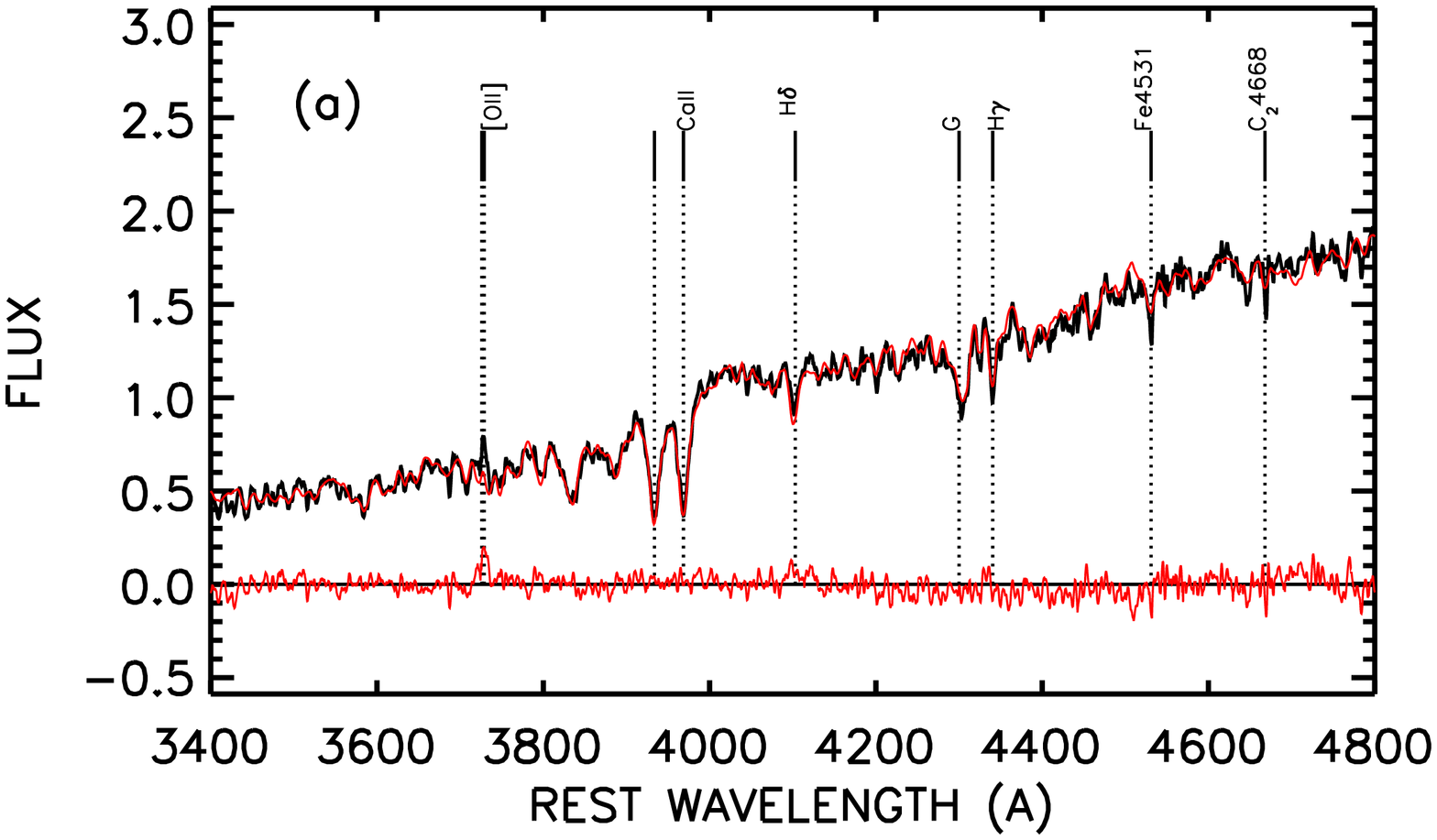,angle=90,width=3.5in}}
\vskip -1.4cm
\centerline{\psfig{figure=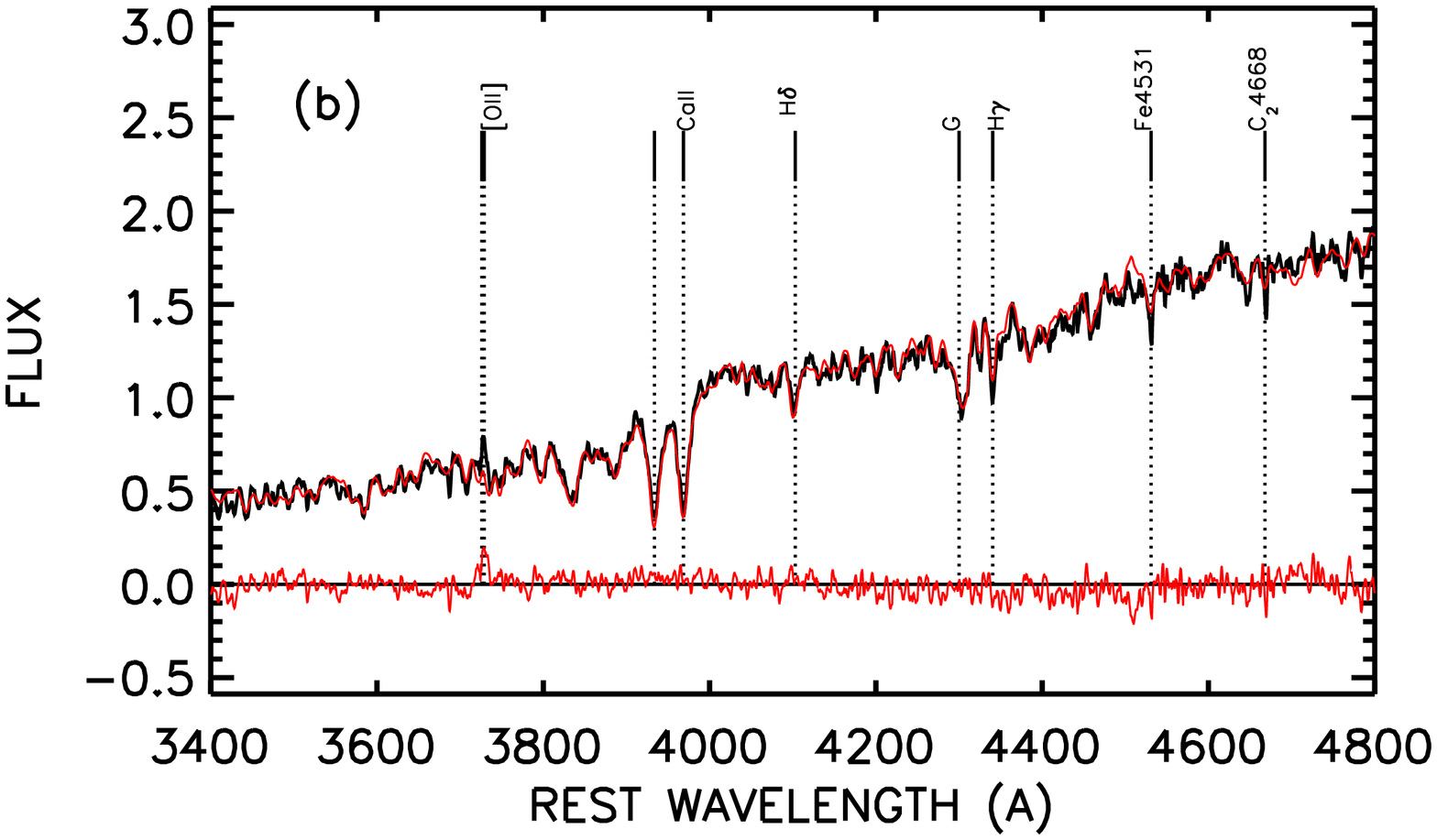,angle=90,width=3.5in}}
\vskip -0.2cm
\figcaption[]{
Stacked spectrum for passive galaxies in the
redshift interval $z=0.475-0.9$ formed by normalizing the individual 
extinction corrected spectra and averaging them {\em (black in each panel)\/}. 
The upper red spectrum in (a) is the best-fit BC03 solar metallicity
model, which is a single burst with an age of 1.9~Gyr. In (b) the
upper red spectrum is the best-fit BC03 $2.5\times$ solar metallicity
model, which is a single burst with an age of 1.0~Gyr.
The overall shapes are matched by applying a second-order 
polynomial over the wavelength range. The lower red spectrum
shows the residual difference between the data and the fit. 
The fits are both excellent and essentially indistinguishable.
\label{stack_passive}
}
\end{inlinefigure}

\subsection{Weakly Active Galaxies}

In contrast to the stacked spectrum for the 
passive galaxies, the stacked spectrum for the weakly active
galaxies clearly shows gaseous emission lines 
(see Fig.~\ref{stack_weak}b). In this section we will derive 
metallicities from the stacked spectrum for the weakly active
galaxies. However, to assess how well we are able to do this,
we will also derive metallicities from the stacked 
spectrum for the strong emission line galaxies and compare 
with the metallicities that we obtained previously for the 
individual sources.

In Figure~\ref{stack_weak}a we show the stacked spectrum
{\em (black)\/} for the strong emission line galaxies
[EW(\hb$)>4$~\AA] in the redshift interval $z=0.475-0.9$ 
and over the mass interval $10^{10.5}-10^{11}$~M$_\odot$. We 
overlay the best-fit BC03 solar metallicity
model, which is exponentially declining with $\tau=5$~Gyr 
and an age of 2.8~Gyr {\em (upper red)\/}. 
(Note again that it does not matter what metallicity we adopt,
as it would only affect the age.)
We also show the residual spectrum {\em (lower red)\/} 
after subtracting the best-fit model from the stacked spectrum.  

In Figure~\ref{stack_weak}b we show the stacked spectrum 
{\em (black)\/} for the weakly active galaxies in the same 
redshift interval and over the same mass interval.
We overlay the best-fit BC03 solar metallicity model, which is
exponentially declining with $\tau=5$~Gyr and an age of 7.5~Gyr
{\em (upper red)\/}. We again also show the residual 
spectrum {\em (lower red)\/}.

%
%
\begin{inlinefigure}
\figurenum{35}
\centerline{\psfig{figure=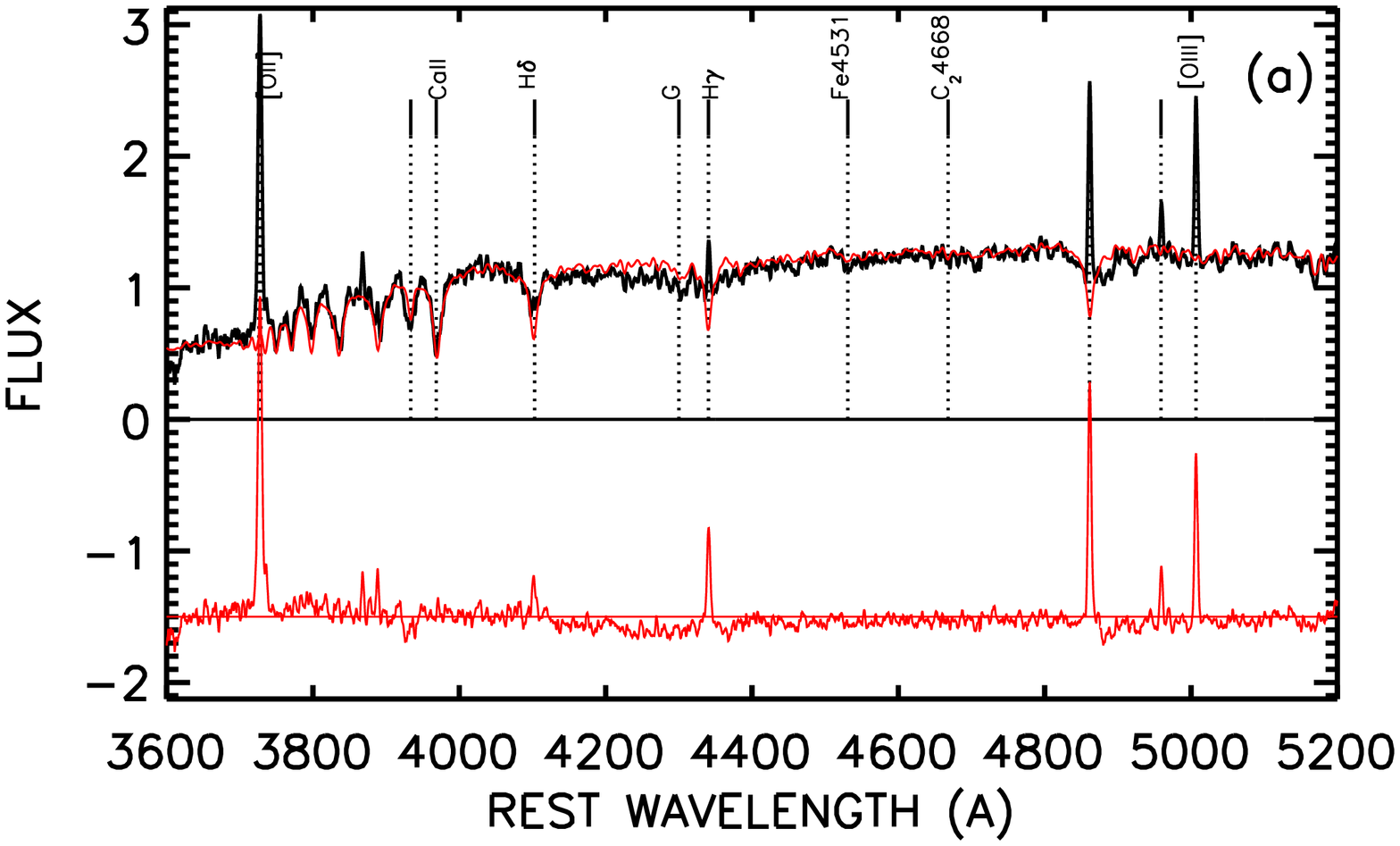,angle=90,width=3.5in}}
\vskip -1.3cm
\centerline{\psfig{figure=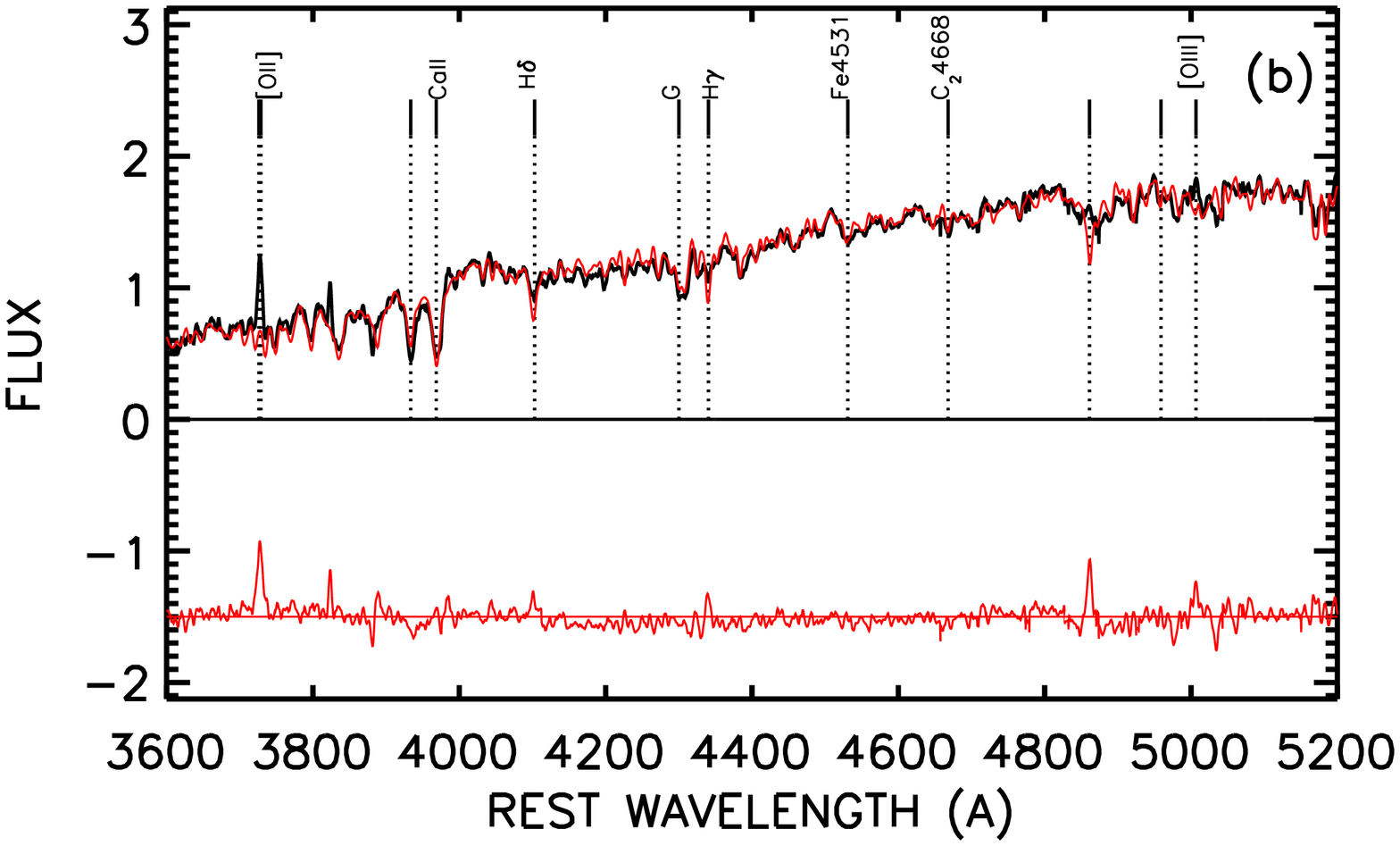,angle=90,width=3.5in}}
\vskip -1.3cm
\centerline{\psfig{figure=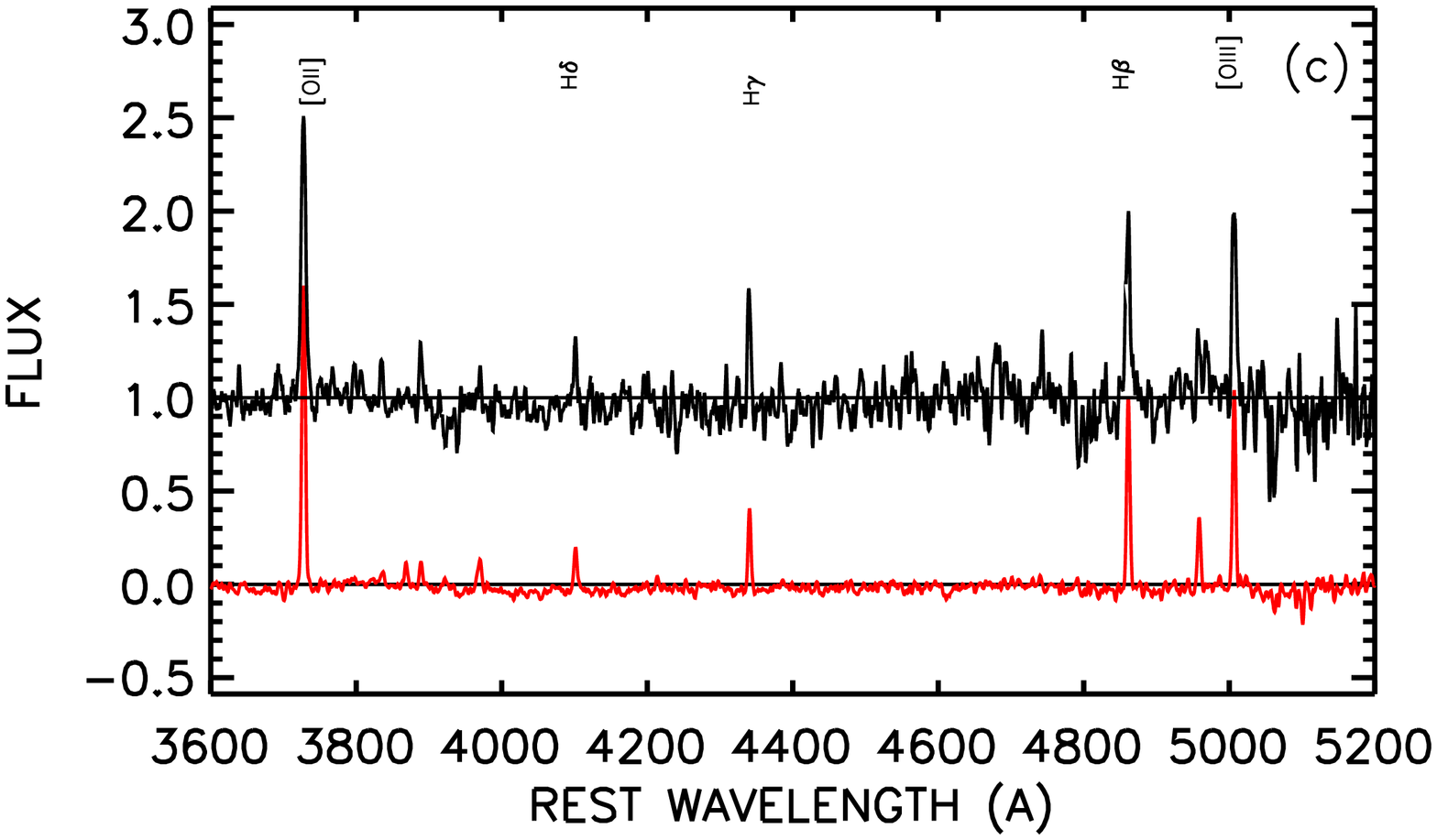,angle=90,width=3.5in}}
\figcaption[]{
Stacked spectra for galaxies with $z=0.475-0.9$ 
and masses $10^{10.5}-10^{11}$~M$_\odot$ formed by normalizing 
the individual extinction corrected spectra and averaging
them. (a) Strong emission line galaxies with EW(\hb$)>4$~\AA, 
all of which were included in the individual metallicity analysis
{\em (black)\/}. The upper red spectrum is the best-fit BC03 
solar metallicity model, which is exponentially declining 
with $\tau=5$~Gyr and has an age of 2.8~Gyr. 
(b) Weakly active galaxies with EW(\hb$)<4$~\AA\ 
and either EW(\oii$)>5$~\AA\ or a $24~\mu$m detection.
The upper red spectrum is the best-fit BC03 solar metallicity
model, which is exponentially declining with $\tau=5$~Gyr and 
has an age of 7.5~Gyr. In (a) and (b) the lower red spectrum 
shows the residual difference between the data and the fit.
(c) A direct comparison of the residual spectra from 
(a) {\em (red)\/} and (b) {\em (displaced black)\/}.
Both have been normalized to unity in the \hb\ line
so that the strengths of the other lines may be directly
compared. The measured R23 is 3.24 for the strong
emission line galaxies and 5.62 for the weakly active galaxies.
\label{stack_weak}
}
\end{inlinefigure}

In Figure~\ref{stack_weak}c we show the two residual
spectra normalized to unity in the \hb\ line so that the
strengths of the other lines may be directly compared.

We can make a rough analysis of the metallicities in the 
stacked spectra using the R23 method. 
The high degree of homogeneity we saw earlier
in the ionization parameters and in the metallicities at a 
given mass gives us some confidence that this procedure may
work reasonably well in analyzing the ensemble of galaxies.
The assumptions are
similar to those we made to determine the metallicities 
for the individual galaxies: namely, that the metallicity 
determined from the composite spectrum of H~II regions in the galaxy
roughly matches the average metallicity that would be derived 
from an independent analysis of the individual H~II regions 
in the galaxy. In general this type of averaging results
in a small underestimate (by about 0.1~dex) in the derived
oxygen abundance for the individual galaxies 
(Kobulnicky et al.\ 1999).

In Figure~\ref{active_stack_o} we compare the metallicities 
derived from the stacked spectrum for the strong emission line
galaxies (using the Tremonti04 calibration) {\em (solid symbols)\/} 
with the best fits to the metallicities of the individual galaxies 
{\em (solid lines)\/} from Equations~\ref{eqr23rel_tr} and 
\ref{eqr23hizrel_tr} (which also use the Tremonti04 calibration). 
The red (black) color denotes the $z=0.05-0.475$ ($z=0.475-0.9$) interval. 
We use the Tremonti04 calibration so that we may compare the
results directly with the local measurements in \S\ref{secdisc}.
The lower redshift points are in good agreement with the fit to the 
individual measurements, while the higher redshift points are about 
0.1~dex higher on average. The open red (black) symbols show the 
measurements from the weakly active galaxies for the redshift 
interval $z=0.05-0.475$ ($z=0.475-0.9$). 
These show substantial scatter, but the measurements are 
in broad agreement with the strong emission line galaxies. 
We will assume in \S\ref{secdisc} that the weakly active galaxies 
parallel the evolution of the strong emission line galaxies.

%
%
\begin{inlinefigure}
\figurenum{36}
\centerline{\psfig{figure=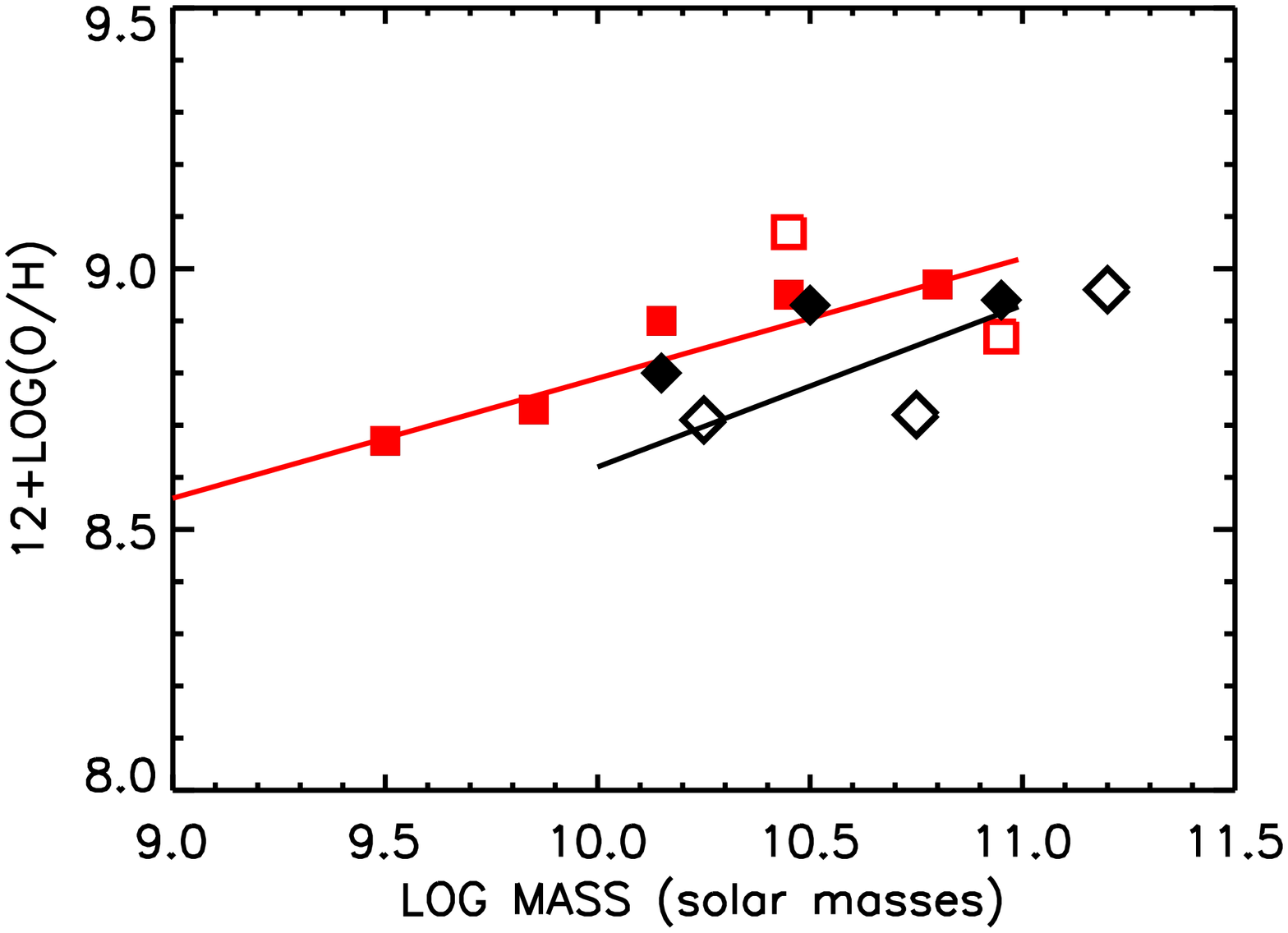,angle=90,width=3.5in}}
\vskip -0.2cm
\figcaption[]{
Metallicities derived from either the stacked spectrum 
for the strong emission line galaxies {\em (solid symbols)\/} or 
the stacked spectrum for the weakly active galaxies 
{\em (open symbols)\/} vs. logarithmic galaxy mass
for the redshift intervals $z=0.05-0.475$ {\em (red)\/}
and $z=0.475-0.9$ {\em (black)\/}. All metallicities were 
computed with the Tremonti04 calibration. The solid lines show
the fits to the individual galaxies in the strong 
emission line sample using the Tremonti04 calibration
[Eq.~\ref{eqr23rel_tr} {\em (red)\/} and 
Eq.~\ref{eqr23hizrel_tr} {\em (black)\/}]. 
\label{active_stack_o}
}
\end{inlinefigure}

\section{Discussion}
\label{secdisc}

We now bring together the results from the previous sections 
to form an integrated picture of galactic stellar mass assembly, 
star formation, and metallicity evolution. In
\S\ref{secgsmf} and \S\ref{secmassassembly} we measure the 
galactic stellar mass functions and the growth of the stellar 
mass densities with redshift. In \S\ref{secsfh} we measure
the star formation rate densities as a function of redshift and
galaxy mass. In \S\ref{secgrowth} we compute the expected 
formed stellar mass density growth rates produced by star formation 
and show that they match those measured from the formed stellar mass 
functions if the IMF is slightly increased from the Salpeter
IMF at intermediate masses ($\sim10~$M$_\odot$). In
\S\ref{secstarspec} we determine the instantaneous specific star 
formation rates, which give a quantitative description 
of the range of behaviors in the galaxies. We show that only 
galaxies below about $10^{11}~$M$_\odot$ are growing substantially 
over $z=0.05-1.5$. In \S\ref{seccolor} we analyze the distributions of
galaxy colors, equivalent widths, and 4000~\AA\ break strengths and 
find that star formation in all but the lowest mass galaxies in 
our sample is occurring in bursts with characteristic time intervals 
of about $4\times10^9$~yr. We also find that most of the growth in 
the mass density
is in the red sequence galaxies, whether these are chosen
by color or from the rest-frame EW(\oii). In \S\ref{secmassmorph} 
we show that as the redshift decreases, the galaxy
types smoothly migrate from spirals to S0s and elliptical
galaxies at all masses. The mass 
build-up is primarily in the E/S0 galaxies, which are also
the red sequence galaxies, and there is little change
in the galactic stellar mass function of the spirals.
In \S\ref{secenv} we find that although the masses are a 
strongly increasing function of environment, there is little 
redshift dependence in this relation. Unlike local results, 
the fraction of galaxies in the red sequence shows little 
environmental dependence and appears to depend primarily on 
the galaxy mass. Finally, in \S\ref{secmetev} we compare the 
metallicity evolution in the present sample with local and 
high-redshift metallicity measurements. In \S\ref{secgasmass} 
we combine the measured increases in metallicity with redshift 
with the metals returned from the measured SFRs to make a crude 
estimate of the galaxy gas mass reservoir, and we compare this 
to the stellar mass density as a function of galaxy mass.

\subsection{Galactic Stellar Mass Functions} 
\label{secgsmf}

In Figure~\ref{fig_mfun} we show the galactic stellar 
mass functions {\em (black squares)\/} in three redshift intervals
[(a) $z=0.9-1.5$, (b) $z=0.475-0.9$, and (c) $z=0.05-0.475$]
computed using the $1/V$ methodology 
(Felten 1976) and compared with the local mass function of
Cole01 adjusted to the present cosmology {\em (purple crosses)\/}. 
Both the Cole01 and our own mass functions are 
computed with the Salpeter IMF assumed throughout.

For each mass function we have computed the errors in two
ways. First, we assigned $1\sigma$ errors based on the number
of sources in each bin {\em (red error bars)\/}. 
These errors dominate at the high-mass end where there 
are small numbers of sources in each bin. Second, we
estimated the effect of cosmic variance in a simple empirical way.
In each redshift interval we eliminated the strongest redshift
sheet from the sample. For example, in the $z=0.475-0.9$
redshift interval we removed all of the sources lying
between $z=0.845$ and $z=0.86$ (see Figure~\ref{mass_byz}).
Typically the strongest single sheet will contain about
10\%$-20$\% of the galaxies in the redshift interval.
We then recomputed the mass function corresponding to the 
new volume and used the difference as an error estimate 
({\em black error bars\/}). These error estimates dominate 
at the low-mass end. The internal error estimate from this
method are similar in size to the analytic estimates of
Somerville et al. 2004 which would give systematic error
of 0.25, 0.15 and 0.1 dex in the $z=0.05-0.475$, $z=0.475-0.9$
and $z=0.9-1.5$ redshift intervals.

%
%
\begin{inlinefigure}
\figurenum{37}
\centerline{\psfig{figure=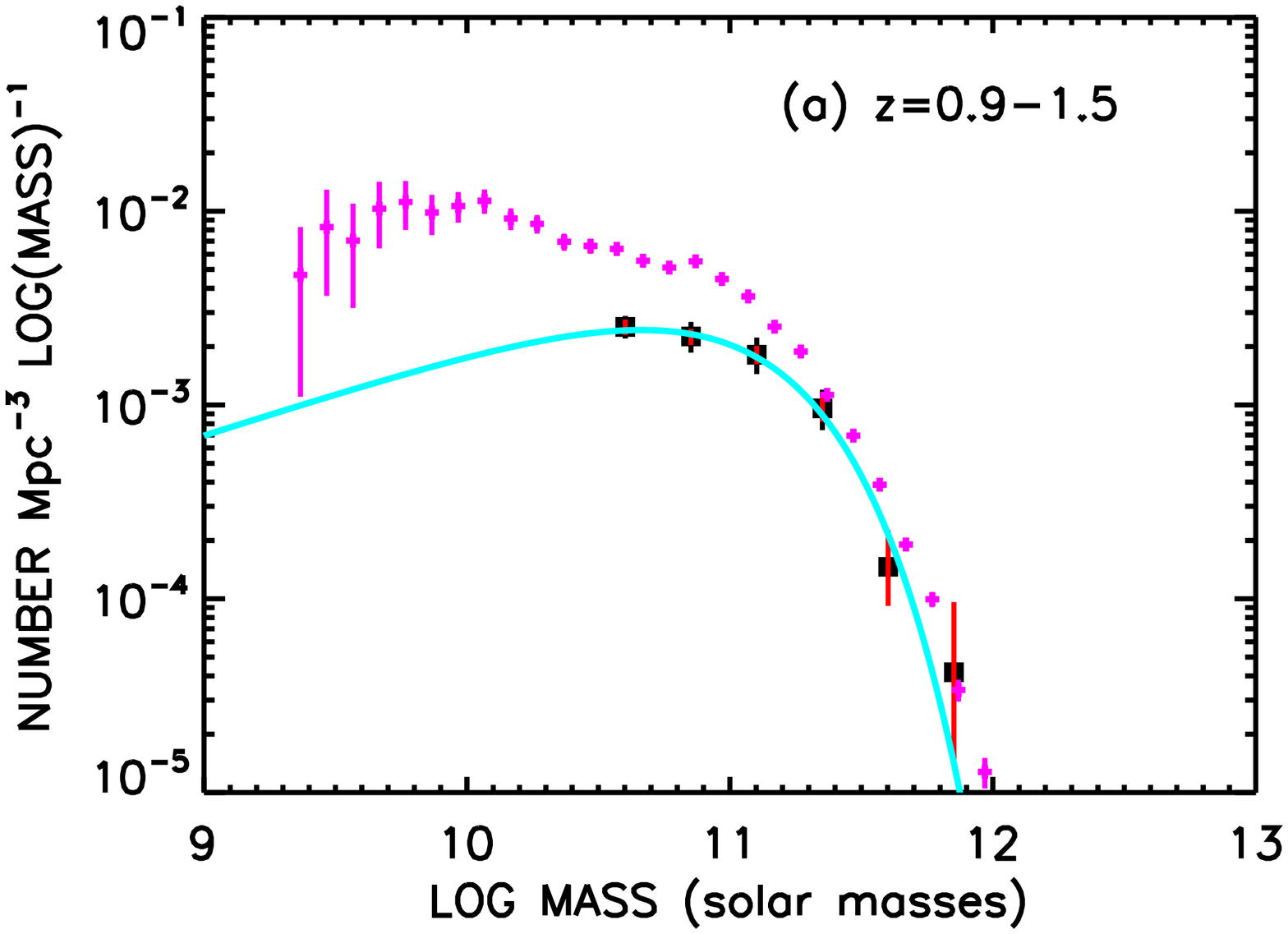,angle=90,width=3.5in}}
\vskip -0.6cm
\centerline{\psfig{figure=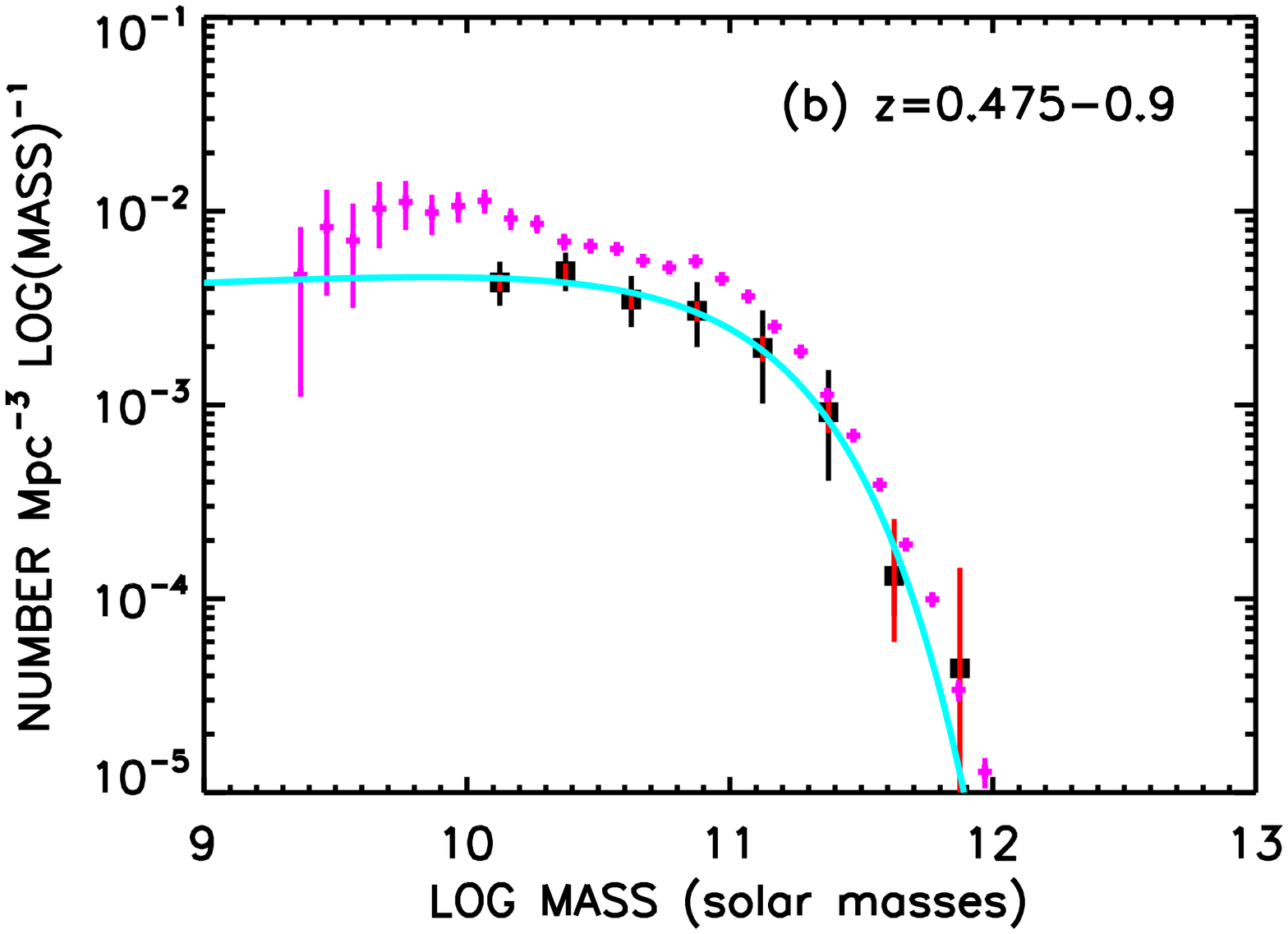,angle=90,width=3.5in}}
\vskip -0.6cm
\centerline{\psfig{figure=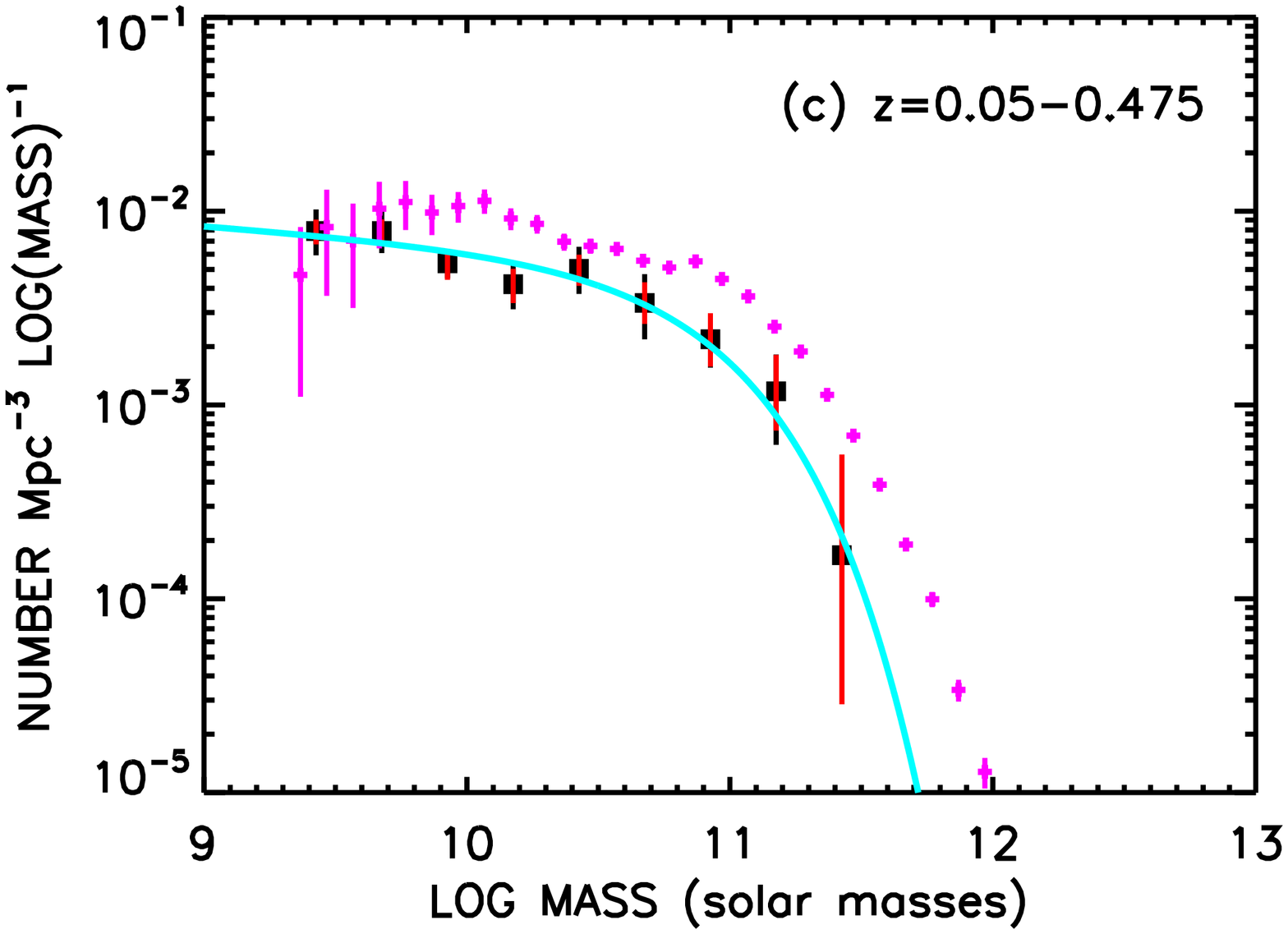,angle=90,width=3.5in}}
\vskip -0.2cm
\figcaption[]{
Galactic stellar mass functions in the redshift
intervals (a) $z=0.9-1.5$, (b) $z=0.475-0.9$, and
(c) $z=0.05-0.475$ {\em (black squares)\/}.
The red error bars are $1\sigma$, based on the
number of sources in each bin, while the black error 
bars are derived from excluding the strongest
velocity sheet in each redshift interval (see text for details).
The purple crosses and associated $1\sigma$ error bars 
are the same in all three redshift intervals and show the 
local mass function of Cole01 for the Salpeter 
IMF adjusted to the present cosmology. The cyan curves
show the best-fit Schechter functions obtained using
the Sandage et al.\ (1979) maximum likelihood method.
\label{fig_mfun}
}
\end{inlinefigure}

In order to provide parametric fits to the data, we have 
assumed a Schechter (1976) form, 
\begin{equation}
\phi(M) = \phi_{\star}\Biggl({M\over {M_{\star}(z)}}\Biggr)^{\alpha(z)}
{e^{-M/{M_{\star}(z)}}\over M_{\star}(z)} \,,
\label{eqnshec}
\end{equation}
for the mass function,
where $\phi(M)$ is the number of galaxies per unit mass per Mpc$^{3}$
at mass $M$. We used the Sandage et al.\ (1979) maximum likelihood
method to determine the power-law index $\alpha(z)$ and the characteristic mass 
$M_\star(z)$ for each of the three redshift intervals. The best-fit 
functions are shown in Figure~\ref{fig_mfun} {\em (cyan curves)\/}, 
and the derived $\alpha$ and $M_\star$ for each redshift 
interval are shown in Figure~\ref{mlf_all} together with the 
68\% and 95\% confidence error ellipses.
We obtained the normalizations $\phi_\star(z)$ by normalizing
to the number of objects in each redshift interval. We 
used our variance estimates (which dominate the error budget) 
in computing the error on this quantity. We summarize the 
maximum likelihood fits in each redshift interval in 
Table~\ref{sty_fits}, together with the 68\% confidence ranges.

%
%
\begin{inlinefigure}
\figurenum{38}
\centerline{\psfig{figure=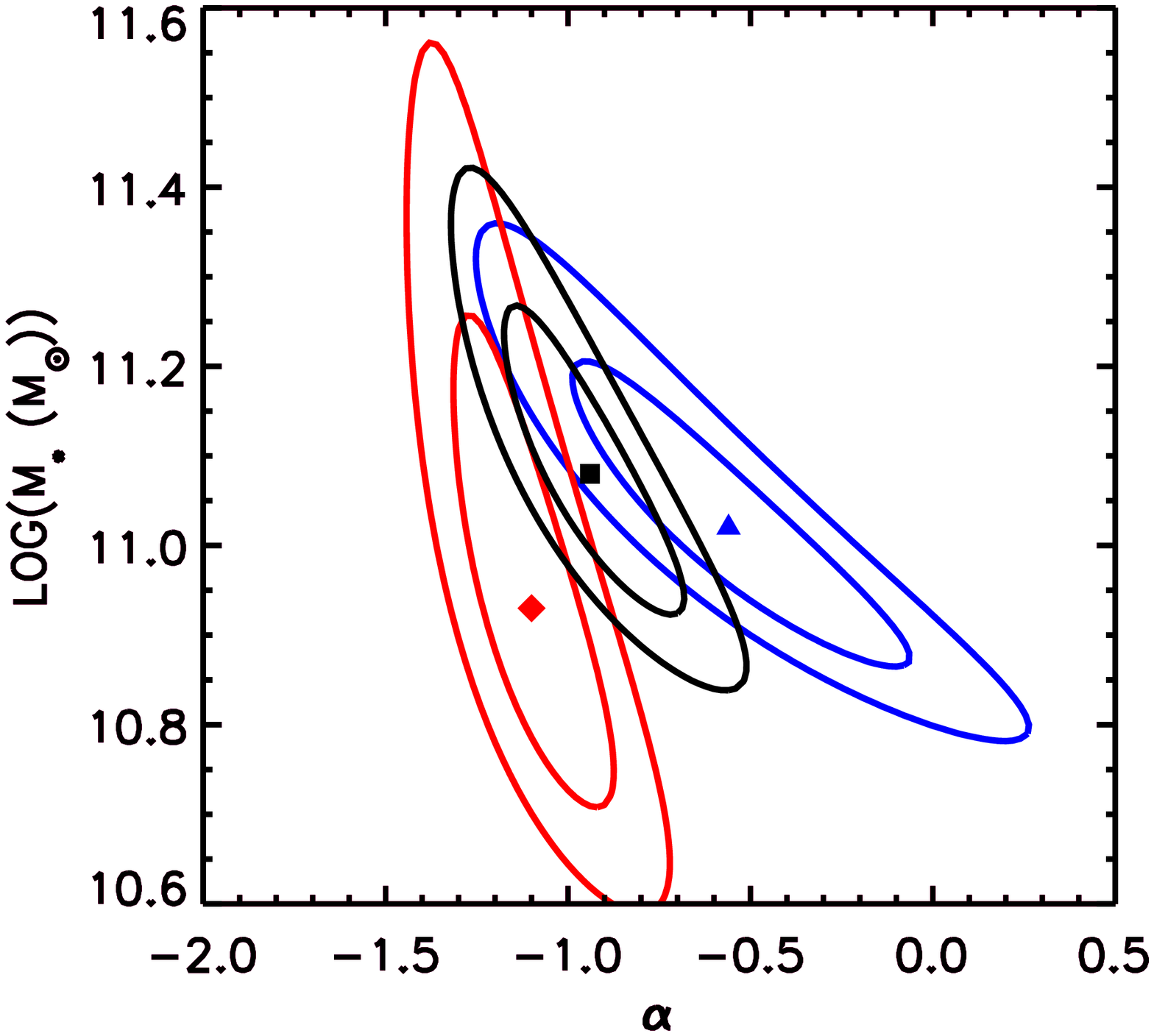,angle=90,width=3.5in}}
\vskip -0.2cm
\figcaption[]{
Sandage et al.\ (1979) fits to $\alpha(z)$ and $M_\star(z)$ for
the three redshift intervals $z=0.05-0.475$ {\em (red diamond)\/},
$z=0.475-0.9$ {\em (black square)\/}, and $z=0.9-1.5$ 
{\em (blue triangle)\/}. In each case the inner contour 
shows the 68\% confidence range and the outer contour shows
the 95\% confidence range for the symbol of the same color.
\label{mlf_all}
}
\end{inlinefigure}

A number of analyses of galactic stellar mass functions
have recently been undertaken (see references in \S\ref{secintro}),
and our results are broadly consistent. We may make
the most straightforward comparison with Fontana et al.\ (2006;
hereafter Fontana06), who analyze their data in a very similar 
way and who also use the Salpeter IMF. While their GOODS-MUSIC 
sample relies heavily on photometric redshifts, in other respects 
it is comparable to the present sample. Their Figure~4 bears a 
striking resemblence to our Figure~\ref{fig_mfun}.
Conselice07 use a much larger field size with almost
twenty times as many objects and also draw a similar conclusion.
The agreement between fields is reassuring 
that our results are robust and are not being dominated by cosmic 
variance.

The effects of downsizing can be clearly seen in both
Fontana06's Figure~4 and our Figure~\ref{fig_mfun}: the galaxy 
number densities at the low-mass 
end are still rising down to the lowest redshift interval, while 
the number densities at the high-mass end ($>10^{11}$~M$_\odot$) 
are changing much more slowly over $z=0.05-1.5$.

Over this redshift range Fontana06 find a relatively invariant
$\alpha(z)\sim -1.2$, but their $M_\star(z)$ rises by about 
0.12~dex from $z=0$, where $\log M_\star =11.16$ from Cole01, 
to $z=1.35$. Our $\alpha(z)$ values are slightly greater than 
this, though consistent with a constant $\alpha(z)=-1.2$ within
the $2\sigma$ errors (see Fig.~\ref{mlf_all}).
Our slightly greater $\alpha(z)$ values are
a consequence of Fontana06 having a turn-up 
at the low-mass end (see their Fig.~4).
It is quite likely that this is a result of 
their use of photometric redshifts, which are more problematic 
for low-mass galaxies because such galaxies are predominantly blue 
and harder to fit, but it could also be a measure of the cosmic 
variance.

However, we do see a significant evolution in the mean mass 
(a rise in either $M_\star$ or $\alpha$ with increasing redshift) 
in our data. In particular, the fitted values in the highest redshift 
interval are not consistent at the $3\sigma$ level with those
in the lowest redshift interval (see Fig.~\ref{mlf_all}). We can see 
this evolution most clearly by adopting a fixed $\alpha(z)=-1.18$ 
from the local Cole01 fit and computing 
$M_\star(z)$ for this fixed slope. This is given as the final 
column in Table~\ref{sty_fits}, and it shows a rise of about 
0.12~dex between $z=0$ (Cole01) and our highest redshift interval, 
an identical result to that of Fontana06.

%
%
\begin{inlinefigure}
\figurenum{39}
\centerline{\psfig{figure=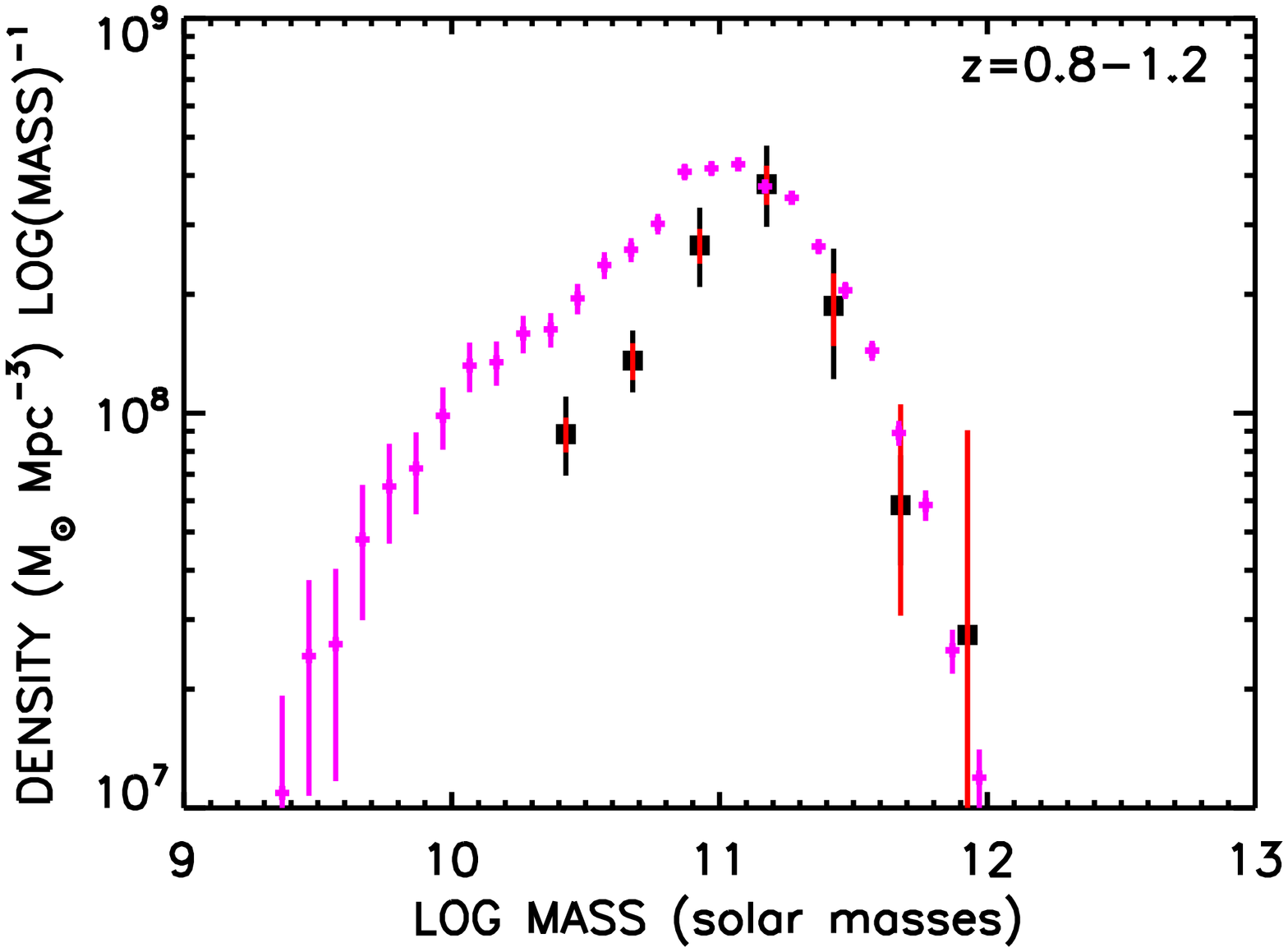,angle=90,width=3.5in}}
\vskip -0.2cm
\figcaption[]{
Stellar mass density per logarithmic galaxy mass interval
in the redshift range $z=0.8-1.2$ {\em (black squares)\/}
compared with the local distribution from Cole01 {\em (purple crosses)\/}.
The error bars are as in Figure~\ref{fig_mfun}.
\label{mass_den_plot}
}
\end{inlinefigure}

The change in $M_\star(z)$ is a numerical consequence of the 
build-up of the low-mass region of the galactic stellar mass
function relative to the high-mass region. 
In Figure~\ref{mass_den_plot} we compare the 
stellar mass density per logarithmic galaxy mass interval
in the redshift range $z=0.8-1.2$ {\em (black squares)\/} with 
that from Cole01 locally {\em (purple crosses)\/}. Both distribution 
functions are strongly peaked with most of the mass density lying in 
galaxies with masses close to the peak value. However, the Cole01
function is broader and extends to lower masses.

The peak of the mass density per logarithmic
galaxy mass, $M_\star(z) (2\alpha(z)+1)$,
provides an alternative way to characterize the Schechter function
(see Baldry et al.\ 2006). The evolution of the peak is 
better defined than the evolution of either $\alpha(z)$ and 
$M_{\star}(z)$ separately. The location of the peak increases
from $11.07\pm0.01$~M$_\odot$ in Cole01 and
10.88 ($10.77-11.09$)~M$_\odot$ in our lowest redshift interval
to $11.18\pm0.03$~M$_\odot$ in our highest redshift interval.

\subsection{Stellar Mass Density Growth with Redshift} 
\label{secmassassembly}

In Figure~\ref{mden_byz} we plot the stellar mass density
versus redshift for various mass
intervals. In both panels we use black solid squares
and 68\% confidence limits to show the mass density
evolution for all of the galaxies in our sample
above $10^{10.5}$~M$_\odot$.
We also obtained the local mass density above $10^{10.5}$~M$_\odot$
by integrating the Cole01 function.
We denote this by a black open square, which
we show extended to all redshifts {\em (black dashed
line)\/} for ease of comparison.
$10^{10.5}$~M$_\odot$ is the lowest mass to which we can measure 
the mass density over our entire $z=0.05-1.5$ redshift range.

%
%
\begin{inlinefigure}
\figurenum{40}
\centerline{\psfig{figure=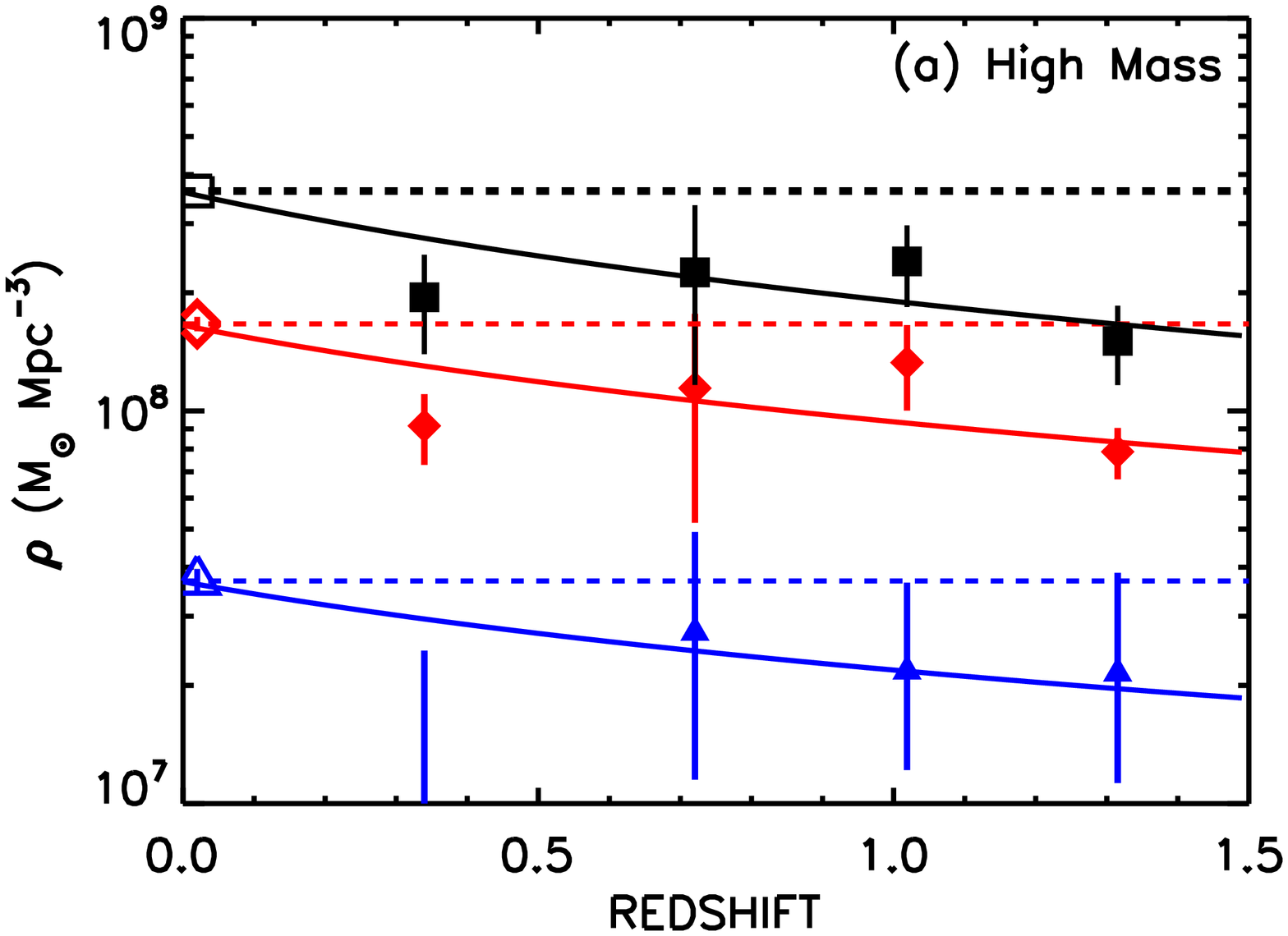,angle=90,width=3.5in}}
\vskip -0.6cm
\centerline{\psfig{figure=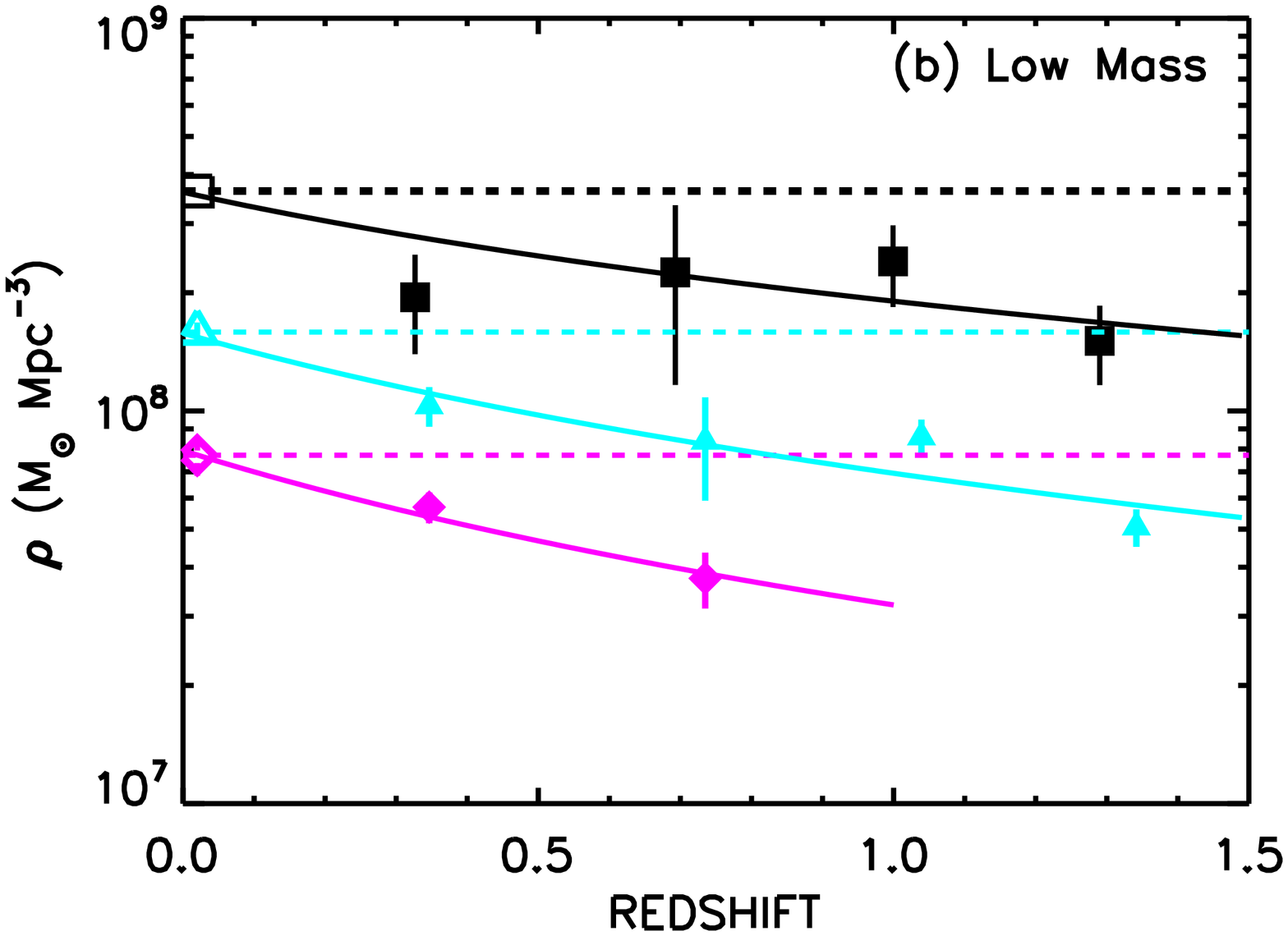,angle=90,width=3.5in}}
\vskip -0.2cm
\figcaption[]{
Universal stellar mass density vs. redshift. In both panels
the black solid squares show the evolution
of the mass density for all sources above $10^{10.5}$~M$_\odot$,
and the black open square (and black dashed line) shows the local 
mass density above $10^{10.5}$~M$_\odot$ obtained by integrating the 
Cole01 function. 
(a) The evolution in the mass intervals $10^{11.5}-10^{12}$~M$_\odot$
{\em (blue triangles, highest-mass interval)\/} and
$10^{11}-10^{11.5}$~M$_\odot$ {\em (red diamonds, high-mass interval)\/}.
(b) The evolution in the mass intervals
$10^{10.5}-10^{11}$~M$_\odot$ {\em (cyan triangles, intermediate-mass 
interval)\/} and $10^{10}-10^{10.5}$~M$_\odot$
{\em (purple diamonds, low-mass interval)\/}.
The low-mass interval is only shown at $z<0.9$
where it is complete. Errors are the maximum of the numerical
and variance errors discussed in the text.
The least-square polynomial fits of $\log \rho$ vs. 
$\log(1+z)$ are shown with the solid curves.
The local mass densities for each of the mass intervals 
obtained by integrating the Cole01 function 
{\em (open symbols, color corresponds to the mass interval)\/} 
are also shown extended to all redshifts {\em (dashed lines)\/}.
\label{mden_byz}
}
\end{inlinefigure}

In Figure~\ref{mden_byz}a
we show the mass density evolution for two high-mass 
intervals: $10^{11.5}-10^{12}$~M$_\odot$ {\em (blue triangles;
hereafter, our highest-mass interval)\/}
and $10^{11}-10^{11.5}$~M$_\odot$ {\em (red diamonds;
our high-mass interval)\/}. 
In Figure~\ref{mden_byz}b we show the mass density evolution
for two lower mass intervals: $10^{10.5}-10^{11}$~M$_\odot$
{\em (cyan triangles; our intermediate-mass interval)\/} 
and $10^{10}-10^{10.5}$~M$_\odot$ {\em (purple diamonds;
our low-mass interval)\/}. We show the evolution
in the low-mass interval only to $z=0.9$ where the sample is complete. 
We obtained the local mass density for each of the mass 
intervals by integrating the Cole01 function over 
those intervals {\em (open symbols)\/}. For ease of comparison,
we show those local mass densities extended to all redshifts
{\em (dashed lines)\/}.

Locally almost equal mass densities lie in the high
{\em (open red diamond in Fig.~\ref{mden_byz}a)\/} and intermediate 
{\em (open cyan triangle in Fig.~\ref{mden_byz}b)\/}
mass intervals. However, 
the high-mass interval material is changing more slowly
than the intermediate-mass interval material, which is growing 
smoothly over the whole $z=0.05-1.5$ range. The low-mass interval 
material is also growing smoothly over $z=0.05-0.9$. 
The highest-mass interval material has a slow
evolution (and indeed is consistent with 
having no evolution) over the entire $z=0.05-1.5$ range, though 
the error bars are large and the amount of material
contained in this mass interval is small. 

The mass-sliced data may be most easily compared with
Conselice07, who present their data in a similar way. 
While the Conselice07 sample 
is a large one (almost 50,000 objects), the data are not of the 
same quality as that in Fontana06 or in the present paper.
Only about 22\% of their sample have spectroscopic
redshifts from the DEEP2 observations,
and their photometric redshifts were calculated using 
only 4 ($BRIK$; half the sample) or 5 ($BRIJK$) band colors. 
These are too few colors to get reliable measurements of all of 
the desired quantities (i.e., extinction, photometric redshifts,
mass, age/metallicity, and evolutionary model).
For comparison, recall that over the redshift range $z=0.05-1.5$
we have spectroscopic redshifts 
for 84\% of the galaxies in our uniform NIR flux-limited 
sample (see \S\ref{secspec}). 
Moreover, for the very small fraction
of sources where we used photometric redshifts, they were derived
from 13  band colors ($UBVRIz'JHK_s$ and the four {\em Spitzer\/} 
IRAC bands). Fontana06 used 14 band colors to determine their 
predominantly photometric redshifts.

Despite these limitations, the agreement is good.
We can compare our Figure~\ref{mden_byz} with Figure~4 of
Conselice07 if we convert our stellar masses and stellar mass 
densities to the Chabrier (2003) IMF used by Conselice07 by 
dividing by 1.70, as discussed in \S\ref{secintro}.
Our results extend to lower masses than Conselice07 since our 
NIR sample is deeper.
However, if we compare the region of mass and redshift where
the two studies overlap, then we find good agreement between the
two results in both absolute 
value and shape. Both analyses show slow growth at the
higher masses and more rapid growth at the lower masses. The one 
quantitative difference is that we do not reproduce the sharp 
drop in the mass density which Conselice07 see in their 
$10^{11}-10^{11.5}$~M$_\odot$ 
interval above $z=1.2$. It is not clear whether this is 
simply a result of cosmic variance (neither of the surveys are 
large enough for cosmic variance not to matter), or whether it
is related to Conselice07's use of 4 or 5 band photometric redshifts.
Since the present work is based on a much deeper and far more 
spectroscopically complete sample than that of Conselice07, our 
results should be more reliable if this is the explanation.

In order to quantify the results of Figure~\ref{mden_byz}, we 
made least-square polynomial fits to the logarithmic stellar 
mass densities [including the local values obtained by integrating 
the Cole01 function] versus the logarithmic cosmic time. 
We show these in Figure~\ref{mden_byz} with solid curves. 
For the total mass above $10^{10.5}$~M$_\odot$ we find
\begin{equation}
\log {\rm \rho_\ast(>10.5~{\rm M_\odot})} = 8.56 + 
(0.73\pm0.16) \log(t/t_{0})  \,,
\label{massint1}
\end{equation}
and for the mass intervals we find
\begin{equation}
\log {\rm \rho_\ast(10^{10}-10^{10.5}~{\rm M_\odot})} = 
7.89 + (1.11\pm0.28) \log(t/t_{0})  \,,
\label{massint2}
\end{equation}
\begin{equation}
\log {\rm \rho_\ast(10^{10.5}-10^{11}~{\rm M_\odot})} = 
8.20 + (0.95\pm0.10) \log(t/t_{0})  \,,
\label{massint3}
\end{equation}
\begin{equation}
\log {\rm \rho_\ast(10^{11}-10^{11.5}~{\rm M_\odot})} = 
8.22 + (0.65\pm0.12) \log(t/t_{0})  \,,
\label{massint4}
\end{equation}
\begin{equation}
\log {\rm \rho_\ast(10^{11.5}-10^{12}~{\rm M_\odot})} = 
7.57 + (0.59\pm0.37) \log(t/t_{0})  \,.
\label{massint5}
\end{equation}
The fit for Equation~\ref{massint2} is over the $z=0.05-0.9$ redshift
range, while the remaining fits are over the $z=0.05-1.5$ range.
The low-mass ranges ($<10^{11}$~M$_\odot$) are growing approximately
linearly with time, while the high-mass ranges are growing more
slowly. Thus, the low-mass galaxies have roughly constant stellar
mass density growth rates, and it is the drop 
in the growth rates in the high-mass galaxies that is responsible for 
the overall drop in the growth rates seen in Equation~\ref{massint1}.

We also computed the total stellar mass density
evolution by integrating to the limiting mass in each redshift interval 
and extrapolating to estimate the contribution from lower mass galaxies.
This correction is not large since
the total mass density is dominated by galaxies
near $M_\star(z)$. Locally about 72\% of the mass density lies
in galaxies above $10^{10.5}$~M$_\odot$ and about
89\% in galaxies above $10^{10}$~M$_\odot$.
If we use our best-fit Schechter function, then in the $z=0.9-1.5$
interval about 88\% of the mass density lies
in galaxies above $10^{10.5}$~M$_\odot$ and about
97\% in galaxies above $10^{10}$~M$_\odot$. If we instead
force-fit to $\alpha=-1.18$, then the percentages are
77\% and 91\%, respectively. When we include these corrections,
the total stellar mass density has a slightly steeper
dependence on redshift than does the mass density above 
$10^{10.5}$~M$_\odot$. Using, respectively, our best-fit Schechter 
function and the fit where $\alpha$ is forced to $-1.18$, we find
\begin{equation}
\log {\rm \rho_\ast(total)} = 8.77 + (0.91\pm0.15) \log(t/t_{0})  \,,
\label{massinttot1}
\end{equation}
and
\begin{equation}
\log {\rm \rho_\ast(total)} = 8.77 + (0.80\pm0.15) \log(t/t_{0})  \,.
\label{massinttot2}
\end{equation}

\subsection{Star Formation History}
\label{secsfh}

We tested our empirical SFR calibrations (\S\ref{seccal})
by calculating the universal SFH from $z=0.05-1$ for 
all of the galaxies in our NIR sample. We show this in 
Figure~\ref{ha_rho_star_bymass},
where we use red open (solid) squares to denote the star formation
rate densities (SFRDs) calculated from the \hb\ (UV) luminosity
densities after applying our extinction corrections.
As is well known, the extinction corrections are
substantial, typically factors of five in the various redshift
bins. At $z=0.9$ about 80\% of the light is dust
reradiated. Our extinction corrected values at $z>0.3$ agree broadly
with the many measurements in the literature (see the references
in \S\ref{secintro}). We explicitly compare
with the values derived from radio and submillimeter
data from the stacking analysis of Wang et al.\ (2006)
{\em (black squares)\/}. These should be a good measure of the 
total star formation history, including highly-obscured sources.
The good agreement suggests that we are seeing most of
the higher redshift star formation in our present sample.

%
%
\begin{inlinefigure}
\figurenum{41}
\centerline{\psfig{figure=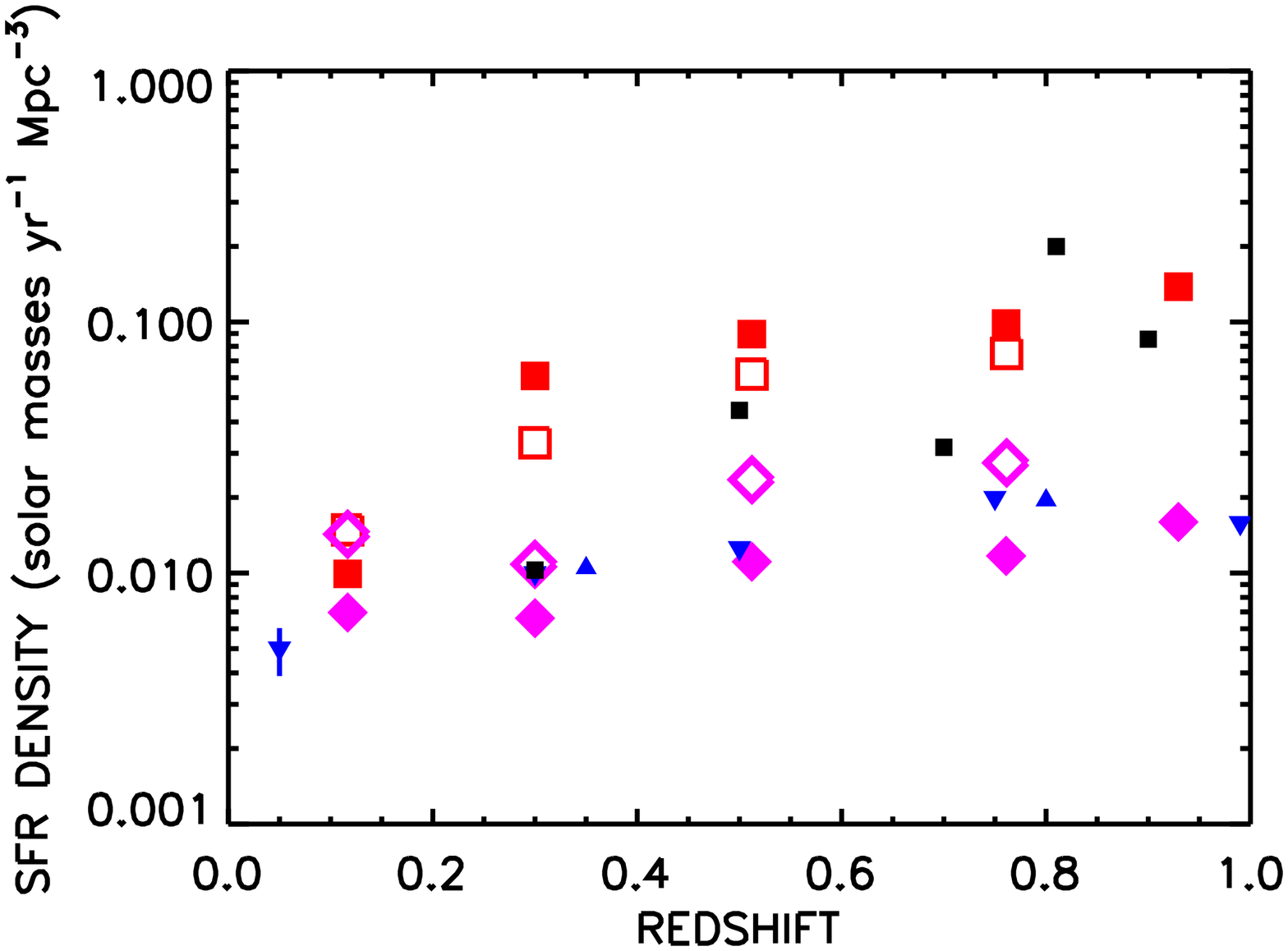,angle=90,width=3.5in}}
\vskip -0.2cm
\figcaption[]{
Star formation histories calculated from the rest-frame 
\hb\ {\em (open)\/} and UV {\em (solid)\/} luminosity densities
of the NIR sample using the calibrations of \S\ref{seccal}. 
The red symbols show the extinction corrected SFRDs.
The purple symbols show the extinction uncorrected SFRDs.
The black solid squares were derived by Wang et al.\ (2006) from 
radio and submillimeter data and agree well with our extinction 
corrected values. The blue symbols show the SFRDs
directly seen at rest-frame UV wavelengths for UV selected galaxies.
We only show the {\em GALEX\/} determinations of
Wyder et al.\ (2005; local) and Schiminovich et al.\ (2005)
{\em (downward pointing triangles)\/} and the ground-based
determinations of Wilson et al.\ (2002)
{\em (upward pointing triangles)\/}, since these
are the most accurate measurements near $z=1$.
All agree reasonably well with our extinction uncorrected values.
The formal errors are mostly smaller than the symbol sizes.
\label{ha_rho_star_bymass}
}
\end{inlinefigure}

However, the SFRDs from the NIR sample drop steeply at
lower redshifts. At these lower redshifts most of the star formation
is instead seen as direct UV emission from lower mass galaxies, which
have very small extinctions. Thus,
in Figure~\ref{ha_rho_star_bymass} we also show the SFRDs derived
from our extinction uncorrected data. Here we use purple open
(solid) diamonds to denote the SFRDs calculated from the \hb\ (UV)
luminosity densities. Since the extinction correction is
smaller at \hb, the \hb\ points lie above the
UV points. We can compare the UV data with the literature results
for rest-frame UV flux measurements of UV selected
galaxies {\em (blue symbols)\/}, such as the ground-based observations
of Wilson et al.\ (2000) {\em (upward pointing triangles)\/} and
the more recent {\em GALEX\/} determinations of Wyder et al.\ (2005; local)
and Schiminovich et al.\ (2005) {\em (downward pointing
triangles)}. We again find close agreement.

\begin{figure*}
\figurenum{42}
\centerline{\psfig{figure=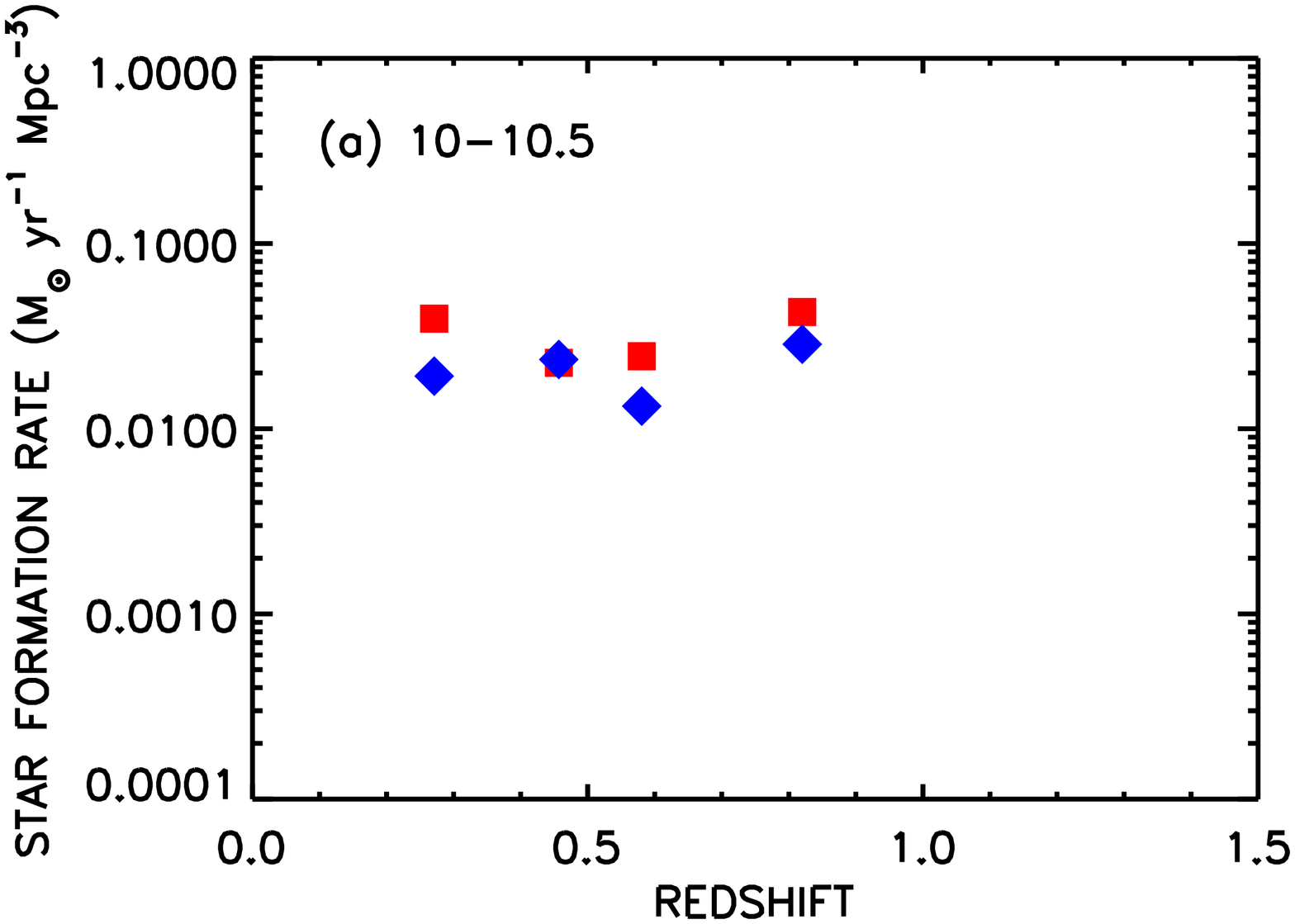,angle=90,width=3.5in}
\psfig{figure=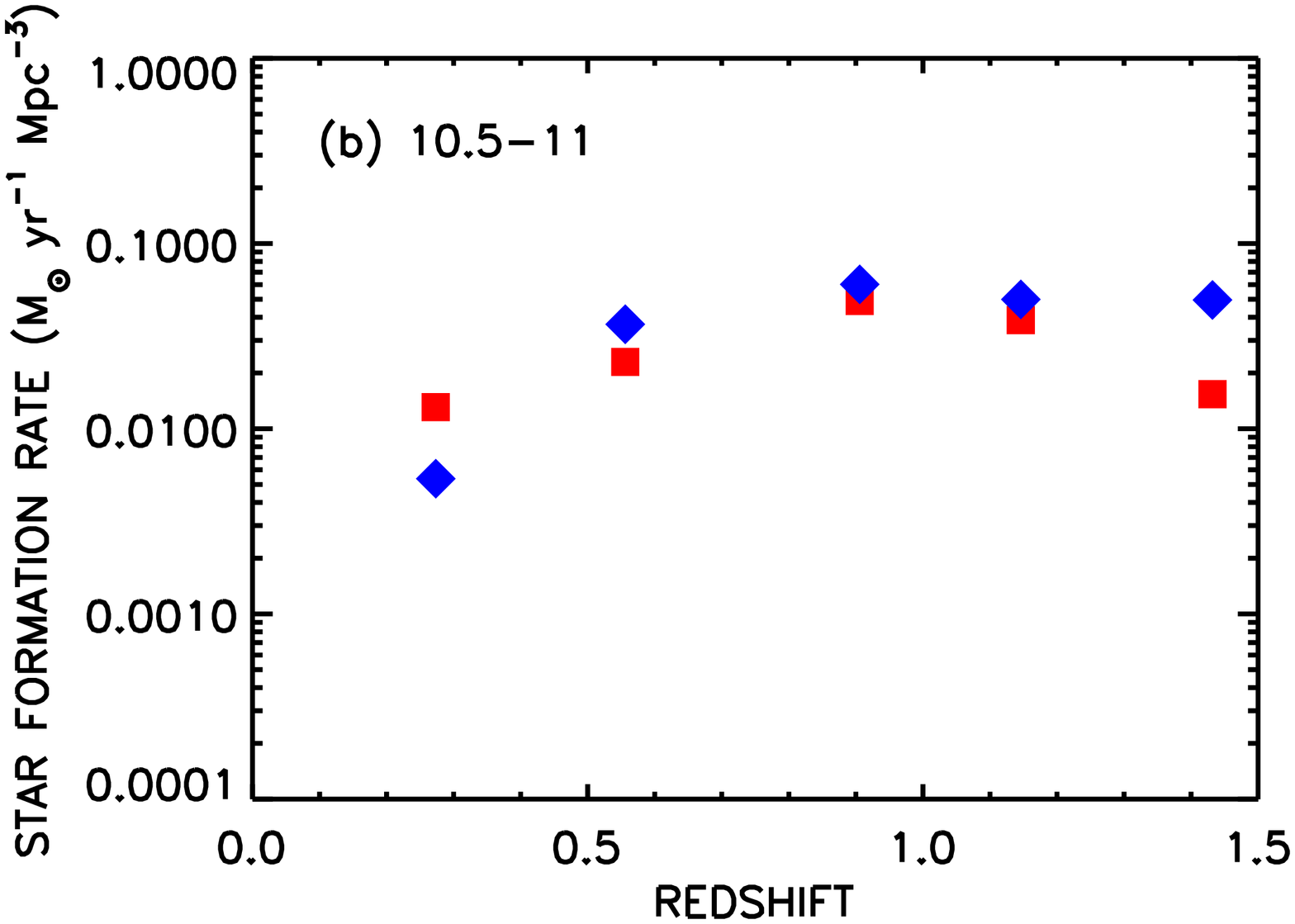,angle=90,width=3.5in}}
\vskip -0.6cm
\centerline{\psfig{figure=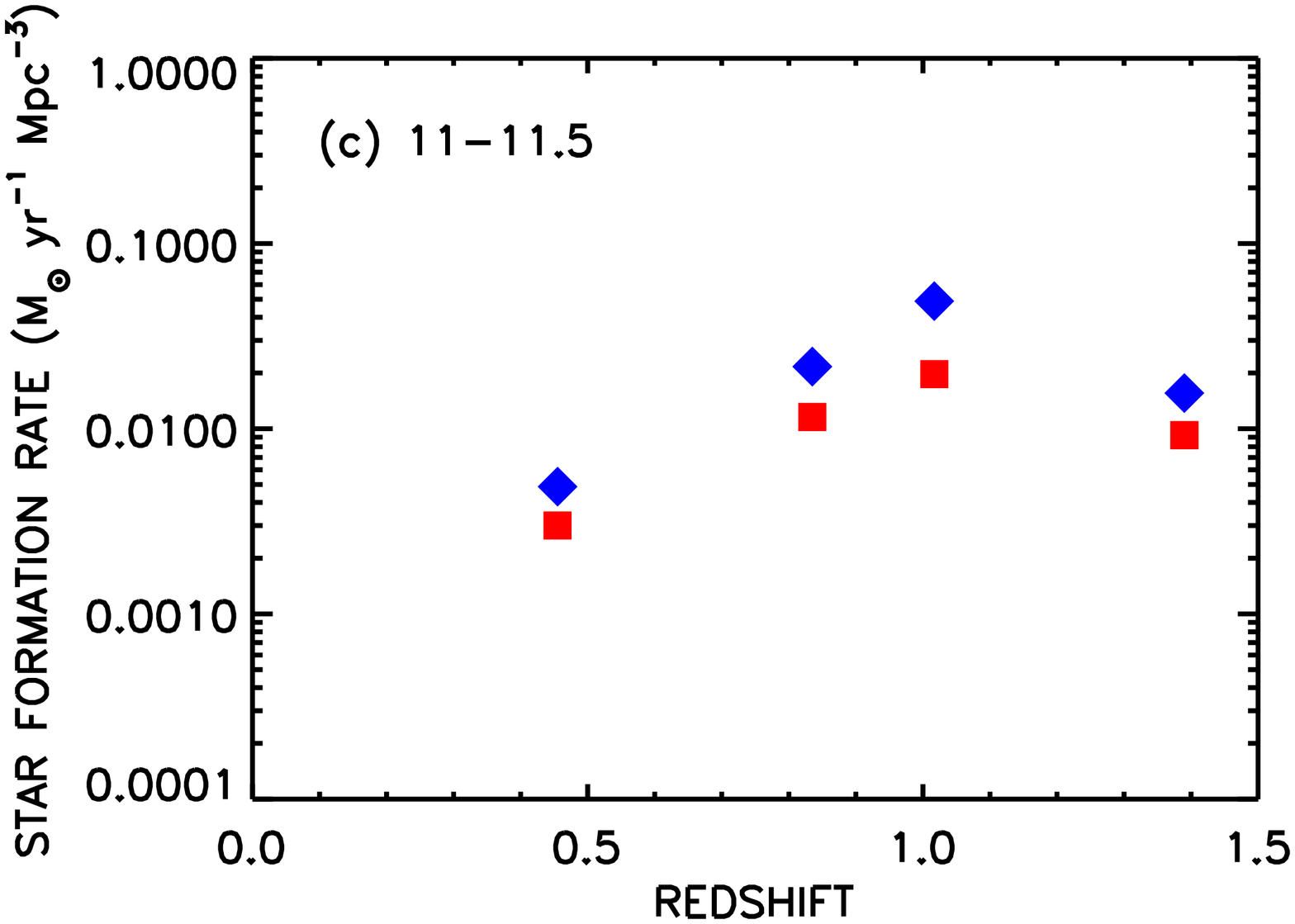,angle=90,width=3.5in}
\psfig{figure=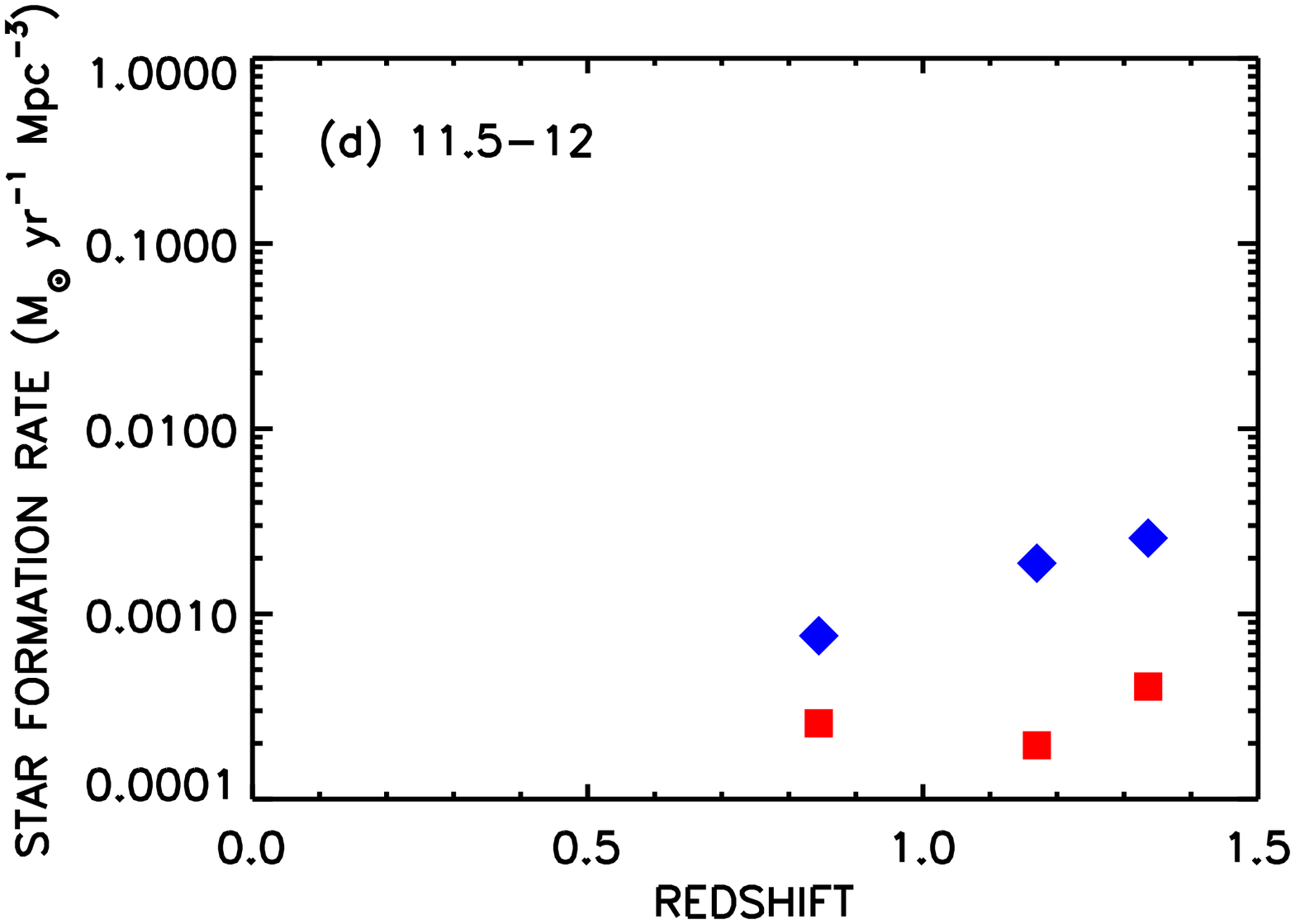,angle=90,width=3.5in}}
\vskip -0.2cm
\figcaption[]{
Star formation rate densities from two different methods vs. 
redshift for the mass intervals (a) $10^{10}-10^{10.5}$~M$_\odot$,
(b) $10^{10.5}-10^{11}$~M$_\odot$,
(c) $10^{11}-10^{11.5}$~M$_\odot$, and
(d) $10^{11.5}-10^{12}$~M$_\odot$.
The red squares show the extinction corrected SFRDs calculated
from the UV luminosity densities using our empirical calibrations.
The blue diamonds show the completely independent calculation
of the SFRDs from the 24~$\mu$m fluxes and the extinction uncorrected 
\oii\ fluxes.
\label{sfr_comp_bymass}
}
\end{figure*}

In Figure~\ref{sfr_comp_bymass} we show our extinction
corrected SFRDs {\em (red squares)\/}
calculated from the rest-frame UV
luminosity densities versus redshift 
for the mass intervals (a) $10^{10}-10^{10.5}$~M$_\odot$,
(b) $10^{10.5}-10^{11}$~M$_\odot$,
(c) $10^{11}-10^{11.5}$~M$_\odot$, and
(d) $10^{11.5}-10^{12}$~M$_\odot$. 
As another check of our UV extinction corrections, we also 
show the SFRDs {\em (blue diamonds)\/}
obtained by adding those computed from the $24~\mu$m fluxes for 
the obscured star formation with those computed from the extinction 
uncorrected \oii\ luminosities for the unobscured star formation 
(see Conselice07). This method has no dependence on the
UV extinction corrections and shows extremely similar 
results to the UV-based method.

%
%

In Figure~\ref{final_sfr_dist} we show our UV-based extinction
corrected SFRDs per unit logarithmic mass versus the logarithmic 
mass in the redshift intervals $z=0.05-0.475$ 
{\em (red diamonds)\/}, $z=0.475-0.9$ {\em (black squares)\/}, 
and $z=0.9-1.5$ {\em (blue triangles)\/}.
In each redshift interval we only show the SFRDs over 
the mass range where the NIR sample is complete. 
Just for a shape comparison, we also show
on the figure the stellar mass density distribution function for 
$z=0.475-0.9$ divided by the age of the universe at $z=0.7$ 
{\em (black curve)\/}. The contrast between this and the SFRD 
distribution is striking, with the latter being much more 
strongly weighted to low-mass galaxies. At all redshifts there is 
very little star formation at masses $>10^{11.1}$~M$_\odot$, but 
at high redshifts the star formation peaks in the interval
$10^{10.5}-10^{11.1}$~M$_\odot$. 
It is the drop in star formation for galaxies in this
mass interval that results in the drop in the overall SFH.
At lower masses there is relatively little change over $z=0.05-0.9$.

%
%
\begin{inlinefigure}
\figurenum{43}
\centerline{\psfig{figure=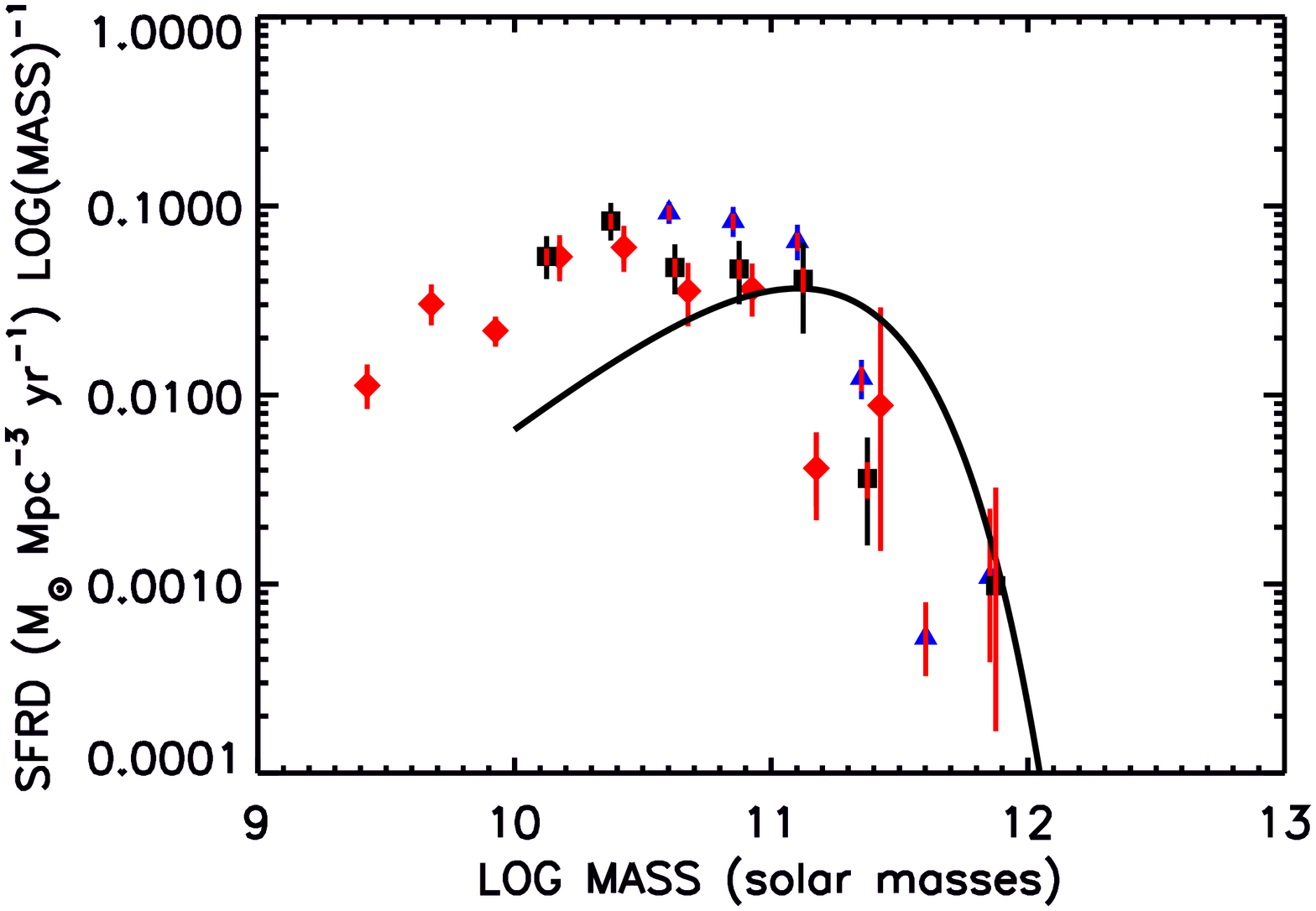,angle=90,width=3.5in}}
\vskip -0.2cm
\figcaption[]{
Extinction corrected star formation rate density per 
unit logarithmic mass vs. logarithmic mass
in the redshift intervals $z=0.05-0.475$ 
{\em (red diamonds)\/}, $z=0.475-0.9$ {\em (black squares)\/}, 
and $z=0.9-1.5$ {\em (blue triangles)\/}.
The solid black curve shows the corresponding shape of the
stellar mass density distribution function for $z=0.475-0.9$
divided by the age of the universe at $z=0.7$.
\label{final_sfr_dist}
}
\end{inlinefigure}

\subsection{Comparison of Growth Rates}
\label{secgrowth}

In this section we want to compare the stellar mass density growth 
rates produced by star formation with those measured from 
the stellar mass functions. Other groups have made this comparison
before (e.g., Borch et al.\ 2006; P{\'e}rez-Gonz{\'a}lez et al.\ 2008),
but generally they have not done so by mass interval, nor have 
they done it self-consistently, since they rely on other groups' 
determinations of the star formation. 
Thus, they have to assume that the galaxies producing the
star formation have masses in the same range as the galaxies
in their stellar mass function analysis. A comparison by mass 
interval is a much more powerful way to analyze this type of data.

\begin{figure*}
\figurenum{44}
\centerline{\psfig{figure=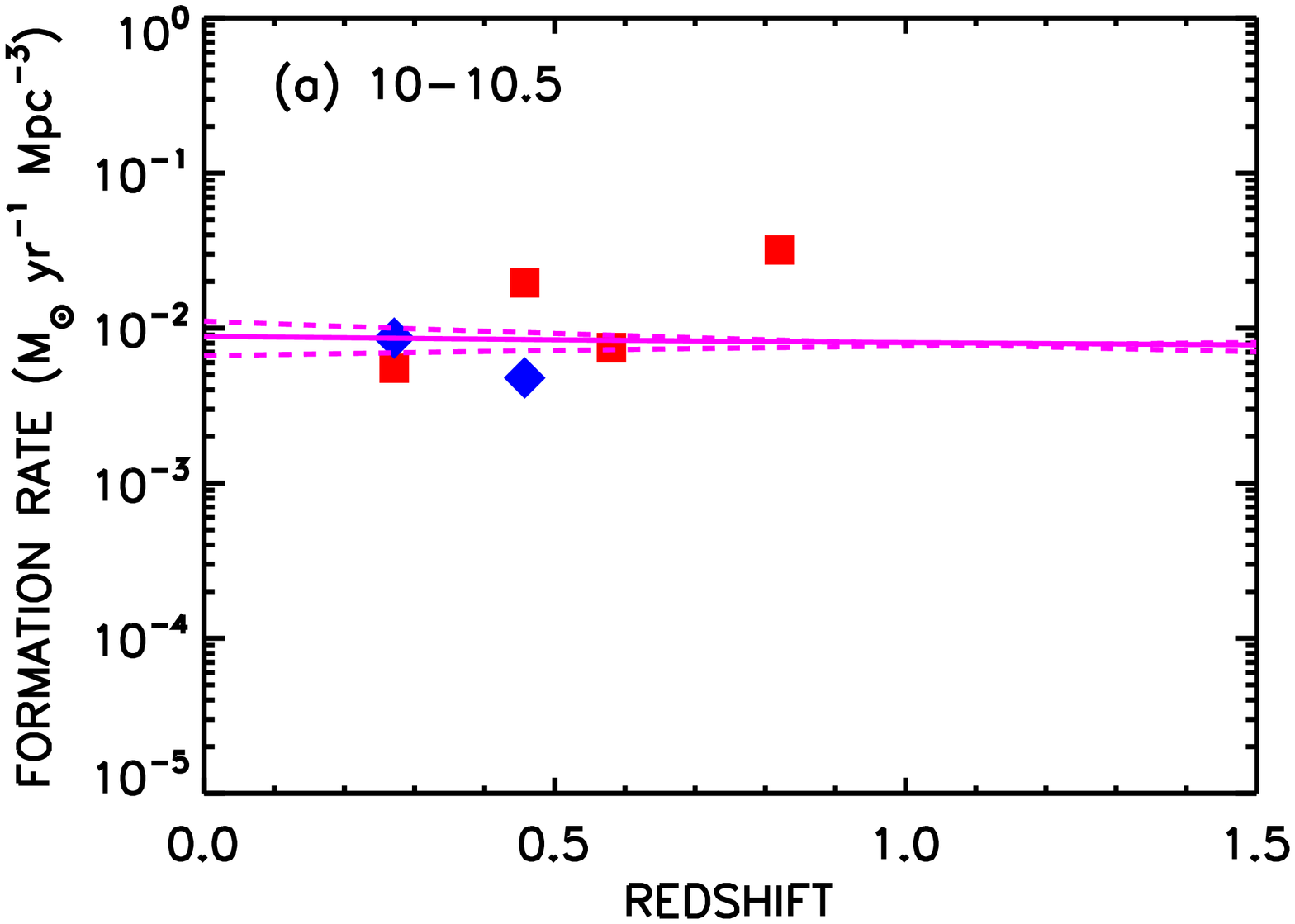,angle=90,width=3.5in}
\psfig{figure=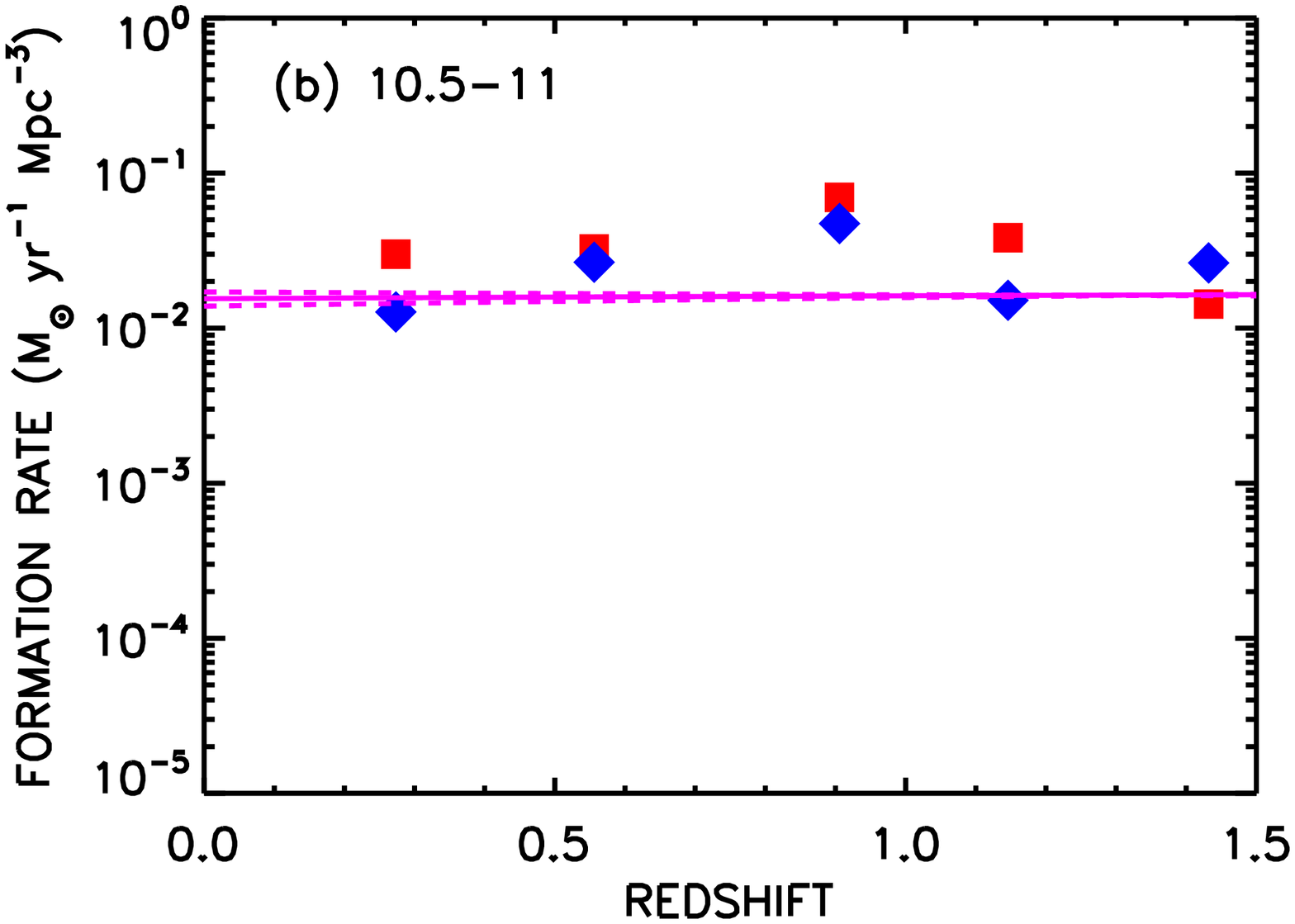,angle=90,width=3.5in}}
\vskip -0.6cm
\centerline{\psfig{figure=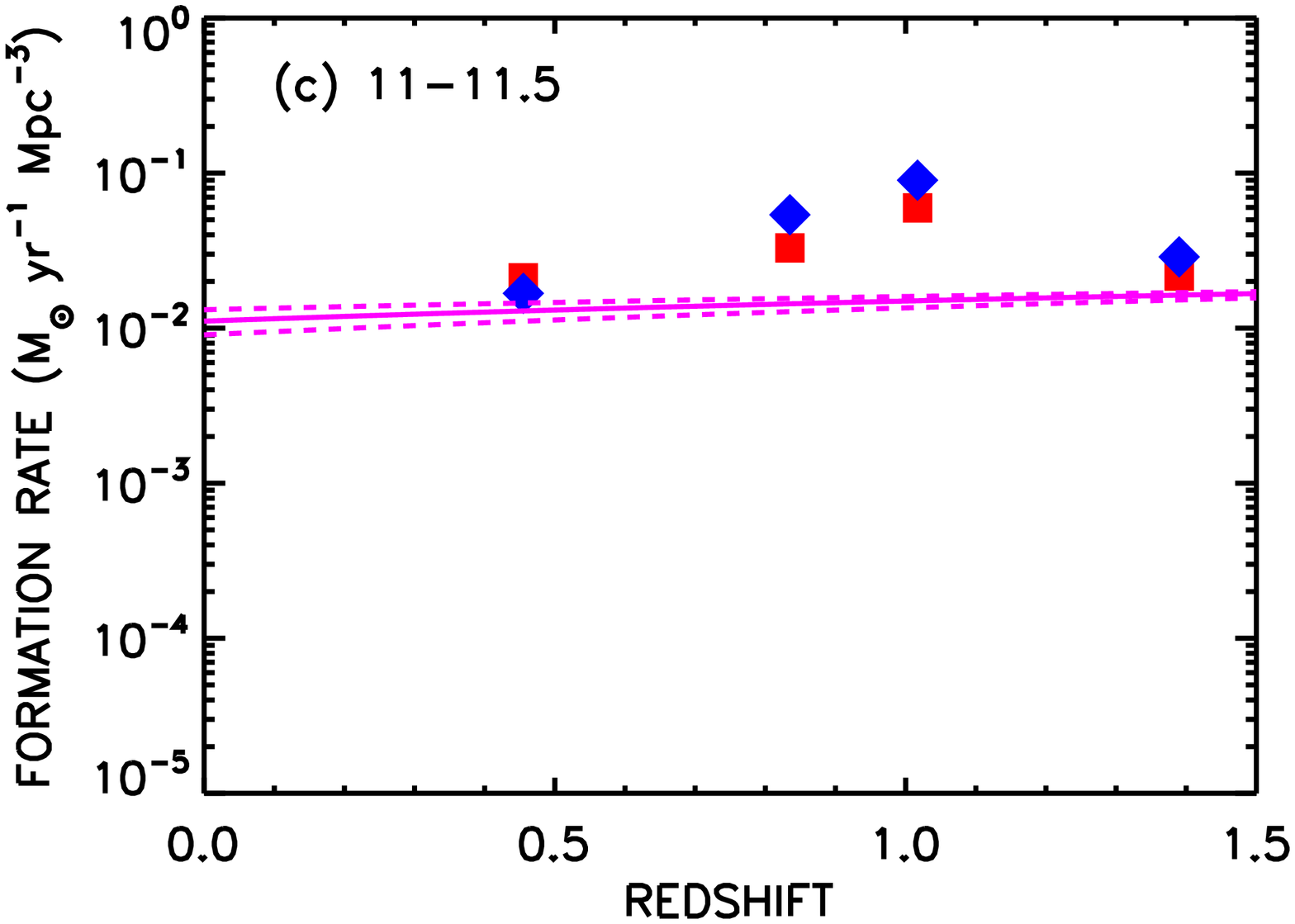,angle=90,width=3.5in}
\psfig{figure=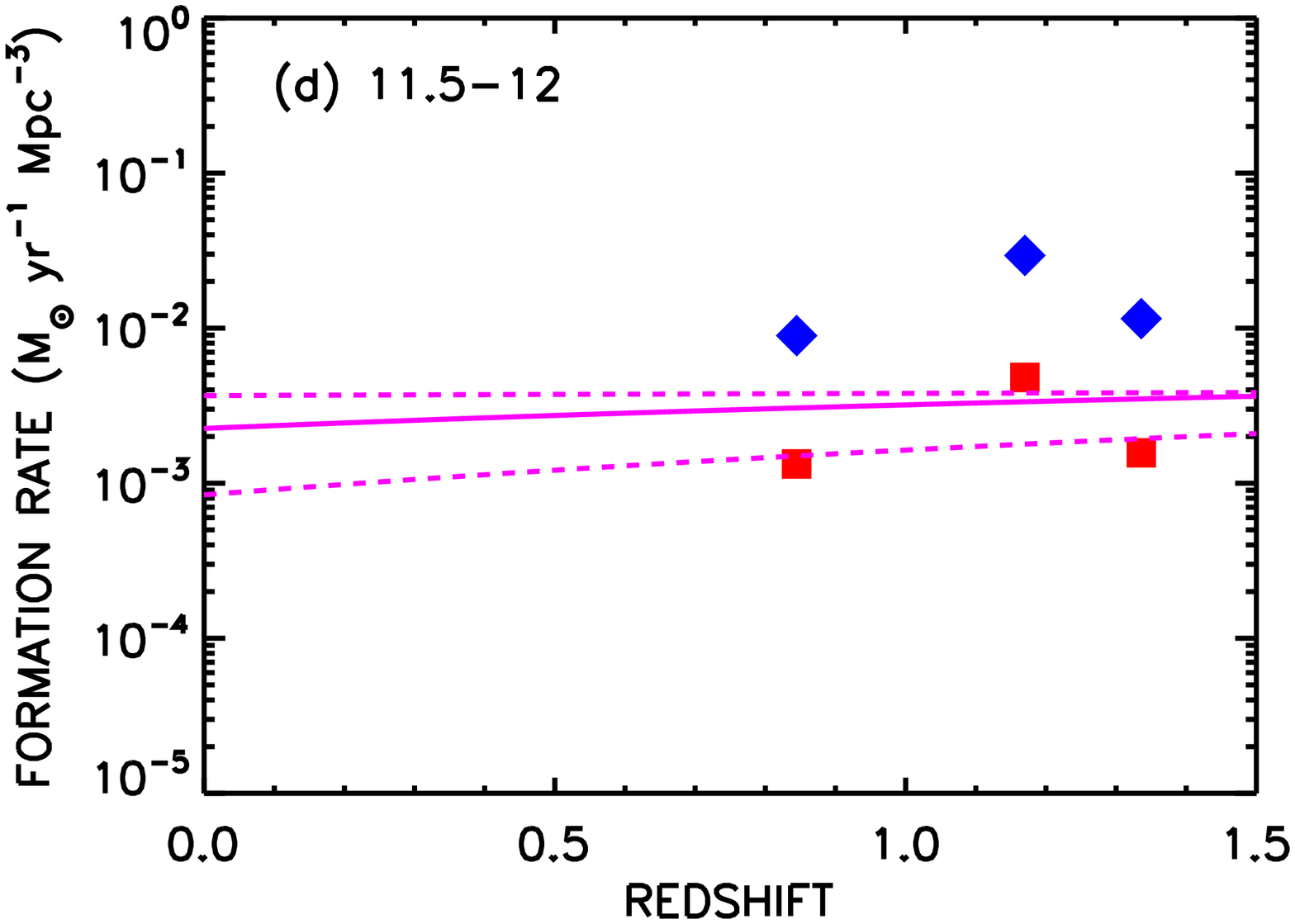,angle=90,width=3.5in}}
\vskip -0.2cm
\figcaption[]{
Mass formation rate densities from two different methods vs. 
redshift for the mass intervals (a) $10^{10}-10^{10.5}$~M$_\odot$,
(b) $10^{10.5}-10^{11}$~M$_\odot$,
(c) $10^{11}-10^{11.5}$~M$_\odot$, and
(d) $10^{11.5}-10^{12}$~M$_\odot$.
The red squares show the formed stellar mass density growth rates 
that the extinction corrected SFRDs calculated from the UV luminosity
densities using our empirical calibrations would produce in each mass 
interval, as calculated from Eq.~\ref{eqnmasscon2}. 
The blue diamonds show the same but this time based on the SFRDs 
calculated from the 24~$\mu$m and extincton uncorrected \oii\ fluxes. 
The purple solid lines show the formed stellar mass density growth 
rates as obtained from the derivative of the least-square polynomial 
fits to the stellar mass density history 
(Eqs.~\ref{massint2}$-$\ref{massint5}) after multiplying
by a factor of 1.35 to convert the present stellar masses to 
formed stellar masses (see \S\ref{secintro}). The purple dashed 
lines show the range given by the $1\sigma$ errors on the fits.
The black dotted lines show the 68\% range in the specific star
formation rates from Noeske et al. 2007.
\label{nuv_rho_star_bymass}
}
\end{figure*}

First, to obtain the growth rates from 
the stellar mass functions, we took the derivative of the least-square 
polynomial fits of Equations~\ref{massint2}$-$\ref{massint5}
after multiplying the mass densities by the average
correction of 1.35 given in \S\ref{secintro} to convert to formed
stellar mass densities (i.e., the total mass density formed into 
stars prior to stellar mass loss; it is the formed 
mass density growth rates which are directly related to the SFRDs). 
We show these growth rates {\em (purple solid lines)\/} in 
Figure~\ref{nuv_rho_star_bymass} versus redshift for the mass intervals 
(a) $10^{10}-10^{10.5}$~M$_\odot$,
(b) $10^{10.5}-10^{11}$~M$_\odot$,
(c) $10^{11}-10^{11.5}$~M$_\odot$, and
(d) $10^{11.5}-10^{12}$~M$_\odot$.

%
%

Next, to compute the expected formed
stellar mass density growth rates produced by star formation,
we need to allow for the fact that the mass of a 
galaxy will grow as it forms stars. Consequently, the mass that 
formed in one mass interval will eventually end up in another mass 
interval. We can describe this movement with the conservation equation
\begin{equation}
\dot{\rho}(M) = s(M) - {d\over dM}(\dot{M}\rho(M)) \,,
\label{eqnmasscon}
\end{equation}
where $\rho(M)$ is the stellar mass density per unit mass 
interval and $s(M)$ is the SFRD per unit mass interval.
Since $\dot{M}$ is related to $s(M)$ through $\dot{M}=s(M)/\phi(M)$,
where $\phi(M)$ is the number of galaxies per unit mass interval,
and $M$ is related to $\rho(M)$ through $M=\rho(M)/\phi(M)$, we
can rewrite the equation as
\begin{equation}
\dot{\rho}(M) = -{ds(M)\over d\ln(M)} \,.
\label{eqnmasscon2}
\end{equation}
Integrated over the total stellar mass function, Equation~\ref{eqnmasscon2}
simply says that the rate of increase in the formed stellar mass
density per unit time is equal to the SFRD.

Merging can also redistribute mass from low to high masses in 
the galactic mass function (e.g., Conselice07), but this is much 
harder to quantify. As we shall see below, 
Equation~\ref{eqnmasscon2} provides a good description of the 
changes seen in the formed stellar mass function 
with mass and time without including substantial merging.
Conselice07 reach a similar conclusion.

As a simple intuitive example of the meaning of 
Equation~\ref{eqnmasscon2}, we may consider the case
where the specific star formation rate is constant 
(i.e., $s\propto \rho$). In
this case the mass distribution function per unit logarithmic
mass, as shown in Figure~\ref{mass_den_plot}, simply moves
to the right in the x-axis. Then the mass density in a given 
interval grows for high masses above the peak in the mass 
distribution function, where $d\rho/dln(M)$ is negative 
($\dot{\rho}$ positive), and the mass density in a given 
interval drops at low masses, where $d\rho/dln(M)$ is 
positive ($\dot{\rho}$ negative).

In Figure~\ref{nuv_rho_star_bymass} we show the formed 
stellar mass density growth rates produced by star formation 
(as calculated from Equation~\ref{eqnmasscon2}) for both the 
UV-based {\em (red squares)\/} and the $24~\mu$m$+$\oii-based 
{\em (blue diamonds)\/} methods. We can
now compare these with the growth rates found earlier from 
the formed stellar mass functions {\em (purple lines)\/}.
The agreement between the shapes with both redshift and mass 
is amazingly good. However, there is a slight normalization 
difference, with the Equation~\ref{eqnmasscon2} UV-based measurements
being higher by about 0.1~dex in the $10^{10}-10^{10.5}$~M$_\odot$ 
interval and higher by about 0.2~dex in the 
$10^{10.5}-10^{11}$~M$_\odot$ 
and $10^{11}-10^{11.5}$~M$_\odot$ intervals. 
(The offset is $-0.2$~dex in the $10^{11.5}-10^{12}$~M$_\odot$ 
interval, but the uncertainties are large.) The offset is 
only slightly reduced if we exclude X-ray selected AGNs
from the star formation calculation. Most previous 
analyses have also found the measurements based on star
formation to be higher than the measurements based on the 
formed stellar mass functions (e.g., Fardal et al.\ 2007 and 
references therein). However, given the completely different 
wavelength ranges used (UV versus NIR), the uncertainties in the
extinction corrections and in the stellar models, and the effects
of cosmic variance, some discrepancy would inevitably
be expected. In this sense, the agreement is remarkably
good, even for the normalization.

The offset is unlikely to be due to relative uncertainties
in the local mass function, which generally lie at the 20\%$-$30\%
level (e.g., compare Cole01, Bell et al.\ 2003, and Eke
et al.\ 2005 with each other). Cole01 lies at the high end of 
the mass density estimates, and reducing the local mass density 
would increase the discrepancy. The offset is also unlikely to
be due to overestimation of the UV extinction corrections, since
our $24~\mu$m$+$\oii-based method for calculating the SFRDs is
completely independent of the UV extinction corrections
and shows extremely similar results to the UV-based method
both in the direct SFRDs (Fig.~\ref{sfr_comp_bymass}) and in the 
formed stellar mass density growth rates computed from 
Equation~\ref{eqnmasscon2} (Fig.~\ref{nuv_rho_star_bymass}).

However, the issue of the uncertainty in the population synthesis 
models and the recent treatments of TP-AGB stars, which we discussed
in \S\ref{secfit}, is more complicated. If the 
BC03 masses which we are using are uniformly too
high, then this would increase the discrepancy. However,
if the local masses are close to the BC03 values
and the high-redshift values are lower than the BC03 values, 
then we will increase the gradients in 
Equations~\ref{massint1}$-$\ref{massint5},
and this will reduce the discrepancy.
We may estimate a maximum effect by assuming that
the mass densities in the $z=1.2-1.5$ interval are only
60\% of the BC03 values (the maximum correction in Bruzual 2007). 
This only slightly increases the formed stellar mass density growth 
rates in the various mass intervals, generally by less than 0.1~dex.
We can see the reason for this by inspecting Figure~\ref{mden_byz}.
Typically we are building a large fraction of the local mass
density over $z=0.05-1.5$ and changing
the starting point mass density downward has a relatively limited
effect on the required growth rate. Thus, it appears that
the mass density growth rates cannot be raised enough to explain
the discrepancy with the rates inferred from star formation.

A possible explanation for the offset
is that the IMF is slightly different from the assumed
Salpeter form. We can resolve the problem by changing the 
index of the IMF to $-1.10$. This is well within the uncertainties 
in the slope of the high-end IMF (Kroupa 2001; Chabrier 2003) and 
close to the Baldry \& Glazebrook (2003) value of $-1.15$.
However, in order to obtain our observed \ha\ to UV ratio, we 
must turn this over at high masses. Using the $-1.10$ index to 
10~M$_\odot$ and an index of $-1.6$ from 10~M$_\odot$ to 
100~M$_\odot$ resolves both problems. Fardal et al.\ (2007) 
argued that this type of mid--mass-weighted IMF, which they 
describe as paunchy, can also help in providing a consistent 
description of the extragalactic background light.

\subsection{Specific Star Formation Rates}
\label{secstarspec}

We now consider the distribution of specific SFRs (SSFRs)
in the individual galaxies. Although in computing the
instantaneous SSFRs we are simplifying the effects of the
time history in the individual galaxies, it is still useful
to have a quantitative description of the range of behaviors
in the galaxies. We shall return to a study of
the time history of the galaxies in \S\ref{seccolor}.

%
%
\begin{inlinefigure}
\figurenum{45}
\centerline{\psfig{figure=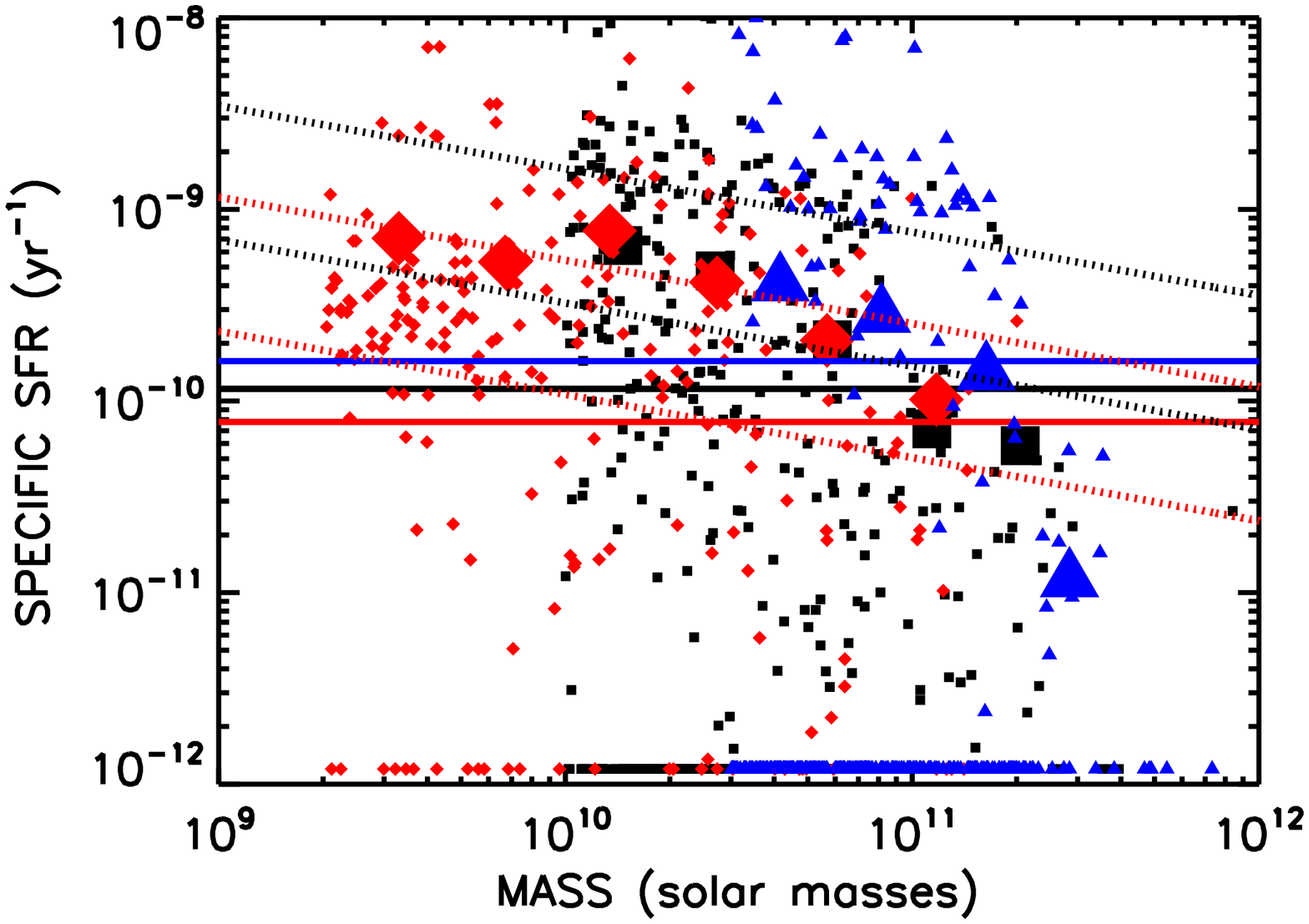,angle=90,width=3.5in}}
\vskip -0.2cm
\figcaption[]{
Specific star formation rates
(SFR per unit mass in the galaxy) vs. mass in the redshift intervals 
$z=0.05-0.475$
{\em (red diamonds)\/}, $z=0.475-0.9$ {\em (black squares)\/},
and $z=0.9-1.5$ {\em (blue triangles)\/}. The high-redshift
sample is only shown above the limiting $3\times10^{10}$~M$_\odot$
to which it is complete. The large symbols show the mean values
for each redshift interval. The solid lines show the
inverse age of the universe at redshifts $z=0.05$ {\em (red)\/},
$z=0.475$ {\em (black)\/}, and $z=0.9$ {\em (blue)\/}.
Only galaxies with an average SSFR
above the inverse age of the universe at the redshift of the
galaxy can undergo a significant change in mass.
The red ($z=0.2-0.7$) and black ($z=0.85-1.1$) dotted lines 
show the ``main sequence'' range of Noeske et al.\ (2007).  
They claim that 68\% of the galaxies should lie within this range
based on their DEEP2 observations.
\label{mass_mstar}
}
\end{inlinefigure}

In Figure~\ref{mass_mstar} we show the SSFRs versus
galaxy mass in the redshift intervals $z=0.05-0.475$
{\em (red diamonds)\/}, $z=0.475-0.9$ {\em (black squares)\/},
and $z=0.9-1.5$ {\em (purple triangles)\/}. 
There is clearly a wide spread at all redshifts and masses. 
However, only galaxies with SSFRs larger than the inverse age 
of the universe at the redshift of the galaxy can change their 
mass significantly if those rates are maintained
over the full time interval. We shall refer to such
galaxies as strong star formers. Note that if the star formation is
episodic, then the mass change in the galaxies will be smaller.
Thus, the number of strong star formers
represents an upper bound on the fraction of galaxies that
may grow significantly at a given redshift.
The mean SSFRs {\em (large symbols)\/}
reproduce the results of \S\ref{secsfh}. That is,
they show that, on average, only galaxies with masses
$\lesssim10^{11}$~M$_\odot$ grow significantly in
any of the redshift intervals. 
Since the high-redshift blue triangles
cross the growth line at slightly higher masses than the black squares
and red diamonds of the lower redshift intervals, the typical
mass at which growth is taking place is downsizing in the later 
redshift intervals. In the lowest redshift interval where we can 
measure the masses below $10^{10}$~M$_\odot$,
we see that the mean SSFRs finally flatten out (at a high level).
There also appears to be a rough maximum to the SSFRs of
about $3\times10^{-9}$~yr$^{-1}$. 

In Figure~\ref{dist_specific_bymass} we show the 
distribution functions of the SSFRs for the mass intervals
(a) $10^{11}-10^{11.5}$~M$_\odot$,
(b) $10^{10.5}-10^{11}$~M$_\odot$,
and (c) $10^{10}-10^{10.5}$~M$_\odot$. In each panel
the redshift intervals $z=0.9-1.5$ {\em (blue triangles)\/},
$z=0.475-0.9$ {\em (black squares)\/}, and
$z=0.05-0.475$ {\em (red diamonds)\/} are shown.
We see little evolution in the distribution functions over
the observed redshift range, but they do have very
different shapes in the different mass intervals.

%
%
\begin{inlinefigure}
\figurenum{46}
\centerline{\psfig{figure=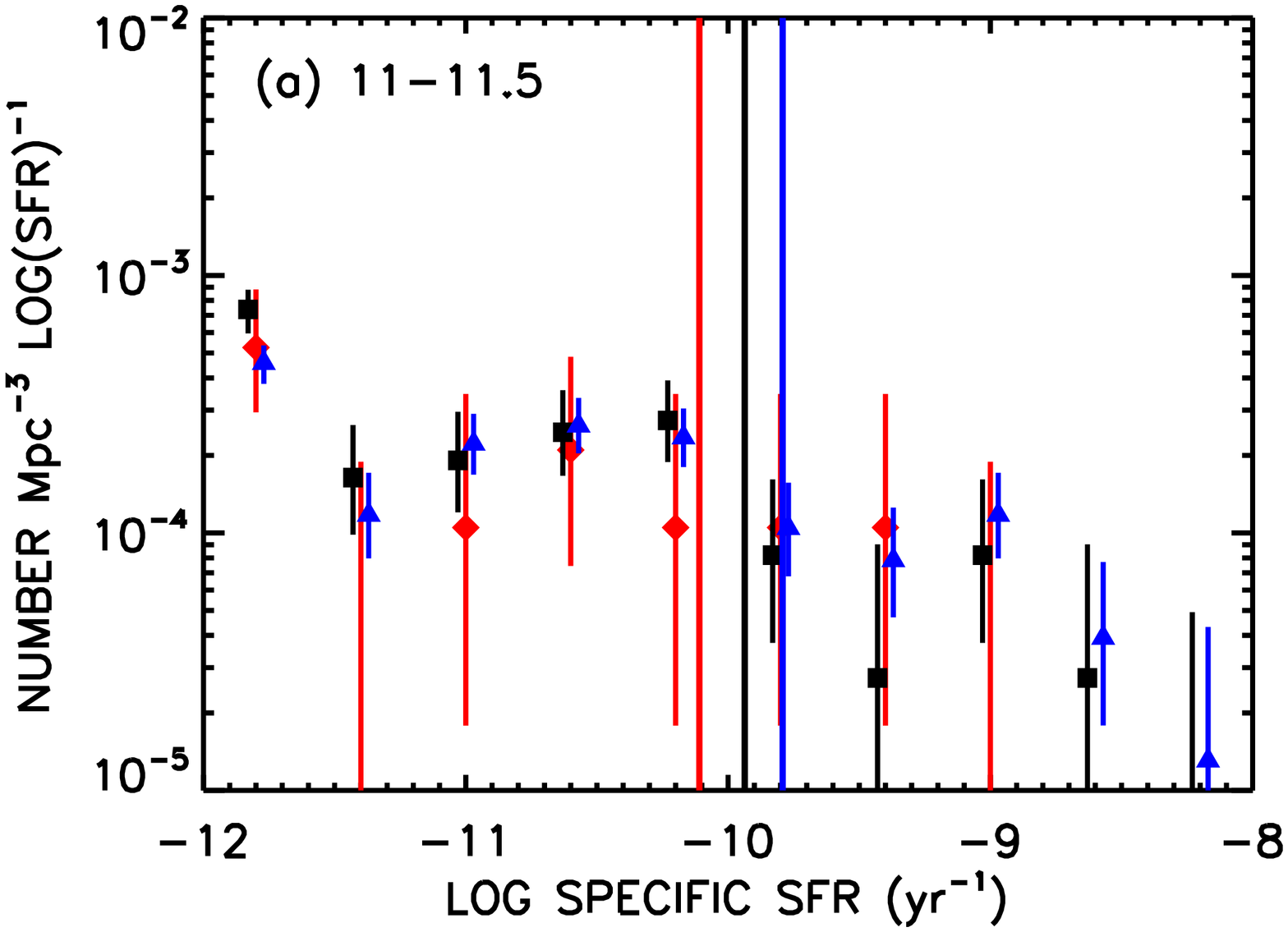,angle=90,width=3.5in}}
\vskip -0.6cm
\centerline{\psfig{figure=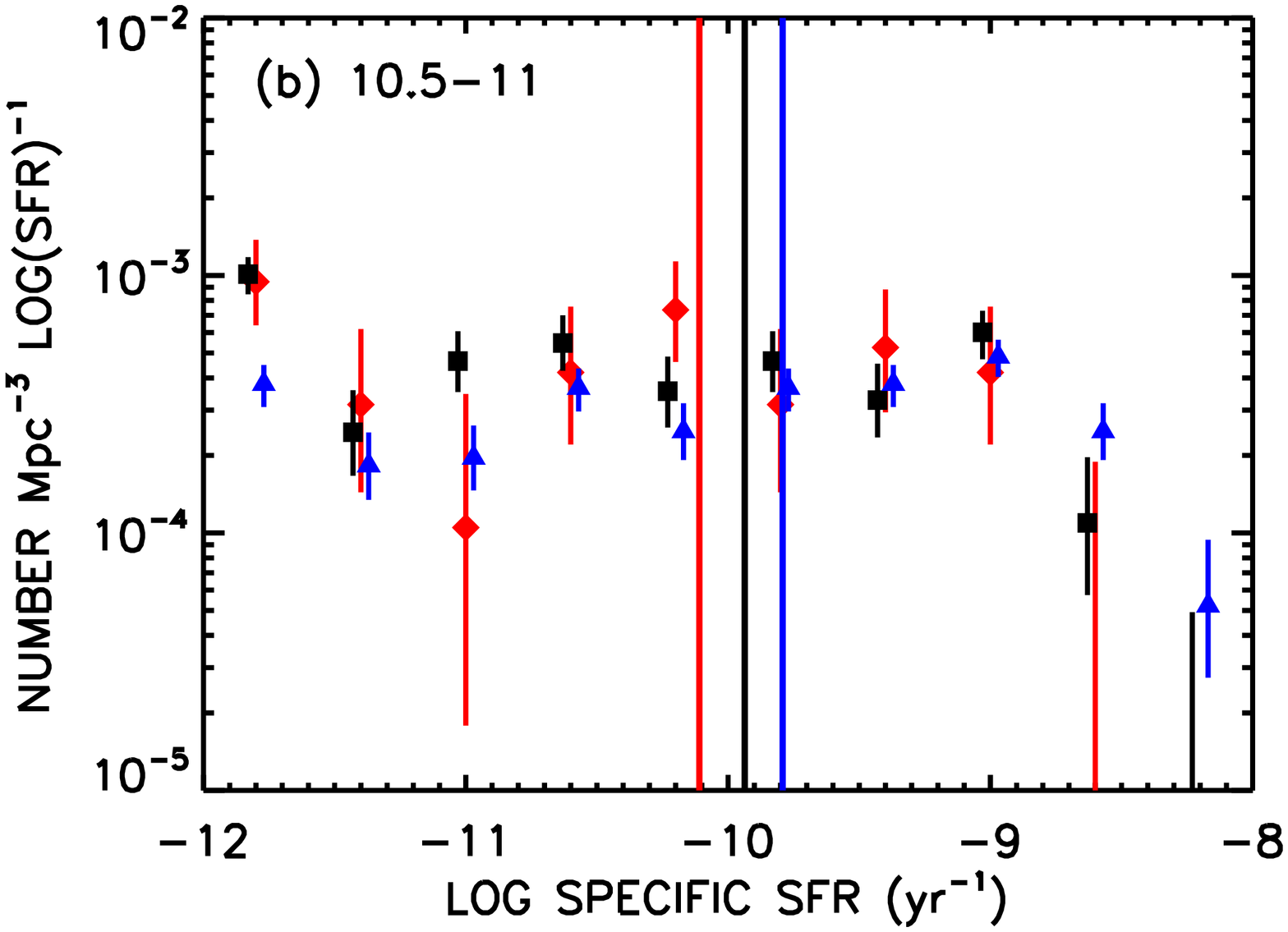,angle=90,width=3.5in}}
\vskip -0.6cm
\centerline{\psfig{figure=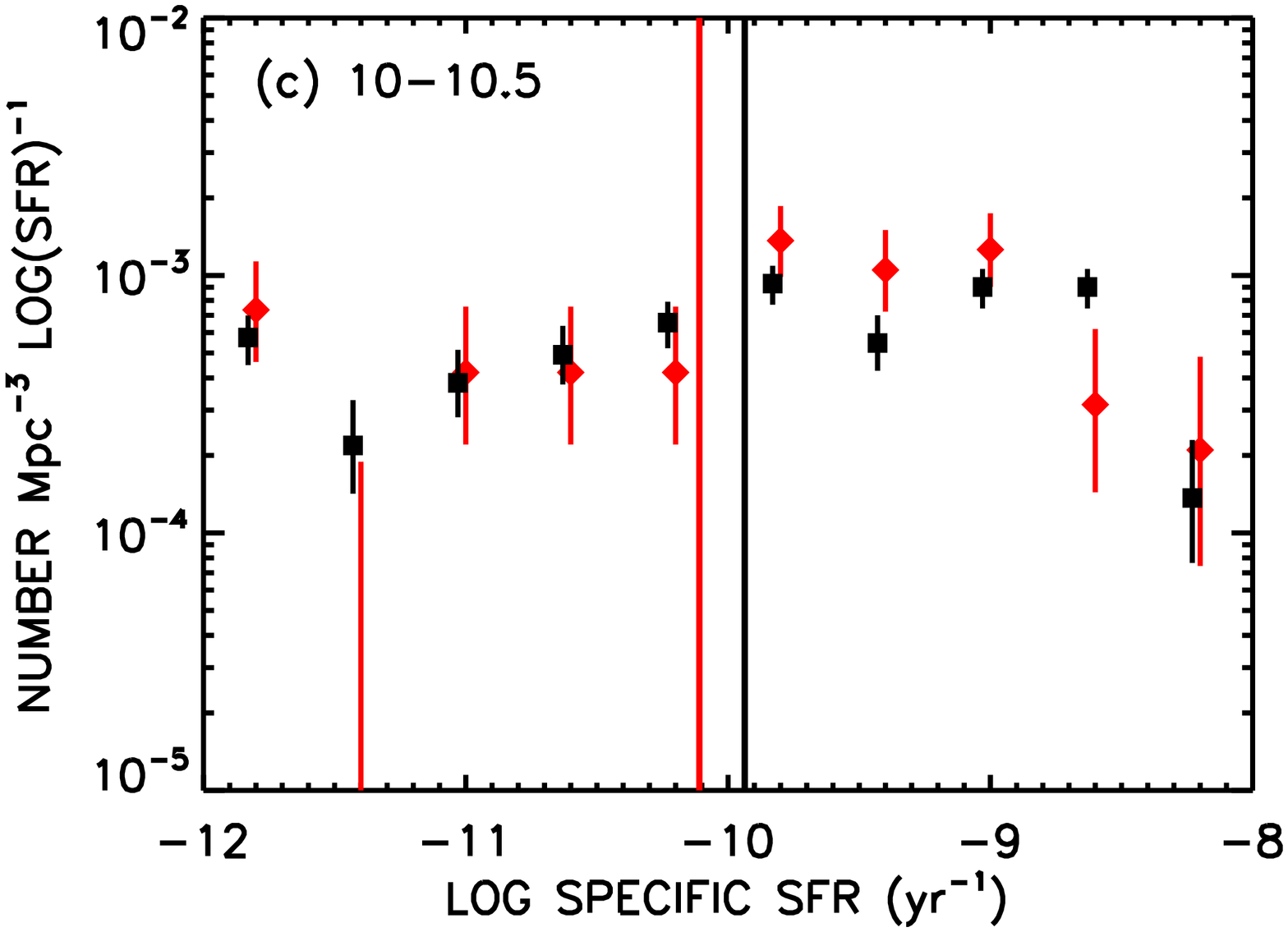,angle=90,width=3.5in}}
\vskip -0.2cm
\figcaption[]{
Distribution functions of the specific star formation
rates for logarithmic mass intervals (a) $11-11.5$~M$_\odot$,
(b) $10.5-11$~M$_\odot$, and (c) $10-10.5$~M$_\odot$.
In each panel the red diamonds denote $z=0.05-0.475$,
the black squares $z=0.475-0.9$, and the blue triangles $z=0.9-1.5$.
The error bars show the 68\% confidence limits. The blue triangles 
and black squares have been slightly displaced in the x-axis (by plus 
and minus 0.03, respectively) to allow the error bars to be distinguished.
The solid vertical lines show the log of the 
inverse age of the universe at redshifts $z=0.05$ {\em (red)\/},
$z=0.475$ {\em (black)\/}, and $z=0.9$ {\em (blue)\/}.
\label{dist_specific_bymass}
}
\end{inlinefigure}

In the highest mass interval (Fig.~\ref{dist_specific_bymass}a) 
most of the galaxies have very low SSFRs. There are only a small 
fraction of stong star formers at any redshift. Overall only about 
10\% of the galaxies can be growing significantly.
In the two lower mass intervals (Figs.~\ref{dist_specific_bymass}b,c) 
the number of sources with strong SSFRs increases. In fact, in the 
lowest mass interval (Fig.~\ref{dist_specific_bymass}c) galaxies 
with SSFRs above the $10^{-10}$~yr$^{-1}$ dominate the population.
The percentages of strong star formers in all three mass
intervals are given in Table~\ref{tabsf}. In the lowest
mass range we give the values only in the two lower redshift
intervals where the sample is complete.

We may conclude from this that what star formation is occurring
in the galaxies in the $10^{11}-10^{11.5}$~M$_\odot$ interval
is spread over many galaxies,
and there are very few galaxies in this mass interval that are
undergoing significant growth. The situation is less clear
in the lower mass intervals. By the time we reach the
$10^{10}-10^{10.5}$~M$_\odot$ interval, the distribution
is roughly evenly split between galaxies
undergoing strong star formation and galaxies with
weak star formation (see Table~\ref{tabsf}).
This could be a distinction between two populations: one with
strong ongoing star formation and one with weak or little growth.
Alternatively, it could be that there is a high frequency of
bursting relative to steady star formation at these redshifts
with all galaxies undergoing significant star formation on average.
Regardless of this point, a substantial number of the low-mass
galaxies have SSFRs that, if maintained over the time
frame, would change their mass significantly.

We cannot easily compare our results with previous analyses
of the evolution of the SSFRs over this redshift interval,
such as Brinchmann \& Ellis (2000) or Bauer et al.\ (2005),
since they did not include extinction corrections, which
make substantial increases in the SSFRs. However, we can
compare our results with a recent analysis by Noeske et al.\ (2007),
who used a portion of the DEEP2 sample with $K$-band and
24~$\mu$m observations to analyze the SSFRs.
They used 24~$\mu$m plus emission line estimates of the
SFRs. Noeske et al.\ (2007) claim that the SSFRs
lie within a rather tightly defined range as a 
function of mass. The normalization of this range increases 
with redshift, with the SSFRs increasing by roughly a factor 
of three from $z=0.3$ to $z=1$. They argue that this
implies a smooth evolution in the galaxy SFRs,
a result which would be inconsistent with our subsequent
analysis of the star formation histories in the more
massive galaxies using Balmer lines and colors. 
We show their ranges in Fig.~\ref{mass_mstar} with the dotted 
lines {\em (red: $z=0.2-0.7$; black: $z=0.85-1.1$)\/},
where we have corrected their
Kroupa masses to Salpeter. It is clear that the present results
are inconsistent with the Noeske et al.\ (2007) analysis. While
their upper bound corresponds roughly to the maximum values
seen in the present SSFRs, we
see a much larger scatter in the values for the high-mass galaxies
above $10^{10}$~M$_\odot$. Our data include many galaxies with
low SSFRs. The result is not dependent on
the method we used to calculate the SFRs. 
We find the same effect using the 24~$\mu$m plus emission line 
estimates of the SFRs. 
The result is also not a simple consequence of the optical 
magnitude selection used in DEEP2 ($R=24.1$), since nearly all 
of the high-mass galaxies would be included by such a selection, 
as Noeske et al.\ (2007) discuss and we self-consistently 
find in the present data. The difference may lie in more subtle 
effects of the spectroscopic completeness versus color and 
optical magnitude or in the limited photometry of the DEEP2 sample.

We can also compare 
our results with the local analysis of Brinchmann et al.\ (2004), 
and, in particular, with their Figure~24. (Their masses are based 
on the Kroupa IMF and must be increased by a factor of 1.54 
to match ours.) Brinchmann et al.\ (2004) only included star 
forming galaxies in their analysis, so their distribution
is truncated at low SSFRs and the means are slightly
higher. Nevertheless, the overall shape and normalization,
including the roughly constant SSFRs at low mass, the
decline in the SSFRs above $10^{10}$~M$_\odot$, the upper bound 
on the SSFRs, and the change in the distribution of SSFRs
at high mass are all in extremely good agreement with the 
present results.

\subsection{Galaxy Colors, Equivalent Widths, and the 4000~\AA\ Break}
\label{seccolor}

Rest-frame galaxy colors and features in the
spectra, such as the equivalent widths of the emission
lines and the strengths of the 4000~\AA\ break,
provide a measure of the SSFRs convolved with the
recent time history of the star formation. Since
the colors and the various spectral features are sensitive 
to different stellar mass 
ranges, they can provide information
on the time history of the star formation and how smooth
or episodic it is. (See, e.g., the Kauffmann et al.\ 2003a 
analysis of the SDSS sample using the 4000~\AA\ break and the
H$\delta$ line.) Thus, the combination of photometric
and spectroscopic information in the present
sample provides a powerful tool to investigate
the nature of the star formation.

Locally the SDSS results have shown that the galaxy colors
are bimodal and divide into a red sequence of galaxies that
are not currently undergoing star formation and a blue cloud
of galaxies with active star formation
(e.g., Strateva et al.\ 2001;
Baldry et al.\ 2004). The red sequence dominates above 
$3\times 10^{10}$~M$_\odot$ (Kauffmann et al.\ 2003b), and the
blue cloud dominates below.
This color bimodality is also seen in higher redshift,
optically-selected samples (e.g., Bell et al.\ 2004;
Weiner et al.\ 2005; Giallonga et al.\ 2005;
Willmer et al.\ 2006).
However, these analyses did not correct for internal
extinction, which, as we shall show below, is important.
Moreover, the bimodality appears to be at least partially
a consequence of the optical selection and is not so strong 
in our mass-selected samples at the higher redshifts.

In Figure~\ref{color_redshift} we show the rest-frame
UV$-$blue (AB3400$-$AB4500) colors uncorrected for
extinction for our full NIR sample versus redshift.
(Note that if we change to the Vega-based magnitudes used
by Willmer et al.\ 2006 in their DEEP2 analysis, then we find
a nearly identical range of colors as they.) We see almost 
no evolution in the color distribution over the $z=0.05-1.5$ 
redshift range. The precise split between
the blue cloud and the red sequence depends on the mass
or luminosity, but we have shown the rough split with the
red dashed line. This is based on the average value of the
relation given by van Dokkum et al.\ (2000) in the appropriate
luminosity range. It is not easy to see evidence for strong
bimodality in this figure.
Rather, we see a uniform spread of colors stretching
from the blue cloud to the red sequence. Nearly all of the
galaxies in the intermediate color range (sometimes referred
to as the green valley) are 24$~\mu$m sources
{\em (green triangles)\/}.

%
%
\begin{inlinefigure}
\figurenum{47}
\centerline{\psfig{figure=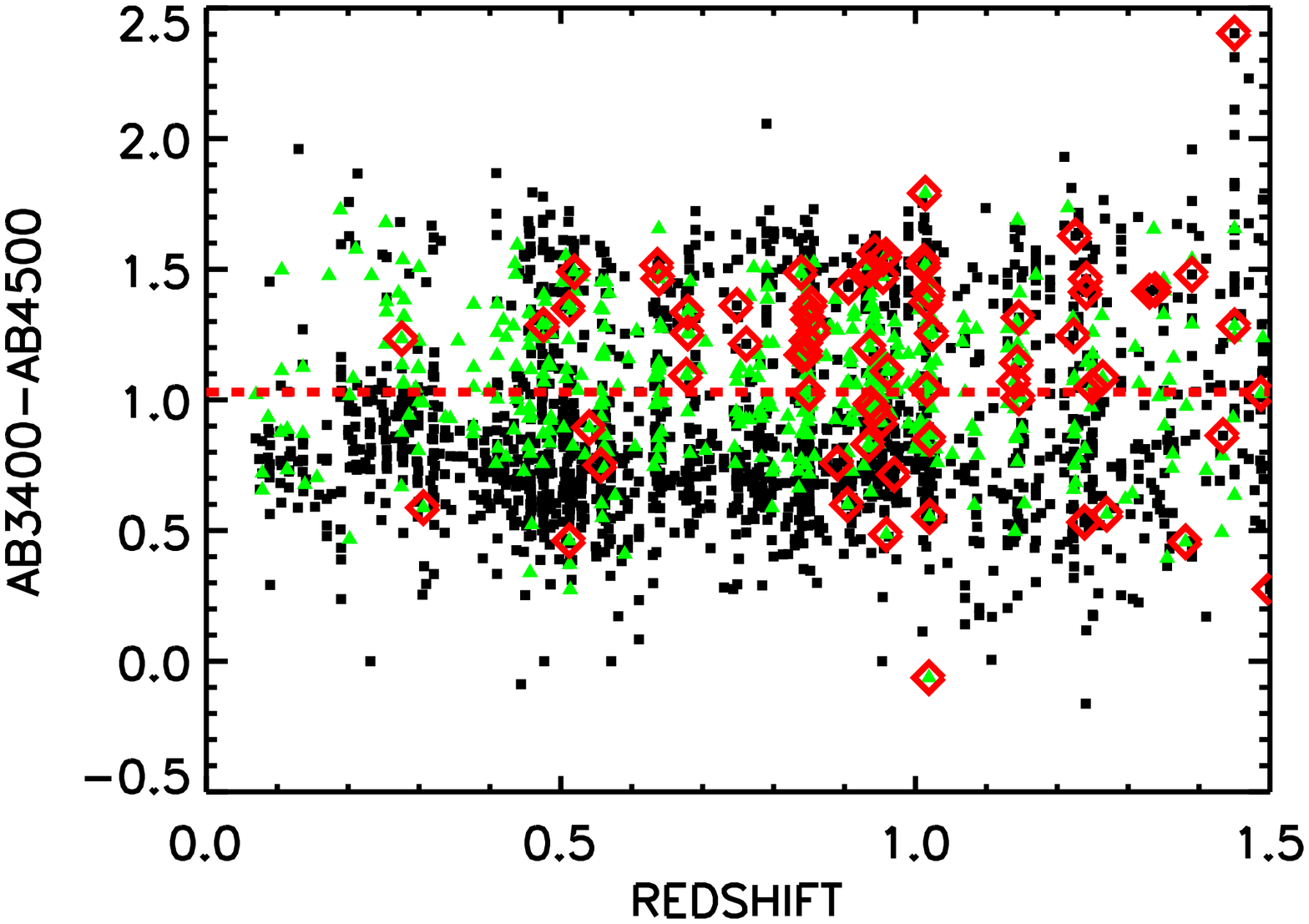,angle=90,width=3.5in}}
\vskip -0.2cm
\figcaption[]{
Rest-frame $3400-4500$~\AA\ color vs. redshift 
for our NIR sample. Green triangles denote $24~\mu$m sources,
and black squares denote sources which are not detected
at $24~\mu$m. Sources with X-ray luminosities implying
the presence of an AGN are enclosed in red diamonds.
The red dashed horizontal line shows the approximate
separation between the blue cloud and the red sequence
based on the average value of the relation given by
van Dokkum et al.\ (2000) in the appropriate luminosity
range.
\label{color_redshift}
}
\end{inlinefigure}

The spread in colors may be more clearly seen
in histogram form. In Figure~\ref{ub_hist}a
we show the colors prior to applying any extinction corrections.
We see that the 24~$\mu$m sources lie in the green valley
{\em (green dashed line)\/}. While there is a hint of bimodality
in the total sample {\em (black solid line)\/}, it is quite weak
with sources present at all colors. In Figure~\ref{ub_hist}b
we show the colors after correcting for
extinction. Now the bulk of the 24~$\mu$m sources lie in the
blue cloud, and there is a more clearly bimodal distribution
in the total sample. Even with the extinction correction there 
are still many sources (both $24~\mu$m and non-$24~\mu$m)
in the intermediate color region, but it is clear that
applying extinction corrections is critical when
analyzing the galaxy colors. Many of the sources
seen in the green valley and the red sequence prior to correcting
for extinction are, in fact, dusty sources with intrinsically blue 
colors. Quantitatively, 801 out of 2254 sources (35\%) are in the red
sequence (defining this as AB3400$-$AB4500$>1.03$) prior to the 
extinction correction, but nearly half of these are dusty blue 
galaxies. After applying the extinction correction, the number in
the red sequence gets reduced to 466 out of 2254 sources, or 
roughly 20\%.

The extinction corrections have a mass dependence since 
higher mass galaxies, with their larger column densities of gas 
and dust, reprocess more of their UV light. (The very highest
mass galaxies will generally have lower extinctions because
they are gas deficient.) The extinction
corrections may also have a redshift dependence due to the 
evolution in the metallicity and gas content of the galaxies. 
Thus, any analyses of the color versus mass or color versus 
luminosity relations that do not apply extinction corrections
will be biased.

In Figure~\ref{color_mass_dered} we show the dereddened
colors versus galaxy mass for the redshift intervals
(a) $z=0.9-1.5$, (b) $z=0.475-0.9$, and (c) $z=0.05-0.475$.
The red sequence is clearly seen in all of the intervals.
Rather than attempt to measure the slope of the color-mass
relation from the present data, we have assumed the locally
determined slope of 0.08~mag per dex in mass determined
by van der Wel et al.\ (2007). We then normalized this
slope to match the red sequence galaxies in the $z=0.9-1.5$
redshift interval with masses above $3\times10^{10}$~M$_\odot$
to obtain the red sequence relation with mass, $M$,
\begin{equation}
{\rm AB3400}-{\rm AB4500} = 1.26+0.08(\log M-9) \,.
\label{sepeq}
\end{equation}
We show this relation with the solid black line in 
Figure~\ref{color_mass_dered}.
In contrast to previous results from optically-selected
and extinction uncorrected data (e.g., Bell et al.\ 2004; 
van der Wel et al.\ 2007), we see no change in the position 
of the red sequence with redshift. This suggests that the 
effect they observed was primarily a consequence of reddening. 
The intrinsic colors of the reddest galaxies are 
not changing over this redshift interval.

%
%
\begin{inlinefigure}
\figurenum{48}
\centerline{\psfig{figure=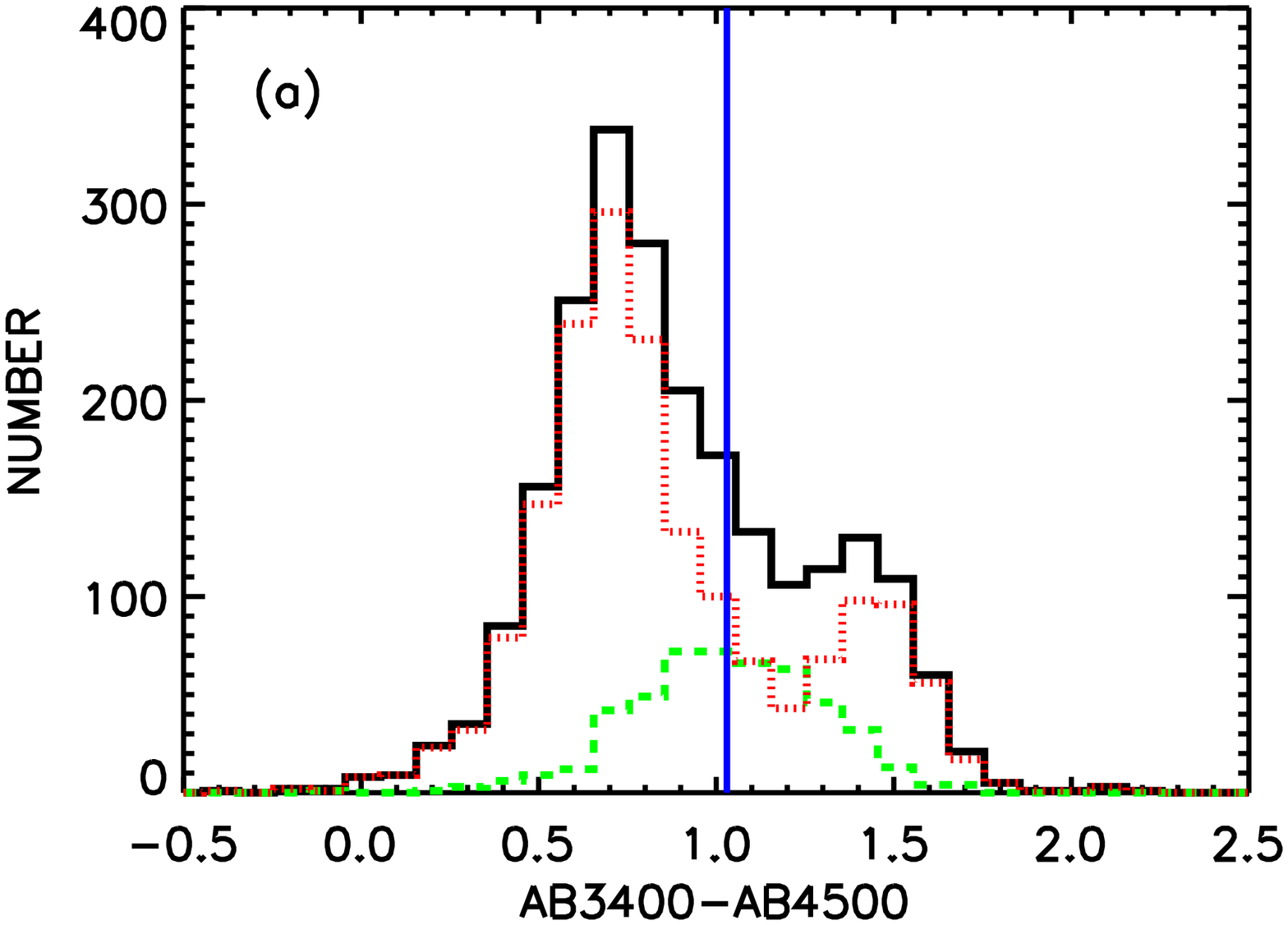,angle=90,width=3.5in}}
\vskip -0.6cm
\centerline{\psfig{figure=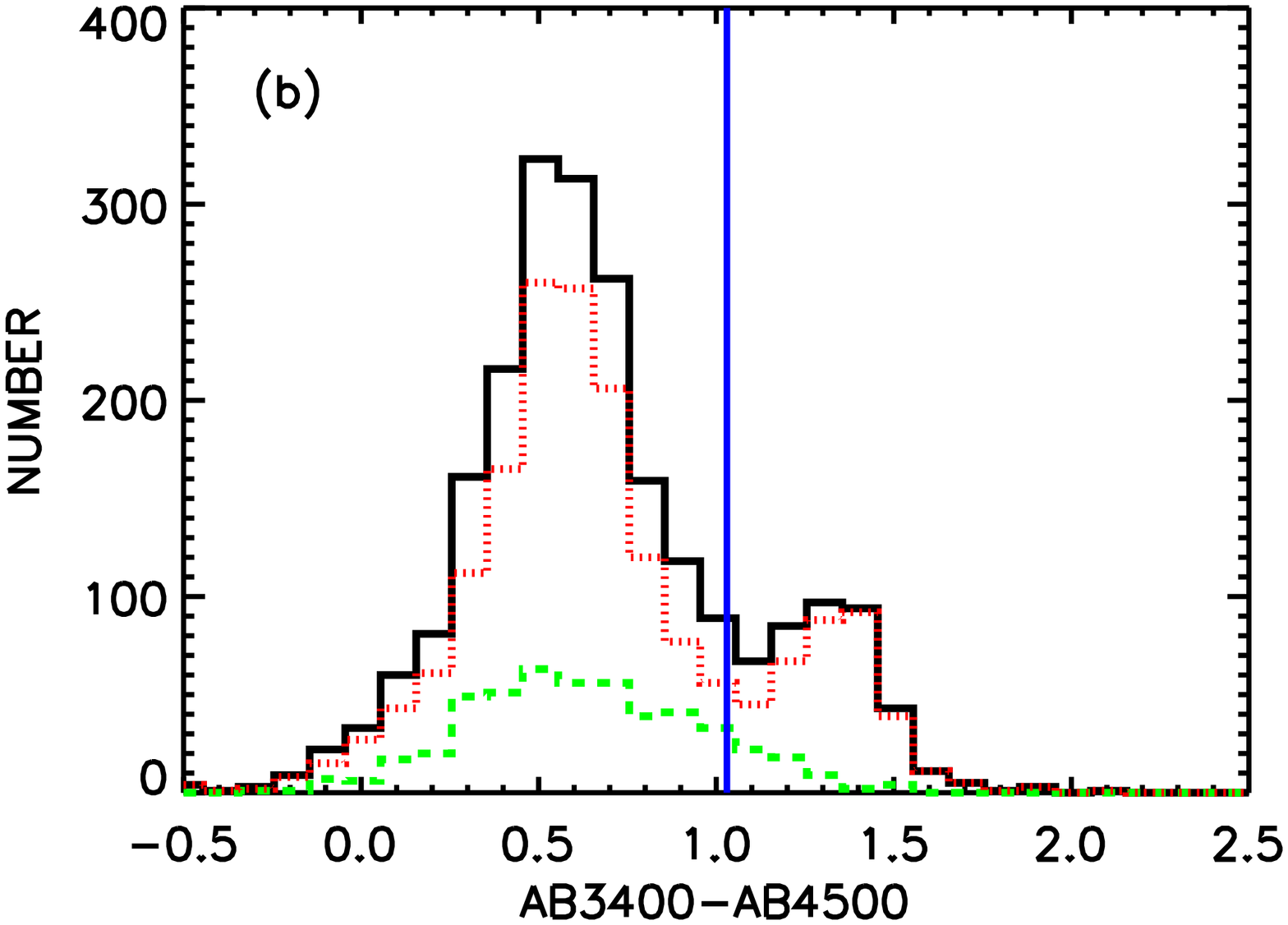,angle=90,width=3.5in}}
\vskip -0.2cm
\figcaption[]{
Distribution of rest-frame $3400-4500$~\AA\ color
for our NIR sample (a) before and (b) after
correcting for internal extinction. The black histogram
shows the distribution of the total sample. The
green dashed (red dotted) line shows the distribution of
$24~\mu$m (non-24~$\mu$m) sources.
The blue vertical line shows the approximate separation
between the blue cloud and the red sequence based on the
average value of the relation given by
van Dokkum et al.\ (2000) in the appropriate luminosity
range.
\label{ub_hist}
}
\end{inlinefigure}

While it is clear from Figure~\ref{ub_hist} that
there is no precise split between the red sequence
and the blue cloud, we may approximately separate
them with a cut lying about 0.25~mag below
the track of the red sequence,
\begin{equation}
{\rm AB3400}-{\rm AB4500} = 1.01+0.08(\log M-9) \,.
\label{sepeqn2}
\end{equation}
This allows for the spread in colors in the red sequence
itself. We show this relation in Figure~\ref{color_mass_dered}
with the solid blue line.
The high-mass, blue cloud galaxies are nearly all $24~\mu$m
sources {\em (green triangles)\/}. Moreover, most of the
$24~\mu$m sources lie in the blue cloud, though at the
highest masses we also see some $24~\mu$m sources that
move into the red sequence region.

%
%
\begin{inlinefigure}
\figurenum{49}
\centerline{\psfig{figure=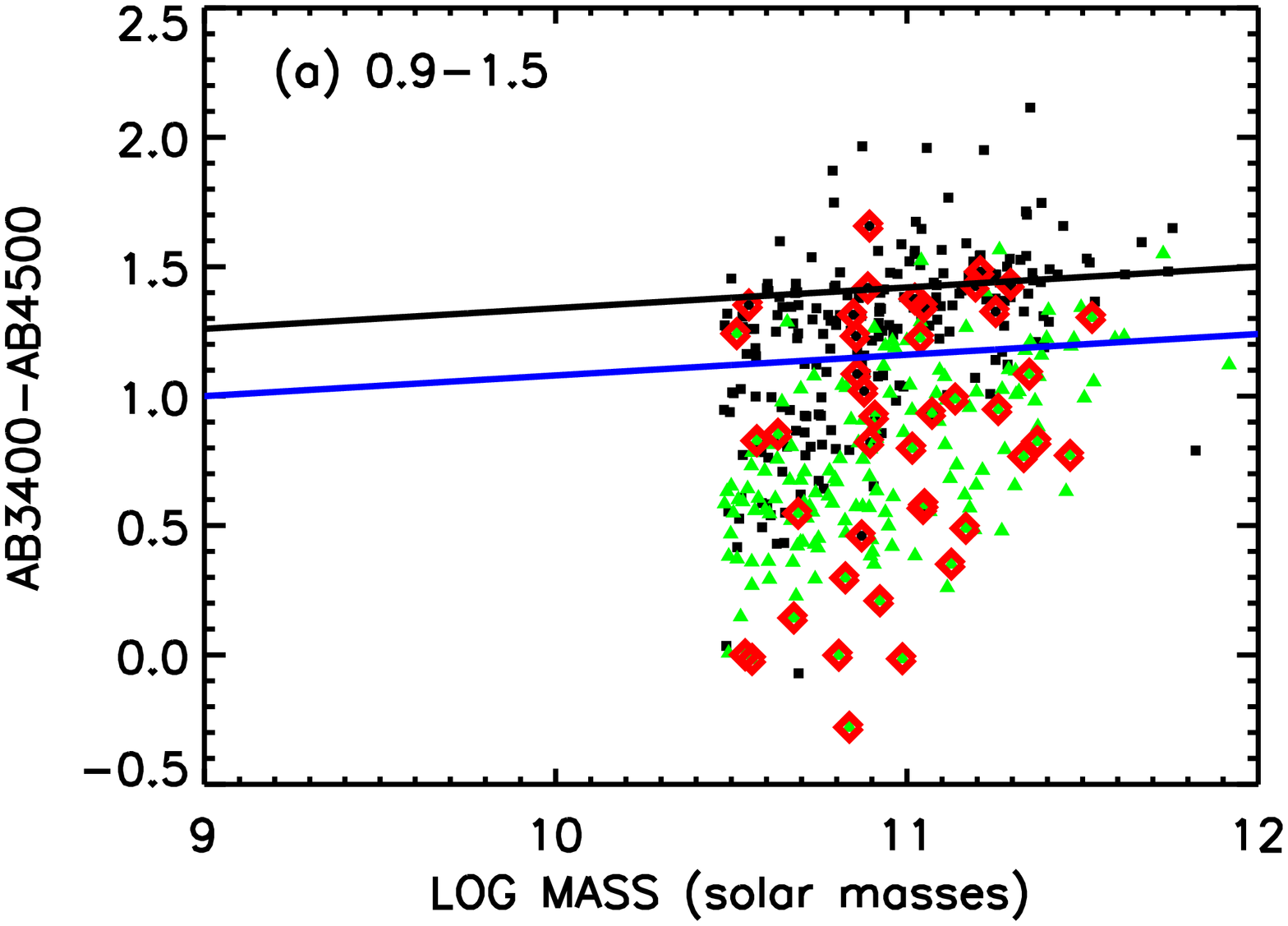,angle=90,width=3.5in}}
\vskip -0.6cm
\centerline{\psfig{figure=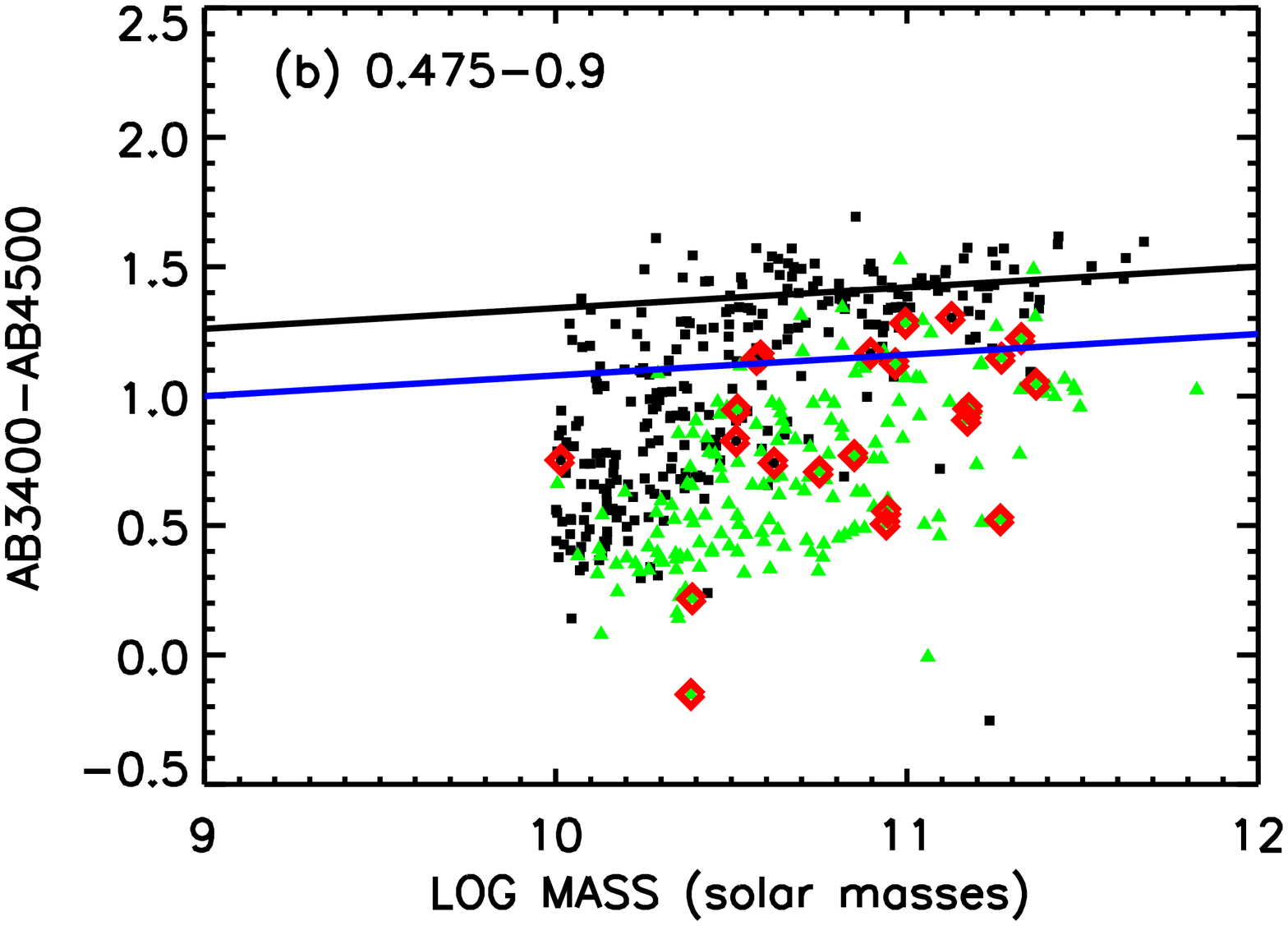,angle=90,width=3.5in}}
\vskip -0.6cm
\centerline{\psfig{figure=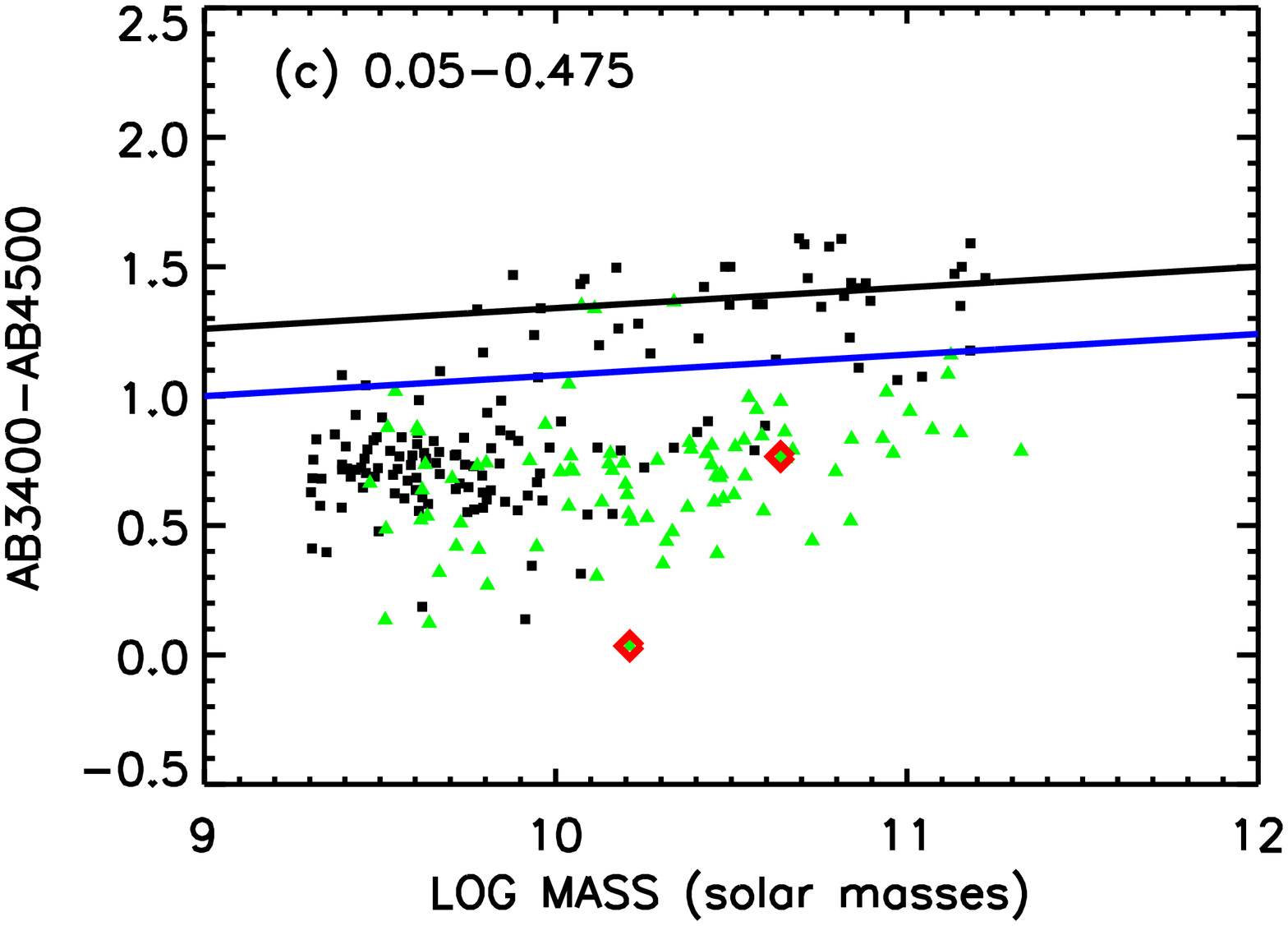,angle=90,width=3.5in}}
\vskip -0.2cm
\figcaption[]{
Rest-frame $3400-4500$~\AA\ color corrected for
extinction vs. logarithmic mass for our NIR sample
in the redshift intervals (a) $z=0.9-1.5$, (b) $z=0.475-0.9$,
and (c) $z=0.05-0.475$. The colors are only shown above the
mass at which the sample in the given redshift interval is complete.
Green triangles denote $24~\mu$m sources, and black squares
denote sources which are not detected at $24~\mu$m. Sources
with X-ray luminosities implying the presence of an AGN
are enclosed in red diamonds. The solid black line shows
the red sequence (with locally determined slope 0.08~mag
per dex in mass) normalized to match the $z=0.9-1.5$
interval. The blue line shows this relation offset by 
0.25~mag, which we adopt as the split between the red 
sequence and the blue cloud.
\label{color_mass_dered}
}
\end{inlinefigure}

Although we can see that the red sources dominate at
the high masses and that nearly all low-mass galaxies
below $10^{10}$~M$_\odot$ are blue cloud galaxies,
there does not appear to be a clear transition
mass in any of the redshift intervals. Rather, the
fraction of galaxies with high SSFRs
drops as we move to higher masses in all the
redshift intervals. We also note that the split
between the blue cloud and the red sequence becomes
more pronounced at low redshifts, while at
higher redshifts there are a considerable number of
intermediate color sources.

%
%
\begin{inlinefigure}
\figurenum{50}
\centerline{\psfig{figure=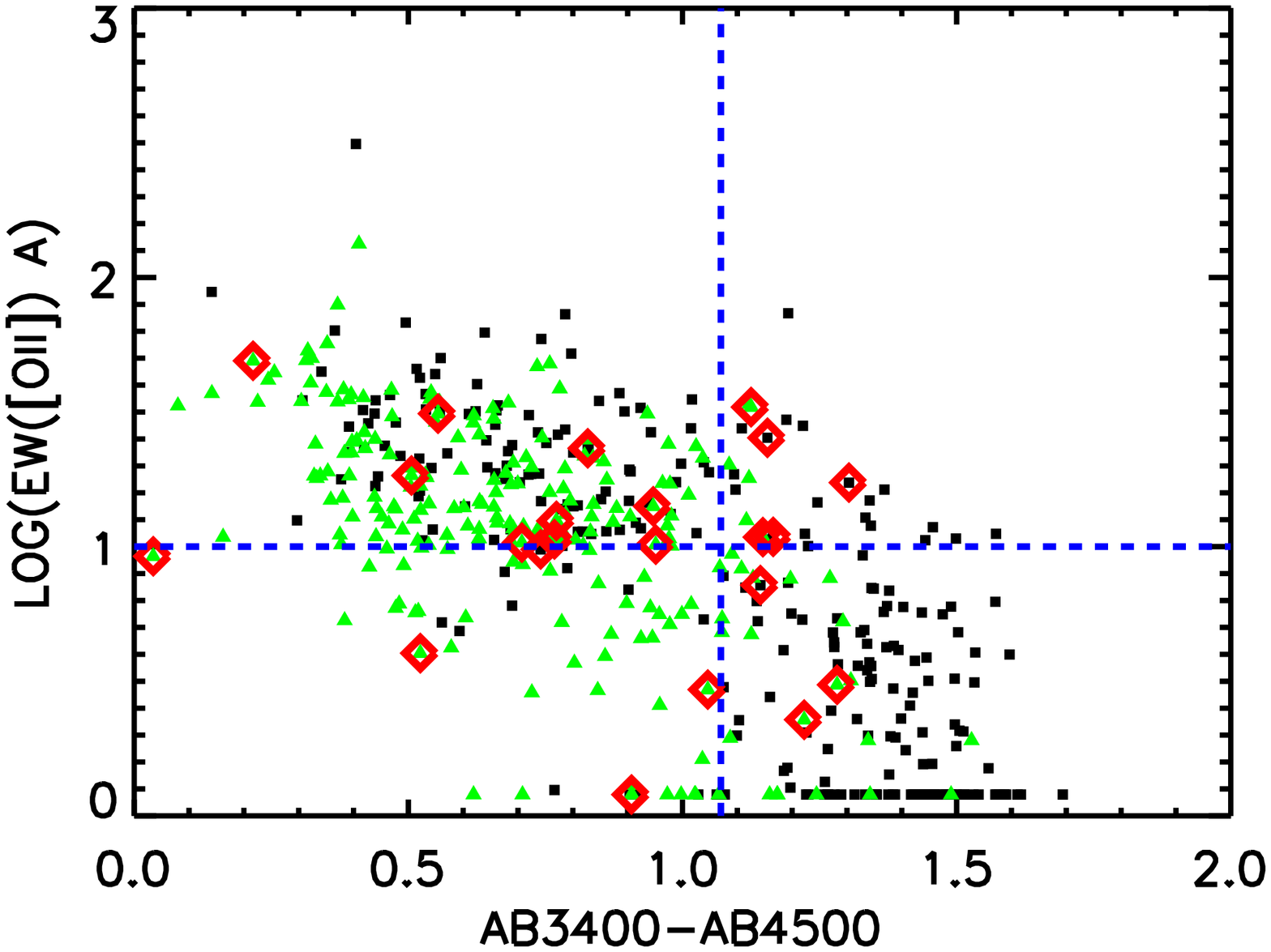,angle=90,width=3.5in}}
\vskip -0.6cm 
\centerline{\psfig{figure=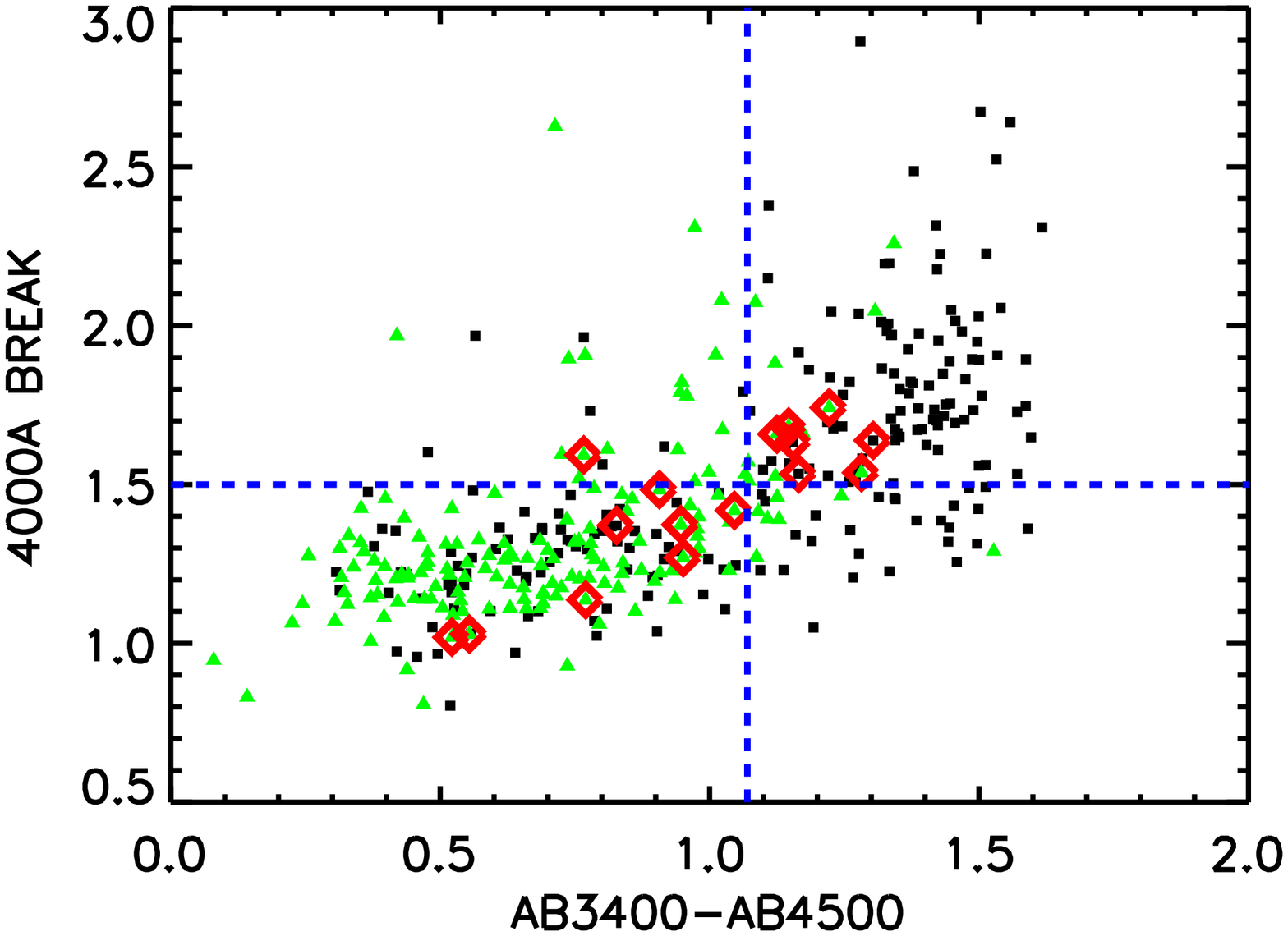,angle=90,width=3.5in}}
\vskip -0.2cm
\figcaption[]{
(a) EW(\oii) and (b) the 4000~\AA\ break vs. the
extinction corrected rest-frame
AB3400$-$AB4500 color for the mid-$z$ sample ($z=0.05-0.9$)
with masses greater than $10^{10}$~M$_\odot$.
The black squares (green triangles) show sources without (with)
$24~\mu$m detections. Sources containing AGNs based on their
X-ray luminosities are enclosed in large red diamonds.
The blue dashed lines show the rough division between
the blue cloud and the red sequence. A color selection
of AB3400$-$AB4500$>1.07$, which would separate the
red sequence from the blue cloud at $10^{11}$~M$_\odot$,
roughly corresponds to an EW(\oii)$<10$~\AA\ or a 
4000~\AA\ break strength greater than 1.5.
\label{tests}
}
\end{inlinefigure}

We may also use the EW(\oii) or the 4000~\AA\ break
to separate the galaxies. The EW(\oii) is independent of
the extinction correction, and the 4000~\AA\ break is nearly
independent so these provide an invaluable check of
our analysis of the colors. In particular this removes
any dependence on our BC03 fitting. In Figure~\ref{tests} we show
(a) the EW(\oii) and (b) the 4000~\AA\ break versus the
extinction corrected rest-frame AB3400$-$AB4500 color.
While there is not a perfect one-to-one relation, a color
selection of AB3400$-$AB$4500>1.07$ {\em (blue dashed
vertical line)\/}, which would separate the red sequence 
from the blue cloud at $10^{11}$~M$_\odot$, roughly corresponds
to an EW(\oii)$<10$~\AA\ {\em (blue dashed horizontal line in a)\/}
or a 4000~\AA\ break strength greater than 1.5 
{\em (blue dashed horizontal line in b)\/}. 
Thus, these cuts may also be used to separate red sequence 
galaxies from blue cloud galaxies.

%
%
\begin{inlinefigure}
\figurenum{51}
\centerline{\psfig{figure=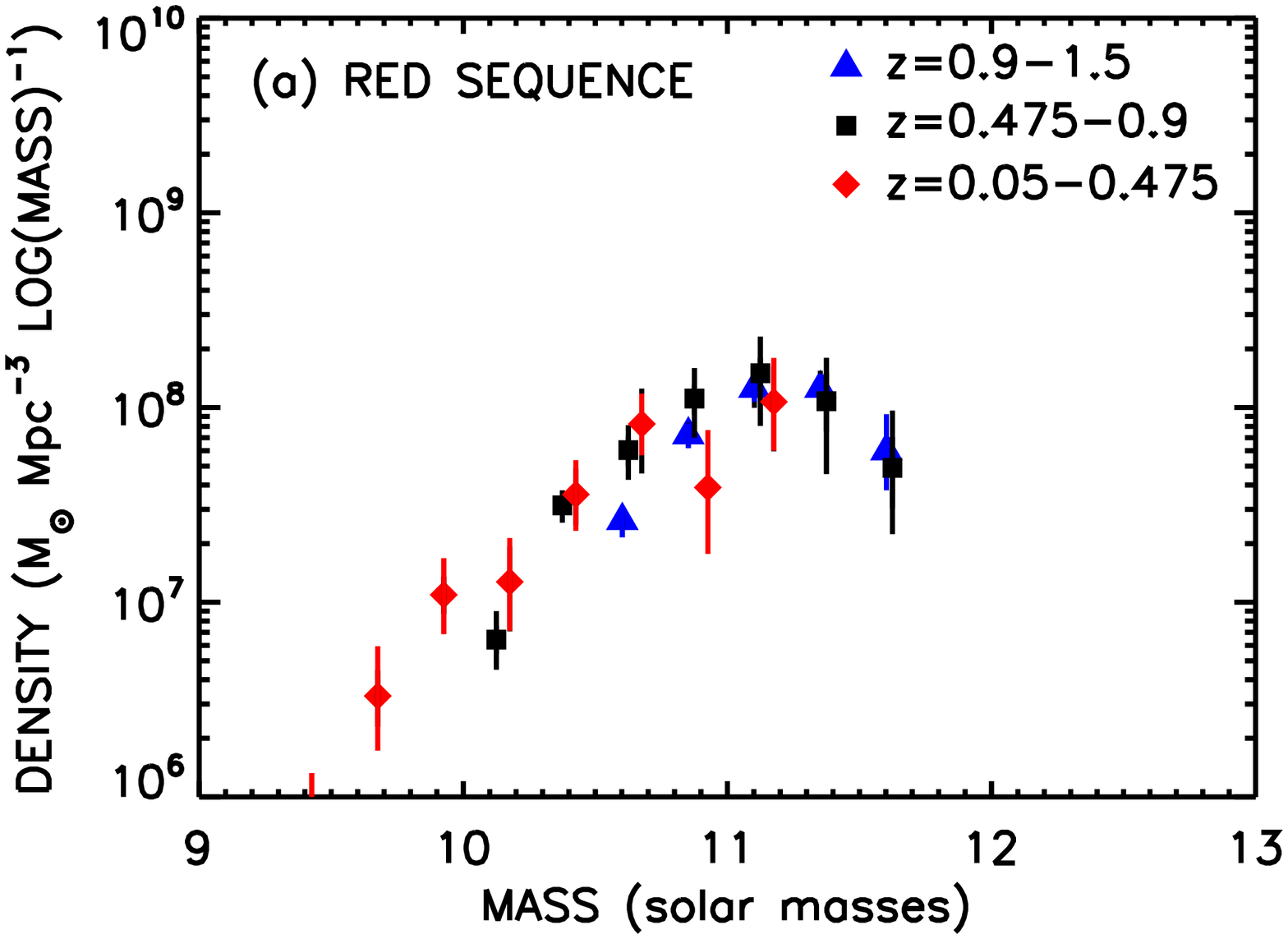,angle=90,width=3.5in}}
\vskip -0.6cm
\centerline{\psfig{figure=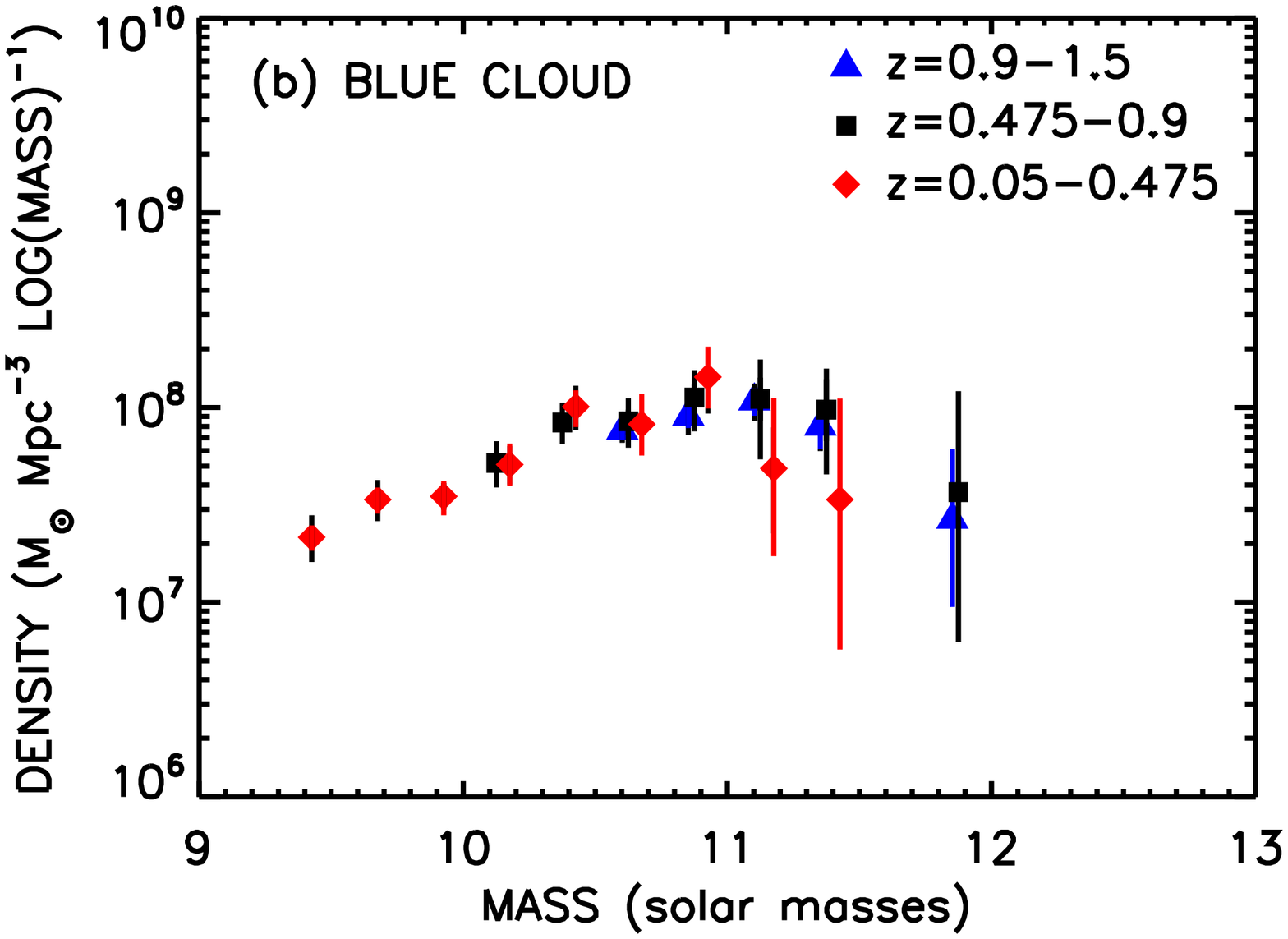,angle=90,width=3.5in}}
\vskip -0.2cm
\figcaption[]{
Mass density per unit log mass functions for
(a) the red sequence galaxies and
(b) the blue cloud galaxies in the redshift intervals
$z=0.9-1.5$ {\em (blue triangles)\/},
$z=0.457-0.9$ {\em (black squares)\/}, and
$z=0.05-0.475$ {\em (red diamonds)\/}.
\label{evol_mass}
}
\end{inlinefigure}

In Figure~\ref{evol_mass} we show the mass density per unit log mass
distribution
functions separated by color using our dividing line between
the blue cloud and the red sequence given in Equation~\ref{sepeqn2}.
There are almost equal amounts of mass in the two color-selected
samples, though the red sequence is highly peaked at $10^{11}$
M$_\odot$ while the blue sequence has a substantial contribution
from lower mass galaxies. In the $z=0.475-0.9$ redshift interval
the red sequence contains $1.29\times10^{8}$ M$_\odot$
Mpc$^{-3}$ and the blue cloud  $1.45\times10^{8}$ M$_\odot$ Mpc$^{-3}$
to the $10^{10}$ M$_\odot$ completeness limit at this redshift.
Figure~\ref{evol_mass}a shows that growth in the mass density
is occurring in red galaxies with masses in the interval
$10^{10.5}-10^{11}$~M$_\odot$.
However, it must be noted that this
result is only based on the difference between the $z=0.475-0.9$
and $z=0.9-1.5$ redhift intervals in the one mass bin and therefore the
conclusion is rather weak.
In contrast, Figure~\ref{evol_mass}b
shows little apparent change in the mass distribution of
the blue cloud with redshift. 
Bell et al.\ (2004), who first noted this effect,
argued that since the star formation, and hence the mass build-up,
is primarily occurring in the blue galaxies,
the blue galaxies must be shifting to the red sequence
at all times in order to leave the blue mass function invariant.
However, Bundy et al. (2006) and Borch et al. (2006) both show
evidence for a decline in the massive blue galaxies with cosmic
time.

%
%
\begin{inlinefigure}
\figurenum{52}
\centerline{\psfig{figure=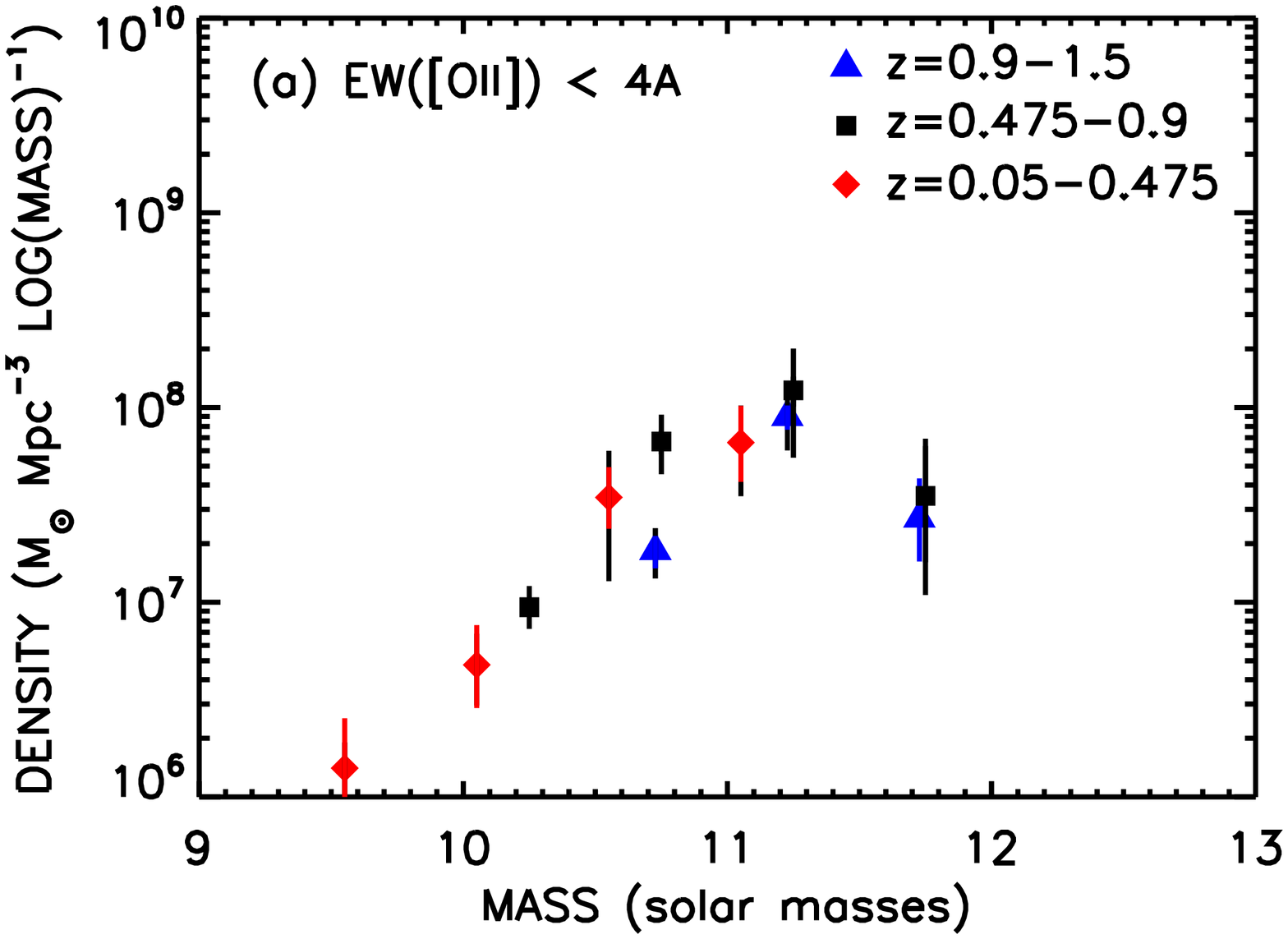,angle=90,width=3.5in}}
\vskip -0.6cm
\centerline{\psfig{figure=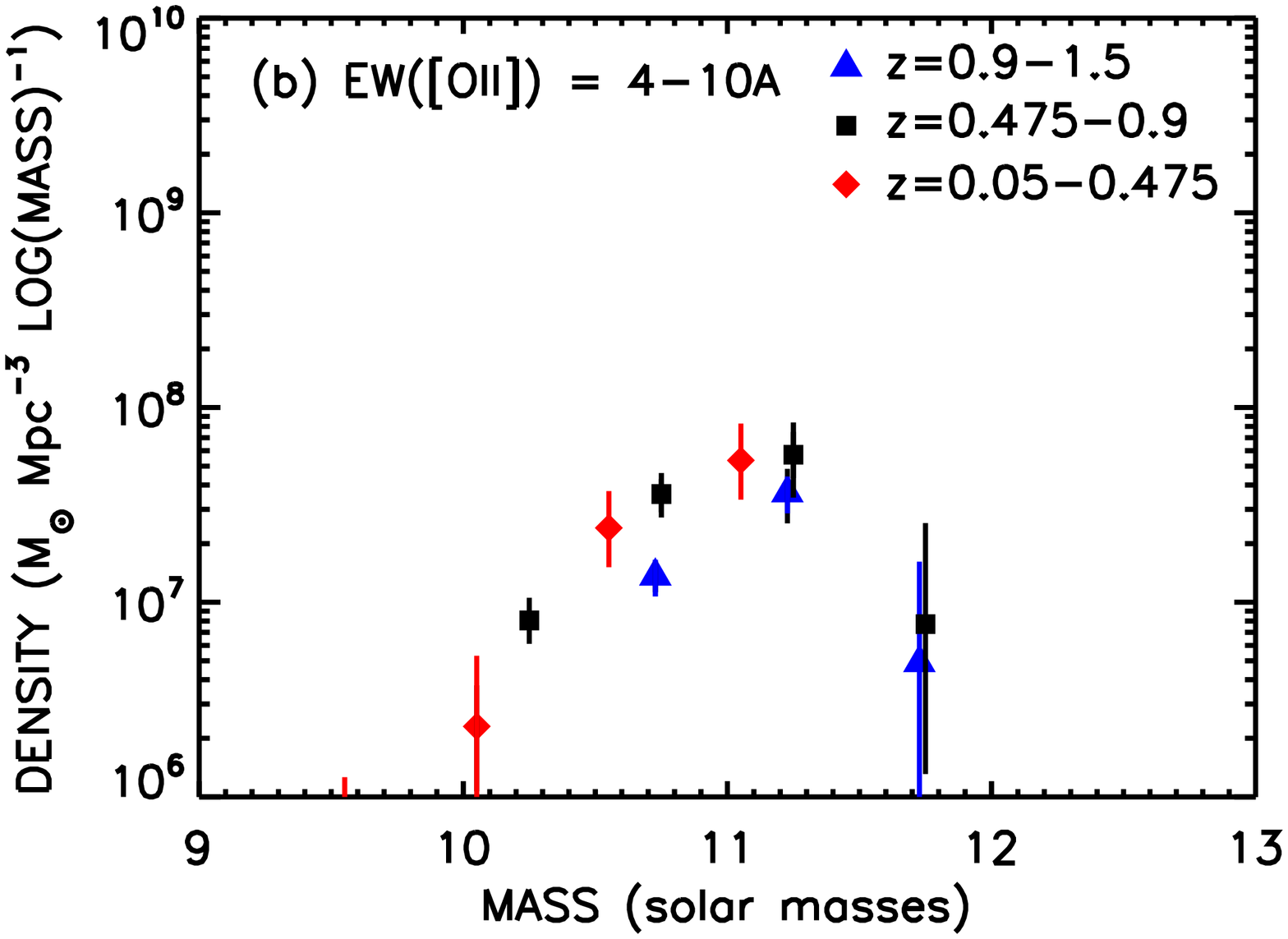,angle=90,width=3.5in}}
\vskip -0.6cm
\centerline{\psfig{figure=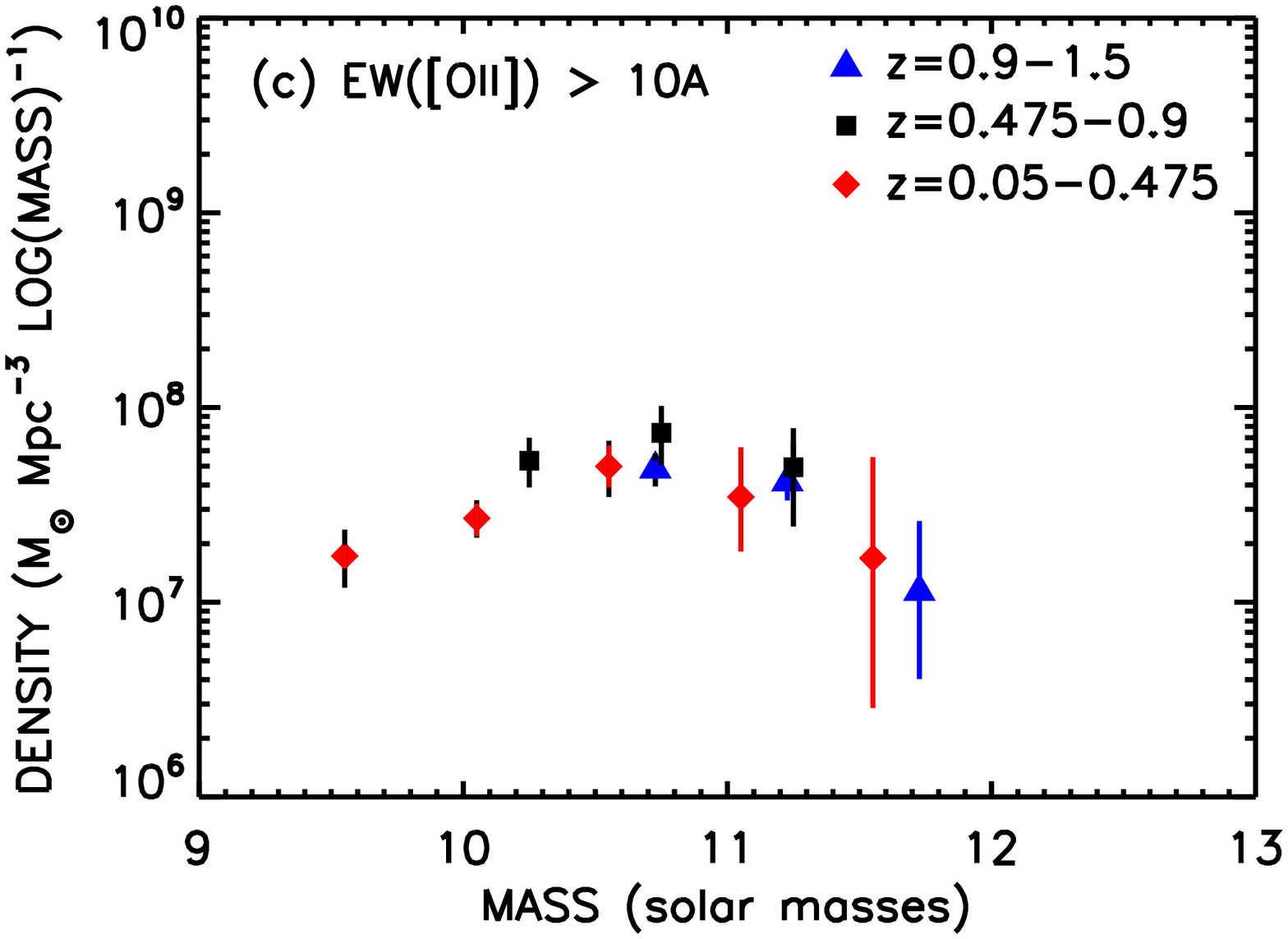,angle=90,width=3.5in}}
\vskip -0.2cm
\figcaption[]{
Mass density per unit log mass functions for rest-frame
(a) EW(\oii$)<4$~\AA, (b) 4~\AA$~<~$EW(\oii$)<10$~\AA,
and (c) EW(\oii$)>10$~\AA\ in the redshift intervals
$z=0.9-1.5$ {\em (blue triangles)\/},
$z=0.457-0.9$ {\em (black squares)\/}, and
$z=0.05-0.475$ {\em (red diamonds)\/}.
\label{evol_o2}
}
\end{inlinefigure}

We can make a finer division using the EW(\oii).
In Figure~\ref{evol_o2} we show the mass function split into 
three classes according to the strength of the \oii\ line. 
In (c) we show sources with rest-frame EW(\oii$)>10$~\AA,
which, as we have discussed above, roughly corresponds to the
blue cloud color selection. As for Figure~\ref{evol_mass}b, we
see few signs of evolution. (Note that cosmic variance may
be causing us some problems with the lowest redshift sources.)
We then split the remaining galaxies 
into (b) weak emitters with $4$~\AA$~<~$EW(\oii$)<10$~\AA\ and
(a) passive sources with EW(\oii$)<4$~\AA. With this division 
we can see growth occurring in both the passive sources and in 
the sources with weak star formation signatures, suggesting that
the red sequence contains both a truly passive population with no 
signs of recent star formation and a population which has more 
recently evolved off the blue cloud and still contains signatures 
of recent star formation.

We may consider this further by plotting the rest-frame
AB3400$-$AB8140 color versus the rest-frame EW(\hb) to
investigate the star formation history. Both are measures of 
the SSFRs in the galaxies and are therefore correlated with 
one another, but the EW(\hb) is produced by higher mass stars 
and thus fades more rapidly than the AB3400$-$AB8140 color, 
providing a well-known age signature.
In Figure~\ref{ewhb_color} we plot the extinction corrected 
rest-frame AB3400$-$AB8140 color versus the rest-frame EW(\hb)
for the $z=0.05-0.9$ sample with masses
(a) $10^{11}-10^{12}$~M$_\odot$,
(b) $10^{10.5}-10^{11}$~M$_\odot$, and
(c) $10^{10}-10^{10.5}$~M$_\odot$.
All but one of the galaxies in the highest mass interval
(Figure~\ref{ewhb_color}a) have little \hb\ emission.
Most of the sources are very red, but there is
a tail which extends to bluer colors. The 24~$\mu$m detected
galaxies {\em (blue triangles)\/} are preferentially bluer
than the non-24~$\mu$m galaxies {\em (black squares)\/}.
We compare the observations with some simple evolutionary
tracks from the BC03 models for galaxies with exponentially
declining SFRs of $10^8$~yrs {\em (black solid)\/} (essentially
a burst model),
$5\times 10^8$~yrs {\em (red dashed)\/}, and
$5\times 10^9$~yrs {\em (green dotted)\/}.
The positions of the galaxies in Figure~\ref{ewhb_color}a
are not consistent with the green dotted line, where the
smoothly declining star formation history
would continue to produce \hb\ emission at intermediate
AB3400$-$AB8140 colors.
Rather, we appear to be seeing episodic bursts of star
formation that have moved the galaxy off
the red sequence, but where the massive stars powering the \hb\
emission have already burned away. Considering the burst model
shown by the black line in Figure~\ref{ewhb_color}a, 
about a quarter of the galaxies
have colors which would require a burst to have occurred in the
last $3\times10^8-10^9$~yrs {\em (thin portion)\/}.
It is these galaxies, which still have substantial
UV flux, that are inferred to have high SSFRs in 
Figure~\ref{dist_specific_bymass}, while the more evolved 
galaxies lie in the low SSFR portion of this diagram.
(This emphasizes again that the SSFRs in 
Figure~\ref{dist_specific_bymass} are not necessarily measures 
of the instantaneous star formation in the galaxy but are a 
time convolution of the SFR history.)
In order to have this fraction of galaxies
in the portion of the evolutionary track
we would require bursts to occur in all galaxies about 
every $4\times10^9$~yrs.
If some galaxies are totally passive and do not
participate in this cycling between the red sequence
and the blue cloud then the remaining galaxies
must have more frequent bursts.

Differential extinction, where the very massive stars 
producing the \hb\ line are more
extinguished than the stars producing the UV continuum, could
increase the EW(\hb) strength. However, this effect would
have to be very large to move the blue sources with very
weak \hb\ to the smooth star formation curves, and, as
we have discussed in \S\ref{secrelext}, we see no signs of this
effect in our comparison of the Balmer line ratios and the
SED derived extinctions. Errors in the UV extinction corrections 
may also introduce scatter in the y-axis and place some sources 
at bluer locations than they should have, but, again, this effect
cannot be large enough to move the sources onto the smooth
star formation curves. Finally, truncation of the SFRs in the smooth
models can move the tracks laterally over onto the burst model
on short timescales, but it would not account for the bluest sources
in the figure, which can only be reproduced with short bursts.

%
%
\begin{inlinefigure}
\figurenum{53}
\centerline{\psfig{figure=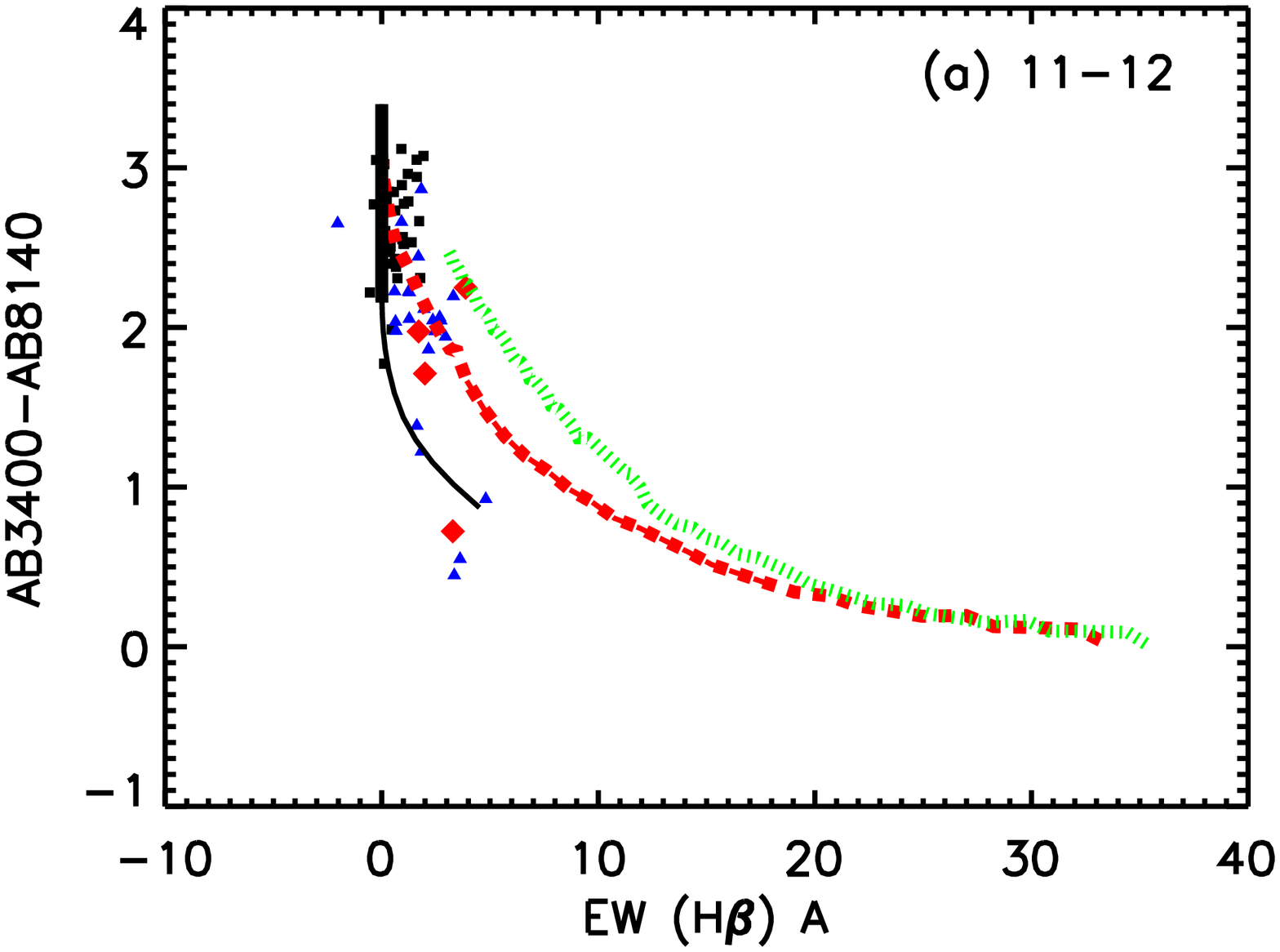,angle=90,width=3.5in}}
\vskip -0.6cm
\centerline{\psfig{figure=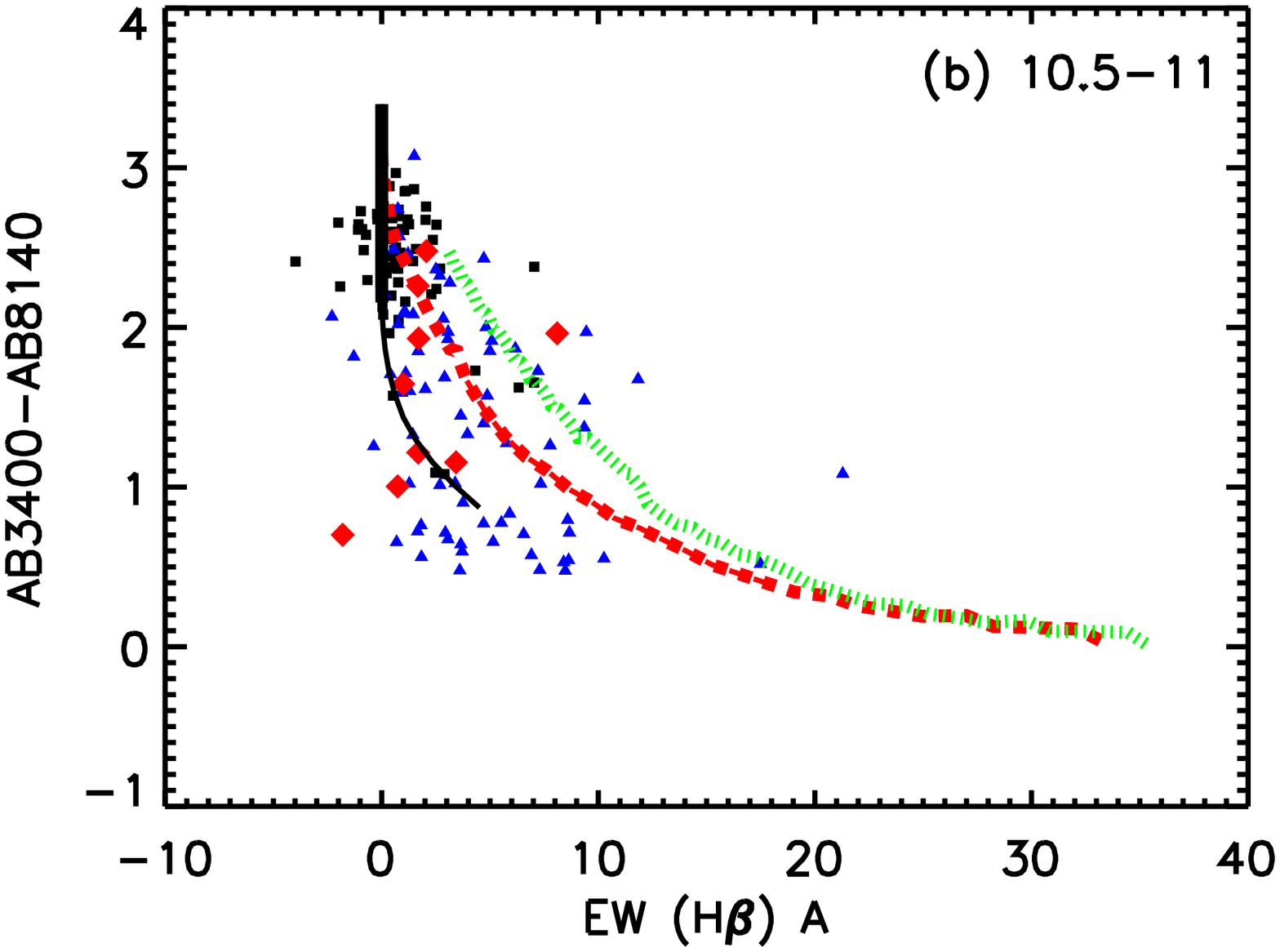,angle=90,width=3.5in}}
\vskip -0.6cm
\centerline{\psfig{figure=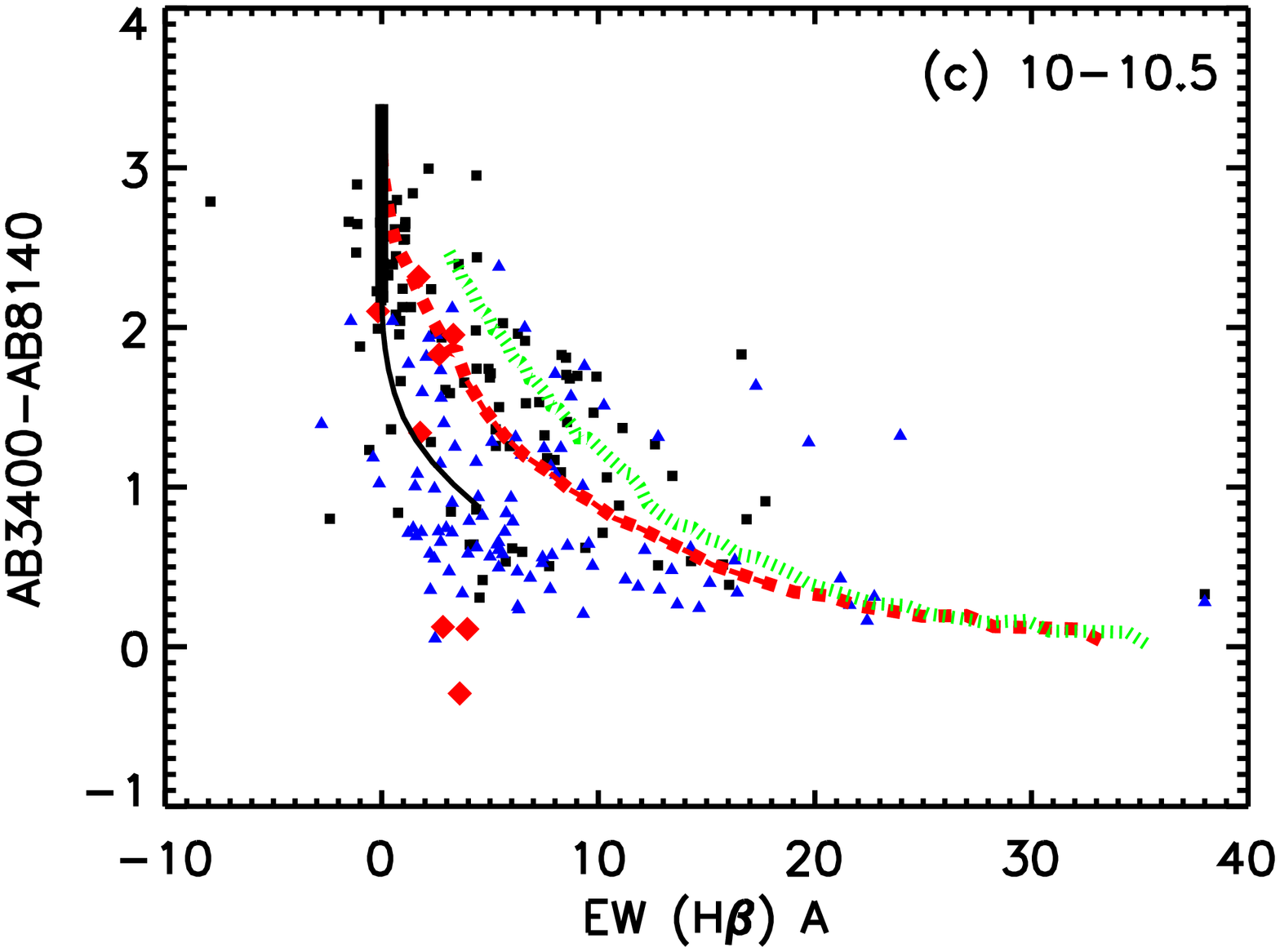,angle=90,width=3.5in}}
\vskip -0.2cm
\figcaption[]{
Extinction corrected rest-frame AB3400$-$AB8140 vs. 
rest-frame EW(\hb)
for the $z=0.05-0.9$ sample in the logarithmic
mass intervals (a) $11-12$~M$_\odot$, (b) $10.5-11$~M$_\odot$,
and (c) $10-10.5$~M$_\odot$.
The black squares (blue triangles) show sources without (with)
$24~\mu$m detections. Sources containing AGNs based on their
X-ray luminosities are denoted by large red diamonds.
The curves show the tracks expected from the BC03 models for
galaxies with exponentially declining star formation rates
of $10^8$~yrs {\em (black solid)\/},
$5\times10^8$~yrs {\em (red dashed)\/},
and $5\times10^9$~yrs {\em (green dotted)\/}.
The black curve is divided into ages of
$3\times10^{8}-10^9$~yrs {\em (thin portion)\/}
and greater than $10^9$~yrs {\em (thick portion)\/}.
\label{ewhb_color}
}
\end{inlinefigure}

The intermediate mass galaxies in Figure~\ref{ewhb_color}b
have a larger fraction of galaxies with blue AB3400$-$AB4500 colors,
suggesting that the burst frequency is higher in these galaxies.
For the lowest mass galaxies shown in Figure~\ref{ewhb_color}c
there appears to be a distinction between the 24~$\mu$m sources
{\em (blue triangles)\/}, 
which still preferentially lie along the burst track, and the 
non-24~$\mu$m sources {\em (black squares)\/}, 
which are more consistent with smooth 
ongoing star formation. This suggests that it is the burst process
which results in the dusty galaxies producing
the 24~$\mu$m emission. In the higher mass galaxies
bursting is the dominant process and all the galaxies
with high SFRs are 24~$\mu$m sources.

\subsection{Galaxy Morphologies}
\label{secmassmorph}

Galaxy morphologies, while closely related to
the colors and spectral properties of the galaxies,
provide an alternative view of the evolution.
In particular, we would like to see in which types
of galaxies the star formation and mass evolution
is occurring and use this information to clarify the
relationship between star formation and
stellar growth in the mass density.

The Ellis morphological classifications that
we are using (see \S\ref{secmorph}) are
based on the {\em HST\/} F850LP images. Therefore, in
order to avoid biasing in type by observing the
galaxies in the rest-frame UV rather than in the rest-frame
optical, we will restrict the NIR sample to only galaxies 
at $z<1.2$ for this section. This ensures that the F850LP 
band corresponds to rest-frame wavelengths above 4000~\AA.

We first compare the morphological typings with the
spectral characteristics of the galaxies. In
Figure~\ref{morph_split} we show the distribution of the
galaxy types in both rest-frame EW(\hb) and 4000~\AA\ break
strength. The black symbols show the E/S0 galaxies (classes $0-2$),
the green symbols show the Sab and S galaxies (classes $3-4$), 
and the small red symbols show the Scd and Irr galaxies
(classes $5-6$). The large red squares show the galaxies
classified as Mergers (class 8). Both the spectral and the
morphological typings place nearly all of the massive galaxies
($10^{11}-10^{12}$~M$_\odot$; Figure~\ref{morph_split}a)
into the E/S0 or Sab-S categories, while the lower mass galaxies
($10^{10}-10^{11}$~M$_\odot$; Figure~\ref{morph_split}b)
show a much wider distribution, including many Irr.
Figure~\ref{morph_split} may be compared with
Figure~4 of Barbaro \& Poggianti (1997) for a local
sample. The positions of the morphological types
in the 4000~\AA\ break-EW(\hb) plane match closely
the positions of the local values. There is no change
in this distribution over the redshift range $z=0.05-0.9$.

In Figure~\ref{ew_morph} we quantitatively show the 
distribution of the EW(\hb) by morphological type 
for the mass intervals (a) $10^{11}-10^{12}$~M$_\odot$,
(b) $10^{10.5}-10^{11}$~M$_\odot$,
and (c) $10^{10}-10^{10.5}$~M$_\odot$.
We divide the galaxies into the broader classes of
E/S0s (classes $0-2$) {\em (solid black line)\/}, 
Spirals (classes $3-5$) {\em (dashed red line)\/},
and Peculiars (classes 6 and 8) {\em (dotted cyan line)\/}.
The distribution of equivalent widths
is very similar for the Spirals and the Peculiars,
but it is a strong function of mass. Most of the massive
Spirals are only weak star formers, while the
lower mass Spirals have a much wider distribution
of equivalent widths, and the mean equivalent
width is much larger.

%
%
\begin{inlinefigure}
\figurenum{54}
\centerline{\psfig{figure=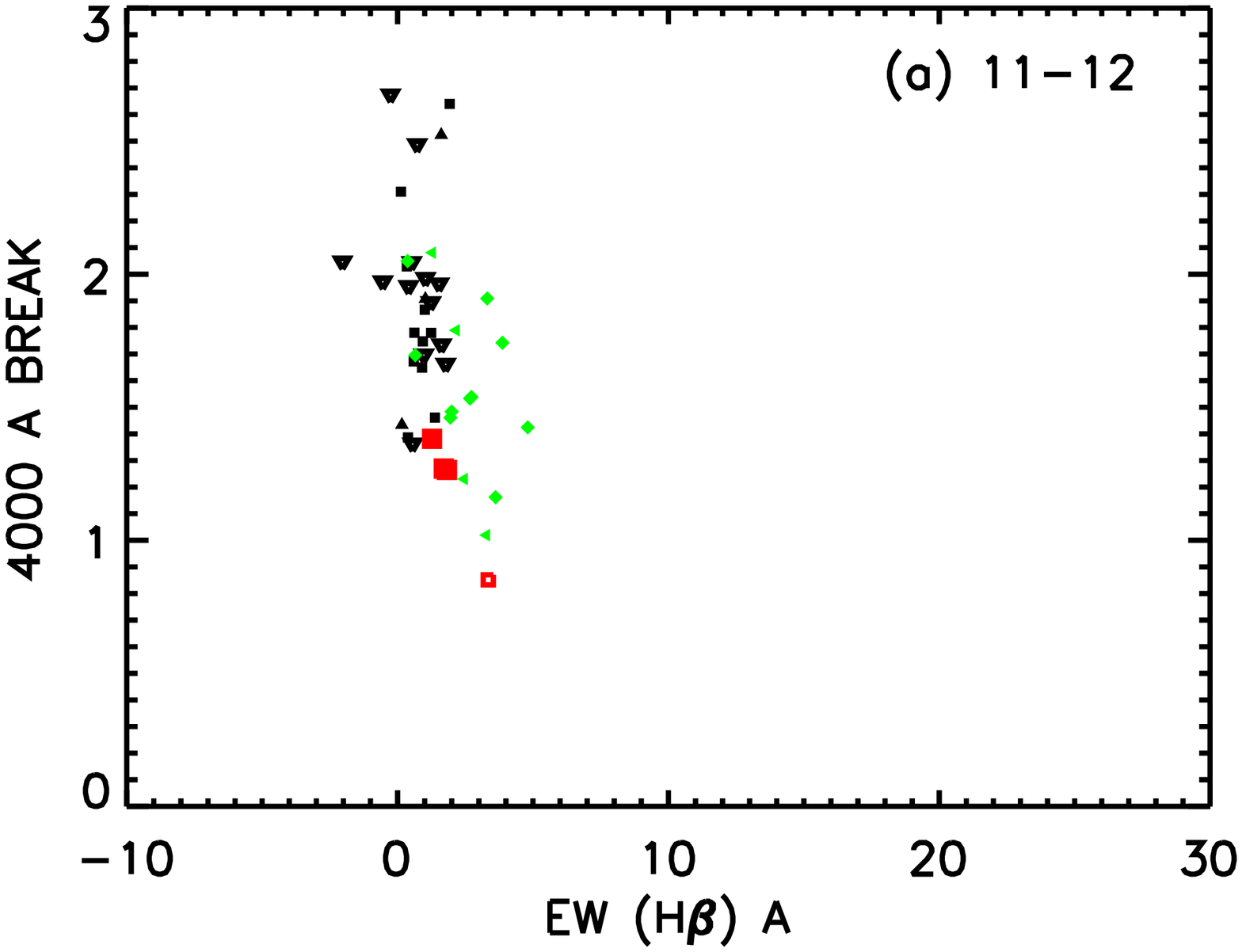,angle=90,width=3.5in}}
\vskip -0.6cm
\centerline{\psfig{figure=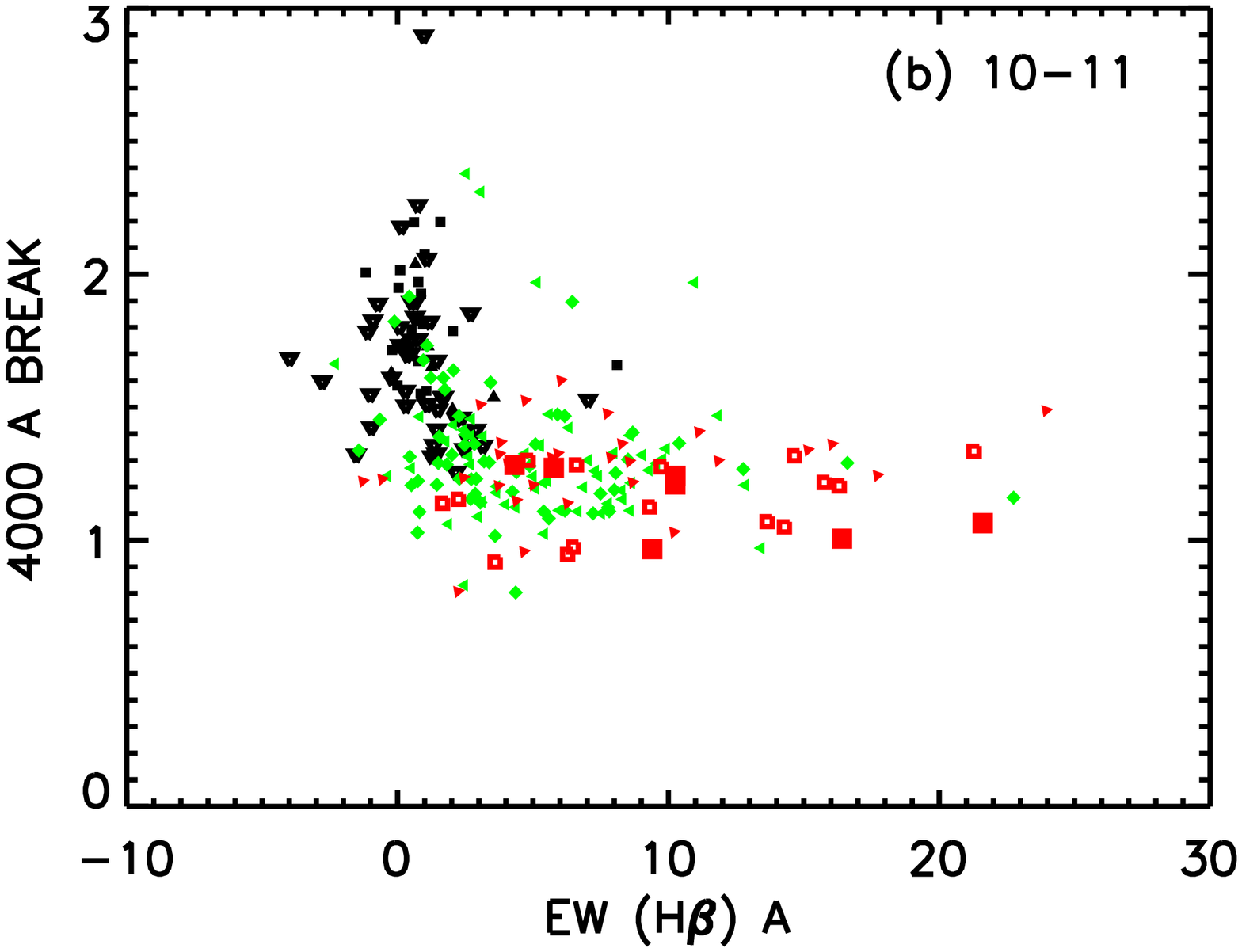,angle=90,width=3.5in}}
\vskip -0.2cm
\figcaption[]{
Distribution of morphological types
in both 4000~\AA\ break strength and
rest-frame EW(\hb) in the mid-$z$ sample
($z=0.05-0.9$) for the logarithmic mass intervals
(a) $11-12$~M$_\odot$ and (b) $10-11$~M$_\odot$.
The black symbols show the E/S0 galaxies
(solid squares = class 0, solid triangles = class 1, and
open downward-pointing triangles = class 2);
the green symbols show the Sab and S
galaxies (diamonds = class 3, leftward-pointing triangles = class 4);
the small red symbols show the Scd and Irr galaxies
(open squares = class 5, rightward-pointing triangles = class 6); 
and the large red squares show the Mergers (class 8).
\label{morph_split}
}
\end{inlinefigure}

In Figure~\ref{mass_morph} we show galaxy morphological 
type versus mass for the redshift intervals $z=0.05-0.475$
{\em (red diamonds)\/}, $z=0.475-0.9$ {\em (black squares)\/},
and $z=0.9-1.2$ {\em (blue triangles)\/}. For each redshift
range we only show galaxies above our mass completeness
level of $2\times10^9$~M$_\odot$, $10^{10}$~M$_\odot$ and
$2\times10^{10}$~M$_\odot$ respectively. Confirming a well-known
result, we see a strong correlation between the galaxy morphology
and the galaxy mass, with many of the most massive galaxies
being E/S0s (classes $0-2$). The large symbols show the median
morphological types by mass and by redshift. A strong evolution
with redshift in the mass-morphology relation is also evident. 
As an example, the typical $10^{11}$~M$_\odot$ galaxy has moved
from being an Sb-like (class 4) galaxy in the $z=0.9-1.2$ redshift
interval {\em (blue)\/} to being an S0 (class 2) galaxy in the 
$z=0.05-0.475$
redshift interval {\em (red)\/}, but this kind of morphological type 
evolution is present across the entire mass range.

%
%
\begin{inlinefigure}
\figurenum{55}
\centerline{\psfig{figure=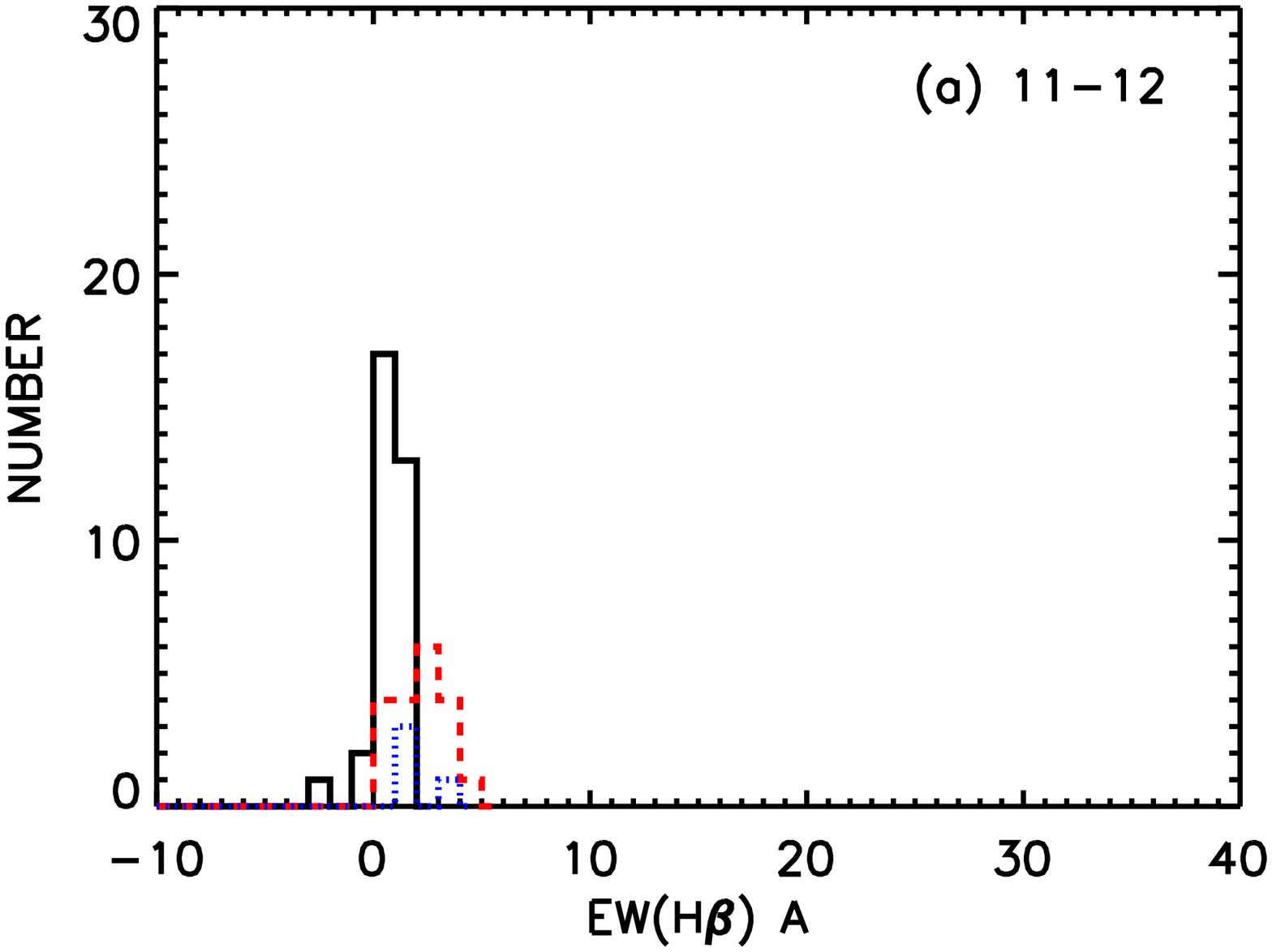,angle=90,width=3.5in}}
\vskip -0.6cm
\centerline{\psfig{figure=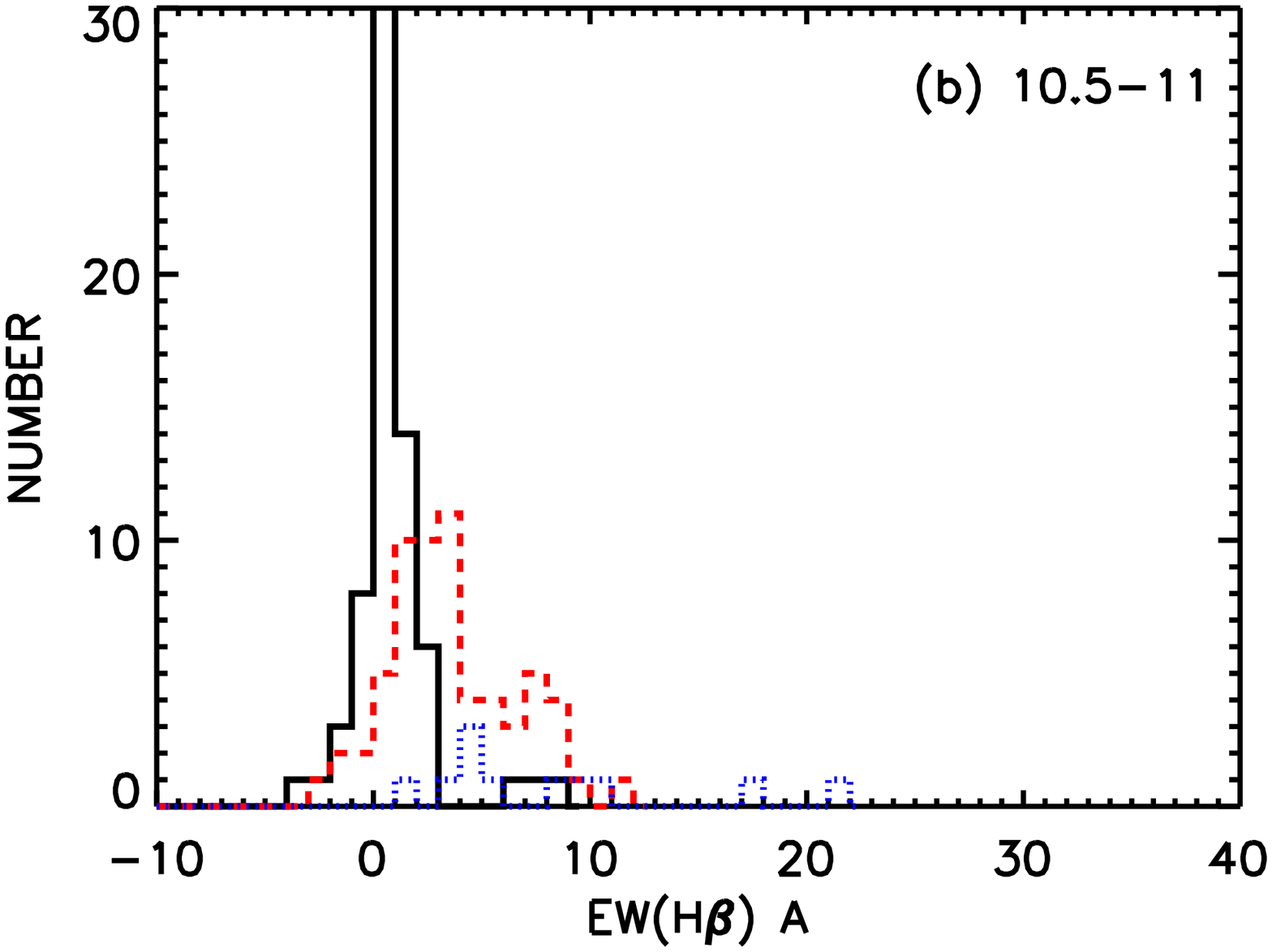,angle=90,width=3.5in}}
\vskip -0.6cm
\centerline{\psfig{figure=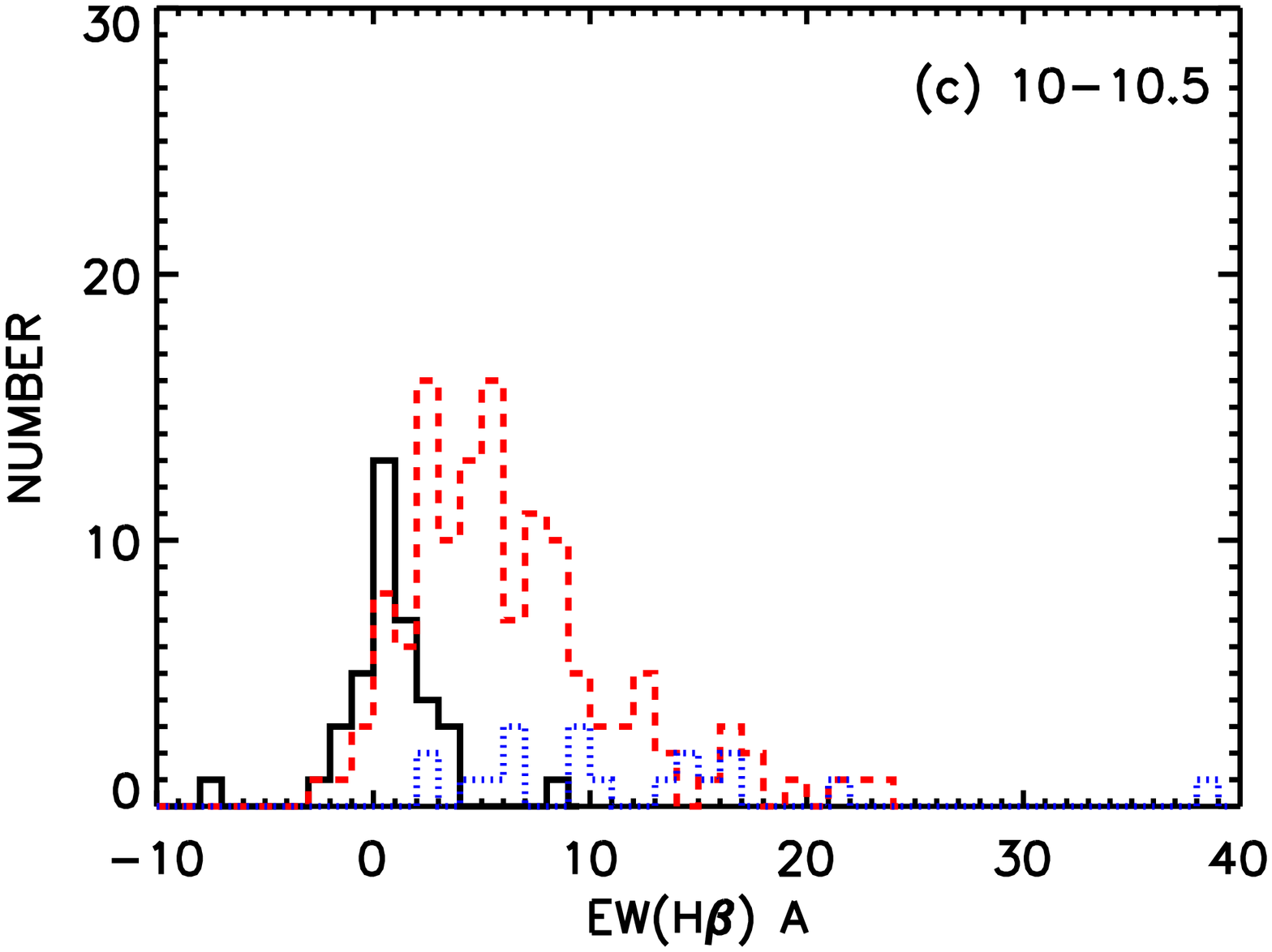,angle=90,width=3.5in}}
\vskip -0.2cm
\figcaption[]{
Distribution of rest-frame EW(\hb) for galaxies
in the redshift interval $z=0.05-0.9$ and in the logarithmic
mass intervals (a) $11-12$~M$_\odot$, (b) $10.5-11$~M$_\odot$,
and (c) $10-10.5$~M$_\odot$. In each panel the solid black line
shows the E/S0 galaxies (classes $0-2$), the red dashed line
shows the Spirals (classes $3-5$), and the dotted blue line
shows the Peculiars (classes 6 and 8).
\label{ew_morph}
}
\end{inlinefigure}

We may now examine the morphological type distribution in 
which the stellar growth in the mass density is occurring. 
As we showed in
\S\ref{seccolor} (Fig.~\ref{evol_o2}), the mass build-up in 
the redshift range $z=0.05-0.9$ is primarily in the passive
(EW(\oii$)<4$~\AA) and weakly
active galaxies (4~\AA$~<~$EW(\oii$)<12$~\AA)
in the $10^{10.5}-10^{11}$~M$_\odot$ interval. 
In Figure~\ref{ewo2_morph_dist} we show the distribution
of morphological types in this redshift and mass interval 
split into passive galaxies {\em (black solid line)\/},
weakly active galaxies {\em (red dashed line)\/},
and strong emission line galaxies {\em (blue dotted line)\/}.
It can be seen that the weakly active galaxies primarily
lie in the spiral and S0 classes, though they are more
strongly weighted to S0s than are the strong emitters,
while the passive galaxies predominantly lie in the E/S0 classes.
(We note in passing that we have visually checked the
emission line galaxies that are morphologically classified as
E, and these classifications are generally robust.)

%
%
\begin{inlinefigure}
\figurenum{56}
\centerline{\psfig{figure=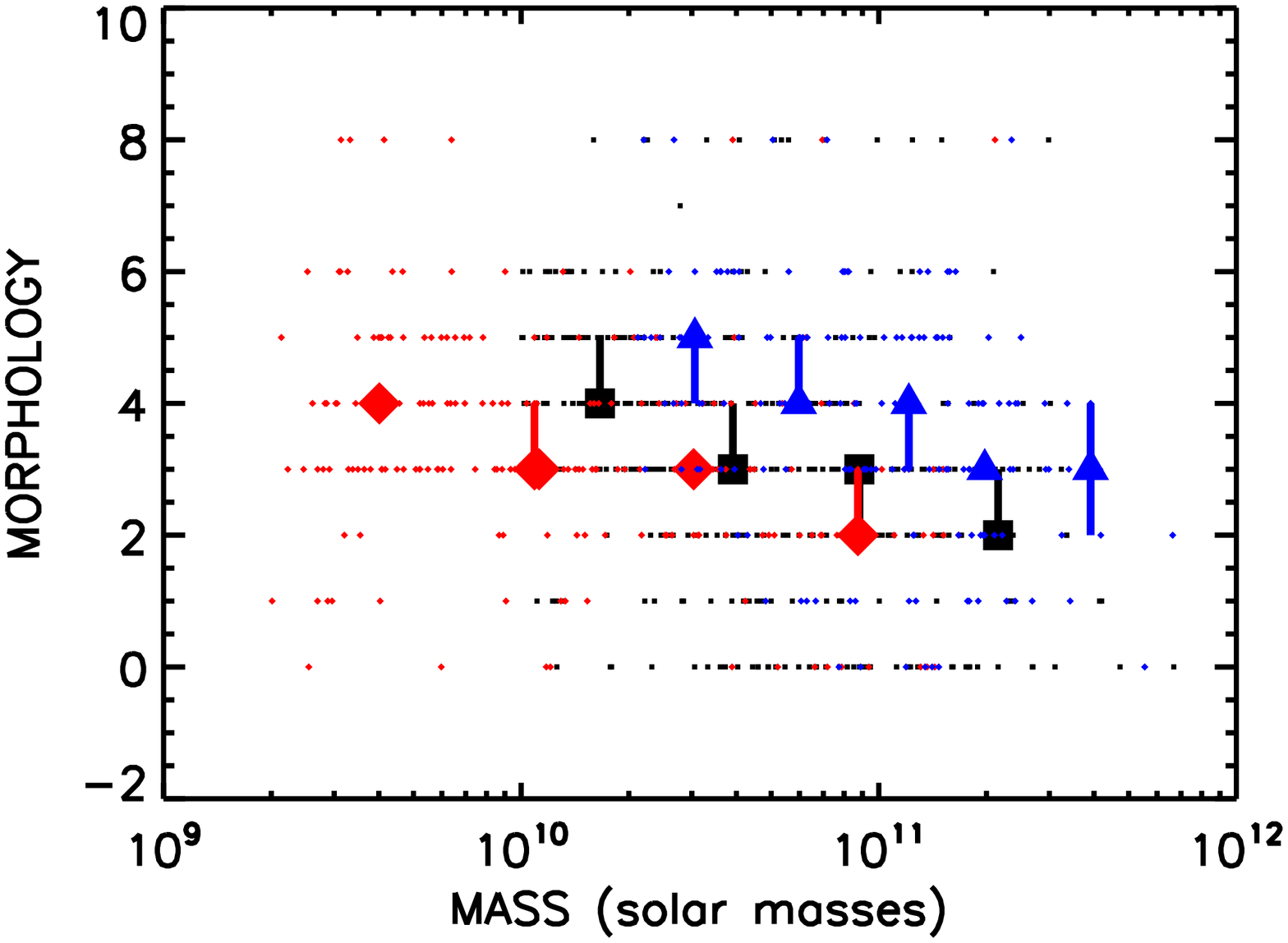,angle=90,width=3.5in}}
\vskip -0.2cm
\figcaption[]{
Galaxy morphological types (classes $0-8$) vs. mass for 
the redshift intervals $z=0.05-0.475$ {\em (red diamonds)\/},
$z=0.475-0.9$ {\em (black squares)\/}, and
$z=0.9-1.2$ {\em (blue triangles)\/}. For each redshift
range we only show galaxies above our mass completeness
level of $2\times10^9$~M$_\odot$, $10^{10}$~M$_\odot$ and
$2\times10^{10}$~M$_\odot$ respectively. In each redshift 
interval the large symbols show the median values for
that mass interval with 68\% confidence limits.
The error bars are generally one morphological class
or less which can result in an asymmetrical appearance.
\label{mass_morph}
}
\end{inlinefigure}

%
%
\begin{inlinefigure}
\figurenum{57}
\centerline{\psfig{figure=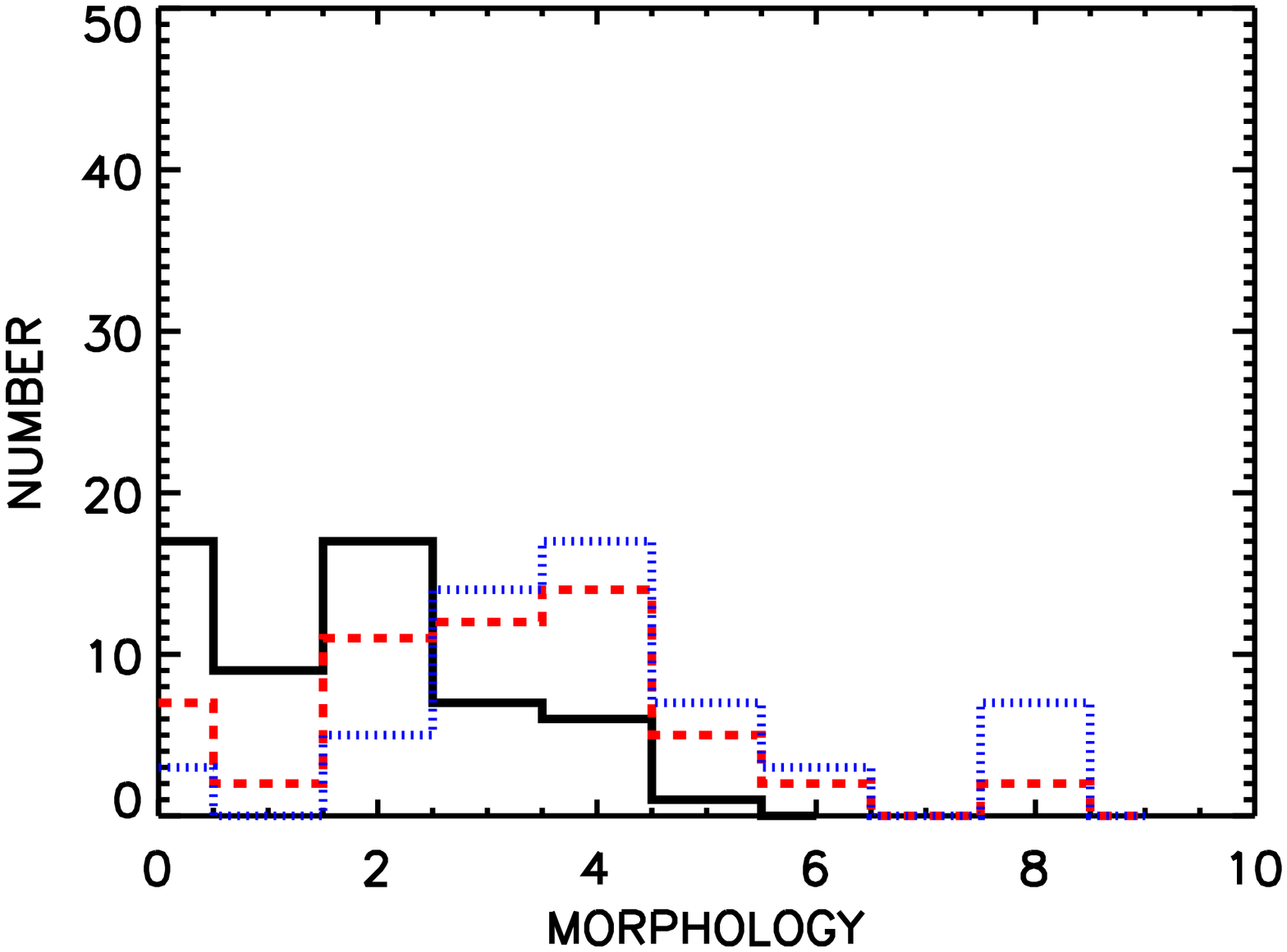,angle=90,width=3.5in}}
\vskip -0.2cm
\figcaption[]{
Distribution of morphological types for galaxies
in the $z=0.05-0.9$ redshift range with logarithmic masses
$10.5-11$~M$_\odot$. The black solid line shows passive
galaxies with EW(\oii$)<4~$\AA, the red dashed line shows
weakly active galaxies with 4~\AA$~<~$EW(\oii$)<12$~\AA,
and the blue dotted line shows strong emission
line galaxies with EW(\oii$)>12$~\AA.
\label{ewo2_morph_dist}
}
\end{inlinefigure}

%
%
\begin{figure*}
\figurenum{58}
\centerline{\psfig{figure=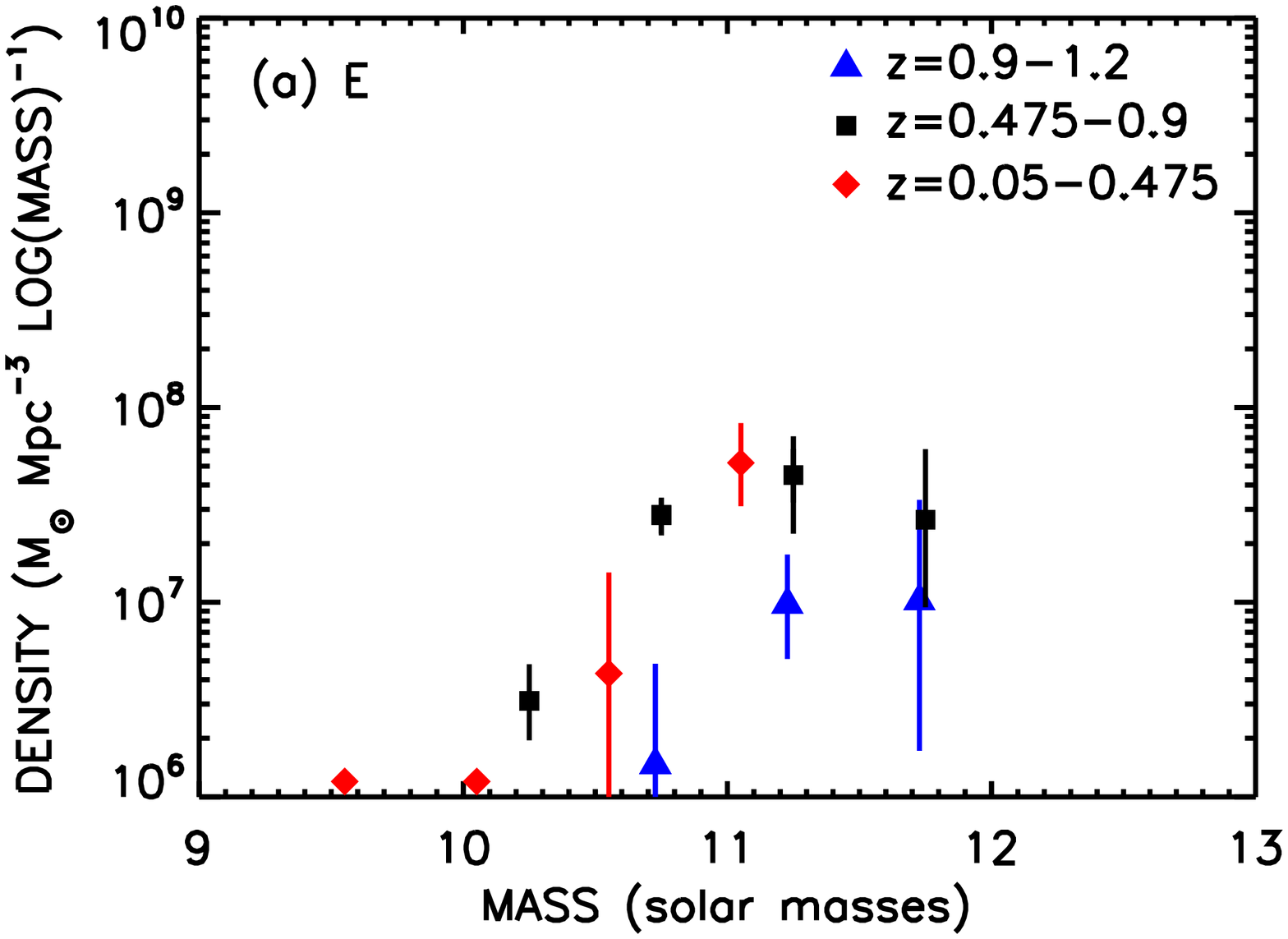,angle=90,width=3.5in}
\psfig{figure=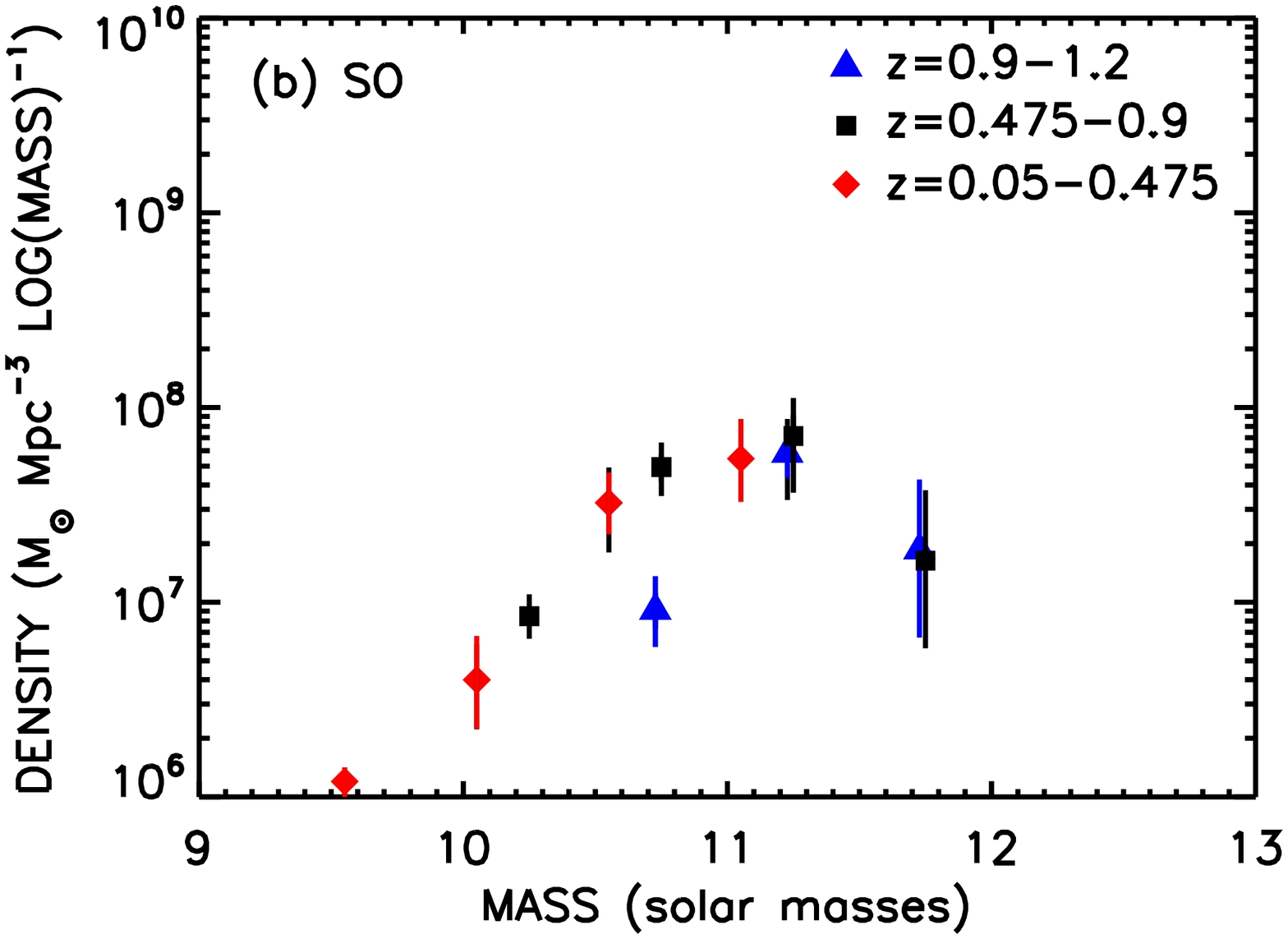,angle=90,width=3.5in}}
\centerline{\psfig{figure=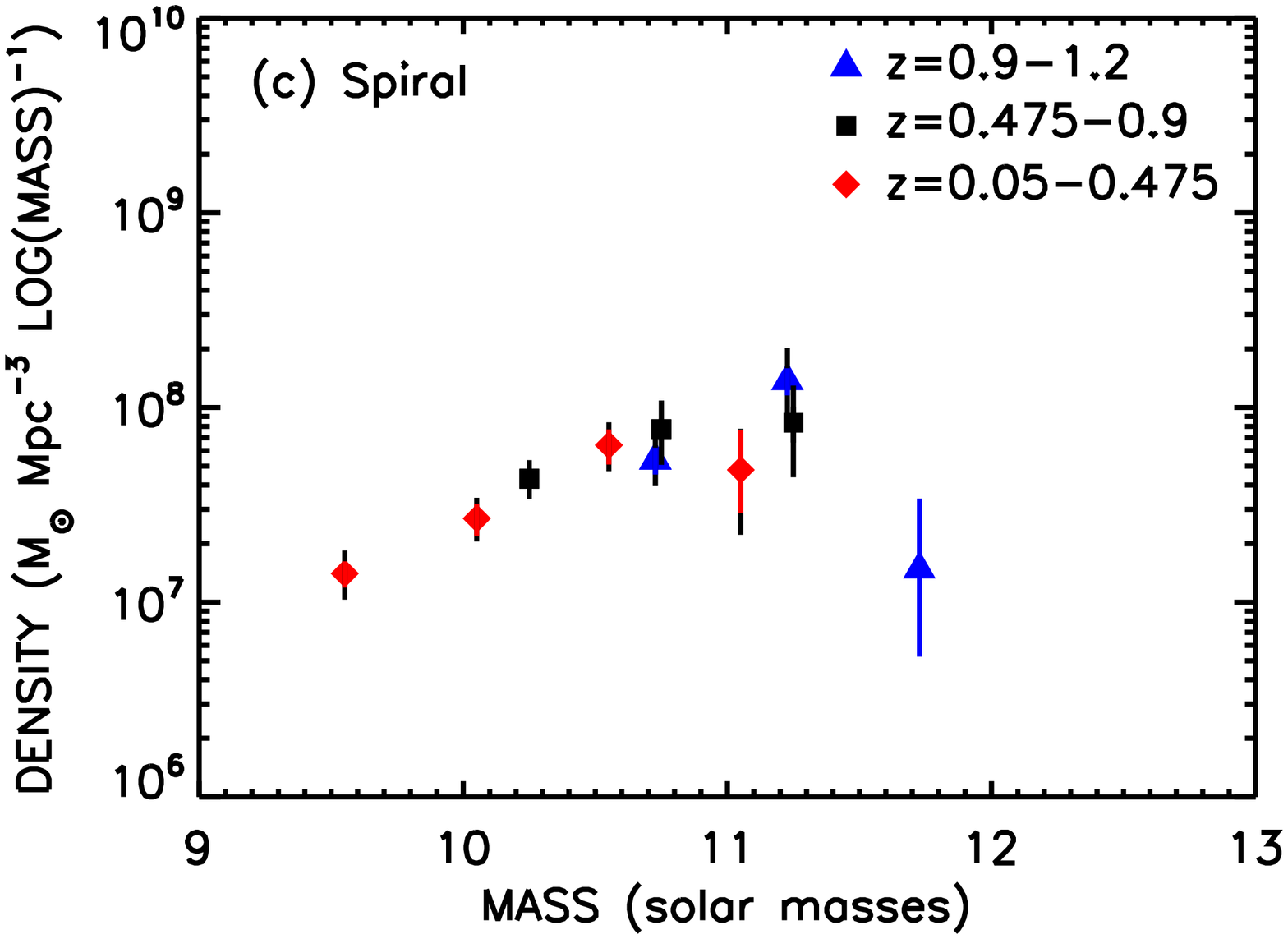,angle=90,width=3.5in}
\psfig{figure=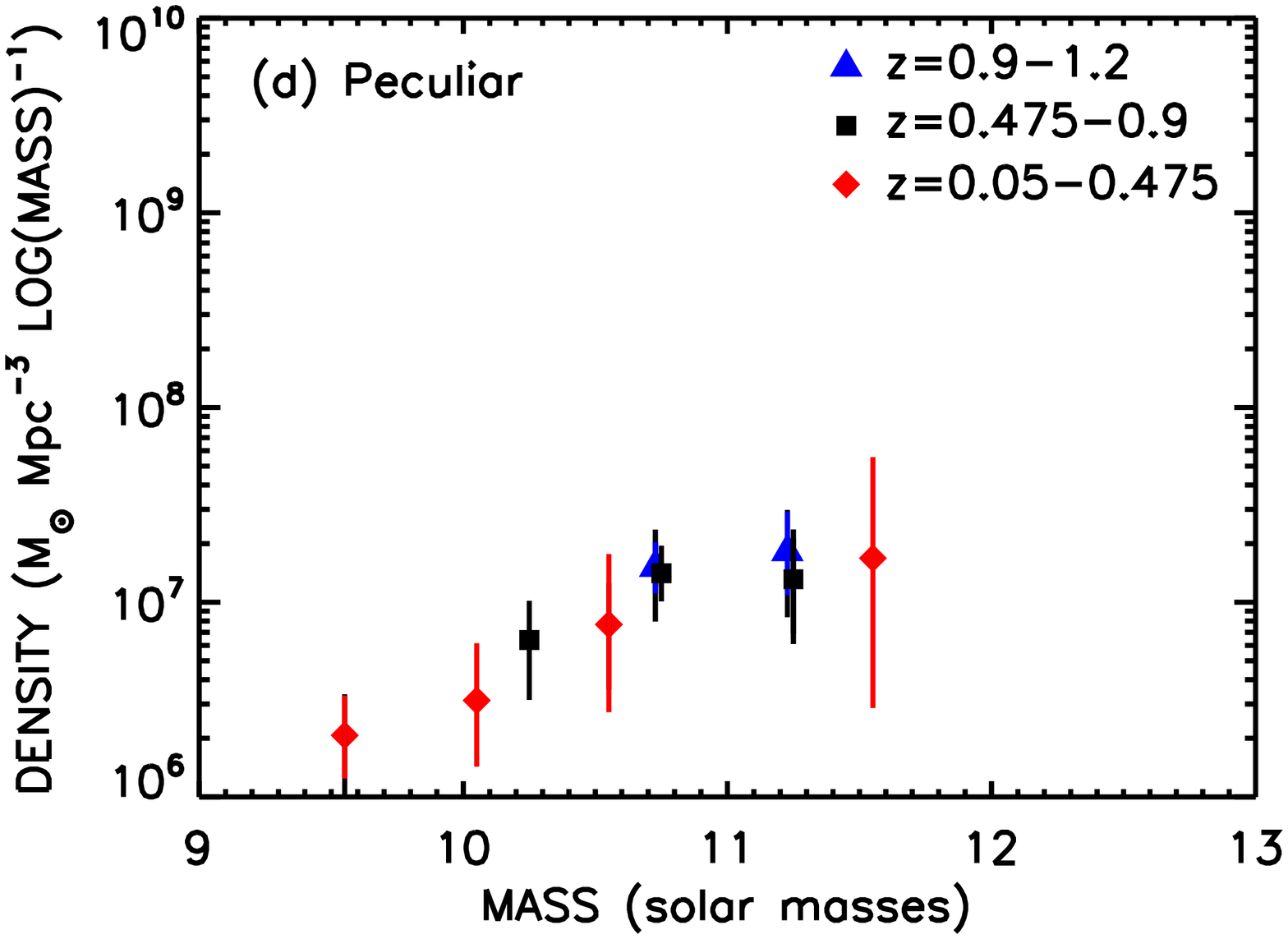,angle=90,width=3.5in}}
\vskip -0.2cm
\figcaption[]{
Build-up of the mass density per unit logarithmic mass
for the redshift intervals $z=0.9-1.2$ {\em (blue triangles)\/}, 
$z=0.475-0.9$ {\em (black squares)\/}, and $z=0.05-0.475$ 
{\em (red diamonds)\/} and for the morphological classes 
(a) E (class 0), (b) S0 (classes $1-2$), (c) Spirals (classes $3-5$), 
and (d) Peculiars (classes 6 and 8).
\label{evol_morph_mass}
}
\end{figure*}

In Figure~\ref{evol_morph_mass} we show the build-up of the 
mass density per unit logarithmic mass as a function of redshift 
and galaxy type. We have separated the E/S0 class into E (class
0) and S0 (classes $1-2$) classes, though if we instead separate
it into E (class $0-1$) and S0 (class 2) classes, it does not 
change our conclusions. Here it can be seen that both the E 
(Fig.~\ref{evol_morph_mass}a) and S0 (Fig.~\ref{evol_morph_mass}b) 
classes are building up strongly in the $10^{10.5}-10^{11}$~M$_\odot$
interval, while the Spirals (Fig.~\ref{evol_morph_mass}c) 
and Peculiars (Fig.~\ref{evol_morph_mass}d) are not changing 
significantly. When we further separate the E/S0 galaxies by
the EW(\oii), we find that all of the growth is in the passive 
or weakly active galaxies.

Thus, it appears that the mass build-up is primarily moving into 
elliptical and S0 galaxies. In other words, mass formation
occurs in the spiral galaxies, and the spiral galaxies
gradually move into the E/S0 class with decreasing redshift as the 
overall star formation dies away. This results in the mass function 
of the strong emitters and spiral galaxies being relatively 
invariant and the mass build-up being primarily seen in the 
passive and weakly active E/S0 galaxies.

\subsection{Galaxy Environments}
\label{secenv}

As described in \S\ref{secgalenv}, we use the projected
nearest neighbor density, $\Sigma_3$, to characterize the
galaxy environment. In order to provide a uniform sample
with a sufficiently high density to minimize edge effects,
we use the galaxy sample with masses greater than 
$2\times10^{10}$~M$_\odot$ to compute $\Sigma_3$.
This restricts our analysis to the redshift range $z=0.3-1.2$.
The lower redshift bound is set by the size of the field and our
edge restriction (objects must be more than 1~Mpc from the
field edge), and the upper redshift bound is set by the mass limit 
of the sample. The average density is $\Sigma_3=0.96$~Mpc$^{-2}$,
and very few objects have densities less than 0.3~Mpc$^{-2}$,
where edge effects begin to enter. (A $4\times10^{10}$~M$_\odot$
sample gives an average $\Sigma_3=0.31$\ Mpc$^{-2}$, where this 
issue would be more significant.)

%
%
\begin{inlinefigure}
\figurenum{59}
\centerline{\psfig{figure=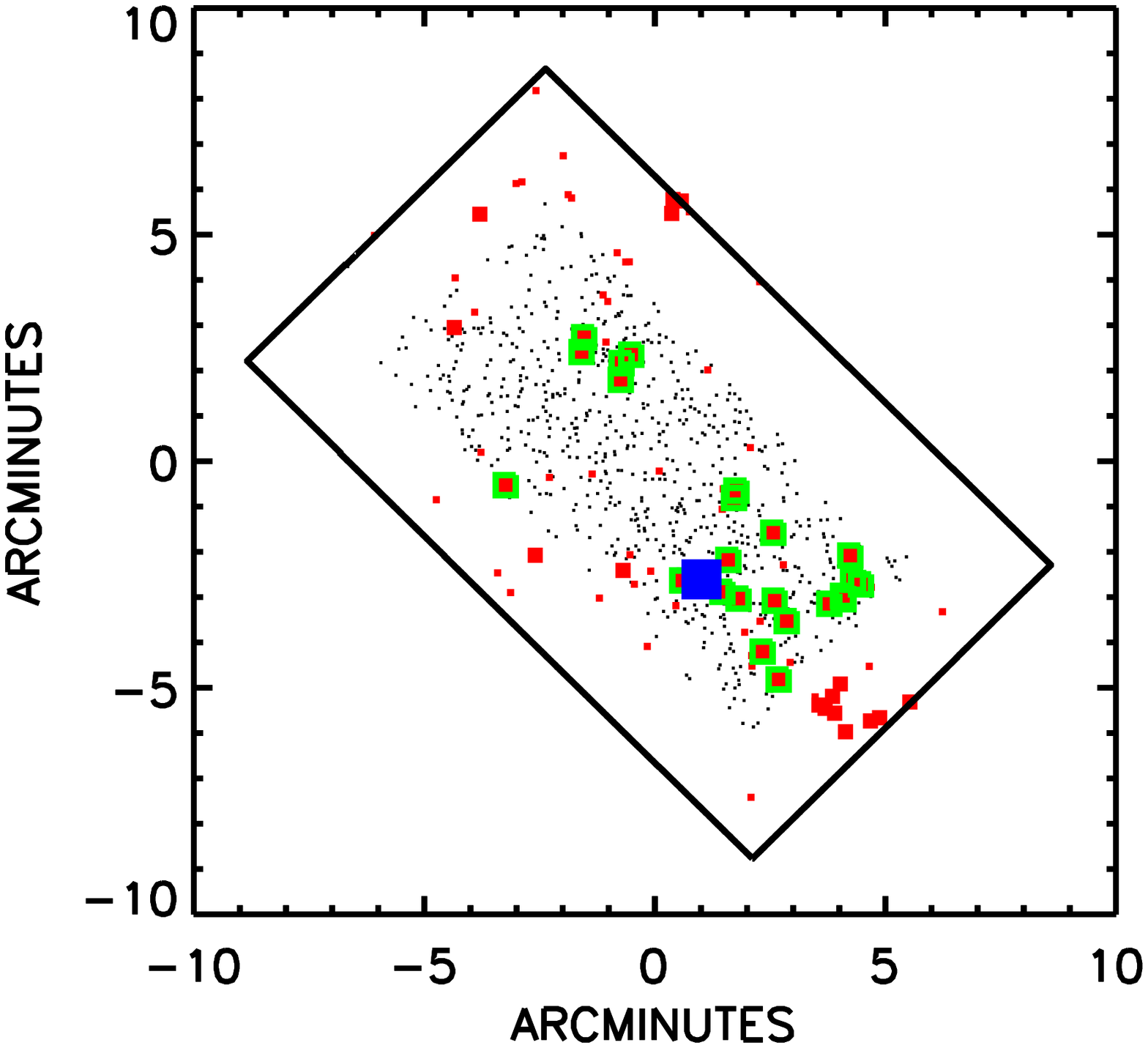,angle=90,width=3.5in}}
\vskip -0.2cm
\figcaption[]{
Galaxies in the $z=1.0156$ sheet in the GOODS-N region are shown
by the red squares (large symbols correspond to galaxies
with masses above $2\times10^{11}$~M$_\odot$; small
symbols to lower mass galaxies). The full region of the
field is shown by the black rectangle. The galaxies
with measured $\Sigma_3$ are shown with small black squares.
This region is smaller than the full field because of the edge
constraint. Galaxies in the velocity slice with $\Sigma_3>3$~Mpc$^{-2}$ 
are shown enclosed in green squares.
The large blue symbol shows the position of
the diffuse X-ray emission from Bauer et al.\ (2002).
\label{goods_group}
}
\end{inlinefigure}

Nearly all of the sources with $\Sigma_3>5$\ Mpc$^{-2}$ lie 
in just four velocity sheets at $z=0.4851$,\ 0.8472,\ 0.9367,\ 
and 1.0156. The strongest of these is the well-known structure 
at $z=1.0156$, which we illustrate in Figure~\ref{goods_group}
{\em (red squares)\/}. This feature is dominated by a fairly
substantial cluster lying at the southern end of the GOODS-N region,
together with a smaller concentrated group to the north. We have
identified 29 galaxies in the southern cluster with masses above 
$2\times10^{10}$~M$_\odot$. The total stellar mass of 
these 29 galaxies alone is $4\times10^{12}$~M$_\odot$.
The velocity dispersion is 470~km~s$^{-1}$. There is
associated diffuse X-ray emission centered on one part of the
cluster (Bauer et al.\ 2002) {\em (blue square)\/}.
Within the region where the density parameter can be measured,
nearly all of the galaxies at $z=1.0156$ lie in substantially 
overdense regions with $\Sigma_3>3~$Mpc$^{-2}$ 
{\em (green squares)\/}. The other sheets are weaker 
and, in some cases (e.g., the $z=0.8472$ structure), more diffuse.

Because the redshift intervals have different mass limits, 
it is most natural to plot the density parameter versus mass. 
In Figure~\ref{mass_dens} we show this relation for the redshift
intervals $z=0.9-1.2$ {\em (blue triangles)\/},
$z=0.475-0.9$ {\em (black squares)\/}, and $z=0.3-0.475$
{\em (red diamonds)\/}. 
The mass-density relation is clearly 
seen in the median values {\em (large symbols)\/}. 
At lower masses (below $10^{11}$~M$_\odot$) the dependence
on the environment is very weak. However, nearly all of the 
most massive galaxies lie in higher density regions. There is 
relatively little evolution in the mass-density relation 
over the $z=0.05-0.9$ redshift range.

%
%
\begin{inlinefigure}
\figurenum{60}
\centerline{\psfig{figure=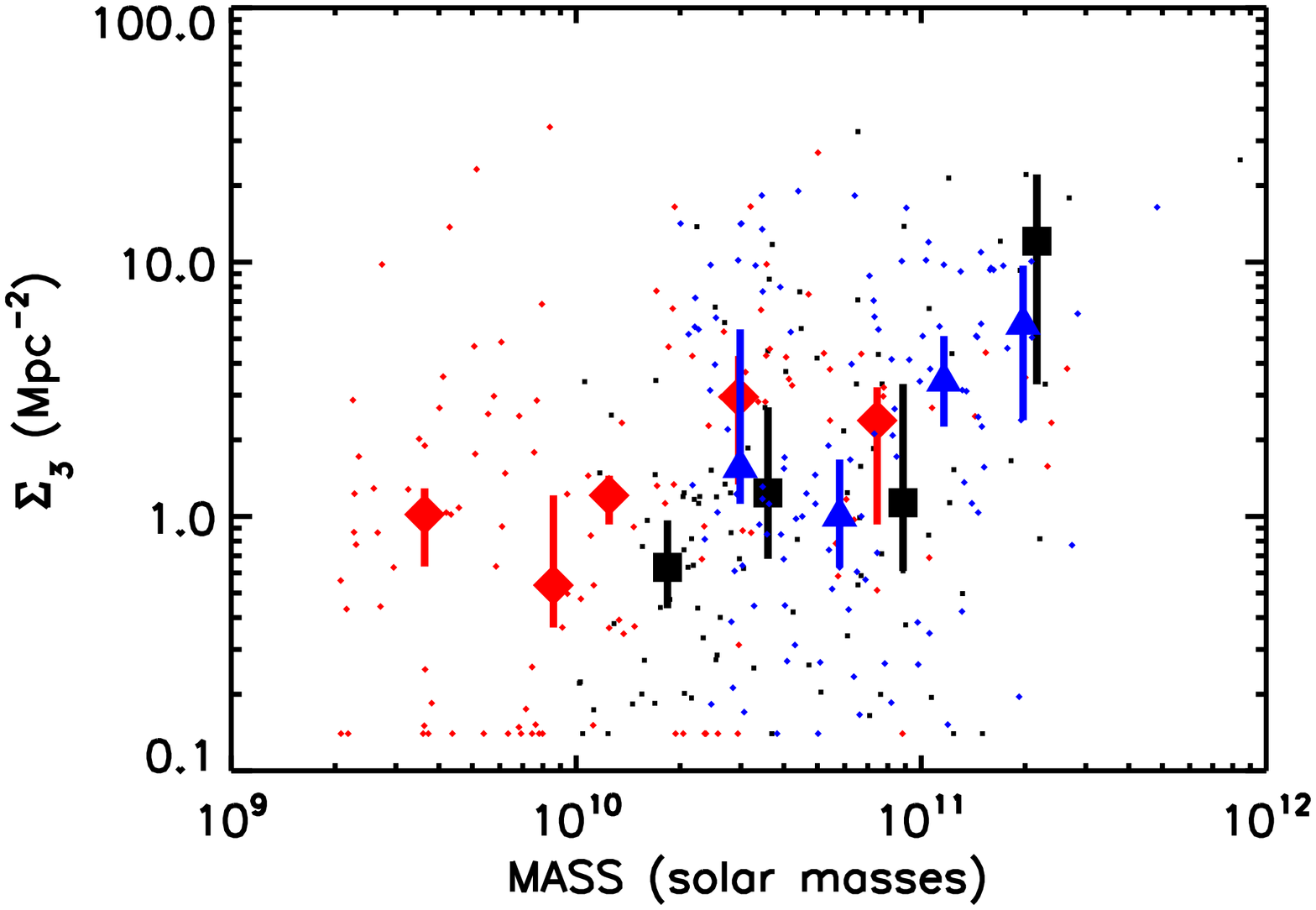,angle=90,width=3.5in}}
\vskip -0.2cm
\figcaption[]{
$\Sigma_3$ density parameter vs. stellar mass for the redshift
intervals $z=0.9-1.2$ {\em (blue triangles)\/},
$z=0.475-0.9$ {\em (black squares)\/}, and $z=0.3-0.475$
{\em (red diamonds)\/}. In each case the large symbols
show the median values for the mass intervals with 68\%
confidence limits.
\label{mass_dens}
}
\end{inlinefigure}

While this relationship is well known in general terms, 
it is not easy to compare it precisely with other results,
either because the environmental parameters are expressed
in different ways, or because the increases are
measured for optical luminosity rather than for mass 
(e.g., Croton et al.\ 2005; Hoyle et al.\ 2005; 
Cooper et al.\ 2007, 2008). However, Baldry et al.\ (2006) 
provide a local analysis of the galaxy masses using
a similar environmental parameter.
They characterize their Schechter function fits to
the galaxy mass functions in different environments 
with the mass at which the contribution to
the local galaxy mass density per dex in galaxy mass peaks.
They show in their Figure~8d how this peak mass depends
on $\Sigma$.
We have measured this quantity for our $z=0.6-1.2$
sample and show the result in Figure~\ref{mass_dens_plot}.
In order to compare with Baldry et al.\ (2006) 
{\em (blue solid line)\/}, we have adjusted their Kroupa 
masses to Salpeter masses, and we have matched their median 
$\Sigma$ parameter to ours. We see from
Figure~\ref{mass_dens_plot} that the mass 
versus $\Sigma_3$ relation has a very similar slope at both
redshifts but that the mass in a given environment
is about 0.3~dex higher at $z=0.9-1.2$ than it is now.
This is expected, since the high-mass galaxies are
already in place at the higher redshifts, while the
low-mass galaxies are still forming. However,
it shows that this relative growth of the low-mass
galaxies is occurring across our measured density
range and is not a strong function of environment.

%
%
\begin{inlinefigure}
\figurenum{61}
\centerline{\psfig{figure=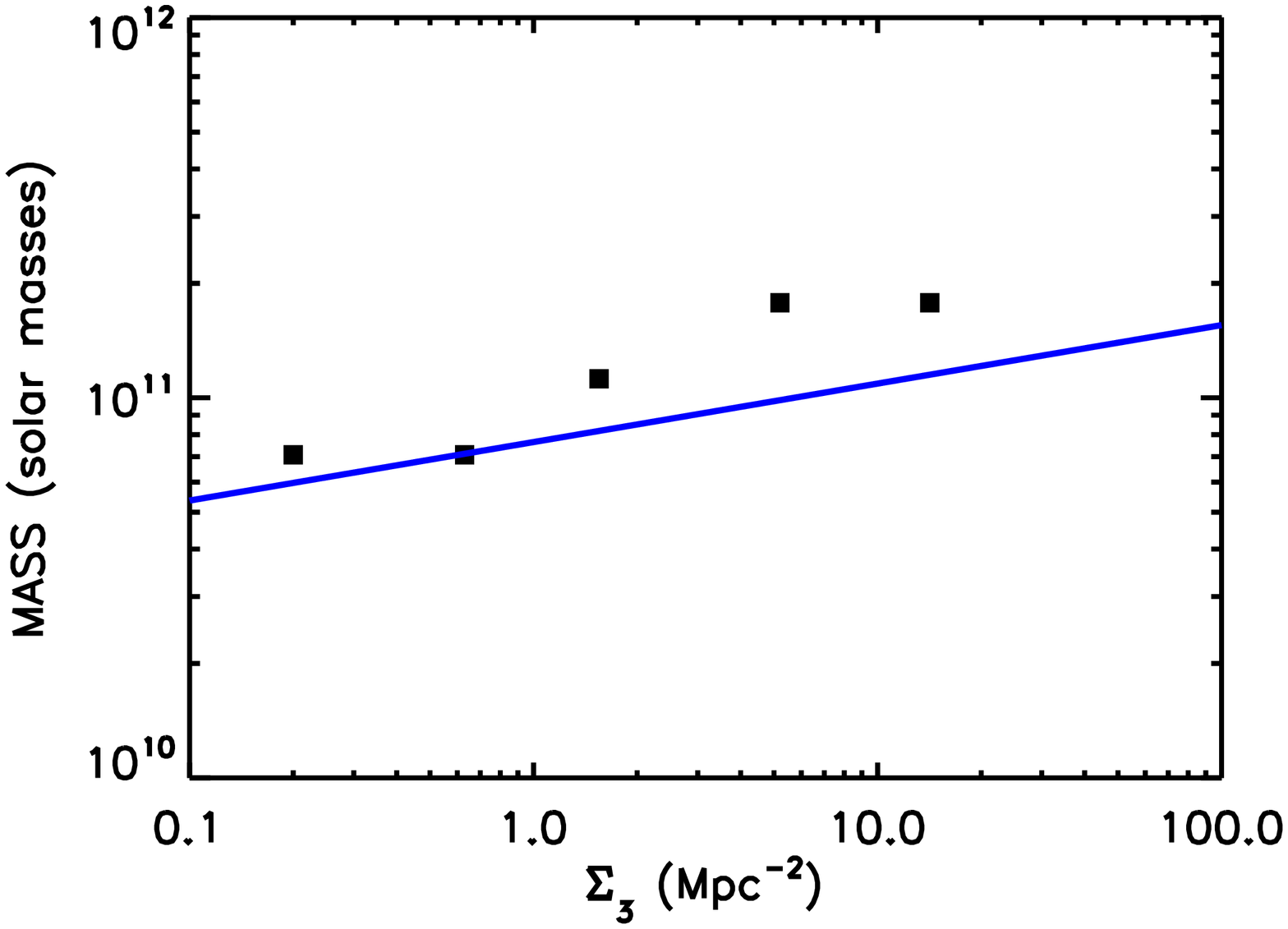,angle=90,width=3.5in}}
\vskip -0.2cm
\figcaption[]{
Mass at which contribution to the mass density per
dex in galaxy mass peaks vs. $\Sigma_3$ density parameter. 
The black squares show the results from the present data 
in the $z=0.9-1.2$ redshift interval. The blue line shows 
the corresponding local result derived by Baldry et al.\ (2006).
\label{mass_dens_plot}
}
\end{inlinefigure}

\begin{figure*}
\figurenum{62}
\centerline{\psfig{figure=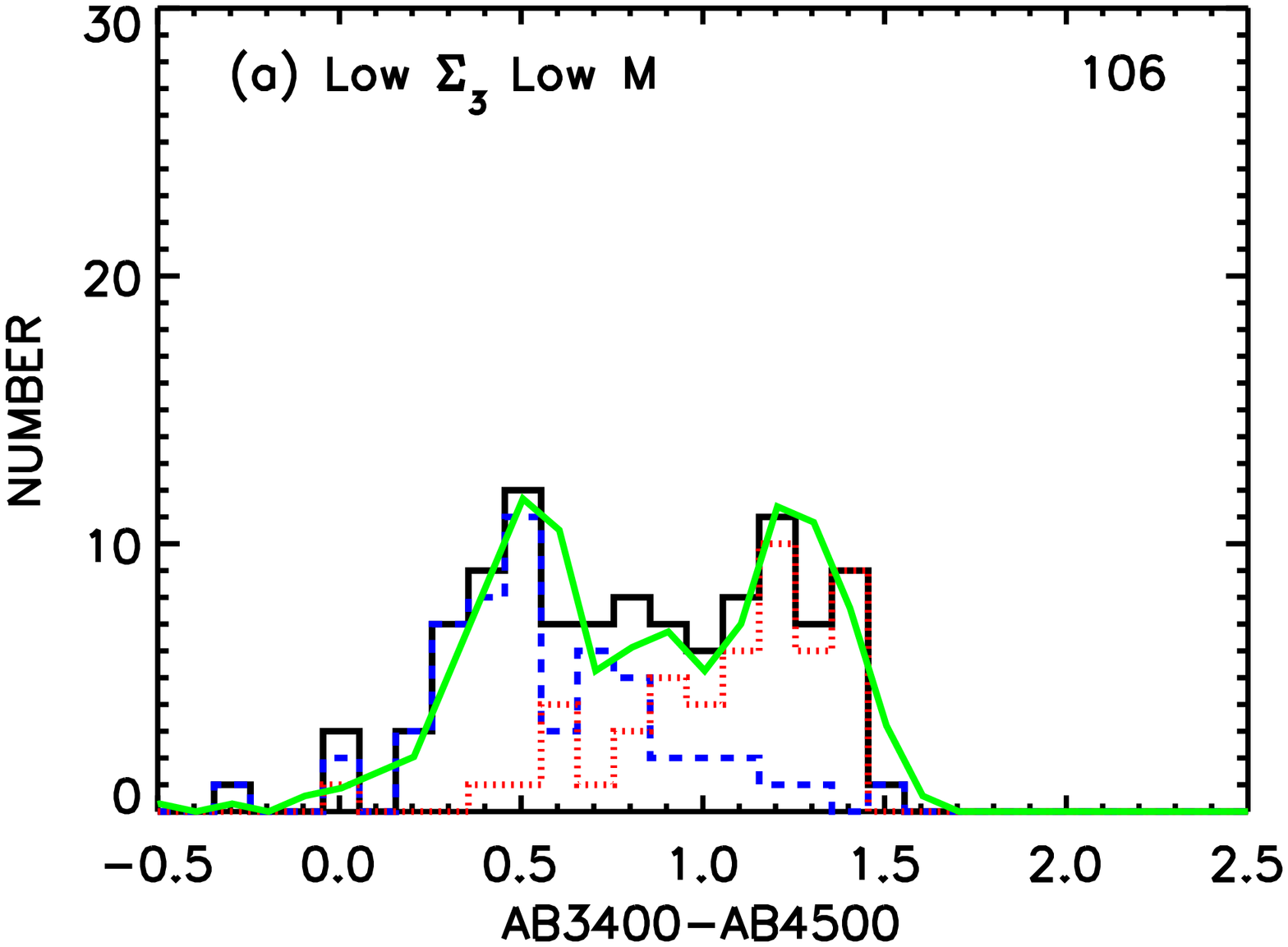,angle=90,width=3.5in}
\psfig{figure=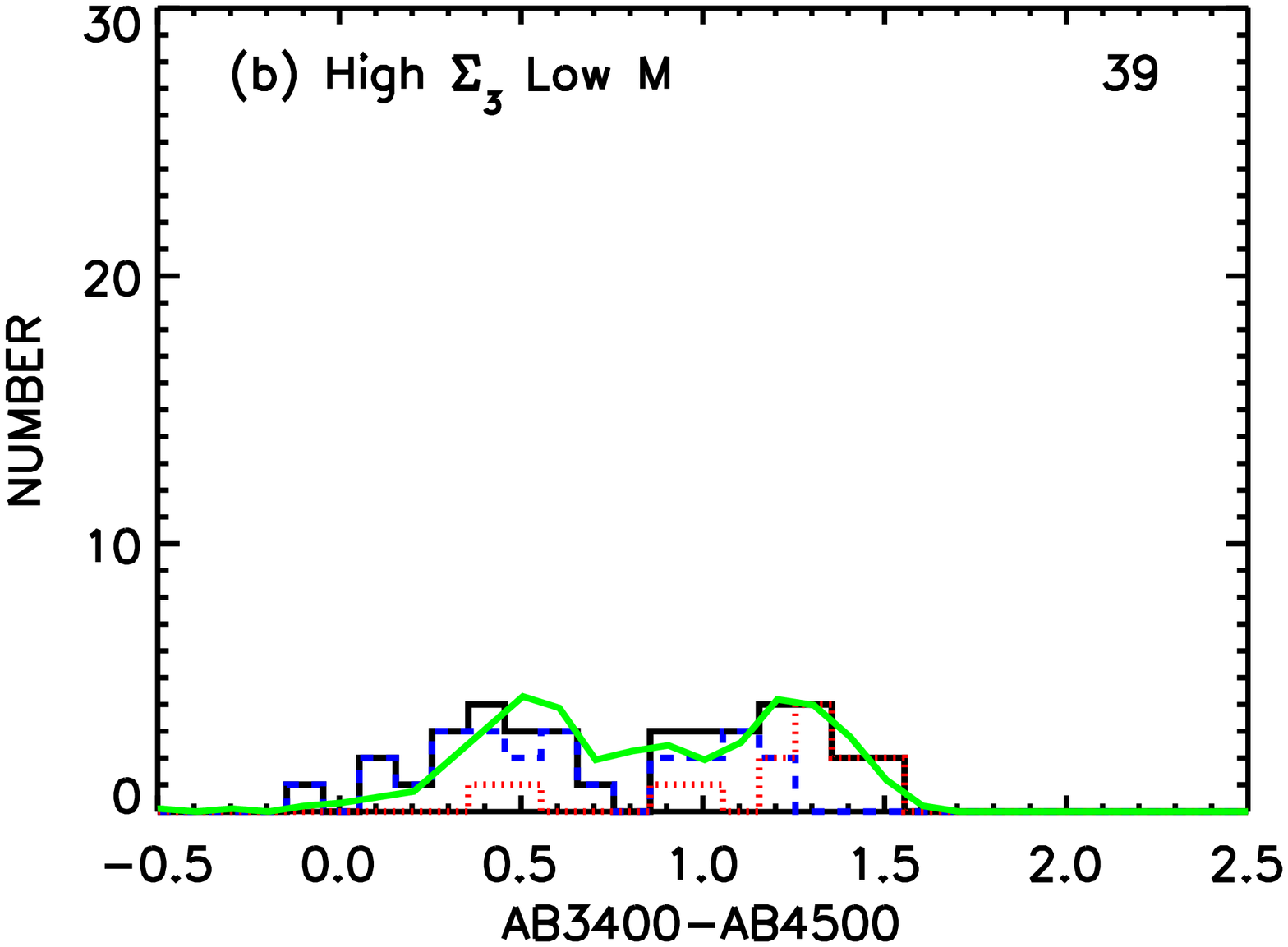,angle=90,width=3.5in}}
\centerline{\psfig{figure=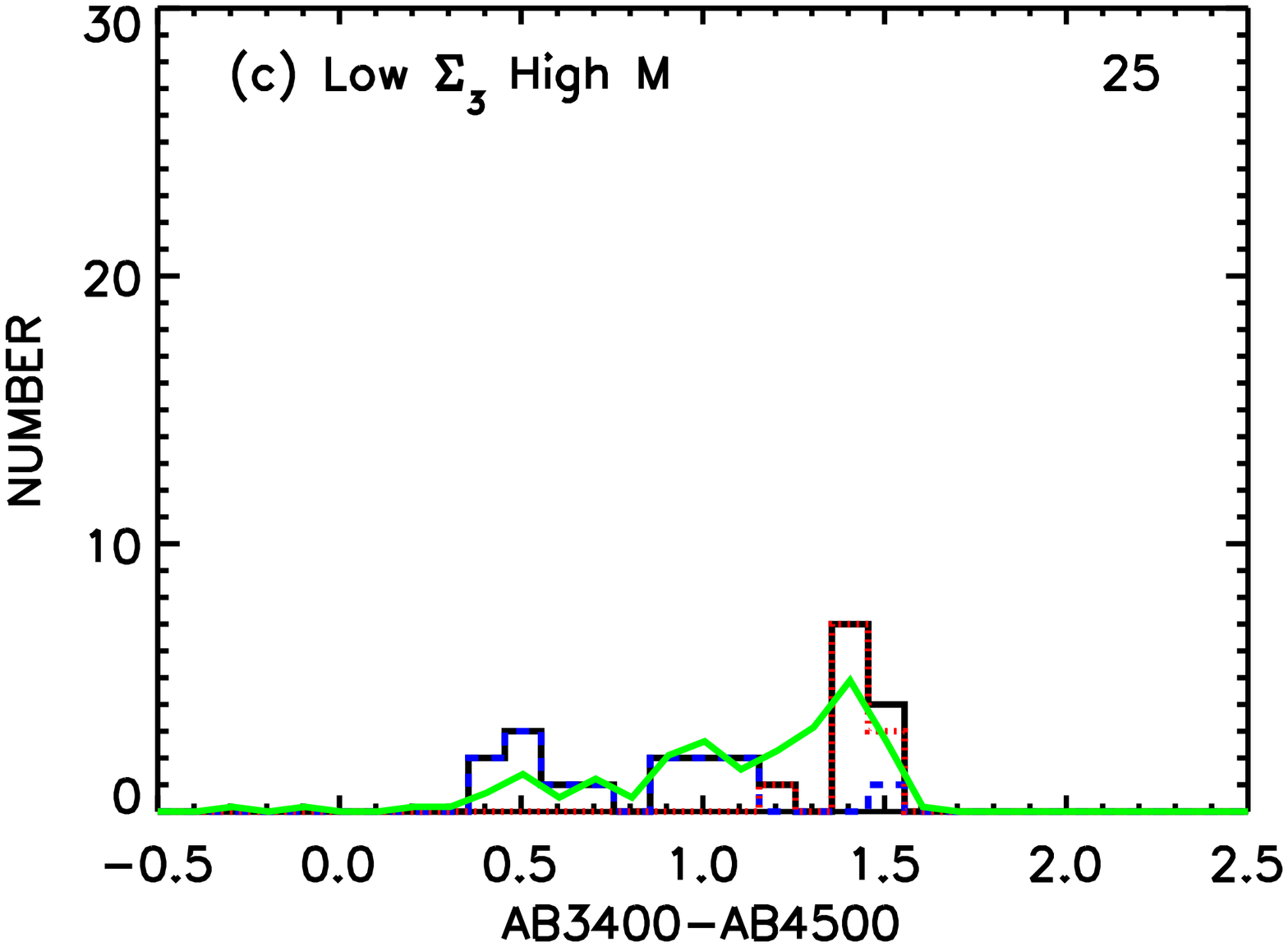,angle=90,width=3.5in}
\psfig{figure=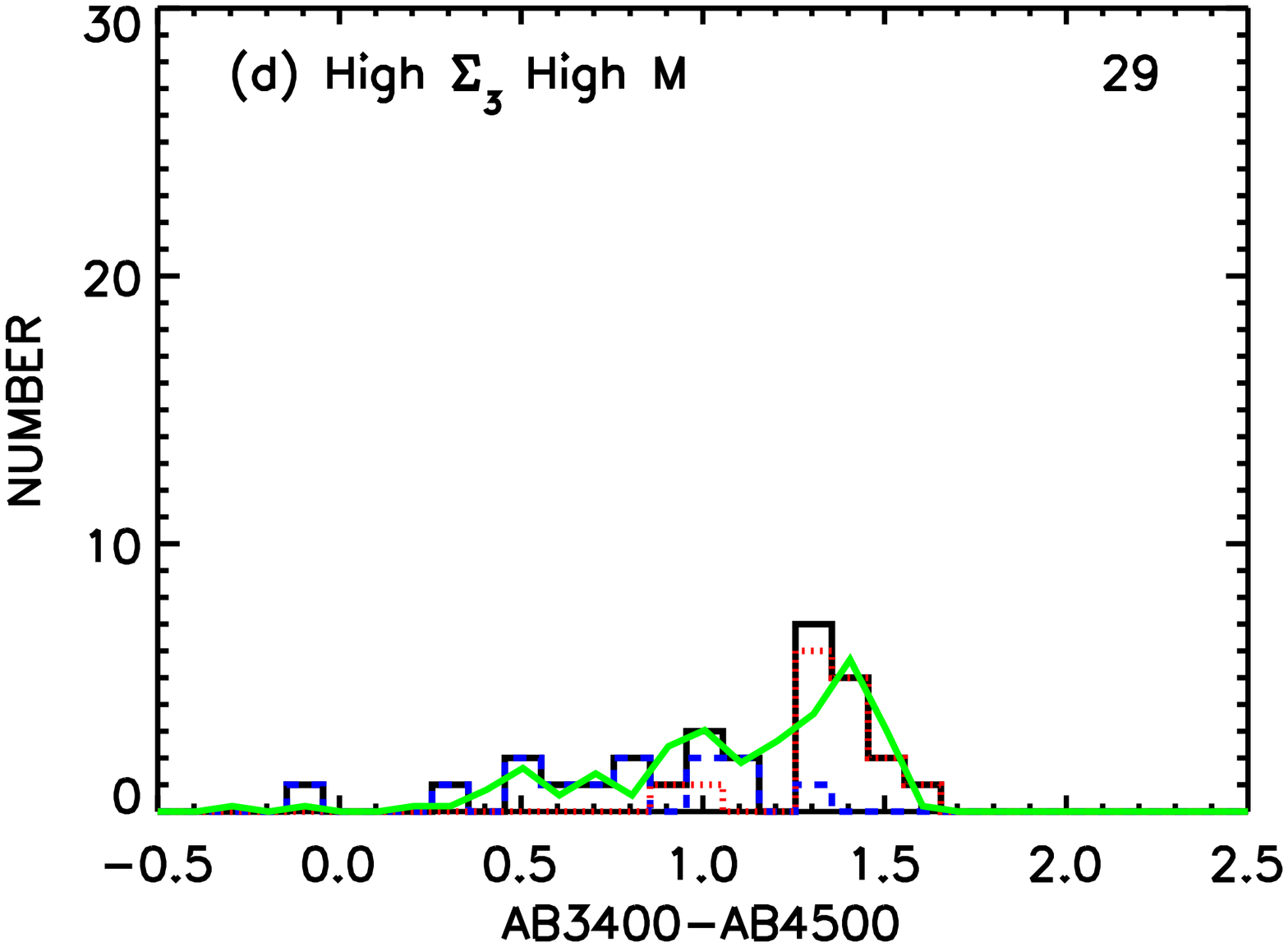,angle=90,width=3.5in}}
\figcaption[]{
Distribution of the AB$3400-$AB4500 colors separated by both mass 
and environment for galaxies with $z=0.6-1.2$ {\em (solid black 
histogram)\/}. In (a) and (b) we show the distributions for
the mass interval $2\times10^{10}-10^{11}$~M$_\odot$.
In (c) and (d) we show the distributions for the mass
interval $10^{11}-10^{12}$~M$_\odot$.
(a) and (c) correspond to $\Sigma_3<1$~Mpc$^{-2}$,
and (b) and (d) correspond to $\Sigma_3$ greater than this value.
The green curves show the distributions of all galaxies in the
given mass interval normalized to the number of galaxies in that
particular sample, which is shown in the upper right corner. The
red (blue) histograms show the distribution of galaxies
detected (undetected) at 24~$\mu$m, respectively.
\label{ub_mass_dens}
}
\end{figure*}

However, unlike the local analysis of Baldry et al.\ (2006),
we do not see an environmental dependence for the distribution
of galaxies between the blue cloud and the red sequence at $z=0.6-1.2$. 
In Figure~\ref{ub_mass_dens} we show the distribution of colors 
for galaxies with $z=0.6-1.2$ separated by both mass and 
environment {\em (black histograms)\/}.
We compare this with the distribution of all galaxies in each
mass interval {\em (green curves)\/} normalized to the number of 
galaxies in that particular sample {\em (number in upper 
right corner)\/}. The blue histograms show sources detected
at $24~\mu$m which dominate the blue cloud at these masses,
and the red histograms show sources which are not detected at
$24~\mu$m which dominate the red sequence. 
It can be seen that the distributions are 
essentially invariant with environment, while the 
fraction of red galaxies increases with mass. Thus,
the environmental dependence of the red fraction seen in the local
sample must have been imprinted over the $z=0-1$ redshift interval. 
This suggests that the star formation switch-off may have been 
more rapid in the higher density environments.

%
%

\subsection{Metal Evolution}
\label{secmetev}

We show the evolution of the metallicity-mass relation with
redshift in Figure~\ref{tremonti}.
In Figure~\ref{tremonti}a we compare the locally derived
metallicity-mass relation of Tremonti04 {\em (black solid curve)\/}
with our relations derived from the R23 method (using the Tremonti04
calibration) for $z=0.05-0.475$ {\em (red)\/}
and $z=0.475-0.9$ {\em (cyan)\/}.
Since the Tremonti04 masses are computed for the Kroupa (2001)
IMF, we had to increase them by a factor of 1.54 to make the comparison
(see \S\ref{secintro}). As we have discussed in \S\ref{secintro},
the adopted IMF does not otherwise affect the results.
We show both the polynomial fits to the metallicity-mass relations
{\em (colored lines)\/} and the median values and errors in various
mass bins {\em (symbols)\/}.

%
%
\begin{inlinefigure}
\figurenum{63}
\centerline{\psfig{figure=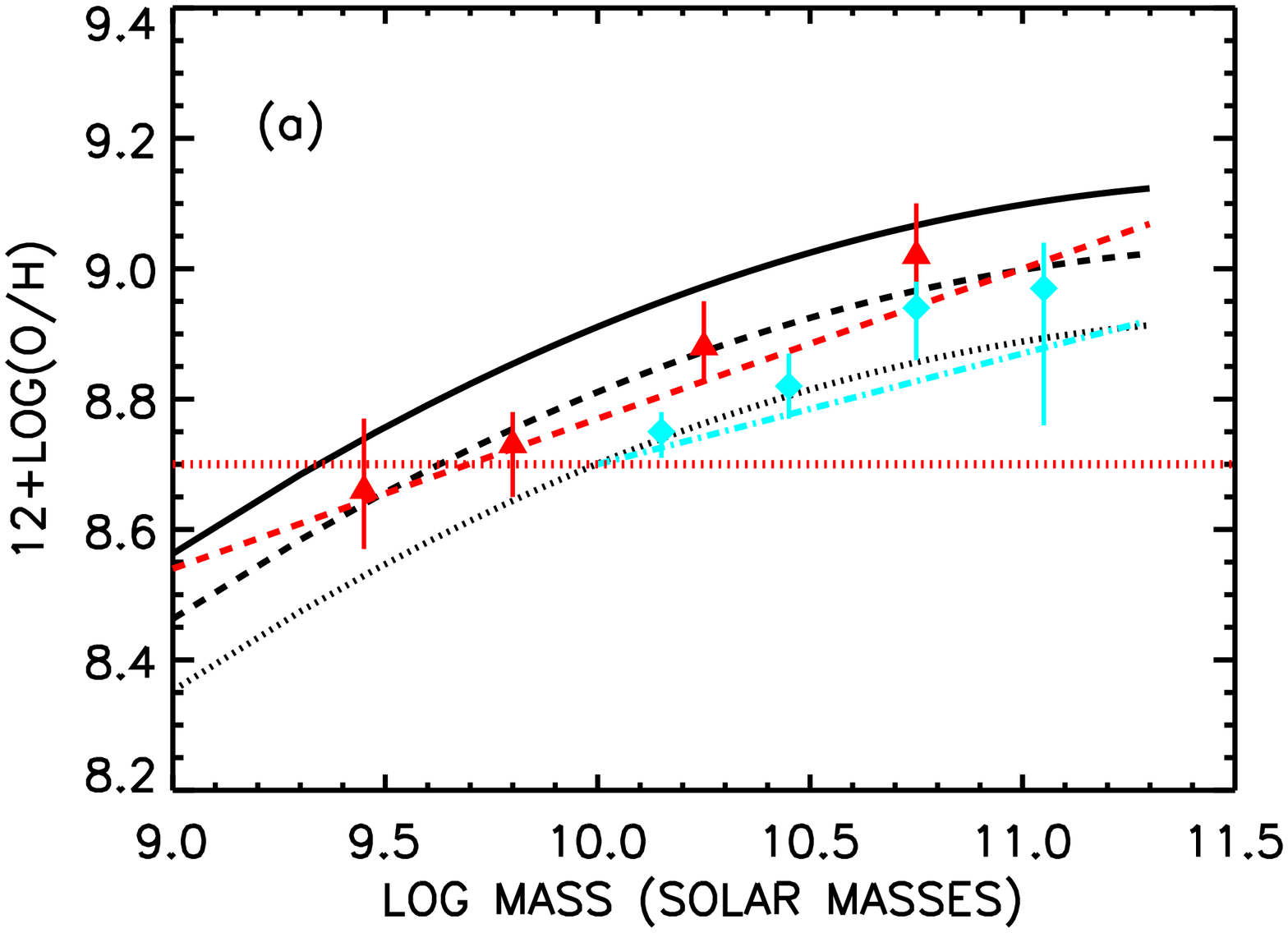,angle=90,width=3.5in}}
\vskip -0.6cm
\centerline{\psfig{figure=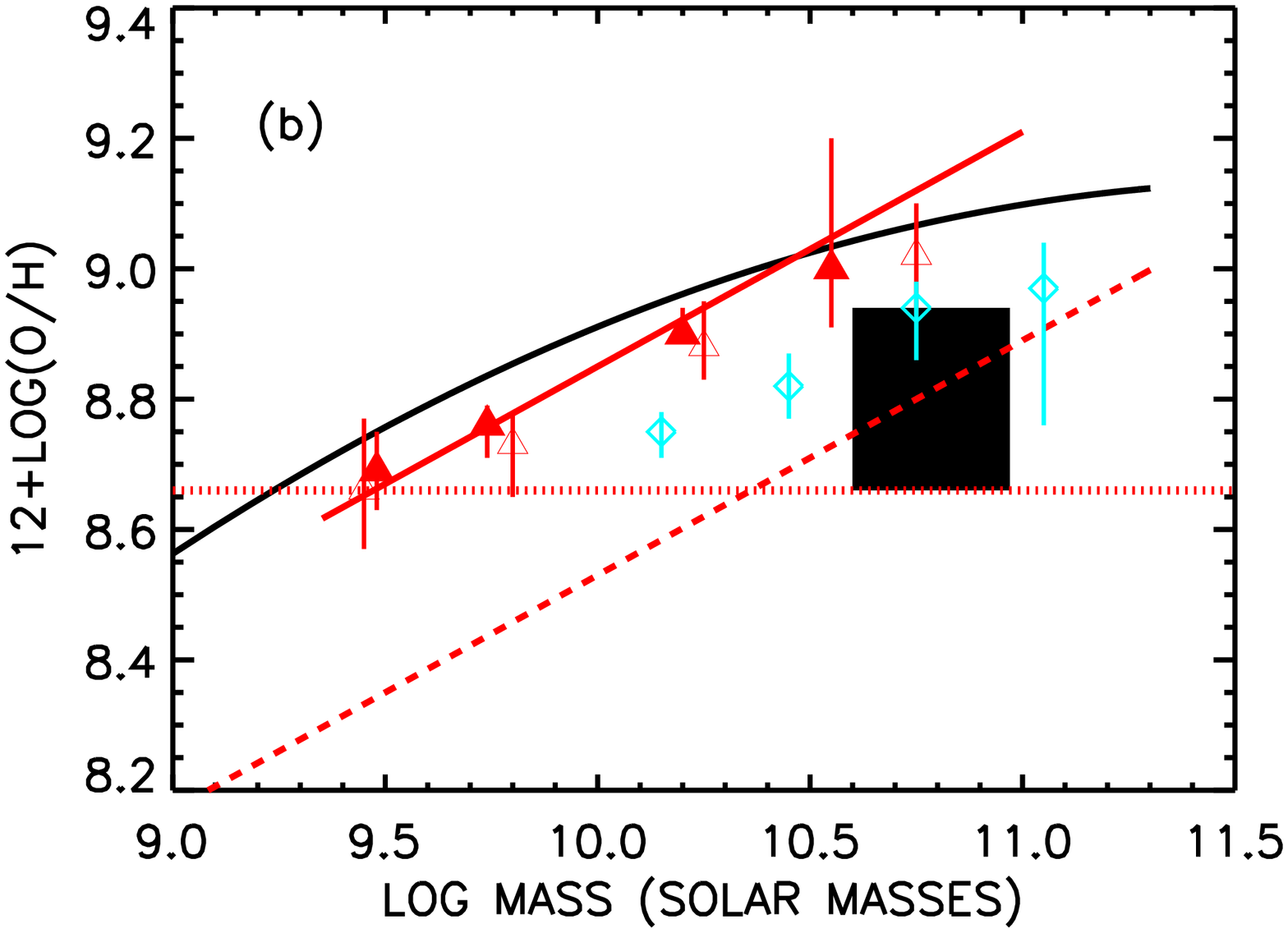,angle=90,width=3.5in}}
\vskip -0.2cm
\figcaption[]{
(a) Metallicity-mass relations in the $z=0.05-0.475$
{\em (red triangles)\/} and $z=0.475-0.9$ {\em (cyan diamonds)\/}
redshift intervals derived from the R23 method using the Tremonti04
calibration and compared with Tremonti04's local metallicity-mass
relation {\em (black curve)\/}. The colored lines show the
polynomial fits to the individual data points. The solid symbols
show the median values in the mass bins with 68\% confidence limits.
The medians lie slightly higher than the average values represented
by the fits. The dotted red line shows the solar abundance.
The black dashed (dotted) line shows the local relation
reduced by 0.14~dex (0.22~dex) to match the data in the
$z=0.05-0.475$ ($z=0.475-0.9$) interval.
(b) Metallicity-mass relation in the $z=0.05-0.475$ redshift
interval {\em (red solid line and solid triangles)\/} computed
using the \nii/\ha\ diagnostic and compared with the
Shapley et al.\ (2004) measurements of the LBG
sample computed using the same method {\em (solid black region)\/}.
The red dashed line shows the red solid line reduced
by 0.32~dex to match the LBGs. The black curve, the red open
triangles, and the cyan open diamonds are all taken from (a) and
show the R23-based results.
\label{tremonti}
}
\end{inlinefigure}

We summarize the decrease in the metallicities with increasing
redshift and decreasing mass in Table~\ref{metal_drop}.
In column~2 we give the average of the Tremonti04 values
for galaxies in the given mass interval, and then in
columns~3 and 4 we give the drop from Tremonti04 to each of
our average values, respectively, for $z=0.05-0.475$ and
$z=0.475-0.9$. The errors are 68\% confidence limits.

The data are not adequate to determine if the shape of the
local relation has changed with redshift. There are hints
at the $2\sigma$ level that there is less evolution
at the high-mass end, but within the accuracy that can be
obtained with our data, the shape of the metallicity-mass
relation over the observed mass range could be invariant
from $z=0.05-0.9$. However, if we normalize the local relation
to our data at $z=0.05-0.475$ {\em (black dashed curve)\/} and
at $z=0.475-0.9$ {\em (black dotted curve)\/}, then the best
fits show that the metallicity in the $10^{10}-10^{11}$~M$_\odot$
interval is lower by $0.10\pm0.04$~dex than the local value at
the median redshift of $0.44$ (low-redshift interval) and 
lower by $0.21\pm0.03$~dex than the local value at the median 
redshift of $0.75$ (high-redshift interval).

The conclusion that there is a decrease in the metallicity
at a given mass with increasing redshift from $z=0.05-0.9$
is consistent with previous work, though the precise values
have varied considerably. The most recent work of
Savaglio et al.\ (2005) gives a considerably larger change
in the normalization and also finds a steeper slope at $z\sim0.7$,
namely $12+\log({\rm O/H})=8.84+(0.48\pm0.06)\log M_{10}$,
which may be compared with our relation of
$12+\log({\rm O/H})=8.70+(0.17\pm0.05)\log M_{10}$
(Eq.~\ref{eqr23hizrel_tr}). The Savaglio et al.\ (2005)
relation is fitted over a much wider mass range (down to masses
below $10^{9}$~M$_\odot$) using a sample that is substantially
incomplete and biased towards star formers at the lower
masses. This weights the lower mass bins to lower
metallicities and steepens the fit.

In Figure~\ref{tremonti}b we compare the median metallicities
{\em (red solid triangles)\/} and the least-square polynomial 
fit {\em (red solid line)\/}
computed from the \nii/\ha\ method in the $z=0.05-0.475$
redshift interval with the $z\sim2.1$ Lyman break galaxy
(LBG) sample of Shapley et al.\ (2004), which was also computed
using this diagnostic {\em (black solid region)\/}.
The R23-based measurements from Figure~\ref{tremonti}a are
also shown {\em (black curve and colored open symbols)\/}. 
The R23-based red open triangles for the same redshift interval
are not significantly different than the \nii/\ha-based red
solid triangles. The LBG galaxies lie about 0.32~dex
lower {\em (dashed red line)\/} than the median $z=0.44$
galaxies {\em (red solid line)\/}
at the same mass and are similar in metallicity to
local galaxies that are almost an order of magnitude lower in mass.

\subsection{Gas Masses}
\label{secgasmass}

In the simplest closed-box model for metal evolution,
the metallicity $Z$(O) (the fraction by mass of O)
in the gas is simply related to the oxygen yield $y$(O)
by the well-known relation
\begin{equation}
Z({\rm O}) = y({\rm O})~\log(M_{g}/M_{T})  \,,
\label{meteq}
\end{equation}
where $M_{g}$ is the gas mass and $M_{T}$ is
the sum of the gas mass and the stellar mass.
The change in metallicity for a change in stellar
mass is
\begin{equation}
\delta Z({\rm O}) = y({\rm O})~\delta M_{star}/M_{g}  \,.
\label{deriveq}
\end{equation}
Thus, the derivative of $Z$(O) with respect to the stellar
mass measures the quantity $y$(O)/$M_{g}$, and, if we
assume a value for the yield, we can derive the
gas mass. Essentially we are measuring the gas reservoir
required to dilute the metals returned from the known
star formation to match the observed metal evolution.

The effective yields
have been empirically measured for local galaxies and are
found to be approximately independent of galaxy mass for
masses above $10^{9.5}$~M$_\odot$ (Garnett 2003).
Tremonti04 find a weak dependence on mass in this mass range,
but their results depend on using star formation as a proxy
for gas mass. We will assume a time-independent and
mass-independent value of $\log y({\rm O})=-1.9$, which
is probably a reasonable approximation given the uncertainties.

We roughly compute the gas reservoir mass densities using the 
mass change between $z=0.05$ and $z=0.77$ from the least-square 
fits of Equations~\ref{massint2}$-$\ref{massint4} and
the corresponding change in $Z$(O) from Table~1.
For the lowest logarithmic mass interval $9.5-10$~M$_\odot$
we made the fit over $z=0.05-0.44$,
where the mass sample is complete. If the local
mass density is lower than the Cole01 estimate this would
reduce $\delta M_{star}$ and hence the inferred gas mass. 
Deriving the values from
the star formation instead gives broadly similar results,
with the largest change being an increase by almost a factor
of two in $M_g$ in the logarithmic mass interval
$10-10.5$~M$_\odot$.

%
%
\begin{inlinefigure}
\figurenum{64}
\centerline{\psfig{figure=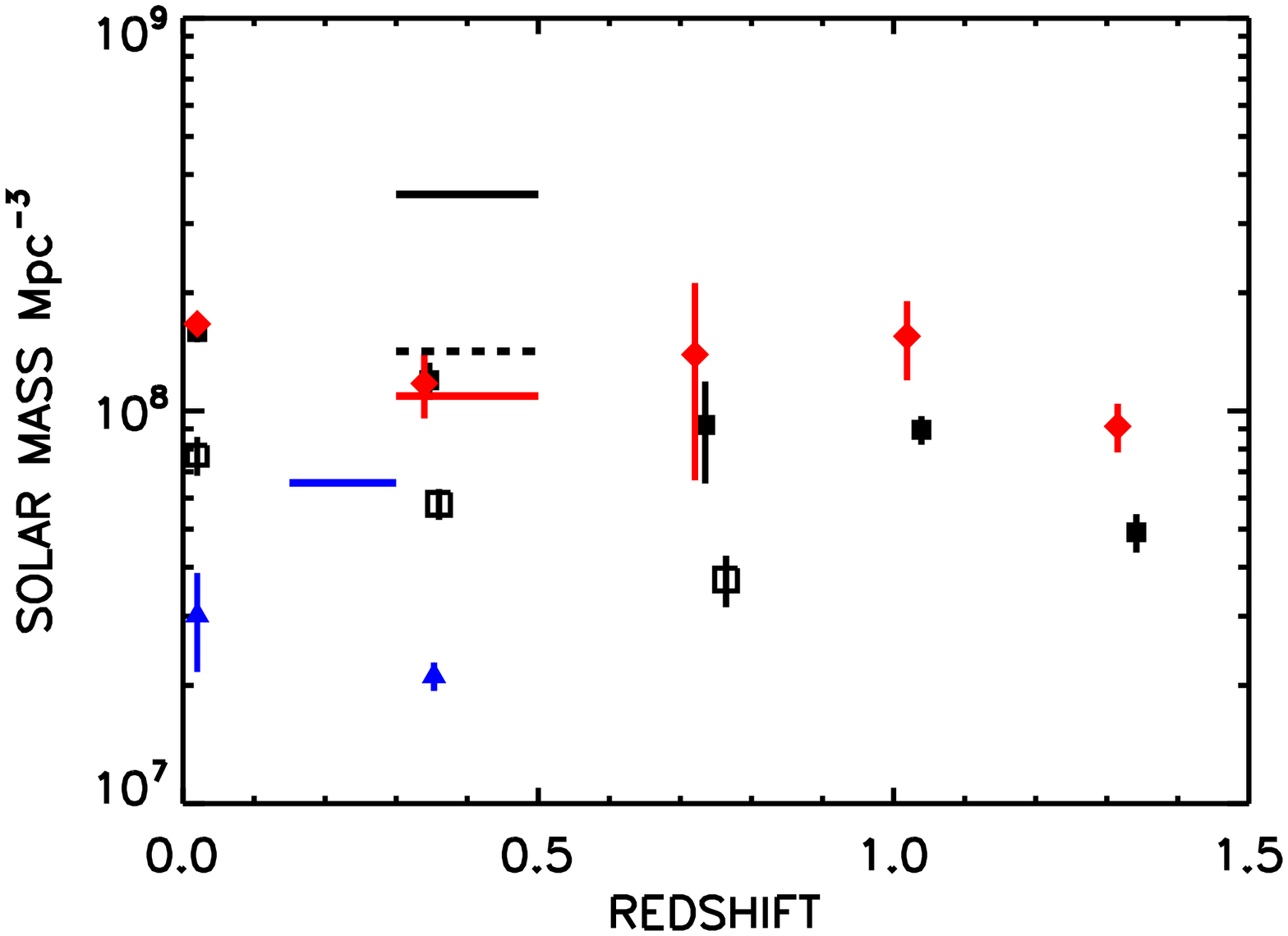,angle=90,width=3.5in}}
\vskip -0.2cm
\figcaption[]{
Mass density of the gas reservoirs
inferred from the metal evolution compared with
the stellar mass density history.
The mass density of the gas reservoirs is shown
with colored lines, and the stellar mass density
history is shown with corresponding colored symbols
in the mass intervals $10^{9.5}-10^{10}$~M$_\odot$
{\em (blue line and triangles)\/}, 
$10^{10}-10^{10.5}$~M$_\odot$ 
{\em (black dashed line and open squares)\/},
$10^{10.5}-10^{11}$~M$_\odot$ {\em (black solid line and solid squares)\/},
and $10^{11}-10^{11.5}$~M$_\odot$ {\em (red line and diamonds)\/}.
The gas mass densities are computed over $z=0.05-0.77$
for the higher mass intervals and over $z=0.05-0.44$
for the lowest mass interval. The local mass densities
for each of the mass intervals were obtained by
integrating the Cole01 function.
\label{mgas_byz}
}
\end{inlinefigure}

We compare the gas mass densities {\em (colored lines)\/} with the
stellar mass density history {\em (corresponding colored
symbols)\/} for the mass intervals $10^{9.5}-10^{10}$~M$_\odot$
{\em (blue line and triangles)\/}, $10^{10}-10^{10.5}$~M$_\odot$ 
{\em (black dashed line and open squares)\/}, 
$10^{10.5}-10^{11}$~M$_\odot$
{\em (black solid line and solid squares)\/},
and $10^{11}-10^{11.5}$~M$_\odot$ 
{\em (red line and diamonds)\/}
in Figure~\ref{mgas_byz}. 
(The gas and stellar mass densities are derived 
using the Salpeter IMF and thus will change consistently if 
we use an alternate IMF.) In the highest mass interval {\em (red)\/}
where the stellar mass is growing only slowly, the sum of
the gas plus stellar mass densities at $z=0.44$
is roughly comparable to the local stellar mass density
obtained by integrating the Cole01 function. 
By contrast, the galaxes in the lower mass intervals 
where assembly is still progressing have much larger gas reservoirs.
Thus, these results suggest that star formation is terminating 
as a consequence of gas depletion.

We estimate the total gas mass density over all galaxies 
larger than $10^{9.5}$~M$_\odot$ at $z=0.44$ to be
$\sim7\times10^{8}$~M$_\odot$~Mpc$^{-3}$. We can compare this
with the current mass density in stars over the same mass range, 
namely $4.7\times10^{8}$~M$_\odot$~Mpc$^{-3}$
obtained by integrating Cole01. This suggests that 
there is still a significant amount of gas mass to be converted 
into stars. From Figure~\ref{mgas_byz} we can see that this 
principally lies in galaxies with masses below $10^{11}$~M$_\odot$.
We may very crudely infer
that at the present time the highest mass galaxies
will have very little gas mass remaining, while the lower
mass galaxies will have comparable amounts of gas and stars.
However, the uncertainties in this are very large, both from
the observations and from the overly simple modeling.

\section{Summary}
\label{seccon}

We have used a very large and highly spectroscopically
complete galaxy sample selected by rest-frame NIR bolometric
flux to conduct an integrated study of star formation
and galactic stellar mass assembly from $z=0.05-1.5$ and
galactic metallicity evolution from $z=0.05-0.9$. We summarize
our results below.

$\bullet$ We constructed a rest-frame NIR ($0.8-2.4~\mu$m) 
bolometric flux sample in the GOODS-N field. 
We have spectroscopic redshifts and high-quality 
spectra for 77\% of the galaxies in the sample, and we measured
13-band photometric redshifts for the remaining sources.
Since most of the sources with only photometric redshifts
lie outside of our two redshift ranges of interest, the 
spectroscopic completeness inside these ranges is extremely 
high ($>91$\% for $z=0.05-0.9$ and $>84$\% for $z=0.05-1.5$). 

$\bullet$ We constructed four uniform NIR luminosity samples 
for our mass assembly analysis (two at lower redshifts,
$z=0.05-0.475$ and $z=0.05-0.9$, which we also use for 
our metallicity analysis, and two at higher redshifts, 
$z=0.9-1.2$ and $z=1.2-1.5$) and fitted BC03
models (assuming a Salpeter IMF, a solar metallicity,
and a Calzetti extinction law) to every galaxy to obtain
galactic stellar masses and extinctions. We tested 
the extinctions by comparing the extinguished UV light, which 
is reradiated into the FIR, with the 24~$\mu$m light, which 
is a completely independent measure of the dust-reradiated 
light. We detected at 24~$\mu$m most of the sources with 
large reradiated fluxes, confirming the assignment of
substantial extinctions. However, some of the 24~$\mu$m sources 
had low reradiated UV fluxes. To test whether we were 
failing to assign extinctions to these galaxies, we also measured,
where possible, $f($\ha$)/f($\hb$)$ for them. In all cases these 
Balmer ratio measurements were also consistent with little extinction.

$\bullet$ We adopted a maximum ratio of the stellar mass to
the observed NIR luminosity of $10^{-33}~$M$_\odot/$ergs~s$^{-1}$.
Multiplying the NIR luminosity limits of the four samples by
this ratio, we found that the samples included all galaxies
with masses above $2\times 10^9$~M$_\odot$ ($z=0.05-0.475$),
$10^{10}$~M$_\odot$ ($z=0.05-0.9$), $2\times 10^{10}$~M$_\odot$
($z=0.9-1.2$), and $3\times 10^{10}$~M$_\odot$ ($z=1.2-1.5$).

$\bullet$ For each galaxy spectrum we measured the equivalent 
widths of a standard set of lines. We found a strong trend
to higher EWs at lower NIR luminosities. We also found a rapid 
drop in EWs at lower redshifts.
We estimated a fixed offset of 1~\AA\ to correct the EW(\hb)
for the effects of underlying stellar absorption, having found 
no strong dependence of the correction on galaxy type.

$\bullet$ We measured the line fluxes (calibrated by broadband 
fluxes) and the 4000~\AA\ break strengths in the galaxies.
We measured the offsets between the rest-frame $3400-4500$~\AA\
magnitudes measured from the spectra and those measured from
the photometry and found no dependence on redshift. Thus,
there are no relative errors in the photometric calibration
of the UV and optical data.

$\bullet$ Although we can only measure Balmer ratio 
extinctions for our line fluxes using our lowest 
redshift sample, we found that when we compared these extinctions 
to the continuum extinctions that we obtained from our BC03 fits, 
the two were consistent, on average, within the errors. Thus, 
we can (and do) use the BC03 extinctions to deredden the line 
fluxes at higher redshifts.

$\bullet$ We calibrated the star formation diagnostics 
internally using our lowest redshift sample. We found that
the UV flux had a somewhat tighter relation to the \ha\ flux
than the \oii\ flux did and hence provided a better estimate
of the SFR. We also saw this in the systematic dependence 
of the \oii\ SFR on galaxy properties, such as 
the \ha\ SFR of the galaxy and the galaxy mass. We adopted 
the UV SFR calibration as our primary calibration.

$\bullet$ Because it has become common in the literature to
use 24~$\mu$m data to estimate SFRs, we also computed SFRs
from our 24~$\mu$m fluxes and compared them with the reradiated
SFRs that we determined from our UV luminosities. (The
latter were obtained by taking the difference between the SFRs 
that we computed after correcting for extinction and the SFRs 
that we computed without making an extinction correction.) 
Although we found a substantial spread for the individual
determinations, the two methods give good agreement when 
applied to the galaxy population as a whole.

$\bullet$ We constructed a number of emission line
diagnostics for our $z=0.05-0.475$ sample, since many of
the spectra cover all of the emission lines from \oii\ to
\sii. We dereddened all of the line fluxes using our BC03
extinctions. Adopting various calibrations from the literature,
we derived metallicities from the N2O2, NH, and R23 diagnostic 
ratios, and we derived ionization parameters from the combination 
of O32, which has a dependence on $q$ and metallicity, and N202 
or R23.
For $z=0.475-0.9$ we could only compute R23 from the optical
spectra. We compared the metallicity-mass relations that we
derived from R23 for the two redshift intervals (restricting 
to masses above $10^{10}$~M$_\odot$, where both samples are 
complete, and below $10^{11}$~M$_\odot$) and found that,
within the wide errors, the slopes were consistent.
We also found that the median metallicity increased by 0.13~dex 
as the redshift decreased from median redshift 0.75 to median 
redshift 0.44.

$\bullet$ We determined that only about 20\%$-$30\% of the
total galaxy mass density is included in our strong emission
line metallicity analysis. We divided the remaining galaxies
into an apparently passive category and a weakly active
category. About 40\% of the total galaxy mass density is
contained in the passive population, and the majority of
the massive galaxies ($>10^{11}$~M$_\odot$) fall into this
category. Due to a lack of \hb\ and \oiii\ features in
the stacked spectrum for the passive galaxies, there was no 
way we could use the gaseous emission to estimate
metallicities. However, we were able to estimate metallicities 
from the stacked spectrum for the weakly active galaxies.
As a check on how well we could do this, we also
estimated metallicities from the stacked spectrum for the
strong emission line galaxies. For $z=0.05-0.475$ these
estimates were in good agreement with the fit that we had made
to the individual measurements, while for $z=0.475-0.9$ they
were about 0.1~dex higher on average. The metallicity estimates
from the stacked spectrum for the weakly active galaxies showed 
substantial scatter but were in broad agreement with the strong 
emission line galaxies, so we subsequently assumed that they 
paralleled the evolution of the strong emission line galaxies.

$\bullet$ We constructed galactic stellar mass functions and
compared them with the local mass function of Cole01.
We found that the galaxy number densities at the low-mass
end are still rising down to the lowest redshift interval,
while the number densities at the high-mass end 
($>10^{11}$~M$_\odot$) are changing much more slowly over
$z=0.05-1.5$. We provided parametric fits to the data by 
assuming a Schechter form for the mass functions. We found
a significant evolution in the mean mass (a rise in either
$M_\star$ or $\alpha$ with increasing redshift) due to the
build-up of the low-mass region of the galactic stellar mass
function relative to the high-mass region.

$\bullet$ We determined the mass density evolution by mass
interval and compared them with the local mass densities 
obtained by integrating the Cole01 function over each mass 
interval. We found slow
growth in the higher mass intervals and more rapid growth 
in the lower mass intervals. To quantify the results, we
made least-square polynomial fits to the logarithmic
stellar mass densities versus the logarithmic cosmic time.
We found that the low-mass ranges ($<10^{11}$~M$_\odot$) 
are growing approximately linearly with time, while the
high-mass ranges are growing more slowly.

$\bullet$ We computed the universal star formation history
from $z=0.05-1$ from our NIR sample. We calculated the SFRDs 
using our empirical calibrations and our extinction corrected
\hb\ and UV luminosities. 
We found good agreement with the values derived from
radio and submillimeter data. The SFRDs for the NIR sample
drop steeply at lower redshifts, where most of the star 
formation is seen as direct UV emission from lower mass 
galaxies, which have very small extinctions. Thus, we also 
calculated the SFRDs from our extinction uncorrected data,
which we found agreed well with rest-frame UV flux 
measurements of UV selected galaxies. Finally, to check
our UV extinction corrections, we added
the SFRDs that we computed from the 24~$\mu$m fluxes
for the obscured star formation to the SFRDs that we computed 
from the extinction uncorrected \oii\ luminosities for the
unobscured star formation. This method has no dependence
on the UV extinction corrections. Reassuringly, when we compared
these SFRDs by mass interval with those calculated from 
the UV-based method, we found very similar results.

$\bullet$ We compared the expected formed stellar mass density 
growth rates produced by star formation (computed with the
conservation equation using both the UV-based and the
24~$\mu$m$+$\oii-based SFRDs) with those measured
from the formed stellar mass functions. Here the
formed stellar mass density is the total mass density 
formed into stars prior to stellar mass loss.
We found amazing agreement over a wide range of redshifts
and masses, though there is a slight normalization
difference between the two measurements (the measurements 
based on the star formation are higher). We explored various 
possibilities to explain the offset, and we concluded that the 
most likely explanation is that we need an IMF which is more 
heavily weighted to mid-mass stars than the Salpeter, Kroupa, 
and Chabrier IMFs. Over the $1-100$~M$_\odot$ range, a broken 
power law with an index of $-1.10$ below 10~M$_\odot$ and 
$-1.60$ above works well. This is well within the range of 
uncertainty in the high-mass IMF and may be favored for 
other reasons (Fardal et al.\ 2007).

$\bullet$ We obtained a quantitative description of the
range of behaviors in the galaxies by computing the
instantaneous SSFRs. There is a wide spread in
the SSFRs at all redshifts and masses. However, from 
the mean SSFRs we found that, on average, only galaxies with 
masses $\lesssim 10^{11}$~M$_\odot$ grow significantly in any 
of the redshift intervals. In the lowest redshift
interval where we can measure the masses below $10^{10}$~M$_\odot$,
we finally saw the flattening of the SSFRs (at a high level).
From the distribution functions of the SSFRs by
mass interval, we concluded that what star formation is
occurring in the galaxies in the $10^{11}-10^{11.5}$~M$_\odot$
interval is spread over many galaxies, and there are very few
galaxies in this mass interval that are undergoing significant
growth. In contrast, in the $10^{10}-10^{10.5}$~M$_\odot$
interval a substantial number of galaxies have SSFRs that,
if maintained over the time frame, would change their
mass significantly.

$\bullet$ We examined the rest-frame UV$-$blue (AB3400$-$AB4500)
colors uncorrected for extinction for our NIR sample and
found almost no evolution with redshift and a uniform spread of 
colors stretching from the blue cloud to the red sequence. Nearly
all of the galaxies in the intermediate color range (sometimes
referred to as the green valley) are 24~$\mu$m sources.
After correcting for extinction, we found that the bulk of the 
24~$\mu$m sources lie in the blue cloud. We also then saw a more 
clearly bimodal distribution in the total sample. 
Applying extinction corrections is critical when analyzing the
galaxy colors, since many of the sources seen in the green
valley and the red sequence prior to correcting for
extinction are dusty sources with intrinsically blue colors.
In contrast to literature results from optically-selected 
and extinction uncorrected data, we saw no change in the position
of the red sequence with redshift in our extinction corrected 
data. Thus, the intrinsic colors of the reddest galaxies are not
changing over $z=0.05-1.5$.

$\bullet$ We approximately separated the red sequence from the
blue cloud using a cut 0.25~mag below the red sequence. We
also separated the red sequence from the blue cloud using the 
EW(\oii), which is independent of the extinction correction, 
and the 4000~\AA\ break strength, which is nearly independent.
Although we found comparable amounts of mass in the red sequence 
and the blue cloud, the growth in the mass density is primarily 
occurring in the red sequence galaxies with masses 
$10^{10.5}-10^{11}$~M$_\odot$.

$\bullet$ We investigated how star formation is occurring
using the extinction corrected rest-frame AB3400$-$AB8140 
colors and the rest-frame EW(\hb). The EW(\hb) is produced by 
higher mass stars and thus fades more rapidly, providing a 
well-known age signature. We compared the data with some 
evolutionary tracks from the BC03 models and found consistency 
with episodic star formation in all but the lowest mass 
($10^{10}-10^{10.5}$~M$_\odot$) galaxies, where there may be a 
mixture of smooth and episodic star formation. It appears
that the burst process results in the dusty galaxies
producing the 24~$\mu$m emission. 

$\bullet$ We compared the morphological types with the
spectral characteristics of the galaxies. We found no change 
in the distribution of the morphological types in the 4000~\AA\ 
break-EW(\hb) plane over $z=0.05-0.9$. We confirmed a strong 
correlation between the galaxy morphology and the galaxy mass, 
with many of the most massive galaxies being E/S0 galaxies.
We saw very little change in the mass function of the 
strong emitters and spiral galaxies. Instead we saw the mass 
build-up primarily occurring in the passive and weakly active E/S0 
galaxies. We concluded that galaxies are moving from spiral 
galaxy types to E/S0 types with decreasing redshift and that 
the decrease in the mass of the blue cloud from this effect 
offsets the growth due to star formation. 

$\bullet$ We found that massive galaxies ($>10^{11}$~M$_\odot$) 
preferentially occur in higher density environments, but below 
this mass we saw relatively little dependence on the environment.
We also found relatively little evolution in this
mass-density relation over the $z=0.05-0.9$ redshift range.

$\bullet$ We found that the metallicities of galaxies are increasing 
with decreasing redshift at all galaxy masses over $z=0-0.9$.
The increase is $0.21\pm0.03$~dex between $z=0.75$ and $z=0$. 
We compared this with the metal release rate from star formation 
to make a crude estimate of the gas mass reservoirs in the galaxies 
using a simple closed box model. We found that for mass intervals
below $10^{11}$~M$_\odot$, where assembly is still progressing,
the gas reservoirs are larger than are needed to assemble the
present-day stellar mass densities.

\acknowledgements
We thank J.~S.~Gallagher for interesting conversations
and a critical reading of the paper. We also thank
the anonymous referee for an extremely helpful report
with many useful suggestions for improving the paper.
We gratefully acknowledge support from NSF grants 
AST 0407374 and AST 0709356 (L.~L.~C.) and 
AST 0239425 and AST 0708793 (A.~J.~B),
the University of Wisconsin Research Committee with funds 
granted by the Wisconsin Alumni Research Foundation (A.~J.~B.),
and the David and Lucile Packard Foundation (A.~J.~B.).

\clearpage

%
%
\begin{deluxetable}{ccccc}
\tablewidth{480pt}
\tablenum{1}
\tablecaption{Schechter Function Fits}
\tablehead{
Redshift Interval & $\alpha(z)$ & Log $M_\star(z)$ & Log $\phi_\star(z)$ & Log $M_\star(z)$ ($\alpha=-1.18$)\cr
 & & (M$_\odot$) & (Mpc$^{-3}$) & (M$_\odot$)
}
\startdata
Local\tablenotemark{a}  &  $-1.18\pm0.03$  &  $11.16\pm0.01$ & $-2.51\pm0.06$ & $11.16\pm0.1$ \nl
$0.05-0.475$  &  $-1.10 (-0.88,-1.30)$  &  $10.93 (10.71,11.25)$ & $-2.63\pm0.08$ & $11.02 (10.85,11.21)$\nl
$0.475-0.9 $  &  $-0.94 (-0.70,-1.16)$  &  $11.08 (10.93,11.26)$ & $-2.60\pm0.10$ & $11.24 (11.14,11.35)$\nl
$0.9-1.5   $  &  $-0.56 (-0.08,-0.98)$  &  $11.02 (10.87,11.20)$ & $-2.63\pm0.04$ & $11.28 (11.19,11.38)$\nl
\enddata
\tablenotetext{a}{Cole et al.\ (2001)}
\label{sty_fits}
\end{deluxetable}

%
%
\begin{deluxetable}{cccc}
\tablewidth{360pt}
\tablenum{2}
\tablecaption{Percentages of Strong Star Formers}
\tablehead{Logarithmic & \multicolumn{3}{c}{Per Redshift Interval (\%)} \cr
Mass Interval (M$_\odot$) & $z=0.9-1.5$ & $z=0.475-0.9$ & $z=0.05-0.475$
}
\startdata
$ 10-10.5$  &  \nodata   &  $55\pm5$ & $67\pm11$ \nl
$10.5-11$  &  $41\pm 4$  &  $29\pm4$ & $32\pm12$ \nl
$11-11.5$  &  $15\pm 4$  &  $8 (4-12)$\tablenotemark{a} & $10 (0-33)$\tablenotemark{a} \nl
\enddata
\label{tabsf}
\tablenotetext{a}{Parentheses show the 68\% confidence limits.}
\end{deluxetable}

%
%
\begin{deluxetable}{ccll}
\tablewidth{360pt}
\tablenum{3}
\tablecaption{R23 Metallicity Evolution\label{metal_drop}}
\tablehead{
Logarithmic & & & \cr
Mass Interval & Tremonti04 & $z=0.05-0.475$ & $z=0.475-0.9$
}
\startdata
$ 9.30-9.75 $  &  $8.77$    &    $-0.10 \pm 0.08$ & \nodata \nl
$ 9.75-10.25$  &  $8.91$    &    $-0.10 \pm 0.07$ & $-0.21 \pm 0.04$ \nl
$10.25-10.75$  &  $9.02$    &    $-0.07 \pm 0.10$ & $-0.20 \pm 0.04$ \nl
$10.75-11.25$  &  $9.10$    &    \nodata & $-0.12 \pm 0.05$ \nl
\enddata
\end{deluxetable}

\end{document}